\documentclass[11pt]{article}
\usepackage[utf8]{inputenc}

\newcounter{ourcount}
\usepackage{amsfonts,amsmath,amssymb,amsthm,graphicx,epsfig,pdflscape,multirow,gensymb,slashed}
\usepackage[matrix,arrow,curve]{xy}
\usepackage{array,enumitem,setspace,color,multirow} 
\usepackage{mathdots} 
\usepackage{arydshln} 
\usepackage{pst-node} 
\usepackage{caption}

\def\arxiv#1#2{\href{http://arxiv.org/abs/#1}{\textsf{arXiv:#1 #2}}}
\usepackage[nosort]{cite}
\usepackage{pstricks}
\usepackage{pst-plot}
\usepackage[backref=false]{hyperref}
\hypersetup{
colorlinks=true,
citecolor=red,
linkcolor=darkblue,
urlcolor=darkblue
}
\numberwithin{equation}{section}
\usepackage[capitalise,noabbrev]{cleveref}

\newtheoremstyle{alexi}% Name
  {}%                  % Space above
  {}%                  % Space below
  {\itshape}%          % Body font
  {}%                  % Indent amount
  {\scshape}%          % Theorem head font
  {.}%                 % Punctuation after theorem head
  { }%                 % Space after theorem head, ' ', or \newline
  {}%                  % Theorem head spec (can be left empty, meaning `normal')
\theoremstyle{alexi}
\newtheorem{Lemme}{Lemma}[section]
\newtheorem{Theorem}[Lemme]{Theorem}

\newcommand{\nc}{\newcommand}
\nc{\bib}{\bibitem}
\nc{\be}{\begin{equation}}
\nc{\ee}{\end{equation}}
\nc{\Aoneone}{\mbox{$A_1^{\textrm{\fontsize{7pt}{7pt}\selectfont $(1)$}}$}}
\nc{\Atwoone}{\mbox{$A_2^{\textrm{\fontsize{7pt}{7pt}\selectfont $(1)$}}$}}
\nc{\Atwotwo}{\mbox{$A_2^{\textrm{\fontsize{7pt}{7pt}\selectfont $(2)$}}$}}
\nc{\dd}{\mathsf{d}}
\nc{\eE}{\mathsf{e}}
\nc{\iI}{\mathrm{i}}
\nc{\wW}{\mathsf{W}}
\nc{\repI}{\mathsf{I}}
\nc{\tl}{\mathsf{TL}}
\nc{\dtl}{\mathsf{dTL}}
\nc{\pic}{\mbox{$\slashed{c}$}}
\nc{\pis}{\mbox{$\slashed{s}$}}
\nc{\db}{\mbox{\boldmath $d$}}
\nc{\xb}{\mbox{\boldmath $x$}}
\nc{\yb}{\mbox{\boldmath $y$}}
\nc{\Db}{\mbox{\boldmath $D$}}
\nc{\Dbh}{\mbox{\boldmath $\widehat D$}}
\nc{\Ab}{\mbox{\boldmath $A$}}
\nc{\Bb}{\mbox{\boldmath $B$}}
\nc{\Ib}{\mbox{\boldmath $I$}}
\nc{\Jb}{\mbox{\boldmath $J$}}
\nc{\Lb}{\mbox{\boldmath $L$}}
\nc{\Kb}{\mbox{\boldmath $K$}}
\nc{\Mb}{\mbox{\boldmath $M$}}
\nc{\Rb}{\mbox{\boldmath $R$}}
\nc{\Sb}{\mbox{\boldmath $S$}}
\nc{\Yb}{\mbox{\boldmath $Y$}}
\nc{\Jbh}{\mbox{\boldmath $\hat J$}}
\nc{\Dbt}{\mbox{\boldmath $\tilde{D}$}}
\nc{\Dbm}{\mbox{\boldmath $\mathcal D$}}
\nc{\fbm}{\mbox{\boldmath $\mathcal F$}}
\nc{\Jbm}{\mbox{\boldmath $\mathcal J$}}
\nc{\Pbm}{\mbox{\boldmath $\mathcal P$}}
\nc{\Qbm}{\mbox{\boldmath $\mathcal Q$}}
\nc{\wb}{\bar{w}}
\nc{\slu}{\slashed{u}}
\nc{\Azero}{\alpha^0}
\nc{\Aun}{\alpha^1}
\nc{\Adeux}{\alpha^2}
\nc{\Atrois}{\alpha^3}
\nc{\Bzero}{\beta^0}
\nc{\Bun}{\beta^1}
\nc{\Bdeux}{\beta^2}
\nc{\Btrois}{\beta^3}
\nc{\leU}{U}
\nc{\leV}{V} 
\DeclareMathOperator{\maxP}{maxP}
\definecolor{lightblue}{rgb}{.7,.7,1}
\definecolor{lightestblue}{rgb}{.95,.95,1}
\definecolor{lightlightblue}{rgb}{.85,.85,1}
\definecolor{midblue}{rgb}{.7,.7,1}
\definecolor{darkblue}{rgb}{0,0,.8}
\definecolor{red}{rgb}{1,0,0}
\newrgbcolor{darkgreen}{0.20, 0.5, 0.35}

\def\facegrid#1#2{
\psframe[fillstyle=solid,fillcolor=lightlightblue,linewidth=0pt]#1#2
\psgrid[gridlabels=0pt,subgriddiv=1]#1#2
}

\def\loopa{
\pspolygon[fillstyle=solid,fillcolor=lightlightblue](0,0)(1,0)(1,1)(0,1)
}
\def\loopb{
\pspolygon[fillstyle=solid,fillcolor=lightlightblue](0,0)(1,0)(1,1)(0,1)
\psarc[linewidth=1.5pt,linecolor=blue](0,1){.5}{-90}{0}
}
\def\loopc{
\pspolygon[fillstyle=solid,fillcolor=lightlightblue](0,0)(1,0)(1,1)(0,1)
\psarc[linewidth=1.5pt,linecolor=blue](1,0){.5}{90}{180}
}
\def\loopd{
\pspolygon[fillstyle=solid,fillcolor=lightlightblue](0,0)(1,0)(1,1)(0,1)
\psarc[linewidth=1.5pt,linecolor=blue](0,0){.5}{0}{90}
}
\def\loope{
\pspolygon[fillstyle=solid,fillcolor=lightlightblue](0,0)(1,0)(1,1)(0,1)
\psarc[linewidth=1.5pt,linecolor=blue](1,1){.5}{180}{270}
}
\def\loopf{
\pspolygon[fillstyle=solid,fillcolor=lightlightblue](0,0)(1,0)(1,1)(0,1)
\psline[linewidth=1.5pt,linecolor=blue](0.5,0)(0.5,1)
}
\def\loopg{
\pspolygon[fillstyle=solid,fillcolor=lightlightblue](0,0)(1,0)(1,1)(0,1)
\psline[linewidth=1.5pt,linecolor=blue](0,0.5)(1,0.5)
}
\def\looph{
\pspolygon[fillstyle=solid,fillcolor=lightlightblue](0,0)(1,0)(1,1)(0,1)
\psarc[linewidth=1.5pt,linecolor=blue](1,0){.5}{90}{180}
\psarc[linewidth=1.5pt,linecolor=blue](0,1){.5}{-90}{0}
}
\def\loopi{
\pspolygon[fillstyle=solid,fillcolor=lightlightblue](0,0)(1,0)(1,1)(0,1)
\psarc[linewidth=1.5pt,linecolor=blue](0,0){.5}{0}{90}
\psarc[linewidth=1.5pt,linecolor=blue](1,1){.5}{180}{270}
}
\def\loopid{
\pspolygon[fillstyle=solid,fillcolor=lightlightblue](0,0)(1,0)(1,1)(0,1)
\psarc[linewidth=1.5pt,linecolor=blue,linestyle=dashed,dash=2pt 2pt](1,0){.5}{90}{180}
\psarc[linewidth=1.5pt,linecolor=blue,linestyle=dashed,dash=2pt 2pt](0,1){.5}{-90}{0}
}
\def\loopej{
\pspolygon[fillstyle=solid,fillcolor=lightlightblue](0,0)(1,0)(1,1)(0,1)
\psarc[linewidth=1.5pt,linecolor=blue,linestyle=dashed,dash=2pt 2pt](0,0){.5}{0}{90}
\psarc[linewidth=1.5pt,linecolor=blue,linestyle=dashed,dash=2pt 2pt](1,1){.5}{180}{-90}
}
\def\losange{
\pspolygon[fillstyle=solid,fillcolor=lightlightblue](2,1)(1,2)(0,1)(1,0)
\psarc[linewidth=0.025]{-}(0,1){0.16}{-45}{45}
}
\def\losangeNoArc{
\pspolygon[fillstyle=solid,fillcolor=lightlightblue](2,1)(1,2)(0,1)(1,0)
}
\def\triangled{
\pspolygon[fillstyle=solid,fillcolor=lightlightblue](0,0)(1,1)(0,2)
}
\def\triangleg{
\pspolygon[fillstyle=solid,fillcolor=lightlightblue](0,1)(1,0)(1,2)
}
\def\triangleCNd{
\pspolygon[fillstyle=solid,fillcolor=lightlightblue](0,2)(1,1)(0,0)
\pspolygon[fillstyle=solid,fillcolor=black](0.30,0.93)(0.45,0.93)(0.45,1.08)(0.30,1.08)
}
\def\triangleCNdw{
\pspolygon[fillstyle=solid,fillcolor=lightlightblue](0,2)(1,1)(0,0)
\pspolygon[fillstyle=solid,fillcolor=white,linewidth=0.01cm](0.29,0.92)(0.46,0.92)(0.46,1.09)(0.29,1.09)
}

\def\triangleCNg{
\pspolygon[fillstyle=solid,fillcolor=lightlightblue](0,1)(1,2)(1,0)
\pspolygon[fillstyle=solid,fillcolor=black](0.55,0.93)(0.70,0.93)(0.70,1.08)(0.55,1.08)
}
\def\triangleArcd{
\triangled
\psarc[linecolor=blue,linewidth=1.5pt,linestyle=dashed,dash=2pt 2pt](0.75,1){0.50}{90}{270}
}
\def\triangleArcg{
\triangleg
\psarc[linecolor=blue,linewidth=1.5pt,linestyle=dashed,dash=2pt 2pt](0,1){0.71}{-45}{45}
}

\def\jauge{
\pspolygon[fillstyle=solid](0,0)(0.75,0)(0.75,0.25)(0,0.25)
}

\def\specialcircle#1{
\pspolygon[fillstyle=solid,fillcolor=black,linewidth=0.5pt](#1,#1)(#1,-#1)(-#1,-#1)(-#1,#1)
}

\def\wobblyarc#1#2#3{
\psparametricplot[linecolor=blue,linewidth=1.2pt,plotpoints=1435]{#2}{#3}{0.05 40 #1 t mul mul sin mul #1 add t cos mul 0.05 40 #1 t mul mul sin mul #1 add t sin mul}
}%#1=radius #2 = start angle #3 end angle

\def\projectorTwo{
\pspolygon[fillstyle=solid,fillcolor=pink](0,0.1)(0,1.9)(0.3,1.9)(0.3,0.1)
\rput(0,0.1){\psline{-}(0.2,0)(0.2,0.1)(0.3,0.1)}
}

\def\projectorBdryTwo{
\pspolygon[fillstyle=solid,fillcolor=pink](0,0.1)(0,2.6)(0.3,2.6)(0.3,0.1)
\rput(0,0.1){\psline{-}(0.2,0)(0.2,0.1)(0.3,0.1)}
}

\renewcommand{\ge}{\geqslant}
\renewcommand{\le}{\leqslant}
\renewcommand{\geq}{\geqslant}
\renewcommand{\leq}{\leqslant}

\makeatletter
\g@addto@macro\bfseries{\boldmath}
\makeatother

\makeatletter
\renewcommand*\env@matrix[1][*\c@MaxMatrixCols c]{%
  \hskip -\arraycolsep
  \let\@ifnextchar\new@ifnextchar
  \array{#1}}
\makeatother

\begin{document}

\topmargin -5mm
\oddsidemargin 5mm
\setcounter{page}{1}
\hyphenpenalty=30000

\begin{title}
{\bfseries Fusion hierarchies, $T$-systems and $Y$-systems\\[0.2cm] for the dilute $A_2^{\textrm{\small{($2$)}}}$ loop models on a strip}
\end{title}
\date{\vspace{-.5truecm}}  

\maketitle

\begin{center}
{\vspace{-11mm}
{\large Florence Boileau$^\S$, Alexi Morin-Duchesne$^{\ast\ddagger}$, Yvan Saint-Aubin$^\ddagger$}}
\\[.8cm]
{\em {}$^\S$D\'epartement de physique, Universit\'e de Montr\'eal}\\
{\em Montr\'eal, Qu\'ebec, Canada, H3C 3J7}
\\[.2cm]
{\em {}$^\ast$Department of Applied Mathematics, Computer Science and Statistics \\ Ghent University, 9000 Ghent, Belgium}
\\[.2cm]
{\em {}$^\ddagger$D\'epartement de math\'ematiques et de statistique, Universit\'e de Montr\'eal}\\
{\em Montr\'eal, Qu\'ebec, Canada, H3C 3J7}
\\[.7cm] 
{\tt florence.boileau\,@\,umontreal.ca} \hfil
{\tt alexi.morin-duchesne\,@\,gmail.com}
\\
{\tt yvan.saint-aubin\,@\,umontreal.ca}
\end{center}

\begin{abstract}
We study the dilute $\Atwotwo$ loop models on the geometry of a strip of width $N$. Two families of boundary conditions are known to satisfy the boundary Yang-Baxter equation. Fixing the boundary condition on the two ends of the strip leads to four models. We construct the fusion hierarchy of commuting transfer matrices for the model as well as its $T$- and $Y$-systems, for these four boundary conditions and with a generic crossing parameter~$\lambda$. For $\lambda/\pi$ rational and thus $q=-\eE^{4\iI\lambda}$ a root of unity, we prove a linear relation satisfied by the fused transfer matrices that closes the fusion hierarchy into a finite system. The fusion relations allow us to compute the two leading terms in the large-$N$ expansion of the free energy, namely the bulk and boundary free energies. These are found to be in agreement with numerical data obtained for small $N$. The present work complements a previous study (A.~Morin-Duchesne, P.A.~Pearce, J.~Stat.~Mech.~(2019)) that investigated the dilute $\Atwotwo$ loop models with periodic boundary conditions.

\bigskip

\noindent Keywords: dilute loop models, dilute Temperley-Lieb algebra, fusion hierarchy, $T$-system, $Y$-system.

\end{abstract}

%%%%%%%%%%%%%%%%%%%%%
%
% TOC
%
%%%%%%%%%%%%%%%%%%%%%

\newpage
\tableofcontents
\clearpage

%%%%%%%%%%%%%%%%%%%%%%%%%%%%%%%%%%%%%%%%%%%%%%%
%
\section{Introduction} \label{sec:intro}
%
%%%%%%%%%%%%%%%%%%%%%%%%%%%%%%%%%%%%%%%%%%%%%%%

Loop models have historically played a central role in deepening our understanding of critical  statistical models on two-dimensional lattices \cite{J09}. The loops are random curves that typically arise in the study of models with local degrees of freedom, by considering the curves drawn on the lattice by collections of these local objects and thus elevating the model to one with non-local observables. Examples include the contour curves of percolation clusters, Ising spin domains or spanning trees. The formulation of these problems in terms of a transfer matrix or Hamiltonian treats one of the two spatial directions as a time direction along which evolution runs in discrete steps. The various quantities of interest are then expressible in terms of the eigenvalues and eigenstates of these operators. If the model is Yang-Baxter integrable~\cite{BAXTER}, then its transfer matrix and Hamiltonian are elements of larger commuting families and various techniques allow one to compute the eigenvalues and eigenstates. 

Loop models are also intimately tied with vertex models, and likewise to Interaction-Round-a-Face models like the Restricted Solid-on-Solid (RSOS) models. Many interesting features of these models emerge in the continuum scaling limit, where the size $N$ of the lattice is sent to infinity. Theoretical and numerical studies show that in this limit, the spectrum of the transfer matrix, after proper scaling, is described by characters of the Virasoro algebra. The central charge $c$ of the underlying conformal field theory (CFT) then depends on the model considered. For instance, it depends on the loop fugacity $\beta$ for the loop models, on the quantum group parameter~$q$ for the vertex models, and on the number $L$ of allowed heights in the RSOS models. If $c$ is a rational number, then the CFT describing the scaling limit of these models is either rational or logarithmic. The latter case is particularly interesting, with indecomposable yet reducible representations and non-trivial Jordan cells for the dilation operator \cite{Gurarie,PPJRJBZ06,SpecialIssue}. These structures are not fully understood, so this provides in itself a good physical motivation to pursue the study of these models. But the models defined on finite lattices are also rich in algebraic structures, some of which still remain to be unraveled.

The elementary building blocks for the loops models are the face operators (or $\check R$ matrices in the language of vertex models), which take the form of linear combinations of diagrams weighted by Boltzmann weights. The model is  integrable if these weights are chosen so that the Yang-Baxter equation is satisfied. This in turn ensures the existence of a one-parameter family of commuting transfer matrices. Solutions to the Yang-Baxter equations are often labeled by affine Lie algebras \cite{B85,J86}, with the most studied cases corresponding to the low-rank algebras $\Aoneone$, $\Atwoone$ and $\Atwotwo$. Thanks to a large body of work on the dense $\Aoneone$ loop model and the related six-vertex models, it is fair to say that their $s\ell(2)$ integrability structures are now well understood \cite{BAXTER,ZP95,BazhMang2007,AMDPPJR14,FrahmMDP2019}. 

This article focuses on the dilute loop model in the $\Atwotwo$ family, a model with a richer structure also often referred to as the dilute $O(n)$ loop model. Its Boltzmann weights were obtained by Nienhuis~\cite{Nienhuis}, and Izergin and Korepin \cite{IK81}. Some of these weights can be negative, curtailing an immediate interpretation within statistical physics, but also leading to unexpected behavior in the scaling limit \cite{VJS2014}. Here we investigate these loop models defined on the geometry of the strip, with boundary face operators satisfying the boundary Yang-Baxter equation. Two solutions were given by Batchelor and Yung \cite{BatchelorYung}, and further studied by Dubail, Jacobsen and Saleur \cite{DJS10}. We label them by S and C (for sine and cosine). Both sides of the strip may have different boundary conditions, so a total of four models will be studied throughout, namely those with identical boundary conditions on the two sides (SS and CC), and those with mixed ones (SC and CS). We also mention that the current work complements an earlier study \cite{AMDPP19} of the dilute $\Atwotwo$ loop model defined with periodic boundary conditions. 

The transfer matrix that we investigate for the loop model on the strip, denoted $\Db(u)$, takes the form of a linear combination of connectivity diagrams. As such, it is an element of the dilute Temperley-Lieb algebra $\dtl_N(\beta)$~\cite{GP93,P94,Grimm96,JBYSA13}, a generalisation of the Temperley-Lieb algebra \cite{TL71,VFRJ99} where the loops are dilute instead of dense. The transfer matrix depends on a spectral parameter~$u$ and a crossing parameter~$\lambda$, with the latter parameterizing the loop fugacity as $\beta = -2 \cos 4 \lambda$. It is a Laurent polynomial in $\eE^{\iI u}$ whose coefficients are elements of $\dtl_N(\beta)$. The commutativity relation \mbox{$[\Db(u),\Db(v)]=0$}, typical of integrable models, allows one to show that these coefficients are commuting conserved quantities. 

In this paper, we construct the fusion hierarchy of transfer matrices $\Db^{m,n}(u)$ for the $\Atwotwo$~loop model on the strip, with $\Db^{1,0}(u) = \Db(u)$. These transfer matrices satisfy relations that allow us to write each $\Db^{m,n}(u)$ as a function of the elementary transfer matrices $\Db(u+k \lambda)$ with $k \in \mathbb Z$. This directly ensures that $[\Db^{m,n}(u),\Db(u)]=0$. Moreover, the construction is done in such a way that each member of the hierarchy is a Laurent polynomial in $\eE^{\iI u}$, which implies that its coefficients are also conserved quantities. This process might seem vacuous as the resulting fused matrices $\Db^{m,n}(u)$ do not create new commuting quantities, since they are algebraically related to those of $\Db(u)$. The power of the hierarchy is instead revealed for $\lambda/\pi \in \mathbb Q$. In this case,  the fused transfer matrices satisfy a linear {\it closure relation} that turns the hierarchy into a finite system. Since the fused matrices of the hierarchy commute, they can be diagonalized simultaneously and the closure relation thus translates into a functional relation satisfied by the eigenvalues of these matrices. Equivalently, this system can be recast in a unique functional polynomial equation satisfied by $\Db(u)$. This identity lives in the algebra $\dtl_N(\beta)$, but is also satisfied by replacing $\Db(u)$ by any of its eigenvalues. This is Baxter's celebrated method \cite{BAXTER} of functional equations, to extract physical quantities from transfer matrices. The techniques that were developed to solve these systems of equations \cite{PK91,KP91,KP92,AMDAKPP21} are intimately related to the thermodynamic Bethe ansatz.

The outline of the paper is as follows. In \cref{sec:dTL_N}, we review the definition of the dilute Temperley-Lieb algebra and of its standard modules. \cref{sec:faceAndBoundary} defines the dilute $\Atwotwo$ loop model in terms of its face and boundary operators and their integrable Boltzmann weights. It also gives a list of the local relations that they satisfy and reviews the definition of some projectors defined in \cite{AMDPP19}. In \cref{sec:transferMatrices}, we define first the fundamental transfer matrix $\Db(u)$, and subsequently the fused transfer matrices $\Db^{m,n}(u)$ using the fusion hierarchy relations that they satisfy. This longer section also discusses several key properties of the transfer matrices, in particular some alternate but equivalent definitions of these objects, the reduction relations that they satisfy at certain special values of $u$, and their behavior in the braid limit $u\to \iI \infty$. Finally, \cref{sec:transferMatrices} also shows that the fusion hierarchy can be recast in terms of a $T$-system and a $Y$-system. The proof that the fused transfer matrices are Laurent polynomials contains new ideas not seen before for other models, and is the topic of \cref{sec:polynomiality}. Then in \cref{sec:closure}, we set $\lambda/\pi \in \mathbb Q$ and prove the closure relation of the fusion hierarchy. With the polynomiality of the fused matrices established, the closure relation takes the form of an identity between Laurent polynomials in $\eE^{\iI u}$ with coefficients in the dilute Temperley-Lieb algebra. The strategy for the proof consists in checking the identity at a finite number of points. This section also discusses the closure for the $Y$-system. Some of the proofs of \cref{sec:transferMatrices,sec:closure} are direct whereas others are longer and are gathered in \cref{app:proofs}. Finally, \cref{sec:free.energies} shows the physical relevance of the functional relations of the hierarchy by computing the bulk free energies (obtained previously by Warnaar, Batchelor and Nienhuis \cite{WBN92}) and the boundary free energies (which to our knowledge are new). \cref{sec:conclusion} offers some concluding remarks.

%%%%%%%%%%%%%%%%%%%%%%%%%%%%%%%%%%%%%%%%%%%%%%%
%
\section{The dilute Temperley-Lieb algebra} \label{sec:dTL_N}
%
%%%%%%%%%%%%%%%%%%%%%%%%%%%%%%%%%%%%%%%%%%%%%%%

The words \textit{planar algebras} \cite{VFRJ99} refer to a broad set of algebras whose elements are diagrams known as \textit{tangles}. One of their main features is that tangles are multiplied by concatenation. Specific rules to be performed upon concatenation will lead to different algebras. The present section is devoted to the description of the rules defining the dilute Temperley-Lieb algebra $\dtl_N(\beta)$ and its standard representations.

%%%%%%%%%%%%%
\subsection{Definition of the algebra}
%%%%%%%%%%%%%

Similarly to the dense loop model and its underlying Temperley-Lieb algebra $\tl_N(\beta)$,
the dilute $\Atwotwo$ loop model is defined using the dilute Temperley-Lieb algebra $\dtl_N(\beta)$. The algebras $\dtl_N(\beta)$ form a family of associative unital algebras labelled by a positive integer $N$ and a parameter $\beta\in\mathbb C$. The present section introduces a diagrammatic presentation of $\dtl_N(\beta)$ and some of its properties. More details and its basic representation theory are given in \cite{JBYSA13}.

The construction of the basic objects, the \textit{connectivities}, goes as follows. Let $N\in\mathbb{N}$. A rectangle with $N$ nodes on the bottom edge and $N$ other ones on the top edge is drawn. A set of non-intersecting planar loop segments is also drawn to connect a subset of the nodes in pairs. The loop segments are constrained to stay within the borders of the rectangle and to have each end tied to one of the $2N$ nodes. A node may be left unconnected and is then said to be \textit{vacant}. Here are two connectivities with $N=6$, with the loop segments depicted as solid blue lines and the vacancies as solid black circles:
\begin{equation} \label{ex:2connectivites}
c_1 = \ \begin{pspicture}[shift=-0.4](3,1)
\pspolygon[fillstyle=solid,fillcolor=lightlightblue](0,0)(3,0)(3,1)(0,1)
\psbezier[linewidth=1.5pt,linecolor=blue](1.75,0)(1.75,0.5)(0.75,0.5)(0.75,1)
\psarc[linewidth=1.5pt,linecolor=blue](2.0,1){.25}{180}{0}
\psbezier[linewidth=1.5pt,linecolor=blue](1.25,1)(1.25,0.3)(2.75,0.3)(2.75,1)
\psarc[linewidth=1.5pt,linecolor=blue](0.5,0){.25}{0}{180}
\pscircle[fillstyle=solid,fillcolor=black](0.25,1){0.05}
\pscircle[fillstyle=solid,fillcolor=black](1.25,0){0.05}
\pscircle[fillstyle=solid,fillcolor=black](2.25,0){0.05}
\pscircle[fillstyle=solid,fillcolor=black](2.75,0){0.05}
\end{pspicture} \ ,
\qquad
c_2 = \ \begin{pspicture}[shift=-0.4](3,1)
\pspolygon[fillstyle=solid,fillcolor=lightlightblue](0,0)(3,0)(3,1)(0,1)
\psbezier[linewidth=1.5pt,linecolor=blue](0.75,0)(0.75,0.5)(1.25,0.5)(1.25,1)
\psarc[linewidth=1.5pt,linecolor=blue](2.0,1){.25}{180}{0}
\psarc[linewidth=1.5pt,linecolor=blue](1.5,0){.25}{0}{180}
\psarc[linewidth=1.5pt,linecolor=blue](2.5,0){.25}{0}{180}
\pscircle[fillstyle=solid,fillcolor=black](0.25,1){0.05}
\pscircle[fillstyle=solid,fillcolor=black](0.75,1){0.05}
\pscircle[fillstyle=solid,fillcolor=black](2.75,1){0.05}
\pscircle[fillstyle=solid,fillcolor=black](0.25,0){0.05}
\end{pspicture} \ .
\end{equation}
The numbers of vacancies on the two sides of a connectivity always have the same
parity by the requirement that loop segments have both ends connected.

As a vector space, the algebra $\dtl_N(\beta)$ is the formal linear span over $\mathbb{C}$ of connectivities. The product $c_1c_2$ of two connectivities is defined via concatenation. To compute it, the connectivity $c_2$ is drawn above $c_1$. The inner edges and nodes of the rectangles are identified and then removed. The resulting loop segments might have loose ends in the middle of the diagram or might be closed to form a loop. The following rules are then applied to obtain a connectivity. If a loop segment ends at a vacancy, the result is set to zero. If loop segments join to form a loop, it is removed and the diagram is multiplied by a factor $\beta$. Otherwise, the loop segments are contracted and the resulting connectivity is read off directly. Here are two examples:
\begin{equation} \label{ex:2produits}
c_1 c_2 = \ \begin{pspicture}[shift=-0.9](3,2)
\rput(0,1){
\pspolygon[fillstyle=solid,fillcolor=lightlightblue](0,0)(3,0)(3,1)(0,1)
\psbezier[linewidth=1.5pt,linecolor=blue](0.75,0)(0.75,0.5)(1.25,0.5)(1.25,1)
\psarc[linewidth=1.5pt,linecolor=blue](2.0,1){.25}{180}{0}
\psarc[linewidth=1.5pt,linecolor=blue](1.5,0){.25}{0}{180}
\psarc[linewidth=1.5pt,linecolor=blue](2.5,0){.25}{0}{180}
\pscircle[fillstyle=solid,fillcolor=black](0.25,1){0.05}
\pscircle[fillstyle=solid,fillcolor=black](0.75,1){0.05}
\pscircle[fillstyle=solid,fillcolor=black](2.75,1){0.05}
\pscircle[fillstyle=solid,fillcolor=black](0.25,0){0.05}
}
\pspolygon[fillstyle=solid,fillcolor=lightlightblue](0,0)(3,0)(3,1)(0,1)
\psbezier[linewidth=1.5pt,linecolor=blue](1.75,0)(1.75,0.5)(0.75,0.5)(0.75,1)
\psarc[linewidth=1.5pt,linecolor=blue](2.0,1){.25}{180}{0}
\psbezier[linewidth=1.5pt,linecolor=blue](1.25,1)(1.25,0.3)(2.75,0.3)(2.75,1)
\psarc[linewidth=1.5pt,linecolor=blue](0.5,0){.25}{0}{180}
\pscircle[fillstyle=solid,fillcolor=black](0.25,1){0.05}
\pscircle[fillstyle=solid,fillcolor=black](1.25,0){0.05}
\pscircle[fillstyle=solid,fillcolor=black](2.25,0){0.05}
\pscircle[fillstyle=solid,fillcolor=black](2.75,0){0.05}
\end{pspicture}
\ = \beta \ 
\begin{pspicture}[shift=-0.4](3,1)
\pspolygon[fillstyle=solid,fillcolor=lightlightblue](0,0)(3,0)(3,1)(0,1)
\psbezier[linewidth=1.5pt,linecolor=blue](1.75,0)(1.75,0.5)(1.25,0.5)(1.25,1)
\psarc[linewidth=1.5pt,linecolor=blue](2.0,1){.25}{180}{0}
\pscircle[fillstyle=solid,fillcolor=black](0.25,1){0.05}
\pscircle[fillstyle=solid,fillcolor=black](0.75,1){0.05}
\pscircle[fillstyle=solid,fillcolor=black](2.75,1){0.05}
\psarc[linewidth=1.5pt,linecolor=blue](0.5,0){.25}{0}{180}
\pscircle[fillstyle=solid,fillcolor=black](0.25,1){0.05}
\pscircle[fillstyle=solid,fillcolor=black](1.25,0){0.05}
\pscircle[fillstyle=solid,fillcolor=black](2.25,0){0.05}
\pscircle[fillstyle=solid,fillcolor=black](2.75,0){0.05}
\end{pspicture} \ ,
\qquad
c_2 c_1 = \ \begin{pspicture}[shift=-0.9](3,2)
\rput(0,1){
\pspolygon[fillstyle=solid,fillcolor=lightlightblue](0,0)(3,0)(3,1)(0,1)
\psbezier[linewidth=1.5pt,linecolor=blue](1.75,0)(1.75,0.5)(0.75,0.5)(0.75,1)
\psarc[linewidth=1.5pt,linecolor=blue](2.0,1){.25}{180}{0}
\psbezier[linewidth=1.5pt,linecolor=blue](1.25,1)(1.25,0.3)(2.75,0.3)(2.75,1)
\psarc[linewidth=1.5pt,linecolor=blue](0.5,0){.25}{0}{180}
\pscircle[fillstyle=solid,fillcolor=black](0.25,1){0.05}
\pscircle[fillstyle=solid,fillcolor=black](1.25,0){0.05}
\pscircle[fillstyle=solid,fillcolor=black](2.25,0){0.05}
\pscircle[fillstyle=solid,fillcolor=black](2.75,0){0.05}
}
\pspolygon[fillstyle=solid,fillcolor=lightlightblue](0,0)(3,0)(3,1)(0,1)
\psbezier[linewidth=1.5pt,linecolor=blue](0.75,0)(0.75,0.5)(1.25,0.5)(1.25,1)
\psarc[linewidth=1.5pt,linecolor=blue](2.0,1){.25}{180}{0}
\psarc[linewidth=1.5pt,linecolor=blue](1.5,0){.25}{0}{180}
\psarc[linewidth=1.5pt,linecolor=blue](2.5,0){.25}{0}{180}
\pscircle[fillstyle=solid,fillcolor=black](0.25,1){0.05}
\pscircle[fillstyle=solid,fillcolor=black](0.75,1){0.05}
\pscircle[fillstyle=solid,fillcolor=black](2.75,1){0.05}
\pscircle[fillstyle=solid,fillcolor=black](0.25,0){0.05}
\end{pspicture}
\ = 0 \ .
\end{equation}
The algebra $\dtl_N(\beta)$ is associative and non-commutative. There is a unit element, $\Ib$, which is the sum of $2^N$ connectivities:
\begin{equation} \label{def:diagIdentite}
\Ib = \ 
\begin{pspicture}[shift=-0.70](0,-0.3)(3,1)
\pspolygon[fillstyle=solid,fillcolor=lightlightblue](0,0)(3,0)(3,1)(0,1)
\psline[linecolor=blue,linewidth=1.5pt,linestyle=dashed,dash=2pt 2pt](0.25,0)(0.25,1)
\psline[linecolor=blue,linewidth=1.5pt,linestyle=dashed,dash=2pt 2pt](0.75,0)(0.75,1)
\psline[linecolor=blue,linewidth=1.5pt,linestyle=dashed,dash=2pt 2pt](1.25,0)(1.25,1)
\psline[linecolor=blue,linewidth=1.5pt,linestyle=dashed,dash=2pt 2pt](2.25,0)(2.25,1)
\psline[linecolor=blue,linewidth=1.5pt,linestyle=dashed,dash=2pt 2pt](2.75,0)(2.75,1)
\rput(1.8,0.5){$\cdots$}
\rput(0,-0.25){
\rput(0.25,0){$_1$}
\rput(0.75,0){$_2$}
\rput(1.25,0){$_3$}
\rput(2.175,0){$_{N-1}$}
\rput(2.825,0){$_N$}}
\end{pspicture} \ ,
\qquad
\begin{pspicture}[shift=-0.4](0.5,1)
\pspolygon[fillstyle=solid,fillcolor=lightlightblue](0,0)(0.5,0)(0.5,1)(0,1)
\psline[linecolor=blue,linewidth=1.5pt,linestyle=dashed,dash=2pt 2pt](0.25,0)(0.25,1)
\end{pspicture}
\ = \ \begin{pspicture}[shift=-0.4](0.5,1)
\pspolygon[fillstyle=solid,fillcolor=lightlightblue](0,0)(0.5,0)(0.5,1)(0,1)
\psline[linecolor=blue,linewidth=1.5pt](0.25,0)(0.25,1)
\end{pspicture}
\ + \ \begin{pspicture}[shift=-0.4](0.5,1)
\pspolygon[fillstyle=solid,fillcolor=lightlightblue](0,0)(0.5,0)(0.5,1)(0,1)
\pscircle[fillstyle=solid,fillcolor=black](0.25,1){0.05}
\pscircle[fillstyle=solid,fillcolor=black](0.25,0){0.05}
\end{pspicture} \ .
\end{equation}
The dimension of $\dtl_N(\beta)$ was computed in \cite{JBYSA13} 
using the resemblance with the Temperley-Lieb algebra $\tl_N(\beta)$:
\be
\dim \dtl_N(\beta) = \sum_{k=0}^N\frac{1}{k+1}\binom{2k}{k}\binom{2n}{2k} = M_{2N} \ ,
\ee
where $M_{2N}$ is the $2N$-th Motzkin number. It can also be written as
\begin{equation}
\dim \dtl_N(\beta) =  \binom{2N}{0}_2 - \binom{2N}{2}_2
\end{equation}
where the trinomial coefficients are defined by
\begin{equation}
(x+1+x^{-1})^N = \sum_{k=-N}^N \binom{N}{k}_2 x^k.
\end{equation}

%%%%%%%%%%%%%
\subsection{Standard modules}
%%%%%%%%%%%%%

Several key results about loop models can be obtained using the standard modules over the dilute Temperley-Lieb algebra. The set of standard modules on $\dtl_N(\beta)$ is $\{\wW_{N}^d,0\leq d\leq N\}$, where the parameters $N$ and $\beta$ are as above and $d$ is the {\it defect number} introduced below.

A basis for the standard module $\wW_{N}^d$ is constituted of {\it link states}. To construct a link state, a horizontal line is drawn with $N$ marked nodes. A set of non-intersecting loop segments is added above the horizontal line, each one either connecting two nodes together or connecting a node to infinity above. A loop segment of the second type is called a \textit{defect}, and the number of defects is denoted $d$. Like those in a connectivity, the nodes of a link state may be left vacant. The formal linear combinations of link diagrams forms a vector space $\wW_{N}$, and the link states with a fixed $d$, with $0\leq d\leq N$, span a subspace that will be the standard module $\wW_{N}^d$ for the action to be defined below. For example, here are the link states generating the vector spaces of the modules $\wW_3^d$ with $d=0,1,2,3$:
\begin{alignat}{2}
&\wW^3_3 : \quad \begin{pspicture}(1.5,0.5)
\psline[linewidth=1.5pt,linecolor=blue](0.25,0)(0.25,0.5)
\psline[linewidth=1.5pt,linecolor=blue](0.75,0)(0.75,0.5)
\psline[linewidth=1.5pt,linecolor=blue](1.25,0)(1.25,0.5)
\psline(0,0)(1.5,0)
\end{pspicture} \ , \qquad
\wW^2_3 : \begin{pspicture}(1.5,0.5)
\psline[linewidth=1.5pt,linecolor=blue](0.25,0)(0.25,0.5)
\psline[linewidth=1.5pt,linecolor=blue](0.75,0)(0.75,0.5)
\pscircle[fillstyle=solid,fillcolor=black](1.25,0){0.05}
\psline(0,0)(1.5,0)
\end{pspicture} \ \
\begin{pspicture}(1.5,0.5)
\psline[linewidth=1.5pt,linecolor=blue](0.25,0)(0.25,0.5)
\pscircle[fillstyle=solid,fillcolor=black](0.75,0){0.05}
\psline[linewidth=1.5pt,linecolor=blue](1.25,0)(1.25,0.5)
\psline(0,0)(1.5,0)
\end{pspicture} \ \
\begin{pspicture}(1.5,0.5)
\pscircle[fillstyle=solid,fillcolor=black](0.25,0){0.05}
\psline[linewidth=1.5pt,linecolor=blue](0.75,0)(0.75,0.5)
\psline[linewidth=1.5pt,linecolor=blue](1.25,0)(1.25,0.5)
\psline(0,0)(1.5,0)
\end{pspicture} \ ,
\nonumber\\
&\wW^1_3 : \quad \begin{pspicture}(1.5,0.5)
\psline[linewidth=1.5pt,linecolor=blue](0.25,0)(0.25,0.5)
\psarc[linewidth=1.5pt,linecolor=blue](1,0){.25}{0}{180}
\psline(0,0)(1.5,0)
\end{pspicture} \ \
\begin{pspicture}(1.5,0.5)
\psarc[linewidth=1.5pt,linecolor=blue](0.5,0){.25}{0}{180}
\psline[linewidth=1.5pt,linecolor=blue](1.25,0)(1.25,0.5)
\psline(0,0)(1.5,0)
\end{pspicture} \ \
\begin{pspicture}(1.5,0.5)
\psline[linewidth=1.5pt,linecolor=blue](0.25,0)(0.25,0.5)
\pscircle[fillstyle=solid,fillcolor=black](0.75,0){0.05}
\pscircle[fillstyle=solid,fillcolor=black](1.25,0){0.05}
\psline(0,0)(1.5,0)
\end{pspicture} \ \
\begin{pspicture}(1.5,0.5)
\pscircle[fillstyle=solid,fillcolor=black](0.25,0){0.05}
\psline[linewidth=1.5pt,linecolor=blue](0.75,0)(0.75,0.5)
\pscircle[fillstyle=solid,fillcolor=black](1.25,0){0.05}
\psline(0,0)(1.5,0)
\end{pspicture} \ \
\begin{pspicture}(1.5,0.5)
\pscircle[fillstyle=solid,fillcolor=black](0.25,0){0.05}
\pscircle[fillstyle=solid,fillcolor=black](0.75,0){0.05}
\psline[linewidth=1.5pt,linecolor=blue](1.25,0)(1.25,0.5)
\psline(0,0)(1.5,0)
\end{pspicture} \ ,
\\
&\wW^0_3 : \quad \begin{pspicture}(1.5,0.5)
\pscircle[fillstyle=solid,fillcolor=black](0.25,0){0.05}
\psarc[linewidth=1.5pt,linecolor=blue](1,0){.25}{0}{180}
\psline(0,0)(1.5,0)
\end{pspicture} \ \
\begin{pspicture}(1.5,0.5)
\psarc[linewidth=1.5pt,linecolor=blue](0.5,0){.25}{0}{180}
\pscircle[fillstyle=solid,fillcolor=black](1.25,0){0.05}
\psline(0,0)(1.5,0)
\end{pspicture} \ \
\begin{pspicture}(1.5,0.5)
\psarc[linewidth=1.5pt,linecolor=blue](0.75,0){.5}{0}{180}
\pscircle[fillstyle=solid,fillcolor=black](0.75,0){0.05}
\psline(0,0)(1.5,0)
\end{pspicture} \ \
\begin{pspicture}(1.5,0.5)
\pscircle[fillstyle=solid,fillcolor=black](0.25,0){0.05}
\pscircle[fillstyle=solid,fillcolor=black](0.75,0){0.05}
\pscircle[fillstyle=solid,fillcolor=black](1.25,0){0.05}
\psline(0,0)(1.5,0)
\end{pspicture} \ .
\nonumber
\end{alignat}

There is a natural action of $\dtl_N(\beta)$ on the vector space $\wW_{N}^d$. If $c$ is a connectivity and $w$ a link state, the action of $c$ on $w$ is computed similarly as the product of two connectivities in $\dtl_N(\beta)$. The link state $w$ is drawn above $c$ and the nodes on the upper edge of the rectangle are identified with those of the link state and then removed. The resulting connectivity is read directly starting from the lower edge of $c$. Some rules apply. First, if a loop segment is connected to a vacancy, the result is zero. Second, if the concatenation reduces the number of defects, the result of the action is set to zero. Third, any closed loop created by the concatenation is removed from the diagram and the remaining link state is multiplied by the factor $\beta^\ell$, where $\ell$ is the number of loops. The following are examples of the action of connectivities of $\dtl_4(\beta)$ on link states in $\wW_4^0$, $\wW_4^3$ and $\wW_4^1$ respectively: 
\begin{equation} \label{ex:2actions}
\begin{pspicture}[shift=-0.4](2,1.5)
\rput(0,1){
\psarc[linewidth=1.5pt,linecolor=blue](0.5,0){.25}{0}{180}
\psarc[linewidth=1.5pt,linecolor=blue](1.5,0){.25}{0}{180}
}
\pspolygon[fillstyle=solid,fillcolor=lightlightblue](0,0)(2,0)(2,1)(0,1)
\psline[linewidth=1.5pt,linecolor=blue](0.25,0)(0.25,1)
\psarc[linewidth=1.5pt,linecolor=blue](1,1){.25}{180}{0}
\psarc[linewidth=1.5pt,linecolor=blue](1.5,0){.25}{0}{180}
\pscircle[fillstyle=solid,fillcolor=black](0.75,0){0.05}
\pscircle[fillstyle=solid,fillcolor=black](1.75,1){0.05}
\end{pspicture} \ = 0 \ ,
\qquad
\begin{pspicture}[shift=-0.4](2,1.5)
\rput(0,1){
\psline[linewidth=1.5pt,linecolor=blue](0.25,0)(0.25,0.5)
\psline[linewidth=1.5pt,linecolor=blue](0.75,0)(0.75,0.5)
\psline[linewidth=1.5pt,linecolor=blue](1.25,0)(1.25,0.5)
}
\pspolygon[fillstyle=solid,fillcolor=lightlightblue](0,0)(2,0)(2,1)(0,1)
\psline[linewidth=1.5pt,linecolor=blue](0.25,0)(0.25,1)
\psarc[linewidth=1.5pt,linecolor=blue](1,1){.25}{180}{0}
\psarc[linewidth=1.5pt,linecolor=blue](1.5,0){.25}{0}{180}
\pscircle[fillstyle=solid,fillcolor=black](0.75,0){0.05}
\pscircle[fillstyle=solid,fillcolor=black](1.75,1){0.05}
\end{pspicture} \ = 0
\qquad \text{and} \qquad
\begin{pspicture}[shift=-0.4](2,1.5)
\rput(0,1){
\psline[linewidth=1.5pt,linecolor=blue](0.25,0)(0.25,0.5)
\psarc[linewidth=1.5pt,linecolor=blue](1,0){.25}{0}{180}
}
\pspolygon[fillstyle=solid,fillcolor=lightlightblue](0,0)(2,0)(2,1)(0,1)
\psline[linewidth=1.5pt,linecolor=blue](0.25,0)(0.25,1)
\psarc[linewidth=1.5pt,linecolor=blue](1,1){.25}{180}{0}
\psarc[linewidth=1.5pt,linecolor=blue](1.5,0){.25}{0}{180}
\pscircle[fillstyle=solid,fillcolor=black](0.75,0){0.05}
\pscircle[fillstyle=solid,fillcolor=black](1.75,1){0.05}
\end{pspicture} 
\ = \beta \
\begin{pspicture}[shift=-0.1](2,0.5)
\psline[linewidth=1.5pt,linecolor=blue](0.25,0)(0.25,0.5)
\psarc[linewidth=1.5pt,linecolor=blue](1.5,0){.25}{0}{180}
\pscircle[fillstyle=solid,fillcolor=black](0.75,0){0.05}
\psline(0,0)(2,0)
\end{pspicture} \ \ .
\end{equation}
The dimension of the module $\wW_N^d$ is
\begin{equation}
\dim \wW_N^d = \binom{N}{k}_2 - \binom{N}{k+2}_2 \ .
\end{equation}

The standard modules $\wW_N^d$ are indecomposable for all $N$ and $d$ with $0\leq d\leq N$. The structure of the module $\wW_N^d$ depends on the value of $\beta = q+q^{-1}$. The parameter $q$ is said to be {\em generic} when it is not a root of unity. For generic values of $q$, each $\wW_N^d$ is irreducible. When $q$ is a root of unity, the module $\wW_N^d$ is indecomposable but not always irreducible. When it is not, it is instead the quotient $\repI_N^d$ of $\wW_N^d$ by its radical that is irreducible. These quotients form a complete set of isomorphism classes of irreducible modules of $\dtl_N(\beta)$. Let us describe the roots of unity case in more detail. For a given value of $q$, an integer $d$ is said to be {\em critical} if $q^{2(d+1)}=1$. For $d$ critical, $\wW_N^d$ is irreducible. Let $d$ be non critical and $d_c$ be the smallest critical integer larger than~$d$. We define $d^+$ as the integer obtained from the reflection of $d$ with respect to $d_c$, namely $d^+=2d_c-d$. Similarly, we define the integer $d^-$ by a reflection of $d$ across the largest critical integer $d'_c$ smaller than $d$, that is $d^-=2d'_c-d$. The sequence $\{\dots, d^{--}, d^-,d,d^+,d^{++},\dots\}$ is then infinite in both directions. To study the structure of $\wW_N^d$ at a root of unity, we truncate this sequence by keeping only the elements in the range $[0,N]$. This truncated sequence is called the {\em orbit} of $d$. Being in the same orbit is an equivalence relation. If both $d$ and $d_+$ are in the truncated sequence, then there is a non-zero morphism $\wW_N^{d^+}\to \wW_N^d$, and $\wW_N^d$ is reducible. It has two composition factors, $\repI_N^d$ and $\repI_N^{d^+}$, and its Loewy diagram is $\repI_N^d \to \repI_N^{d^+}$. The algebra $\dtl_N(\beta)$ can be seen as a module over itself. This is the {\em regular} representation and it can be decomposed into a direct sum of indecomposable modules. All the elements in this direct sum that contain composition factors that arise in the set of standard modules $\wW_N^e$, with $e$ in the orbit of a given $d$, are said to form a {\it block}. Similarly, an indecomposable $\dtl_N$-module may only have composition factors $\repI_N^{e}$ with their labels~$e$ belonging to the same orbit. They are then naturally associated to the block labeled by this orbit. Non-trivial homomorphism and extension groups exist only between the indecomposables associated to the same block. (See \cite{JBYSA13} for more details.)

%%%%%%%%%%%%%%%%%%%%%%%%%%%%%%%%%%%%%%%%%%%%%%%
%
\section{Integrable face and boundary operators} \label{sec:faceAndBoundary}
%
%%%%%%%%%%%%%%%%%%%%%%%%%%%%%%%%%%%%%%%%%%%%%%%

The transfer matrix of the dilute $\Atwotwo$ loop models is constructed out of two building blocks: the face operator and the boundary operator. The present section introduces these operators and lists some of their properties.

%%%%%%%%%%%%%
\subsection{Definition of the operators}
%%%%%%%%%%%%%

An \textit{elementary tile} is a square with a single node on each edge. Like for a connectivity, these nodes may be connected in pairs by non-intersecting loop segments or stay vacant. There are then nine elementary tiles. The \textit{face operator} is a linear combination of these nine tiles:
\begin{equation}
\begin{gathered} \label{def:tuileDeBase}
\begin{pspicture}[shift=-0.4](1,1)
\facegrid{(0,0)}{(1,1)}
\psarc[linewidth=0.025]{-}(0,0){0.16}{0}{90}
\rput(.5,.5){$u$}
\end{pspicture}
\ = \rho_1(u) \
\begin{pspicture}[shift=-0.4](1,1)
\loopa 
\pscircle[fillstyle=solid,fillcolor=black](0.5,0){0.05}
\pscircle[fillstyle=solid,fillcolor=black](0,0.5){0.05}
\pscircle[fillstyle=solid,fillcolor=black](0.5,1){0.05}
\pscircle[fillstyle=solid,fillcolor=black](1,0.5){0.05}
\end{pspicture}
\ + \rho_2(u) \
\begin{pspicture}[shift=-0.4](1,1)
\loopb
\pscircle[fillstyle=solid,fillcolor=black](0.5,0){0.05}
\pscircle[fillstyle=solid,fillcolor=black](1,0.5){0.05}
\end{pspicture}
\ + \rho_3(u) \
\begin{pspicture}[shift=-0.4](1,1)
\loopc
\pscircle[fillstyle=solid,fillcolor=black](0,0.5){0.05}
\pscircle[fillstyle=solid,fillcolor=black](0.5,1){0.05}
\end{pspicture}
\ + \rho_4(u) \
\begin{pspicture}[shift=-0.4](1,1)
\loopd
\pscircle[fillstyle=solid,fillcolor=black](0.5,1){0.05}
\pscircle[fillstyle=solid,fillcolor=black](1,0.5){0.05}
\end{pspicture}
\ + \rho_5(u) \
\begin{pspicture}[shift=-0.4](1,1)
\loope
\pscircle[fillstyle=solid,fillcolor=black](0.5,0){0.05}
\pscircle[fillstyle=solid,fillcolor=black](0,0.5){0.05}
\end{pspicture} \\[1ex]
+ \rho_6(u) \
\begin{pspicture}[shift=-0.4](1,1)
\loopf
\pscircle[fillstyle=solid,fillcolor=black](0,0.5){0.05}
\pscircle[fillstyle=solid,fillcolor=black](1,0.5){0.05}
\end{pspicture}
\ + \rho_7(u) \
\begin{pspicture}[shift=-0.4](1,1)
\loopg
\pscircle[fillstyle=solid,fillcolor=black](0.5,0){0.05}
\pscircle[fillstyle=solid,fillcolor=black](0.5,1){0.05}
\end{pspicture}
\ + \rho_8(u) \
\begin{pspicture}[shift=-0.4](1,1)
\looph \end{pspicture}
\ + \rho_9(u) \
\begin{pspicture}[shift=-0.4](1,1)
\loopi \end{pspicture} \ ,
\end{gathered}
\end{equation}
where black circles indicate vacancies, as in \cref{sec:dTL_N}. The small quarter arc in the lower-left corner of the left-hand side indicates the orientation to be given to the diagrams. For instance, a face operator with the arc in the lower-right corner would have all the diagrams rotated counter-clockwise by 90\degree{}. The factors $\rho_i(u)$ of the elementary tiles are interpreted as local Boltzmann weights. They are parameterized by $u$, the \textit{spectral parameter}, and $\lambda$, the \textit{crossing parameter}. They are
\begin{alignat}{3}
\rho_1(u) &= \sin(2\lambda) \sin(3\lambda) + \sin(u) \sin(3\lambda-u) \ ,
&\qquad \rho_{2,3}(u) &= \sin(2\lambda) \sin(3\lambda-u) \ , \notag\\
\rho_{4,5}(u) &= \sin(2\lambda) \sin(u) \ ,
&\rho_{6,7}(u) &= \sin(u) \sin(3\lambda-u) \ , \label{def:poidsTuileDeBase} \\
\rho_8(u) &= \sin(2\lambda-u) \sin(3\lambda-u) \ , 
&\rho_9(u) &= -\sin(u) \sin(\lambda-u) \ . \notag
\end{alignat}
The crossing parameter also parameterizes the fugacity of contractible loops as
\begin{equation} \label{def:beta}
\beta = -2 \cos(4\lambda) \ .
\end{equation}
The weights $\rho_i(u)$ are real for $u$, $\lambda$ $\in\mathbb{R}$. Note however that the positivity constraint is relaxed as they can be negative. Our normalization of these local weights is different than that of \cite{AMDPP19}. The present choice removes the singularities at $\lambda=\frac{\pi}{3},\frac{\pi}{2},\frac{2\pi}{3}$. 
The crossing parameter is said to be \textit{generic} if $\frac{\lambda}{\pi}\notin\mathbb{Q}$, whereas otherwise $\frac{\lambda}{\pi}\in\mathbb{Q}$ and thus $q=-\eE^{4\iI \lambda}$ is a root of unity.

The face operator can be seen as an element of $\dtl_2(\beta)$ by fixing a direction of action, that is, from two adjacent nodes that would appear at the top of the connectivities, to the remaining two adjacent nodes, that would be at the bottom \cite{PPJRJBZ06}. The face operator can also be seen as an element of $\dtl_N(\beta)$ acting non-trivially on the adjacent nodes $i$ and $i+1$, with $1\leq i< N$. Then for each $j$ different from $i$ and $i+1$, the $j$-th nodes on the top and bottom of the connectivity are connected by a unit link (represented by a dashed line in \eqref{def:diagIdentite}).

The \textit{boundary operator} is a linear combination of two elementary triangular tiles:
\begin{equation} \label{def:tuileFrontiere}
\begin{pspicture}[shift=-0.65](0.9,1.5)
\psset{unit=0.75cm,linewidth=0.75\pslinewidth}
\psline[linecolor=blue,linewidth=1.5pt,linestyle=dashed,dash=2pt 2pt](0,1.5)(1.25,1.5)
\psline[linecolor=blue,linewidth=1.5pt,linestyle=dashed,dash=2pt 2pt](0,0.5)(1.25,0.5)
\triangled \rput(0.35,1){$u$}
\end{pspicture}
= \delta \left(\tfrac{3\lambda}{2}-u\right) \
\begin{pspicture}[shift=-0.65](0.9,1.5)
\psset{unit=0.75cm,linewidth=0.75\pslinewidth}
\triangled
\psline[linecolor=blue,linewidth=1.5pt](0.74,1.5)(1.25,1.5)
\psline[linecolor=blue,linewidth=1.5pt](0.74,0.5)(1.25,0.5)
\psarc[linecolor=blue,linewidth=1.5pt](0.75,1){0.50}{90}{270}
\end{pspicture}
+ \delta \left(\tfrac{3\lambda}{2}+u\right) \
\begin{pspicture}[shift=-0.65](0.9,1.5)
\psset{unit=0.75cm,linewidth=0.75\pslinewidth}
\triangled
\pscircle[fillstyle=solid,fillcolor=black](0.5,0.5){0.0625}
\pscircle[fillstyle=solid,fillcolor=black](0.5,1.5){0.0625}
\end{pspicture} \ .
\end{equation}
With the triangles pointing rightward as in the equation above, the boundary operator has one node on each diagonal edge but none on the vertical one. We consider two possible choices for the function $\delta(u)$, which we refer to as the \textit{sine} (S) and \textit{cosine} (C) \textit{boundary conditions}. For simplicity and to underline the links between the S and C cases, we also introduce the function $\bar{\delta}(u)$ for each case:
\be \label{def:frontieres}
\begin{array}{clll}
\text{S:} &\quad\delta(u)=\sin(u) \ , &&\bar{\delta}(u)=\cos(u) \ , \\[0.1cm]
\text{C:} &\quad\delta(u)=\cos(u) \ , &&\bar{\delta}(u)=\sin(u) \ .
\end{array} \ee
In the following sections, when no ambiguity is possible, we use the short-hand notations
\begin{equation} \label{def:notation}
u_k = u+k\lambda \ , \qquad
s(\alpha u_k) = \sin(\alpha(u+k\lambda)) \ , \qquad
c(\alpha u_k) = \cos(\alpha(u+k\lambda)) \ .
\end{equation}
We note that the normalisation of the function $s(\alpha u_k)$ used here differs from the one used in \cite{AMDPP19}.

The weights \eqref{def:poidsTuileDeBase} in the face operator and those \eqref{def:frontieres} in the boundary one are not arbitrary. They are chosen to satisfy algebraic constraints, among them the Yang-Baxter equations, that lead to the integrability of the models defined in the next section. They were studied in \cite{Nienhuis,IK81} for the weight of the face operator and in \cite{BatchelorYung,DJS10} for the boundary ones.

%%%%%%%%%%%%%
\subsection{Diagrammatic properties of the face operator}
%%%%%%%%%%%%%

Many useful properties are satisfied by the face operator or a concatenation of a few of them. Some are valid only at given values of the spectral parameter $u$. Those reviewed here are well-known (see for example \cite{AMDPP19}) and thus stated without proof.
These identities are derived in the planar version of the dilute Temperley-Lieb algebra \cite{VFRJ99}. In this context, diagrammatic objects like face operators are glued together, sometimes with the dashed loop segment \eqref{def:diagIdentite} that acts as the identity, and the outside of the diagram has a number of free nodes. These relations can then be used to simplify larger diagrams where a member of one of the identities appears. They can also be converted into identities in $\dtl_N(\beta)$.

The \textit{crossing symmetry} 
\begin{equation} \label{rel:symCroiseTuile}
\begin{pspicture}[shift=-0.4](1,1)
\loopa \psarc[linewidth=0.025]{-}(0,0){0.16}{0}{90}
\rput(.5,.5){$u$}
\end{pspicture}
\ = \
\begin{pspicture}[shift=-0.4](1,1)
\loopa \psarc[linewidth=0.025]{-}(1,0){0.16}{90}{180}
\rput(.5,.5){\footnotesize${3\lambda\!-\!u}$}
\end{pspicture}
\end{equation}
links the face operator evaluated at $u$ to the one rotated by 90\degree{} and evaluated at $3\lambda-u$. The \textit{inversion relation}
\begin{equation} \label{rel:inversionTuile}
\begin{pspicture}[shift=-0.6](2.9,1.4)
\psline[linecolor=blue,linewidth=1.5pt,linestyle=dashed,dash=2pt 2pt](1,1.06)(2,1.06)
\psline[linecolor=blue,linewidth=1.5pt,linestyle=dashed,dash=2pt 2pt](1,0.35)(2,0.35)
\rput(0.71,0){\rput{45}{\loopa \psarc[linewidth=0.025]{-}(0,1){0.16}{-90}{0}}
\rput(0,0.71){$u$}}
\rput(2.12,0){\rput{45}{\loopa \psarc[linewidth=0.025]{-}(0,1){0.16}{-90}{0}}
\rput(0,0.71){$-u$}}
\end{pspicture}
= \rho_8(u) \rho_8(-u) \
\begin{pspicture}[shift=-0.6](1.5,1.4)
\rput{45}(0.71,0){\loopej}
\end{pspicture}
\end{equation}
shows that the concatenation of two face operators, evaluated at $u$ and $-u$ respectively, is a multiple of the identity. This operator is thus invertible as long as $\rho_8(u)\rho_8(-u)\neq 0$, that is, when $u\neq\pm2\lambda$ or $\pm3\lambda$. 

The \textit{Yang-Baxter equation} plays a crucial role in the proof of the integrability of loop models and is given by
\begin{equation} \label{rel:YBTuile}
\begin{pspicture}[shift=-0.9](3,2)
\psline[linecolor=blue,linewidth=1.5pt,linestyle=dashed,dash=2pt 2pt](1.5,0.5)(2,0.5)
\psline[linecolor=blue,linewidth=1.5pt,linestyle=dashed,dash=2pt 2pt](1.5,1.5)(2,1.5)
\losange \rput(1,1){$u-v$}
\rput(2,0){
\facegrid{(0,0)}{(1,2)}
\psarc[linewidth=0.025]{-}(0,1){0.16}{0}{90}
\psarc[linewidth=0.025]{-}(0,0){0.16}{0}{90}
\rput(.5,1.5){$v$}
\rput(.5,0.5){$u$}}
\end{pspicture}
\ = \
\begin{pspicture}[shift=-0.9](3,2)
\psline[linecolor=blue,linewidth=1.5pt,linestyle=dashed,dash=2pt 2pt](1,0.5)(1.5,0.5)
\psline[linecolor=blue,linewidth=1.5pt,linestyle=dashed,dash=2pt 2pt](1,1.5)(1.5,1.5)
\rput(1,0){\losange \rput(1,1){$u-v$}}
\facegrid{(0,0)}{(1,2)}
\psarc[linewidth=0.025]{-}(0,1){0.16}{0}{90}
\psarc[linewidth=0.025]{-}(0,0){0.16}{0}{90}
\rput(.5,1.5){$u$}
\rput(.5,0.5){$v$}
\end{pspicture} \ \ .
\end{equation}

At some values of the spectral parameter, the face operator factorizes into the product of two tangles. Those values are referred to as \textit{degeneration points} in \cite{AMDPP19}. At $u=0$ and $u=3\lambda$, the two tangles are triangles that carry the identity arc:
\begin{equation} \label{rel:evalRemarquable0et3}
\begin{pspicture}[shift=-0.38](1,1)
\loopa \psarc[linewidth=0.025]{-}(0,0){0.16}{0}{90}
\rput(.5,.5){$0$}
\end{pspicture}
\ = s(2\lambda) s(3\lambda) \
\begin{pspicture}[shift=-0.38](1,1)
\loopid \psline[linewidth=1pt](0,0)(1,1)
\end{pspicture} \ ,
\qquad
\begin{pspicture}[shift=-0.38](1,1)
\loopa \psarc[linewidth=0.025]{-}(0,0){0.16}{0}{90}
\rput(.5,.5){$3\lambda$}
\end{pspicture}
\ = s(2\lambda) s(3\lambda) \
\begin{pspicture}[shift=-0.38](1,1)
\loopej \psline[linewidth=1pt](0,1)(1,0)
\end{pspicture} \ .
\end{equation}
At $u=\lambda$ and $u=2\lambda$, these tangles are 
\begin{equation} \label{rel:evalRemarquable1et2}
\begin{pspicture}[shift=-0.4](1,1)
\loopa \psarc[linewidth=0.025]{-}(0,0){0.16}{0}{90}
\rput(.5,.5){$\lambda$}
\end{pspicture}
\ = s(\lambda) s(2\lambda) \
\begin{pspicture}[shift=-0.4](1,1)
\rput{90}(1,0){ \loopa \psline[linewidth=1pt](0,1)(1,0)
\pspolygon[fillstyle=solid,fillcolor=black](0.65,0.65)(0.75,0.65)(0.75,0.75)(0.65,0.75)
\pspolygon[fillstyle=solid,fillcolor=black](0.35,0.35)(0.25,0.35)(0.25,0.25)(0.35,0.25)}
\end{pspicture} \ ,
\qquad
\begin{pspicture}[shift=-0.4](1,1)
\loopa \psarc[linewidth=0.025]{-}(0,0){0.16}{0}{90}
\rput(.5,.5){$2\lambda$}
\end{pspicture}
\ = s(\lambda) s(2\lambda) \
\begin{pspicture}[shift=-0.4](1,1)
\loopa \psline[linewidth=1pt](0,1)(1,0)
\pspolygon[fillstyle=solid,fillcolor=black](0.65,0.65)(0.75,0.65)(0.75,0.75)(0.65,0.75)
\pspolygon[fillstyle=solid,fillcolor=black](0.35,0.35)(0.25,0.35)(0.25,0.25)(0.35,0.25)
\end{pspicture} \ ,
\end{equation}
where the black triangle is
\be
\label{def:triangleNoir}
\psset{unit=0.8cm}
\begin{pspicture}[shift=-0.2cm](0,0)(2,0.8)
\rput{90}(2,0)\triangleCNd
\end{pspicture}
= 2\, c(\lambda) \ \begin{pspicture}[shift=-0.2cm](0,0)(2,0.8)
\rput{90}(2,0)\triangled
\pscircle[fillstyle=solid,fillcolor=black](0.5,.5){0.0625}
\pscircle[fillstyle=solid,fillcolor=black](1.5,.5){0.0625}
\pscircle[fillstyle=solid,fillcolor=black](1,0){0.0625}
\end{pspicture}
+ \begin{pspicture}[shift=-0.2cm](0,0)(2,0.8)
\rput{90}(2,0)\triangled
\psarc[linewidth=1.5pt,linecolor=blue](1,1){.707}{-135}{-45}
\pscircle[fillstyle=solid,fillcolor=black](1,0){0.0625}
\end{pspicture}
+ \begin{pspicture}[shift=-0.2cm](0,0)(2,0.8)
\rput{90}(2,0)\triangled
\psline[linewidth=1pt,linecolor=blue,linewidth=1.5pt](0.5,0.5)(1,0)
\pscircle[fillstyle=solid,fillcolor=black](1.5,.5){0.0625}
\end{pspicture}
+ \begin{pspicture}[shift=-0.2cm](0,0)(2,0.8)
\rput{90}(2,0)\triangled
\psline[linewidth=1pt,linecolor=blue,linewidth=1.5pt](1.5,0.5)(1,0)
\pscircle[fillstyle=solid,fillcolor=black](0.5,.5){0.0625}
\end{pspicture} \ .
\ee
In contrast with the boundary operator \eqref{def:tuileFrontiere} that has two nodes, one on each shorter side, the black triangle tangle has three nodes, one per side. The crossing symmetry ($u\leftrightarrow 3\lambda -u$) links these degeneration points into pairs. 

There are also other local relations where a collection of face operators attached together are evaluated at values of the spectral parameter that differ by $2\lambda$ or $3\lambda$, and then have a simpler form in terms of fewer tangles. In particular, the next two relations are known as \textit{push-through} properties, as they both involve the propagation from right to left of a triangular tile through a pair of face operators. These tangles are those defined above, namely the identity arc and the black triangle, respectively. Using the short-hand notation \eqref{def:notation}, these relations are
\begin{subequations}
\begin{alignat}{2} 
\label{rel:PTbulkTCN}
&\begin{pspicture}[shift=-0.9](2,2)
\psline[linecolor=blue,linewidth=1.5pt,linestyle=dashed,dash=2pt 2pt](1,0.5)(1.5,0.5)
\psline[linecolor=blue,linewidth=1.5pt,linestyle=dashed,dash=2pt 2pt](1,1.5)(1.5,1.5)
\facegrid{(0,0)}{(1,2)}
\rput(0,1){\psarc[linewidth=0.025]{-}(0,0){0.16}{0}{90}\rput(.5,0.5){$u_2$}}
\rput(0,0){\psarc[linewidth=0.025]{-}(0,0){0.16}{0}{90}\rput(.5,0.5){$u_0$}}
\rput(1,0){\triangleCNg}
\end{pspicture}
\ = -s(u_{-3})s(u_2) \
\begin{pspicture}[shift=-0.4](1.5,1)
\pspolygon[fillstyle=solid,fillcolor=lightlightblue](0,0.5)(0.475,0.99)(0.475,0.01)
\psscalebox{0.49}{
\pspolygon[fillstyle=solid,fillcolor=black](0.55,0.93)(0.70,0.93)(0.70,1.08)(0.55,1.08)}
\rput(0.48,0){\loopa \psarc[linewidth=0.025]{-}(0,0){0.16}{0}{90} \rput(.5,0.5){$u_1$}}
\end{pspicture}\ ,
\\[0.25cm] 
\label{rel:PTbulkArc}
&\begin{pspicture}[shift=-0.9](1.5,2)
\psarc[linecolor=blue,linewidth=1.5pt,linestyle=dashed,dash=2pt 2pt](1,1){0.5}{-90}{90}
\facegrid{(0,0)}{(1,2)}
\rput(0,1){\psarc[linewidth=0.025]{-}(0,0){0.16}{0}{90}\rput(.5,0.5){$u_3$}}
\rput(0,0){\psarc[linewidth=0.025]{-}(0,0){0.16}{0}{90}\rput(.5,0.5){$u_0$}}
\end{pspicture}
\ = s(u_2) s(u_{-2}) s(u_3) s(u_{-3}) \
\begin{pspicture}[shift=-0.9](1.5,2)
\facegrid{(0,0)}{(1,2)}
\rput(0,1){\psarc[linewidth=1.5pt,linecolor=blue,linestyle=dashed,dash=2pt 2pt](0,0){.5}{0}{90}
\psarc[linewidth=1.5pt,linecolor=blue,linestyle=dashed,dash=2pt 2pt](1,1){0.5}{180}{-90}}
\psarc[linewidth=1.5pt,linecolor=blue,linestyle=dashed,dash=2pt 2pt](1,0){0.5}{90}{180}
\psarc[linewidth=1.5pt,linecolor=blue,linestyle=dashed,dash=2pt 2pt](0,1){0.5}{-90}{0}
\psarc[linecolor=blue,linewidth=1.5pt,linestyle=dashed,dash=2pt 2pt](1,1){0.5}{-90}{90}
\end{pspicture} \ .
\end{alignat}
\end{subequations}
To obtain the similar push-through properties for a propagation from left to right, one must simply rotate the whole stacks of tangles by 180\degree{}.

A last property links the face operators at $u$ and $u+\pi$. Following \cite{AMDPPJR19}, we define a gauge operator on one site. It is depicted as a white box:
\begin{equation} \label{def:jauge}
\begin{pspicture}[shift=-0.27](-0.1,0)(0.75,0.5)
\rput(0,0.25){\pspolygon(0,0)(0.75,0)(0.75,0.25)(0,0.25)}
\end{pspicture}
\ = (-1)
\begin{pspicture}[shift=-0.27](-0.2,0)(1,0.5)
\rput(0,0.25){\pspolygon[fillstyle=solid,fillcolor=lightlightblue](0,0)(0.75,0)(0.75,0.25)(0,0.25)
\psline[linewidth=1.5pt,linecolor=blue](0.37,0)(0.37,0.25)}
\end{pspicture}
+
\begin{pspicture}[shift=-0.27](-0.2,0)(0.75,0.5)
\rput(0,0.25){\pspolygon[fillstyle=solid,fillcolor=lightlightblue](0,0)(0.75,0)(0.75,0.25)(0,0.25)
\pscircle[fillstyle=solid,fillcolor=black](0.37,0.25){0.05}
\pscircle[fillstyle=solid,fillcolor=black](0.37,0){0.05}}
\end{pspicture} \ .
\end{equation}
The following periodicity property then holds:
\begin{equation} \label{rel:jaugePi}
\begin{pspicture}[shift=-0.4](1,1)
\loopa \psarc[linewidth=0.025]{-}(0,0){0.16}{0}{90}
\rput(.5,.5){\footnotesize$u+\pi$}
\end{pspicture}
\ = \ \begin{pspicture}[shift=-0.65](1,1.5)
\rput(0.125,1.25){\pspolygon(0,0)(0.75,0)(0.75,0.25)(0,0.25)}
\rput(0.125,0.00){\pspolygon(0,0)(0.75,0)(0.75,0.25)(0,0.25)}
\rput(0,0.25){\loopa \psarc[linewidth=0.025]{-}(0,0){0.16}{0}{90}
\rput(.5,.5){$u$}}
\end{pspicture}
\ = \ \begin{pspicture}[shift=-0.4](1.5,1)
\rput{90}(0.25,0.125){\pspolygon(0,0)(0.75,0)(0.75,0.25)(0,0.25)}
\rput{90}(1.50,0.125){\pspolygon(0,0)(0.75,0)(0.75,0.25)(0,0.25)}
\rput(0.25,0){\loopa \psarc[linewidth=0.025]{-}(0,0){0.16}{0}{90}
\rput(.5,.5){$u$}}
\end{pspicture} \ .
\end{equation}

Finally, we recall that some projectors for the dilute $\Atwotwo$ loop model were constructed in \cite{AMDPP19}. Here we only introduce the first two non-trivial ones, $P^{2,0}$ and $P^{1,1}$, defined as
\begin{equation} \label{def:P20P11explicite}
\psset{unit=0.9}
\begin{pspicture}[shift=-0.9](0.5,2)
\projectorTwo\rput(0.18,1){\rput{90}(0,0){$_{2,0}$}}
\end{pspicture}
\ = \ \begin{pspicture}[shift=-0.9](1,2)
\pspolygon[fillstyle=solid,fillcolor=lightlightblue](0,0)(0,2)(1,2)(1,0)
\psline[linecolor=blue,linewidth=1.5pt,linestyle=dashed,dash=2pt 2pt](0,1.5)(1,1.5)
\psline[linecolor=blue,linewidth=1.5pt,linestyle=dashed,dash=2pt 2pt](0,0.5)(1,0.5)
\end{pspicture}
\ - \ \begin{pspicture}[shift=-0.9](2,2)
\psline[linecolor=blue,linewidth=1.5pt,linestyle=dashed,dash=2pt 2pt](0,1.5)(2,1.5)
\psline[linecolor=blue,linewidth=1.5pt,linestyle=dashed,dash=2pt 2pt](0,0.5)(2,0.5)
\triangleCNg 
\rput(1,0){\triangleCNdw
\psarc[linewidth=0.025]{-}(1,1){0.16}{135}{225}}
\end{pspicture}\ \ ,
\qquad \qquad
\begin{pspicture}[shift=-0.9](0.5,2)
\projectorTwo\rput(0.18,1){\rput{90}(0,0){$_{1,1}$}}
\end{pspicture}
\ = \ \begin{pspicture}[shift=-0.9](1,2)
\pspolygon[fillstyle=solid,fillcolor=lightlightblue](0,0)(0,2)(1,2)(1,0)
\psline[linecolor=blue,linewidth=1.5pt,linestyle=dashed,dash=2pt 2pt](0,1.5)(1,1.5)
\psline[linecolor=blue,linewidth=1.5pt,linestyle=dashed,dash=2pt 2pt](0,0.5)(1,0.5)
\end{pspicture}
\ - \ \begin{pspicture}[shift=-0.9](2,2)
\psline[linecolor=blue,linewidth=1.5pt,linestyle=dashed,dash=2pt 2pt](0,1.5)(2,1.5)
\psline[linecolor=blue,linewidth=1.5pt,linestyle=dashed,dash=2pt 2pt](0,0.5)(2,0.5)
\rput(0,0){\triangleArcg}
\rput{90}(2,0){
\pspolygon[fillstyle=solid,fillcolor=lightlightblue](0,1)(1,0)(2,1)
\rput(1,0){\wobblyarc{0.71}{45}{135}}
}
\end{pspicture}
\ \ ,
\end{equation}
where
\begin{subequations}
\label{eq:triangles.m} % triangle blanc
\begin{alignat}{2}
\psset{unit=0.9}
\begin{pspicture}[shift=-.9](0,-0.5)(1,1.5)
\rput{90}(1,-0.5){
\pspolygon[fillstyle=solid,fillcolor=lightlightblue](0,1)(1,0)(2,1)
\rput{-90}(0,1)\triangleCNdw
\psarc[linewidth=0.025]{-}(1,0){0.16}{45}{135}
}
\end{pspicture} 
&
\ = \frac{s^2(2\lambda)}{s(4\lambda)s(5 \lambda)} \ \
\psset{unit=0.9}
\begin{pspicture}[shift=-.9](0,-0.5)(1,1.5)
\rput{90}(1,-0.5){
\pspolygon[fillstyle=solid,fillcolor=lightlightblue](0,1)(1,0)(2,1)
\pscircle[fillstyle=solid,fillcolor=black](0.5,0.5){0.0625}
\pscircle[fillstyle=solid,fillcolor=black](1.5,0.5){0.0625}
\pscircle[fillstyle=solid,fillcolor=black](1,1){0.0625}
}
\end{pspicture}
\ - \frac{s(2\lambda)s(3\lambda)}{s(4\lambda)s(5 \lambda)}\ \
\begin{pspicture}[shift=-.90](0,-0.5)(1,1.5)
\rput{90}(1,-0.5){
\pspolygon[fillstyle=solid,fillcolor=lightlightblue](0,1)(1,0)(2,1)
\pscircle[fillstyle=solid,fillcolor=black](1,1){0.0625}
\psarc[linewidth=1.5pt,linecolor=blue](1,0){.707}{45}{135}
}
\end{pspicture} 
\ - \frac{s(2\lambda)}{s(4\lambda)}\ \
\begin{pspicture}[shift=-.90](0,-0.5)(1,1.5)
\rput{90}(1,-0.5){
\pspolygon[fillstyle=solid,fillcolor=lightlightblue](0,1)(1,0)(2,1)
\pscircle[fillstyle=solid,fillcolor=black](1.5,0.5){0.0625}
\psline[linewidth=1.5pt,linecolor=blue](0.5,0.5)(1,1)
}
\end{pspicture} 
\ + \frac{s(2\lambda)s(3\lambda)}{s(\lambda)s(4\lambda)}\ \
\begin{pspicture}[shift=-.90](0,-0.5)(1,1.5)
\rput{90}(1,-0.5){
\pspolygon[fillstyle=solid,fillcolor=lightlightblue](0,1)(1,0)(2,1)
\pscircle[fillstyle=solid,fillcolor=black](0.5,0.5){0.0625}
\psline[linewidth=1.5pt,linecolor=blue](1.5,0.5)(1,1)
}
\end{pspicture}
\ ,
\\[0.3cm]
\psset{unit=0.9} % triangle fris\'e
\begin{pspicture}[shift=-.90](0,-0.5)(1,1.5)
\rput{90}(1,-0.5){
\pspolygon[fillstyle=solid,fillcolor=lightlightblue](0,1)(1,0)(2,1)
\rput(1,0){\wobblyarc{0.71}{45}{135}}
}
\end{pspicture} 
&\psset{unit=0.9}
\ =  \frac{s(2\lambda)s(3\lambda)s(7\lambda)}{s(\lambda)s(5\lambda)s(6\lambda)} \ \
\begin{pspicture}[shift=-.90](0,-0.5)(1,1.5)
\rput{90}(1,-0.5){
\pspolygon[fillstyle=solid,fillcolor=lightlightblue](0,1)(1,0)(2,1)
\pscircle[fillstyle=solid,fillcolor=black](0.5,0.5){0.0625}
\pscircle[fillstyle=solid,fillcolor=black](1.5,0.5){0.0625}
}
\end{pspicture}
\ + \frac{s(2\lambda)s(3\lambda)}{s(5\lambda)s(6\lambda)}\ \ 
\begin{pspicture}[shift=-.90](0,-0.5)(1,1.5)
\rput{90}(1,-0.5){
\pspolygon[fillstyle=solid,fillcolor=lightlightblue](0,1)(1,0)(2,1)
\psarc[linewidth=1.5pt,linecolor=blue](1,0){.707}{45}{135}
}
\end{pspicture} \ .
\end{alignat}
\end{subequations}
We have the relations
\be
\label{eq:spider.ids}
\psset{unit=0.9cm}
\begin{pspicture}[shift=-.90](0,-1)(2,1)
\rput(0,-1)\triangleCNdw
\pspolygon[fillstyle=solid,fillcolor=lightlightblue](2,1)(1,0)(2,-1)
\psarc[linewidth=0.025]{-}(1,0){0.16}{135}{225}
\rput(1.6,0){\specialcircle{0.075}}
\psarc[linewidth=1.5pt,linecolor=blue,linestyle=dashed,dash=2pt 2pt](1,0){.707}{45}{135}
\psarc[linewidth=1.5pt,linecolor=blue,linestyle=dashed,dash=2pt 2pt](1,0){.707}{-135}{-45}
\end{pspicture} \ = \
\begin{pspicture}[shift=-0.275](0,0)(1.5,0.75)
\pspolygon[fillstyle=solid,fillcolor=lightlightblue](0,0)(0,0.75)(1.5,0.75)(1.5,0)
\psline[linewidth=1.5pt,linecolor=blue,linestyle=dashed,dash=2pt 2pt](0,0.375)(1.5,0.375)
\end{pspicture}\ \ ,
\qquad\quad
\begin{pspicture}[shift=-.90](0,-1)(2,1)
\pspolygon[fillstyle=solid,fillcolor=lightlightblue](0,1)(1,0)(0,-1)
\rput(1,0){\wobblyarc{0.71}{135}{225}}
\pspolygon[fillstyle=solid,fillcolor=lightlightblue](2,1)(1,0)(2,-1)
\psarc[linewidth=1.5pt,linecolor=blue,linestyle=dashed,dash=2pt 2pt](1,0){.707}{-135}{135}
\end{pspicture} \ = 1,
\qquad \quad
\begin{pspicture}[shift=-0.9](0,0)(1.3,2)
\projectorTwo\rput(0.18,1){\rput{90}(0,0){$_{2,0}$}}
\rput(-0.7,1){\pspolygon[fillstyle=solid,fillcolor=lightlightblue](2,1)(1,0)(2,-1)
\rput(1.6,0){\specialcircle{0.075}}}
\psline[linewidth=1.5pt,linecolor=blue,linestyle=dashed,dash=2pt 2pt](0.3,0.5)(0.8,0.5)
\psline[linewidth=1.5pt,linecolor=blue,linestyle=dashed,dash=2pt 2pt](0.3,1.5)(0.8,1.5)
\end{pspicture}
\ = 0,
\qquad \quad
\begin{pspicture}[shift=-0.9](0,0)(1.3,2)
\projectorTwo\rput(0.18,1){\rput{90}(0,0){$_{1,1}$}}
\rput(-0.7,1){\pspolygon[fillstyle=solid,fillcolor=lightlightblue](2,1)(1,0)(2,-1)}
\psarc[linewidth=1.5pt,linecolor=blue,linestyle=dashed,dash=2pt 2pt](0.3,1){.707}{-45}{45}
\psline[linewidth=1.5pt,linecolor=blue,linestyle=dashed,dash=2pt 2pt](0.3,0.5)(0.8,0.5)
\psline[linewidth=1.5pt,linecolor=blue,linestyle=dashed,dash=2pt 2pt](0.3,1.5)(0.8,1.5)
\end{pspicture} \ = 0.
\ee
The first two identities are used to show that $P^{2,0}$ and $P^{1,1}$ are indeed projectors.

%%%%%%%%%%%%%
\subsection{Diagrammatic properties of the boundary operators}\label{sub:propertiesBoundary}
%%%%%%%%%%%%%

Similar relations involving boundary operators \eqref{def:tuileFrontiere} also exist. Some of them involve only boundary operators, whereas others apply to particular combinations of boundary operators and face operators. The first relation is the \textit{crossing symmetry at the boundary}. It is a diagrammatic relation that links boundary operators evaluated at $u$ and $3\lambda-u$:
\begin{equation} \label{rel:symCroiseFrontiere}
\begin{pspicture}[shift=-0.9](3,2)
\psline[linecolor=blue,linewidth=1.5pt,linestyle=dashed,dash=2pt 2pt](0.5,1.5)(1.5,1.5)
\psline[linecolor=blue,linewidth=1.5pt,linestyle=dashed,dash=2pt 2pt](0.5,0.5)(1.5,0.5)
\triangled \rput(0.45,1){$_{3\lambda-u}$}
\rput(1,0){\losange \rput(1,1){$3\lambda-2u$}}
\end{pspicture}
\ = \frac{\rho_8(2u-3\lambda)\bar{\delta}(u-\tfrac{\lambda}{2})}
{\bar{\delta}(\tfrac{5\lambda}{2}-u)} \
\begin{pspicture}[shift=-0.9](1,2)
\triangled \rput(0.4,1){$u$}
\end{pspicture} \ .
\end{equation}
In these equations and in others below, the boundary operators are chosen to be of S or C types and the functions $\delta(u)$ and $\bar \delta(u)$ are set accordingly from \eqref{def:frontieres}. The boundary operator type S or C stays the same when an equality between tangles is presented. We also note that the function in the right-hand side in fact does not have any pole in $u$, as the zero in the denominator is cancelled by a zero in the numerator for both the S and C functions.
The next relation is the \textit{inversion relation at the boundary}:
\begin{equation}
\begin{pspicture}[shift=-0.4](4,1) \label{rel:inversionFrontiere}
\psline[linecolor=blue,linewidth=1.5pt,linestyle=dashed,dash=2pt 2pt](1,0.5)(3,0.5)
\rput(2,0){\rput{90}(2,0){\triangled} \rput(1,0.4){$-u$}}
\rput{90}(2,0){\triangled} \rput(1,0.4){$u$}
\end{pspicture}
\ = \ \delta(\tfrac{3\lambda}{2}-u) \delta(\tfrac{3\lambda}{2}+u)
\begin{pspicture}[shift=-0.4](1.8,1)
\psline[linecolor=blue,linewidth=1.5pt,linestyle=dashed,dash=2pt 2pt](0.2,0.5)(1.8,0.5)
\end{pspicture} \ .
\end{equation}
The boundary operators also satisfy \textit{boundary Yang-Baxter equations}. For example, the left one is
\begin{equation} \label{rel:YBFrontiere}
\begin{pspicture}[shift=-1.9](3,4)
\psline[linecolor=blue,linewidth=1.5pt,linestyle=dashed,dash=2pt 2pt](0,3.5)(2,3.5)
\rput(1,2){\losange \rput(1,1){$u-v$}}
\rput(0,2){\triangled \rput(0.35,1){$u$}}
\rput(0,1){\losange \rput(1,1){$_{3\lambda-(u+v)}$}}
\rput(0,0){\triangled \rput(0.35,1){$v$}}
\end{pspicture}
\ = \ \ 
\begin{pspicture}[shift=-1.9](3,4)
\psline[linecolor=blue,linewidth=1.5pt,linestyle=dashed,dash=2pt 2pt](0,0.5)(2,0.5)
\rput(1,0){\losange \rput(1,1){$u-v$}}
\rput(0,2){\triangled \rput(0.35,1){$v$}}
\rput(0,1){\losange \rput(1,1){$_{3\lambda-(u+v)}$}}
\rput(0,0){\triangled \rput(0.35,1){$u$}}
\end{pspicture} \ \ .
\end{equation}
It is by solving this equation that the expressions \eqref{def:tuileFrontiere} for the boundary tile were obtained in \cite{BatchelorYung,DJS10}.

The previous identities are similar to those appearing in other models, for instance in the dense loop models (see for example \cite{PPJRJBZ06}). In contrast, the following relations are particular to the dilute models. They are analogous to the factorization and push-through properties of the face operator. They do not apply to the boundary operator itself but to specific combinations of tangles which include boundary operators. These combinations will be interpreted as boundary conditions for the first fused transfer matrices in \cref{sub:fused.with.diagrams}. The combinations are
\begin{subequations} \label{def:produitFrontiere}
\begin{alignat}{2}
&\begin{pspicture}[shift=-1.9](2,4) \label{def:produitFrontiere_a}
\psline[linecolor=blue,linewidth=1.5pt,linestyle=dashed,dash=2pt 2pt](0,3.5)(2,3.5)
\psline[linecolor=blue,linewidth=1.5pt,linestyle=dashed,dash=2pt 2pt](0,2.5)(2,2.5)
\psline[linecolor=blue,linewidth=1.5pt,linestyle=dashed,dash=2pt 2pt](0,1.5)(2,1.5)
\psline[linecolor=blue,linewidth=1.5pt,linestyle=dashed,dash=2pt 2pt](0,0.5)(2,0.5)
\rput(0,2){\triangled \rput(0.45,1){$_{3\lambda-u_2}$}}
\rput(0,1){\losange \rput(1,1){$2u-\lambda$}}
\triangled \rput(0.45,1){$_{3\lambda-u_0}$}
\end{pspicture} \ \ ,
&&\begin{pspicture}[shift=-1.9](2,4)
\psline[linecolor=blue,linewidth=1.5pt,linestyle=dashed,dash=2pt 2pt](0,3.5)(2,3.5)
\psline[linecolor=blue,linewidth=1.5pt,linestyle=dashed,dash=2pt 2pt](0,2.5)(2,2.5)
\psline[linecolor=blue,linewidth=1.5pt,linestyle=dashed,dash=2pt 2pt](0,1.5)(2,1.5)
\psline[linecolor=blue,linewidth=1.5pt,linestyle=dashed,dash=2pt 2pt](0,0.5)(2,0.5)
\rput(1,2){\triangleg \rput(0.6,1){$u_2$}}
\rput(0,1){\losange \rput(1,1){$\lambda-2u$}}
\rput(1,0){\triangleg \rput(0.6,1){$u_0$}}
\end{pspicture} \ \ , \\
&\begin{pspicture}[shift=-1.9](2,4) \label{def:produitFrontiere_b}
\psline[linecolor=blue,linewidth=1.5pt,linestyle=dashed,dash=2pt 2pt](0,3.5)(2,3.5)
\psline[linecolor=blue,linewidth=1.5pt,linestyle=dashed,dash=2pt 2pt](0,2.5)(2,2.5)
\psline[linecolor=blue,linewidth=1.5pt,linestyle=dashed,dash=2pt 2pt](0,1.5)(2,1.5)
\psline[linecolor=blue,linewidth=1.5pt,linestyle=dashed,dash=2pt 2pt](0,0.5)(2,0.5)
\rput(0,2){\triangled \rput(0.45,1){$_{3\lambda-u_3}$}}
\rput(0,1){\losange \rput(1,1){$2u$}}
\triangled \rput(0.45,1){$_{3\lambda-u_0}$}
\end{pspicture} \ \ ,
&\hspace{3cm}&\begin{pspicture}[shift=-1.9](2,4)
\psline[linecolor=blue,linewidth=1.5pt,linestyle=dashed,dash=2pt 2pt](0,3.5)(2,3.5)
\psline[linecolor=blue,linewidth=1.5pt,linestyle=dashed,dash=2pt 2pt](0,2.5)(2,2.5)
\psline[linecolor=blue,linewidth=1.5pt,linestyle=dashed,dash=2pt 2pt](0,1.5)(2,1.5)
\psline[linecolor=blue,linewidth=1.5pt,linestyle=dashed,dash=2pt 2pt](0,0.5)(2,0.5)
\rput(1,2){\triangleg \rput(0.6,1){$u_3$}}
\rput(0,1){\losange \rput(1,1){$-2u$}}
\rput(1,0){\triangleg \rput(0.6,1){$u_0$}}
\end{pspicture} \ \ .
\end{alignat}
\end{subequations}
The combinations are grouped in pairs linked by crossing symmetry. Starting from any member of a pair, changing the evaluation point respectively as $u\mapsto\lambda-u$ for \eqref{def:produitFrontiere_a} and as $u\mapsto-u$ for \eqref{def:produitFrontiere_b} and then applying a 180\degree{} rotation gives the second member. The combinations are also invariant under $u\mapsto u+\pi$. An important property of these combinations is that, similarly to \eqref{rel:evalRemarquable0et3} and \eqref{rel:evalRemarquable1et2} for the face operator, they factorize at some specific values of $u$ into products of tangles that involve either a black triangle or an identity arc. The remarkable values of $u$ at which this occurs depend on the type S or C of the boundary operators.

The following remarkable evaluations apply to the combinations in \eqref{def:produitFrontiere_a}, with these objects rewritten as a collection of operators involving black triangles. The evaluations are presented for the rightward pointing triangle, and only the tangles necessary for the factorization are drawn. There are four different values of $u$ for each boundary operator of type S or C. The type of boundary is always specified, and SC indicates that both are allowed. These evaluations are
\begingroup
\allowdisplaybreaks
\begin{subequations}
\label{eq:bdy.identities.type1}
\begin{alignat}{2} \label{eq:firstTangleInBST}
&\text{S} \quad u=\frac{\lambda}{2}:
\hspace{2.5cm}&&\text{C} \quad u=\frac{\lambda}{2} + \frac{\pi}{2}: \nonumber \\
&\begin{pspicture}[shift=-1.4](2,3)
\psline[linecolor=blue,linewidth=1.5pt,linestyle=dashed,dash=2pt 2pt](0.5,2.5)(1.85,2.5)
\psline[linecolor=blue,linewidth=1.5pt,linestyle=dashed,dash=2pt 2pt](0.5,1.5)(1.85,1.5)
\psline[linecolor=blue,linewidth=1.5pt,linestyle=dashed,dash=2pt 2pt](0.5,0.5)(1.85,0.5)
\rput(0,1){\triangled \rput(0.4,1){$\frac{\lambda}{2}$}}
\rput(0,0){\losange \rput(1,1){$0$}}
\end{pspicture}
= p_{1} \ \begin{pspicture}[shift=-1.4](3,3)
\psline[linecolor=blue,linewidth=1.5pt,linestyle=dashed,dash=2pt 2pt](0.5,2.5)(2.85,2.5)
\psline[linecolor=blue,linewidth=1.5pt,linestyle=dashed,dash=2pt 2pt](2,1.5)(2.85,1.5)
\psline[linecolor=blue,linewidth=1.5pt,linestyle=dashed,dash=2pt 2pt](1.58,0.5)(2.85,0.5)
\rput(0,1){\triangled \rput(0.35,1){$\frac{5\lambda}{2}$}}
\rput(0,0){\losangeNoArc}
\psarc[linecolor=blue,linewidth=1.5pt,linestyle=dashed,dash=2pt 2pt](1,2){0.71}{225}{315}
\psarc[linecolor=blue,linewidth=1.5pt,linestyle=dashed,dash=2pt 2pt](1,0){0.71}{45}{135}
\rput(1,1)\triangleCNg
\rput(2,1)\triangleCNd
\end{pspicture}
\hspace{2.5cm}&&\begin{pspicture}[shift=-1.4](2,3)
\psline[linecolor=blue,linewidth=1.5pt,linestyle=dashed,dash=2pt 2pt](0.5,2.5)(1.85,2.5)
\psline[linecolor=blue,linewidth=1.5pt,linestyle=dashed,dash=2pt 2pt](0.5,1.5)(1.85,1.5)
\psline[linecolor=blue,linewidth=1.5pt,linestyle=dashed,dash=2pt 2pt](0.5,0.5)(1.85,0.5)
\rput(0,1){\triangled \rput(0.45,1){\tiny$\frac{\lambda}{2}\!-\!\frac{\pi}{2}$}}
\rput(0,0){\losange \rput(1,1){$\pi$}}
\end{pspicture}
= p_{1} \ \begin{pspicture}[shift=-1.4](3,3)
\psline[linecolor=blue,linewidth=1.5pt,linestyle=dashed,dash=2pt 2pt](0.5,2.5)(2.85,2.5)
\psline[linecolor=blue,linewidth=1.5pt,linestyle=dashed,dash=2pt 2pt](2,1.5)(2.85,1.5)
\psline[linecolor=blue,linewidth=1.5pt,linestyle=dashed,dash=2pt 2pt](1.58,0.5)(2.85,0.5)
\rput(0,1){\triangled \rput(0.45,1){\tiny$\frac{5\lambda}{2}\!-\!\frac{\pi}{2}$}}
\rput(0,0){\losangeNoArc}
\psarc[linecolor=blue,linewidth=1.5pt,linestyle=dashed,dash=2pt 2pt](1,0){0.71}{45}{135}
\psarc[linecolor=blue,linewidth=1.5pt,linestyle=dashed,dash=2pt 2pt](1,2){0.71}{225}{315}
\rput(1,1)\triangleCNg
\rput(2,1)\triangleCNd
\rput{45}(1.44,0.08)\jauge
\rput{45}(0.44,1.08)\jauge
\end{pspicture}
\\
&\text{S} \quad u=\frac{3\lambda}{2}:\label{eq:secondTangleInBST}
\hspace{2.5cm}&&\text{C} \quad u=\frac{3\lambda}{2} + \frac{\pi}{2}: \nonumber \\
&\begin{pspicture}[shift=-1.9](2,4)
\psline[linecolor=blue,linewidth=1.5pt,linestyle=dashed,dash=2pt 2pt](0.5,3.5)(2,3.5)
\psline[linecolor=blue,linewidth=1.5pt,linestyle=dashed,dash=2pt 2pt](0.5,2.5)(2,2.5)
\psline[linecolor=blue,linewidth=1.5pt,linestyle=dashed,dash=2pt 2pt](0.5,1.5)(2,1.5)
\psline[linecolor=blue,linewidth=1.5pt,linestyle=dashed,dash=2pt 2pt](0.5,0.5)(2,0.5)
\rput(0,2){\triangled \rput(0.4,1){$-\!\frac{\lambda}{2}$}}
\rput(0,1){\losange \rput(1,1){$2\lambda$}}
\triangled \rput(0.4,1){$\frac{3\lambda}{2}$}
\end{pspicture}
\ = p_{2} \ \begin{pspicture}[shift=-1.9](2,4)
\psline[linecolor=blue,linewidth=1.5pt,linestyle=dashed,dash=2pt 2pt](0.62,3.5)(2,3.5)
\psline[linecolor=blue,linewidth=1.5pt,linestyle=dashed,dash=2pt 2pt](0.75,2.5)(2,2.5)
\psline[linecolor=blue,linewidth=1.5pt,linestyle=dashed,dash=2pt 2pt](0.75,1.4)(2,1.4)
\pscircle[fillstyle=solid,fillcolor=black](2,0.5){0.05}
\psarc[linecolor=blue,linewidth=1.5pt,linestyle=dashed,dash=2pt 2pt](0.75,2.75){0.75}{95}{270}
\rput(0.75,1)\triangleCNd
\end{pspicture}
\hspace{2.5cm}&&\begin{pspicture}[shift=-1.9](2,4)
\psline[linecolor=blue,linewidth=1.5pt,linestyle=dashed,dash=2pt 2pt](0.5,3.5)(2,3.5)
\psline[linecolor=blue,linewidth=1.5pt,linestyle=dashed,dash=2pt 2pt](0.5,2.5)(2,2.5)
\psline[linecolor=blue,linewidth=1.5pt,linestyle=dashed,dash=2pt 2pt](0.5,1.5)(2,1.5)
\psline[linecolor=blue,linewidth=1.5pt,linestyle=dashed,dash=2pt 2pt](0.5,0.5)(2,0.5)
\rput(0,2){\triangled \rput(0.47,1){\tiny$-\!\frac{\lambda}{2}\!-\!\frac{\pi}{2}$}}
\rput(0,1){\losange \rput(1,1){$2\lambda+\pi$}}
\rput(0,0){\triangled \rput(0.45,1){\tiny$\frac{3\lambda}{2}\!-\!\frac{\pi}{2}$}}
\end{pspicture}
\ = p_{2} \ \begin{pspicture}[shift=-1.9](2,4)
\psline[linecolor=blue,linewidth=1.5pt,linestyle=dashed,dash=2pt 2pt](0.62,3.5)(2,3.5)
\psline[linecolor=blue,linewidth=1.5pt,linestyle=dashed,dash=2pt 2pt](0.75,2.5)(2,2.5)
\psline[linecolor=blue,linewidth=1.5pt,linestyle=dashed,dash=2pt 2pt](0.75,1.4)(2,1.4)
\pscircle[fillstyle=solid,fillcolor=black](2,0.5){0.05}
\psarc[linecolor=blue,linewidth=1.5pt,linestyle=dashed,dash=2pt 2pt](0.75,2.75){0.75}{95}{270}
\rput(0.75,1)\triangleCNd
\rput{45}(1.18,1.08)\jauge
\end{pspicture}
\\
&\text{SC} \quad u=\lambda:\label{eq:thirdTangleInBST}
\hspace{2.5cm}&&\text{SC} \quad u=\lambda+\frac{\pi}{2}: \nonumber \\
&\begin{pspicture}[shift=-1.4](2,3)
\psline[linecolor=blue,linewidth=1.5pt,linestyle=dashed,dash=2pt 2pt](0.5,2.5)(2,2.5)
\psline[linecolor=blue,linewidth=1.5pt,linestyle=dashed,dash=2pt 2pt](0.5,1.5)(2,1.5)
\psline[linecolor=blue,linewidth=1.5pt,linestyle=dashed,dash=2pt 2pt](0.5,0.5)(2,0.5)
\rput(0,1){\triangled \rput(0.4,1){$0$}}
\rput(0,0){\losange \rput(1,1){$\lambda$}}
\end{pspicture}
\ = p_{3} \
\begin{pspicture}[shift=-1.4](2,3)
\psline[linecolor=blue,linewidth=1.5pt,linestyle=dashed,dash=2pt 2pt](0.5,2.5)(2,2.5)
\psline[linecolor=blue,linewidth=1.5pt,linestyle=dashed,dash=2pt 2pt](0.5,1.5)(2,1.5)
\psline[linecolor=blue,linewidth=1.5pt,linestyle=dashed,dash=2pt 2pt](1,0.5)(2,0.5)
\psarc[linecolor=blue,linewidth=1.5pt,linestyle=dashed,dash=2pt 2pt](0.5,2){0.5}{95}{270}
\rput{90}(2,1){\triangleCNd}
\rput{90}(2,0){\triangleCNg}
\end{pspicture}
\hspace{2.5cm} &&\begin{pspicture}[shift=-1.4](2,3)
\psline[linecolor=blue,linewidth=1.5pt,linestyle=dashed,dash=2pt 2pt](0.5,2.5)(2,2.5)
\psline[linecolor=blue,linewidth=1.5pt,linestyle=dashed,dash=2pt 2pt](0.5,1.5)(2,1.5)
\psline[linecolor=blue,linewidth=1.5pt,linestyle=dashed,dash=2pt 2pt](0.5,0.5)(2,0.5)
\rput(0,1){\triangled \rput(0.4,1){$-\frac{\pi}{2}$}}
\rput(0,0){\losange \rput(1,1){$\lambda+\pi$}}
\end{pspicture}
\ = p_{4} \
\begin{pspicture}[shift=-1.4](2,3)
\psline[linecolor=blue,linewidth=1.5pt,linestyle=dashed,dash=2pt 2pt](0.5,2.5)(2,2.5)
\psline[linecolor=blue,linewidth=1.5pt,linestyle=dashed,dash=2pt 2pt](0.5,1.5)(2,1.5)
\psline[linecolor=blue,linewidth=1.5pt,linestyle=dashed,dash=2pt 2pt](1,0.4)(2,0.4)
\psarc[linecolor=blue,linewidth=1.5pt,linestyle=dashed,dash=2pt 2pt](0.5,2){0.5}{95}{270}
\rput{90}(2,1){\triangleCNd}
\rput{90}(2,0){\triangleCNg}
\rput{45}(1.44,0.08)\jauge
\end{pspicture}\ \ .
\end{alignat}
\end{subequations}
\begin{subequations}
The weights are
\begin{alignat}{3}
p_{1} &= \frac{s(\lambda) s(2\lambda) s(3\lambda)}{2s(5\lambda)}\ , \qquad
&p_{2} &= s(\lambda) s^2(2\lambda) s(3\lambda)\ , \\[0.1cm]
p_{3} &= s(\lambda) s(2\lambda) \delta\!\left(\tfrac{3\lambda}{2}\right)\ ,
&p_{4} &= s(\lambda) s(2\lambda) \delta\!\left(\tfrac{3\lambda}{2}-\tfrac{\pi}{2}\right)\ .
\end{alignat}
\end{subequations}
As an example, we give the steps leading to the first identity in \eqref{eq:firstTangleInBST}, 
marking each equal sign by the identity used:
\begin{alignat}{2}
\begin{pspicture}[shift=-1.4](2,3)
\psline[linecolor=blue,linewidth=1.5pt,linestyle=dashed,dash=2pt 2pt](0.5,2.5)(2,2.5)
\psline[linecolor=blue,linewidth=1.5pt,linestyle=dashed,dash=2pt 2pt](0.5,1.5)(2,1.5)
\psline[linecolor=blue,linewidth=1.5pt,linestyle=dashed,dash=2pt 2pt](0.5,0.5)(2,0.5)
\rput(0,1){\triangled \rput(0.4,1){$\frac{\lambda}{2}$}}
\rput(0,0){\losange \rput(1,1){$0$}}
\end{pspicture}
% deuxieme diagramme
\ &\overset{\eqref{rel:evalRemarquable0et3}} 
=\ s(2\lambda) s(3\lambda)\
\begin{pspicture}[shift=-1.4](2,3)
\psline[linecolor=blue,linewidth=1.5pt,linestyle=dashed,dash=2pt 2pt](0.5,2.5)(2,2.5)
\psline[linecolor=blue,linewidth=1.5pt,linestyle=dashed,dash=2pt 2pt](1.58,1.5)(2,1.5)
\psline[linecolor=blue,linewidth=1.5pt,linestyle=dashed,dash=2pt 2pt](1.58,0.5)(2,0.5)
\rput(0,1){\triangled \rput(0.35,1){$\frac{\lambda}{2}$}}
\rput(0,0){\losangeNoArc}
\psline(0,1)(2,1)
\psarc[linecolor=blue,linewidth=1.5pt,linestyle=dashed,dash=2pt 2pt](1,2){0.71}{225}{315}
\psarc[linecolor=blue,linewidth=1.5pt,linestyle=dashed,dash=2pt 2pt](1,0){0.71}{45}{135}
\end{pspicture}\nonumber\\ 
% troisieme diagramme
&\ \overset{\eqref{rel:symCroiseFrontiere}}
=\frac{s(2\lambda) s(3\lambda) \bar{\delta}(2\lambda)}{\bar{\delta}(0)\rho_8(-2\lambda)}\ 
\begin{pspicture}[shift=-1.4](3,3)
\psline[linecolor=blue,linewidth=1.5pt,linestyle=dashed,dash=2pt 2pt](0.5,2.5)(3,2.5)
\psline[linecolor=blue,linewidth=1.5pt,linestyle=dashed,dash=2pt 2pt](1.58,1.5)(3,1.5)
\psline[linecolor=blue,linewidth=1.5pt,linestyle=dashed,dash=2pt 2pt](1.58,0.5)(3,0.5)
\rput(0,1){\triangled \rput(0.35,1){$\frac{5\lambda}{2}$}}
\rput(1,1){\losange \rput(1,1){$2\lambda$}}
\rput(0,0){\losangeNoArc}
\psarc[linecolor=blue,linewidth=1.5pt,linestyle=dashed,dash=2pt 2pt](1,2){0.71}{225}{315}
\psarc[linecolor=blue,linewidth=1.5pt,linestyle=dashed,dash=2pt 2pt](1,0){0.71}{45}{135}
\end{pspicture}
% quatrieme diagramme
\ \overset{\eqref{rel:evalRemarquable1et2}}=p_{1} \ 
\begin{pspicture}[shift=-1.4](3,3)
\psline[linecolor=blue,linewidth=1.5pt,linestyle=dashed,dash=2pt 2pt](0.5,2.5)(3,2.5)
\psline[linecolor=blue,linewidth=1.5pt,linestyle=dashed,dash=2pt 2pt](2,1.5)(3,1.5)
\psline[linecolor=blue,linewidth=1.5pt,linestyle=dashed,dash=2pt 2pt](1.58,0.5)(3,0.5)
\rput(0,1){\triangled \rput(0.35,1){$\frac{5\lambda}{2}$}}
\rput(0,0){\losangeNoArc}
\psarc[linecolor=blue,linewidth=1.5pt,linestyle=dashed,dash=2pt 2pt](1,2){0.71}{225}{315}
\psarc[linecolor=blue,linewidth=1.5pt,linestyle=dashed,dash=2pt 2pt](1,0){0.71}{45}{135}
\rput(1,1)\triangleCNg
\rput(2,1)\triangleCNd
\end{pspicture} \ .
\end{alignat}
\endgroup

The next remarkable evaluations apply to the objects in \eqref{def:produitFrontiere_b}, and the resulting diagrams involve an identity arc. There are only two such values of $u$ for each type of boundary operator. These relations are
\begin{alignat}{2}
&\text{SC} \quad u=0:
\hspace{2cm}&&\text{SC} \quad u=\frac{\pi}{2}: \nonumber \\\label{eq:bdy.identities.type2a}
&\begin{pspicture}[shift=-1.4](2,3)
\psline[linecolor=blue,linewidth=1.5pt,linestyle=dashed,dash=2pt 2pt](0.5,2.5)(2,2.5)
\psline[linecolor=blue,linewidth=1.5pt,linestyle=dashed,dash=2pt 2pt](0.5,1.5)(2,1.5)
\psline[linecolor=blue,linewidth=1.5pt,linestyle=dashed,dash=2pt 2pt](0.5,0.5)(2,0.5)
\rput(0,1){\triangled \rput(0.4,1){$0$}}
\rput(0,0){\losange \rput(1,1){$0$}}
\end{pspicture}
\ = q_{1} \
\begin{pspicture}[shift=-1.4](2.25,3)
\rput(0,1){\triangled}
\rput(0,0){\losangeNoArc}
\psline[linecolor=blue,linewidth=1.5pt,linestyle=dashed,dash=2pt 2pt](0.58,2.5)(2.25,2.5)
\psarc[linecolor=blue,linewidth=1.5pt,linestyle=dashed,dash=2pt 2pt](1,2){0.71}{135}{315}
\psline[linecolor=blue,linewidth=1.5pt,linestyle=dashed,dash=2pt 2pt](1.58,1.5)(2.25,1.5)
\psarc[linecolor=blue,linewidth=1.5pt,linestyle=dashed,dash=2pt 2pt](1,0){0.71}{45}{135}
\psline[linecolor=blue,linewidth=1.5pt,linestyle=dashed,dash=2pt 2pt](1.58,0.5)(2.25,0.5)
\end{pspicture}
\hspace{2cm}&&\begin{pspicture}[shift=-1.4](2,3)
\psline[linecolor=blue,linewidth=1.5pt,linestyle=dashed,dash=2pt 2pt](0.5,2.5)(2,2.5)
\psline[linecolor=blue,linewidth=1.5pt,linestyle=dashed,dash=2pt 2pt](0.5,1.5)(2,1.5)
\psline[linecolor=blue,linewidth=1.5pt,linestyle=dashed,dash=2pt 2pt](0.5,0.5)(2,0.5)
\rput(0,1){\triangled \rput(0.4,1){$-\frac{\pi}{2}$}}
\rput(0,0){\losange \rput(1,1){$\pi$}}
\end{pspicture}
\ = q_{2} \
\begin{pspicture}[shift=-1.4](2.25,3)
\rput(0,1){\triangled}
\rput(0,0){\losangeNoArc}
\psline[linecolor=blue,linewidth=1.5pt,linestyle=dashed,dash=2pt 2pt](0.58,2.5)(2.25,2.5)
\psarc[linecolor=blue,linewidth=1.5pt,linestyle=dashed,dash=2pt 2pt](1,2){0.71}{135}{315}
\psline[linecolor=blue,linewidth=1.5pt,linestyle=dashed,dash=2pt 2pt](1.58,1.5)(2.25,1.5)
\psarc[linecolor=blue,linewidth=1.5pt,linestyle=dashed,dash=2pt 2pt](1,0){0.71}{45}{135}
\psline[linecolor=blue,linewidth=1.5pt,linestyle=dashed,dash=2pt 2pt](1.58,0.4)(2.25,0.4)
\rput{45}(1.44,0.08)\jauge
\end{pspicture}
\end{alignat}
A third evaluation point is worth mentioning even if no identity arc is factored. It cancels the combination \eqref{def:produitFrontiere_b} because a loop segment connects to a vacancy:
\begin{alignat}{2}
&\text{S} \quad u=\frac{3\lambda}{2}:
\hspace{2cm}&&\text{C} \quad u=\frac{3\lambda}{2} + \frac{\pi}{2}: \nonumber \\
&\psset{unit=1.1}\label{eq:bdy.identities.type2b}
\begin{pspicture}[shift=-1.9](2,4)
\psline[linecolor=blue,linewidth=1.5pt,linestyle=dashed,dash=2pt 2pt](0.5,3.5)(2,3.5)
\psline[linecolor=blue,linewidth=1.5pt,linestyle=dashed,dash=2pt 2pt](0.5,2.5)(2,2.5)
\psline[linecolor=blue,linewidth=1.5pt,linestyle=dashed,dash=2pt 2pt](0.5,1.5)(2,1.5)
\psline[linecolor=blue,linewidth=1.5pt,linestyle=dashed,dash=2pt 2pt](0.5,0.5)(2,0.5)
\rput(0,2){\triangled \rput(0.4,1){$-\frac{3\lambda}{2}$}}
\rput(0,1){\losange \rput(1,1){$3\lambda$}}
\rput(0,0){\triangled \rput(0.4,1){$\frac{3\lambda}{2}$}}
\end{pspicture}
\ = q_{3} \
\begin{pspicture}[shift=-1.9](2,4)
\rput(0,2){\triangled
\psarc[linecolor=blue,linewidth=1.5pt](1,1){0.71}{137}{225}}
\rput(0,1){\losangeNoArc
\psarc[linecolor=blue,linewidth=1.5pt,linestyle=dashed,dash=2pt 2pt](0,1){0.71}{315}{45}
\psarc[linecolor=blue,linewidth=1.5pt,linestyle=dashed,dash=2pt 2pt](2,1){0.71}{135}{225}}
\rput(0,0){\triangled
\pscircle[fillstyle=solid,fillcolor=black](0.5,1.5){0.05}
\pscircle[fillstyle=solid,fillcolor=black](0.5,0.5){0.05}}
\psline[linecolor=blue,linewidth=1.5pt,linestyle=dashed,dash=2pt 2pt](0.5,3.5)(2,3.5)
\psline[linecolor=blue,linewidth=1.5pt,linestyle=dashed,dash=2pt 2pt](1.58,2.5)(2,2.5)
\psline[linecolor=blue,linewidth=1.5pt,linestyle=dashed,dash=2pt 2pt](1.58,1.5)(2,1.5)
\psline[linecolor=blue,linewidth=1.5pt,linestyle=dashed,dash=2pt 2pt](0.5,0.5)(2,0.5)
\end{pspicture}
\ = 0
\hspace{2cm}&&\psset{unit=1.1}\begin{pspicture}[shift=-1.9](2,4)
\psline[linecolor=blue,linewidth=1.5pt,linestyle=dashed,dash=2pt 2pt](0.5,3.5)(2,3.5)
\psline[linecolor=blue,linewidth=1.5pt,linestyle=dashed,dash=2pt 2pt](0.5,2.5)(2,2.5)
\psline[linecolor=blue,linewidth=1.5pt,linestyle=dashed,dash=2pt 2pt](0.5,1.5)(2,1.5)
\psline[linecolor=blue,linewidth=1.5pt,linestyle=dashed,dash=2pt 2pt](0.5,0.5)(2,0.5)
\rput(0,2){\triangled \rput(0.46,1){\tiny$-\!\frac{3\lambda}{2}\!-\!\frac{\pi}{2}$}}
\rput(0,1){\losange \rput(1,1){$3\lambda+\pi$}}
\rput(0,0){\triangled \rput(0.46,1){\tiny$\frac{3\lambda}{2}\!-\!\frac{\pi}{2}$}}
\end{pspicture}
\ = q_{3} \
\begin{pspicture}[shift=-1.9](2,4)
\rput(0,2){\triangled
\psarc[linecolor=blue,linewidth=1.5pt](1,1){0.71}{137}{225}}
\rput(0,1){\losangeNoArc
\psarc[linecolor=blue,linewidth=1.5pt,linestyle=dashed,dash=2pt 2pt](0,1){0.71}{315}{45}
\psarc[linecolor=blue,linewidth=1.5pt,linestyle=dashed,dash=2pt 2pt](2,1){0.71}{135}{225}}
\rput(0,0){\triangled
\pscircle[fillstyle=solid,fillcolor=black](0.5,1.5){0.05}
\pscircle[fillstyle=solid,fillcolor=black](0.5,0.5){0.05}}
\psline[linecolor=blue,linewidth=1.5pt,linestyle=dashed,dash=2pt 2pt](0.5,3.5)(2,3.5)
\psline[linecolor=blue,linewidth=1.5pt,linestyle=dashed,dash=2pt 2pt](1.58,2.5)(2,2.5)
\psline[linecolor=blue,linewidth=1.5pt,linestyle=dashed,dash=2pt 2pt](1.58,1.4)(2,1.4)
\psline[linecolor=blue,linewidth=1.5pt,linestyle=dashed,dash=2pt 2pt](0.5,0.5)(2,0.5)
\rput{45}(1.44,1.08)\jauge
\end{pspicture}
\ = 0 \ .
\end{alignat}
The weights in the previous identities are:
\begin{equation}
q_{1} = s(2\lambda) s(3\lambda) \delta\!\left(\tfrac{3\lambda}{2}\right) \ , \qquad
q_{2} = s(2\lambda) s(3\lambda) \delta\!\left(\tfrac{3\lambda}{2}-\tfrac{\pi}{2}\right) \ , \qquad
q_{3} = s(2\lambda) s^3(3\lambda) \ .
\end{equation}
We note that all these identities for the boundary operators of type C can be recovered from those of type S from the relation 
\begin{equation}
\label{eq:SC.periodicity}
\begin{pspicture}[shift=-0.9](1.25,2)
\psline[linecolor=blue,linewidth=1.5pt,linestyle=dashed,dash=2pt 2pt](0,1.5)(1.25,1.5)
\psline[linecolor=blue,linewidth=1.5pt,linestyle=dashed,dash=2pt 2pt](0,0.5)(1.25,0.5)
\triangled \rput(0.35,1){$u$}
\rput(0.75,2){\scriptsize(C)}
\end{pspicture}
\ = \ 
\begin{pspicture}[shift=-0.9](1.25,2)
\psline[linecolor=blue,linewidth=1.5pt,linestyle=dashed,dash=2pt 2pt](0,1.5)(1.25,1.5)
\psline[linecolor=blue,linewidth=1.5pt,linestyle=dashed,dash=2pt 2pt](0,0.5)(1.25,0.5)
\triangled \rput(0.46,1){\scriptsize $u\!-\!\frac \pi 2$}
\rput(0.75,2){\scriptsize(S)}
\rput{45}(0.44,0.08)\jauge
\end{pspicture}\ \ .
\end{equation} 

There are also push-through properties associated to the combinations \eqref{def:produitFrontiere}, which we refer to as \textit{boundary reflection properties}. Let us recall that the push-through properties \eqref{rel:PTbulkTCN} and \eqref{rel:PTbulkArc} both include a pair of face operators whose evaluation points differ by $2\lambda$ or $3\lambda$. The boundary reflection properties include similar shifts, but now in the evaluation points of the boundary operators. The first combination in \eqref{def:produitFrontiere_a} contains two boundary operators shifted by $2\lambda$ and satisfies a reflection property for the black triangle for both S and C cases:
\begin{equation} \label{rel:PTfrontiereTCN}
\begin{pspicture}[shift=-1.9](2.5,4)
\psline[linecolor=blue,linewidth=1.5pt,linestyle=dashed,dash=2pt 2pt](0.5,3.5)(2.25,3.5)
\psline[linecolor=blue,linewidth=1.5pt,linestyle=dashed,dash=2pt 2pt](0.5,2.5)(2.25,2.5)
\psline[linecolor=blue,linewidth=1.5pt,linestyle=dashed,dash=2pt 2pt](0.5,0.5)(2,0.5)
\psline[linecolor=blue,linewidth=1.5pt,linestyle=dashed,dash=2pt 2pt](2,1)(2.25,1)
\rput(0,2){\triangled \rput(0.46,1){$_{3\lambda-u_2}$}}
\rput(0,1){\losange \rput(1,1){$2u-\lambda$}}
\triangled \rput(0.46,1){$_{3\lambda-u_0}$}
\rput(1,0){\triangleCNg}
\end{pspicture}
= 2 s(2u_{-3}) \delta(u_{-5/2}) \bar{\delta}(\tfrac{\lambda}{2}-u) \delta(u_{1/2}) \
\begin{pspicture}[shift=-1.9](2.5,4)
\psline[linecolor=blue,linewidth=1.5pt,linestyle=dashed,dash=2pt 2pt](1.5,3.5)(2.25,3.5)
\psline[linecolor=blue,linewidth=1.5pt,linestyle=dashed,dash=2pt 2pt](1.5,2.5)(2.25,2.5)
\psline[linecolor=blue,linewidth=1.5pt,linestyle=dashed,dash=2pt 2pt](0.5,2.25)(1.25,3)
\psline[linecolor=blue,linewidth=1.5pt,linestyle=dashed,dash=2pt 2pt](0.2,1.25)(2.25,1.25)
\rput(1.25,2){\triangleCNd}
\rput(0,0.75){\triangled \rput(0.45,1){$_{3\lambda-u_1}$}}
\end{pspicture} \ .
\end{equation}
The second reflection property applies to the first object \eqref{def:produitFrontiere_b}, to which an identity arc factor is attached. Similarly to \eqref{rel:PTbulkArc}, this combination contains two boundary operators shifted by $3\lambda$ and holds for both S and C cases:
\begin{equation}\label{rel:PTfrontiereArc}
\begin{pspicture}[shift=-1.9](2.5,4)
\psline[linecolor=blue,linewidth=1.5pt,linestyle=dashed,dash=2pt 2pt](0.5,3.5)(2,3.5)
\psline[linecolor=blue,linewidth=1.5pt,linestyle=dashed,dash=2pt 2pt](0.5,2.5)(2,2.5)
\psline[linecolor=blue,linewidth=1.5pt,linestyle=dashed,dash=2pt 2pt](0.5,0.46)(1.46,0.46)
\rput(0,2){\triangled \rput(0.46,1){$_{3\lambda-u_3}$}}
\rput(0,1){\losange \rput(1,1){$2u$}}
\rput(0,0){\triangled \rput(0.46,1){$_{3\lambda-u_0}$}}
\rput(1,0){\triangleArcg}
\end{pspicture}
= -2 s(2u_{-3}) \delta(u_{-5/2}) \delta(u_{-3/2}) \bar{\delta}(u_{-1/2}) \delta(u_{3/2})
\begin{pspicture}[shift=-1.9](2,4)
\rput(0.25,2){\triangleArcd}
\psline[linecolor=blue,linewidth=1.5pt,linestyle=dashed,dash=2pt 2pt](1.05,3.5)(1.5,3.5)
\psline[linecolor=blue,linewidth=1.5pt,linestyle=dashed,dash=2pt 2pt](1.05,2.5)(1.5,2.5)
\end{pspicture} \ .
\end{equation}

A straightforward method to prove \eqref{rel:PTfrontiereTCN} and \eqref{rel:PTfrontiereArc} is to expand each operator in terms of its elementary tiles, and then compute the resulting connectivities using the rules stated in \cref{sec:dTL_N} for the concatenation of diagrams. The coefficients of identical connectivities are added together and simplified. Then the proof consists of checking that identical connectivities on the two sides of the equation have equal coefficients. This method is direct but cumbersome and cannot be easily applied to calculations implying a large number of tangles.

A second method to prove the previous properties will also play an important role in the following sections. A quick presentation is in order. The weights appearing in the face and boundary operators are trigonometric functions, and the spectral parameter $u$ always appears linearly in their arguments. Seen as functions of the new variable $z=\eE^{\iI u}$, these operators are thus Laurent polynomials of the form $\sum_{-m\leq i\leq m}C_iz^i$ for some linear combination of tangles $C_i$ and some integer $m$. Since the weights in the boundary operator contain a single trigonometric function, the boundary triangles in the tangles on the left-hand side of \eqref{rel:PTfrontiereTCN} and \eqref{rel:PTfrontiereArc} account for polynomials in $z$ of maximal degree $m=1$. The weights $\rho_i$ have either one or two trigonometric functions whose arguments are linear in $u$. The face operator in the two tangles is evaluated at $2u-\lambda$ or $2u$ and its coefficients will be polynomials of maximal degree up to~$m=4$. The other faces in the tangles, namely the black triangles and the triangles containing the identity strand, do not depend on $u$ (or $z$). The tangles in the boundary reflection properties are thus polynomials of the form $\sum_{-m\leq i\leq m}C_iz^i$ with $m=6$. To determine the coefficients $C_i$, it is sufficient to check the equality at $2m+1$ distinct values of $z$.

As stated earlier, the combinations \eqref{def:produitFrontiere} are invariant under $u\mapsto u+\pi$. Because the black triangle and the identity arc do not depend on $u$, the left-hand side of \eqref{rel:PTfrontiereTCN} and \eqref{rel:PTfrontiereArc} are also invariant under this transformation. However, $u$ and $u+\pi$ respectively lead to $z_1=\eE^{\iI u}$ and $z_2=\eE^{\iI(u+\pi)}=-z_1$, which are not equal. Each evaluation in $u$ with $\textrm{Re}(u) \in [0,\pi)$ thus gives us two values of $z$, and it is sufficient to check $m+1$ values of $u$ in this domain.  

Combinations of tangles are easier to compute if the value of the spectral parameter $u$ is such that (i) one of the sides of the identity vanishes trivially or (ii) diagrammatic properties can be used. As an example, one can consider the following evaluation points for \eqref{rel:PTfrontiereTCN} for the boundary of type S:
\be 
\text{(i):}\quad -\frac{\lambda}{2}, \, \frac{\lambda}{2}+\frac{\pi}{2}, \, \frac{5\lambda}{2},
\, 3\lambda, \, 3\lambda+\frac{\pi}{2}, 
\qquad \qquad
\text{(ii):}\quad \frac{\lambda}{2}, \, \frac{3\lambda}{2}, \,
\lambda, \, \lambda+\frac{\pi}{2} \ .
\ee
The right hand side of \eqref{rel:PTfrontiereTCN} vanishes when evaluated at the points (i), and the identities \eqref{eq:firstTangleInBST}, \eqref{eq:secondTangleInBST} and \eqref{eq:thirdTangleInBST} can be used for the evaluations at points (ii). We do not present the remaining explicit computations. All those points are distinct for $\lambda$ generic and thus give more than the seven verifications needed. The push-through property \eqref{rel:PTfrontiereTCN} then extends to all $\lambda$ by continuity.

This technique for proving tangle identities is based on a {\em polynomial argument} and was introduced in \cite{AMDPP19}, replacing the usual diagrammatic method that consists in expanding all the diagrams and verifying the identity for all $u$. 

%%%%%%%%%%%%%%%%%%%%%%%%%%%%%%%%%%%%%%%%%%%%%%%
%
\section{Commuting transfer matrices}\label{sec:transferMatrices}
%
%%%%%%%%%%%%%%%%%%%%%%%%%%%%%%%%%%%%%%%%%%%%%%%

The previous sections prepared the ground for introducing the family of commuting transfer matrices of the $\Atwotwo$ loop models on the geometry of the strip. In \cref{sub:fundamental}, we introduce the fundamental transfer matrix of the model $\Db(u) = \Db^{1,0}(u)$. In \cref{sub:fused.with.relations}, we give a recursive definition of the fused transfer matrices $\Db^{m,n}(u)$, based on the fusion hierarchy relations and valid for all $m,n$. Then \cref{sub:fused.with.diagrams} gives a second equivalent definition for $\Db^{2,0}(u)$ and $\Db^{1,1}(u)$ that is diagrammatic. In \cref{sub:det}, we discuss a formulation of $\Db^{m,n}(u)$ in terms of the determinant of a matrix of size $m+n$ whose entries are proportional to $\Ib$ or $\Db(u+k\lambda)$, with $k \in \mathbb Z$. In \cref{sub:reductions}, we show that the fused transfer matrices $\Db^{m,0}(u)$ with the spectral parameter $u$ specialized to certain special values satisfy reduction relations. In \cref{sub:braid}, we discuss the braid limits $u\to \pm \iI \infty$ of the face and boundary operators and of the fused transfer matrices. Finally, \cref{sub:TY.systems} derives the $T$-systems and $Y$-systems of equations for this integrable hierarchy of commuting transfer matrices. The technical proofs of some properties of the fused transfer matrices are relegated to \cref{app:proofs}. In the proofs here and in the appendix, we assume when needed that the parameter $\lambda$ is generic, that is, $\eE^{\iI\lambda}$ is not a root of unity. The results then extend to roots of unity as the identities under study are between Laurent polynomials in both the variables $\eE^{\iI u}$ and $\eE^{\iI\lambda}$.

%%%%%%%%%%%%%
\subsection{Fundamental transfer tangle} \label{sub:fundamental}
%%%%%%%%%%%%%

In this section, we define the fundamental transfer matrix $\Db(u)$ and discuss some of its important properties. As a first step towards this goal, let us define the following tangle built from $2N$ face operators and two boundary operators:
\begin{equation} \label{def:D10tilde}
\psset{unit=1.1}
\Dbt(u) = \
\begin{pspicture}[shift=-0.9](-1,0)(5,2)
\facegrid{(0,0)}{(4,2)}
\rput(2.5,1.5){$\ldots$}
\rput(2.5,0.5){$\ldots$}
\rput(0,1){\rput(0.5,0.5){$_{u+\xi_{(1)}}$}\psarc[linewidth=0.025]{-}(1,0){0.16}{90}{180}}
\rput(1,1){\rput(0.5,0.5){$_{u+\xi_{(2)}}$}\psarc[linewidth=0.025]{-}(1,0){0.16}{90}{180}}
\rput(3,1){\rput(0.5,0.5){$_{u+\xi_{(N)}}$}\psarc[linewidth=0.025]{-}(1,0){0.16}{90}{180}}
\rput(0,0){\rput(0.5,0.5){$_{u-\xi_{(1)}}$}\psarc[linewidth=0.025]{-}(0,0){0.16}{0}{90}}
\rput(1,0){\rput(0.5,0.5){$_{u-\xi_{(2)}}$}\psarc[linewidth=0.025]{-}(0,0){0.16}{0}{90}}
\rput(3,0){\rput(0.5,0.5){$_{u-\xi_{(N)}}$}\psarc[linewidth=0.025]{-}(0,0){0.16}{0}{90}}
\psline[linecolor=blue,linewidth=1.5pt,linestyle=dashed,dash=2pt 2pt](4,1.5)(4.5,1.5)
\psline[linecolor=blue,linewidth=1.5pt,linestyle=dashed,dash=2pt 2pt](4,0.5)(4.5,0.5)
\rput(4,0){\triangleg \rput(0.65,1){$u$}}
\psline[linecolor=blue,linewidth=1.5pt,linestyle=dashed,dash=2pt 2pt](0,1.5)(-1,1.5)
\psline[linecolor=blue,linewidth=1.5pt,linestyle=dashed,dash=2pt 2pt](0,0.5)(-1,0.5)
\rput(-1,0){\triangled \rput(0.45,1){\scriptsize${3\lambda-u}$}}
\end{pspicture}
\ .
\end{equation}
The tangle $\Dbt(u)$ is an element of $\dtl_N(\beta)$: there are $N$ nodes on its upper edge and $N$ on its lower edge. The tangle depends on the crossing parameter $\lambda$, the spectral parameter $u$ and the number of sites $N$. These parameters $\lambda$ and $N$ will be fixed throughout, so we only write the dependence on the spectral parameter $u$ explicitly. Each column is assigned an arbitrary inhomogeneity $\xi_{(j)}$. We note that the subscript $j$ in parenthesis is a column label and not a shift in the evaluation of the spectrum parameter. Note that, using crossing symmetry, the arcs in the upper tiles can be moved to the left lower corners at the expense of changing the evaluation points from $u+\xi_{(j)}$ to $3\lambda-u-\xi_{(j)}$. The proofs of certain results below will assume that $\xi_{(j_1)}\neq\xi_{(j_2)}$, however by continuity these results will also hold when two inhomogeneities (or more) coincide. The homogeneous transfer tangle is obtained from $\Dbt(u)$ by setting $\xi_{(j)}\rightarrow 0$ for all $j$.

The boundary operators are chosen in \eqref{def:frontieres} to be either of S or C type, and this choice can be different for the two boundaries. This leads to four possible boundary conditions for the transfer tangles, that we gather in two groups: (i) \textit{identical}, for SS and CC, and (ii) \textit{mixed}, for SC and CS. For instance, SC refers to the choice S for the left boundary and to C for the right boundary.

The diagrammatic properties introduced in \cref{sec:faceAndBoundary} result in important properties satisfied by $\Dbt(u)$. The first one is a crossing symmetry which ties the evaluations of the spectral parameter at $u$ and $3\lambda-u$. The proof follows closely the similar proof given in \cite{Behrend} for Interaction-Round-a-Face models and is a direct consequence of \eqref{rel:symCroiseTuile}, \eqref{rel:inversionTuile} and \eqref{rel:symCroiseFrontiere}. The crossing symmetry takes different forms depending on the choice of boundaries:
\begin{subequations} \label{eq:symCroiseMixte}
\begin{alignat}{2}
&\text{SS and CC} \qquad &\Dbt(u)&=\Dbt(3\lambda-u) \ ,
\\
&\hspace{6mm} \text{SC} \label{eq:symCroiseSC}
&s\!\left(u-\tfrac{\lambda}{2}\right) c\!\left(\tfrac{5\lambda}{2}-u\right) \Dbt(u)
&= c\!\left(u-\tfrac{\lambda}{2}\right)s\!\left(\tfrac{5\lambda}{2}-u\right)\Dbt(3\lambda-u) \ ,
\\
&\hspace{6mm} \text{CS} \label{eq:symCroiseCS}
&c\!\left(u-\tfrac{\lambda}{2}\right) s\!\left(\tfrac{5\lambda}{2}-u\right) \Dbt(u)
&= s\!\left(u-\tfrac{\lambda}{2}\right) c\!\left(\tfrac{5\lambda}{2}-u\right)\Dbt(3\lambda-u) \ .
\end{alignat}
\end{subequations}
Note that the pairs of trigonometric factors on the left and right sides of \eqref{eq:symCroiseSC} and \eqref{eq:symCroiseCS} are themselves tied by the crossing symmetry $u \mapsto 3\lambda-u$.

Because $\Dbt(u)$ is a Laurent polynomial in $z = \eE^{\iI u}$, it has no poles for $z \in \mathbb C^\times$. We then deduce from \eqref{eq:symCroiseSC} and \eqref{eq:symCroiseCS} that $\Dbt(u)$ has overall trigonometric factors in the mixed cases. It is convenient to define a renormalized tangle where these factors are removed. The reduced matrices $\Db(u)$ are defined in the four cases as\footnote{The minus sign for SS and CC is included for convenience, so that the fusion hierarchy relations of \cref{sub:fused.with.relations} are written in a uniform way.}
\begin{subequations}\label{def:D10}
\begin{alignat}{4}
&\text{SS:} \qquad &&\Db(u) = -\Dbt(u) \ ,
&&\text{CC:} \qquad &&\Db(u) = -\Dbt(u) \ , \\
&\text{SC:} &&\Db(u) = \frac{\Dbt(u)}
{c\!\left(u-\tfrac{\lambda}{2}\right) s\!\left(\tfrac{5\lambda}{2}-u\right)} \ , \qquad
&&\text{CS:} &&\Db(u) = \frac{\Dbt(u)}
{s\!\left(u-\tfrac{\lambda}{2}\right) c\!\left(\tfrac{5\lambda}{2}-u\right)} \ .
\end{alignat}
\end{subequations}
We shall refer to $\Db(u)$ as the {\em fundamental transfer tangle} (or transfer matrix). Its symmetry properties can be expressed uniformly for all choices of boundary conditions:
\begin{subequations}
\begin{alignat}{3}
\label{eq:symCroiseD10}
\text{crossing symmetry} &\qquad\qquad &&\Db(u) = \Db(3\lambda-u) \ , \\
\label{eq:commutD10}
\text{commutativity} &\qquad\qquad &&[\Db(u),\Db(v)] = 0 \ , \\
\label{eq:piD10}
\text{periodicity} &\qquad\qquad &&\Db(u) = \Db(u+\pi)\ .
\end{alignat}
\end{subequations}
The relation \eqref{eq:commutD10} is the commutation of two transfer tangles with different spectral parameters. It embodies the integrability of the model and can be proved diagrammatically with \eqref{rel:inversionTuile} and the Yang-Baxter equations \eqref{rel:YBTuile} and \eqref{rel:YBFrontiere}. The arguments follow those in section 3.4 of \cite{Behrend}. The periodicity property \eqref{eq:piD10} is a consequence of the gauge transformation \eqref{rel:jaugePi}.

Another feature of the fundamental transfer matrix is that it evaluates to a multiple of the identity at certain special values of $u$:
\begin{subequations}
\label{eq:DI}
\begin{alignat}{2}
\Db(0) &=  f(2\lambda) f(3\lambda) \Ib
\times \frac{s(6\lambda)}{s(2\lambda)} \times \left\{\begin{array}{cc}
\displaystyle -\frac{s(3\lambda/2)c(\lambda/2)s(5\lambda/2)}{c(3\lambda/2)} & \textrm{SS},
\\[0.4cm]
\displaystyle\frac{c(3\lambda/2)s(\lambda/2)c(5\lambda/2)}{s(3\lambda/2)} & \textrm{CC},
\\[0.4cm]
1 & \textrm{SC and CS},
\end{array}\right.
\\
\Db(\tfrac{\pi}2) &= f(2\lambda+\tfrac{\pi}2)f(3\lambda+\tfrac{\pi}2) \Ib
\times \frac{s(6\lambda)}{s(2\lambda)} \times \left\{\begin{array}{cc}
\displaystyle\frac{c(3\lambda/2)s(\lambda/2)c(5\lambda/2)}{s(3\lambda/2)} & \textrm{SS},
\\[0.4cm]
-\displaystyle\frac{s(3\lambda/2)c(\lambda/2)s(5\lambda/2)}{c(3\lambda/2)} & \textrm{CC},
\\[0.4cm]
1 & \textrm{SC and CS},
\end{array}\right.
\end{alignat}
\end{subequations}
where
\begin{equation} \label{def:f_k} 
f(u) = \prod_{j=1}^N s\!\left(u-\xi_{(j)}\right)s\!\left(u+\xi_{(j)}\right) \ .
\end{equation}
To obtain these results, one applies the push-through property \eqref{rel:PTbulkArc} subsequently $N$ times to \eqref{def:D10tilde} specified at $u = 0, \frac \pi 2$, and then evaluates the constant diagrams
\begin{equation}
\begin{pspicture}[shift=-0.9](2,2)
\psline[linecolor=blue,linewidth=1.5pt,linestyle=dashed,dash=2pt 2pt](0,1.5)(2,1.5)
\psline[linecolor=blue,linewidth=1.5pt,linestyle=dashed,dash=2pt 2pt](0,0.5)(2,0.5)
\triangled \rput(0.35,1){$3\lambda$}
\rput(1,0){\triangleg \rput(0.65,1){$0$}}
\end{pspicture}
\ \qquad \textrm{and} \qquad \
\begin{pspicture}[shift=-0.9](2,2)
\psline[linecolor=blue,linewidth=1.5pt,linestyle=dashed,dash=2pt 2pt](0,1.5)(2,1.5)
\psline[linecolor=blue,linewidth=1.5pt,linestyle=dashed,dash=2pt 2pt](0,0.5)(2,0.5)
\triangled \rput(0.45,1){\scriptsize$3\lambda\!-\!\frac \pi 2$}
\rput(1,0){\triangleg \rput(0.65,1){$\frac \pi 2$}}
\end{pspicture}\ \ .
\end{equation}
Of course, because of the crossing symmetry and periodicity, $\Db(u)$ also evaluates to a multiple of the identity at $u = 3 \lambda$, $\pi$, $3 \lambda + \frac \pi 2$, etc.

It directly follows from the definition \eqref{def:D10tilde} that $\Dbt(u)$ is a Laurent polynomial in $z=\eE^{\iI u}$:
\begin{equation} \label{eq:D10tildePoly}
\Dbt(u) = \sum_{j=-(4N+2)}^{4N+2} C_j z^j \ ,
\end{equation} 
with coefficients $C_j$ that are elements of $\dtl_N(\beta)$ that depend on $\lambda$. As discussed in \cref{sub:propertiesBoundary}, the degree is fixed by the number of tiles used in the diagrammatic definition. A centered Laurent polynomial such as the one in \eqref{eq:D10tildePoly} is said to have \textit{maximal power} $\maxP=4N+2$. A consequence of the definitions \eqref{def:D10} is that the maximal power of $\Db(u)$ depends on the type of boundary:
\begin{equation}
\text{identical:} \qquad\maxP(\Db(u)) = 4N+2, \qquad\qquad
\text{mixed:} \qquad\maxP(\Db(u)) = 4N.
\end{equation}
We also remark that, because $\text{maxP\,}(s(u_k-\xi_{(j)}))=1$ for all $j$, the maximal power of the function $f(u)$ is $2N$. Moreover, all zeros of $f(u)$ are distinct for generic inhomogeneities, an observation that will be useful in the following.

%%%%%%%%%%%%%
\subsection{Fusion hierarchy of fused transfer tangles $\Db^{m,n}(u)$} \label{sub:fused.with.relations}
%%%%%%%%%%%%%

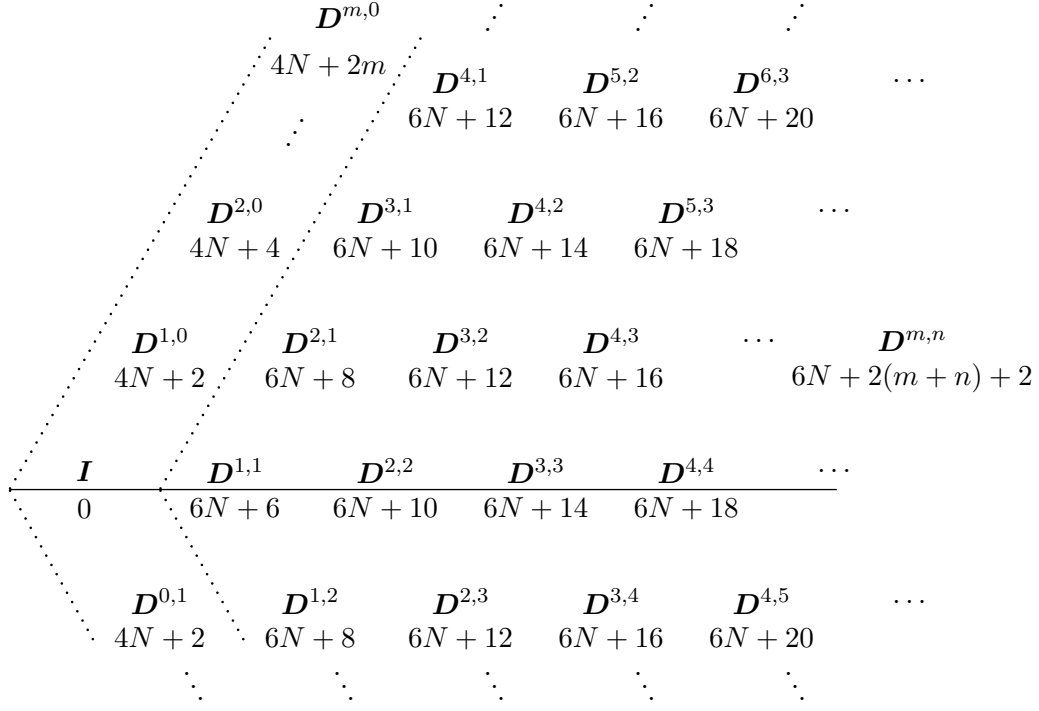
\begin{figure}[t]
\begin{center}
\begin{pspicture}(-1,-3)(12,6)
% ligne 4
\rput(5.45,6.028){$\rput(-0.09,-0.162){.}\rput(0,0){.}\rput(0.09,0.162){.}$}
\rput(7.45,6.028){$\rput(-0.09,-0.162){.}\rput(0,0){.}\rput(0.09,0.162){.}$}
\rput(9.45,6.028){$\rput(-0.09,-0.162){.}\rput(0,0){.}\rput(0.09,0.162){.}$}
% ligne 3.5
\rput(3.5,6.062){$\Db^{m,0}$ \rput(-0.7,-0.5){$4N+2m$}}
% ligne 3
\rput(5,5.196){$\Db^{4,1}$}\rput(5,4.696){$6N+12$}
\rput(7,5.196){$\Db^{5,2}$}\rput(7,4.696){$6N+16$}
\rput(9,5.196){$\Db^{6,3}$}\rput(9,4.696){$6N+20$}
\rput(11,5.196){$\dots$}
% ligne 2.60
\rput(2.80,4.5032){$\rput(-0.09,-0.162){.}\rput(0,0){.}\rput(0.09,0.162){.}$}
% ligne 2
\rput(2,3.464){$\Db^{2,0}$}\rput(2,2.964){$4N+4$}
\rput(4,3.464){$\Db^{3,1}$}\rput(4,2.964){$6N+10$}
\rput(6,3.464){$\Db^{4,2}$}\rput(6,2.964){$6N+14$}
\rput(8,3.464){$\Db^{5,3}$}\rput(8,2.964){$6N+18$}
\rput(10,3.464){$\dots$}
% ligne 1
\rput(1,1.732){$\Db^{1,0}$}\rput(1,1.232){$4N+2$}
\rput(3,1.732){$\Db^{2,1}$}\rput(3,1.232){$6N+8$}
\rput(5,1.732){$\Db^{3,2}$}\rput(5,1.232){$6N+12$}
\rput(7,1.732){$\Db^{4,3}$}\rput(7,1.232){$6N+16$}
\rput(9,1.732){$\dots$}
\rput(11,1.732){$\Db^{m,n}$}\rput(11,1.232){$6N+2(m+n)+2$}
% ligne 0
\rput(0,0){$\Ib$}\rput(0,-0.5){0}
\rput(2,0){$\Db^{1,1}$}\rput(2,-0.5){$6N+6$}
\rput(4,0){$\Db^{2,2}$}\rput(4,-0.5){$6N+10$}
\rput(6,0){$\Db^{3,3}$}\rput(6,-0.5){$6N+14$}
\rput(8,0){$\Db^{4,4}$}\rput(8,-0.5){$6N+18$}
\rput(10,0){$\dots$}
% ligne -1
\rput(1,-1.732){$\Db^{0,1}$}\rput(1,-2.232){$4N+2$}
\rput(3,-1.732){$\Db^{1,2}$}\rput(3,-2.232){$6N+8$}
\rput(5,-1.732){$\Db^{2,3}$}\rput(5,-2.232){$6N+12$}
\rput(7,-1.732){$\Db^{3,4}$}\rput(7,-2.232){$6N+16$}
\rput(9,-1.732){$\Db^{4,5}$}\rput(9,-2.232){$6N+20$}
\rput(11,-1.732){$\dots$}
% ligne -2
\rput(1.45,-2.864){$\rput(-0.09,0.162){.}\rput(0,0){.}\rput(0.09,-0.162){.}$}
\rput(3.45,-2.864){$\rput(-0.09,0.162){.}\rput(0,0){.}\rput(0.09,-0.162){.}$}
\rput(5.45,-2.864){$\rput(-0.09,0.162){.}\rput(0,0){.}\rput(0.09,-0.162){.}$}
\rput(7.45,-2.864){$\rput(-0.09,0.162){.}\rput(0,0){.}\rput(0.09,-0.162){.}$}
\rput(9.45,-2.864){$\rput(-0.09,0.162){.}\rput(0,0){.}\rput(0.09,-0.162){.}$}
%\rput(4,-2.864){$\ddots$}
%\rput(6,-2.864){$\ddots$}
%\rput(8,-2.864){$\ddots$}
% the boundary
\psline[linewidth=1.pt,linestyle=dotted](1.,-0.25)(4.464,5.75)
\psline[linewidth=1.pt,linestyle=dotted](-1.,-0.25)(2.464,5.75)
\psline[linewidth=1.pt,linestyle=dotted](1.,-0.25)(2.115,-2.25)
\psline[linewidth=1.pt,linestyle=dotted](-1.,-0.25)(0.115,-2.25)
\psline[linewidth=.5pt](-1.,-0.25)(10,-0.25)
\end{pspicture}
\end{center}
\caption{The fused transfer matrices and their maximal power ($\maxP$) for identical boundary conditions (SS and CC).
The elements near the boundary of the wedge, between the dotted lines, are the fused transfer tangles $\Db^{m,0}(u)$ and $\Db^{0,n}(u)$ and their $\maxP$ behaves differently to those of the elements $\Db^{m,n}(u)$, for $m,n>1$. For mixed boundary conditions (SC and CS), the maximal powers are $4N$ for the fused tangles $\Db^{m,0}(u)$ or $\Db^{0,n}(u)$ and $6N$ for $\Db^{m,n}(u)$ with $m,n\geq 1$. The index of the tangles $\Db^{m,n}(u)$ are exchanged upon reflection with respect to the horizontal line. 
\label{fig:maxP}}
\end{figure}

The \textit{fused transfer tangles} $\Db^{m,n}(u)$ for the $\Atwotwo$ loop models form a commuting family of $\dtl_N(\beta)$ elements. The fused tangles are labelled by pairs of integers $(m,n)$ and depicted on a lattice in \cref{fig:maxP}. The fundamental transfer matrix considered as an element of this larger set is labelled as
\begin{equation}
\Db(u) = \Db^{1,0}(u) \ .
\end{equation}
The \textit{conjugate fundamental transfer tangle} is defined by
\begin{equation} \label{def:D01}
\Db^{0,1}(u) = \Db^{1,0}(u+\lambda) \ .
\end{equation}
Let us introduce the additional short-hand notations
\begin{equation} \label{def:Db_k}
\Db^{m,n}_k = \Db^{m,n}(u+k\lambda) \ , \qquad f_k = f(u_k)\ .
\end{equation}

Two techniques were previously used to introduce transfer tangles of loop models with higher fusion indices. For some models, a diagrammatic definition is possible in terms of the appropriate projectors. The recursive relations can then be directly derived by diagrammatic manipulations that use the recursive definition of these projectors. This was achieved for the $A^{(1)}_1$ and $A^{(1)}_2$ loop models, in \cite{AMDPPJR14} and \cite{AMDPPJR19} respectively. 

For the $\Atwotwo$ loop models, the appropriate projectors are not all known. An alternative technique was used in \cite{AMDPP19}: a set of fusion hierarchy relations are written down, these are used as a recursive definition of the fused transfer tangles, and then the desired properties are shown to hold. In this section, we employ the second technique to define the transfer matrices $\Db^{m,n}(u)$, with the desired properties given in \eqref{eq:symCroiseDmn} -- \eqref{eq:Dm0D0m}. 

The fused matrices depend on the choice of boundary conditions for $\Db^{1,0}(u)$ and therefore on the corresponding functions defined in \eqref{def:frontieres}. To express the upcoming equations \eqref{eq:HF} in a unified manner for all boundary conditions, we introduce four functions in  \cref{tab:lesPoids}, namely $w(u)$, $\bar w(u)$, $w^{(m)}(u_k)$ and $\bar w^{(n)}(u_k)$, that differ according to the boundary conditions chosen for $\Db^{1,0}(u)$. We stress in particular that the functions $\delta(u)$ and $w(u)$ are not the same. Indeed, the transfer matrix contains two boundary operators, which may or may not be assigned the same choice of $\delta(u)$. In contrast, the functions $w(u)$ and $\bar w(u)$ are associated to a pair of boundary conditions for $\Db(u)$.
\begin{figure}[h!]
\begin{center}
\begin{tabular}{c|cccc}
 & $w(u)$ & $\bar w(u)$ & $w^{(m)}(u)$ & $\bar w^{(n)}(u)$ 
 \\[0.10cm]
\hline
&&& & \\[-0.3cm]
SS & $s(u)$ & $c(u)$ & $s(u+m\frac{\pi}2)$ & $c(u+n\frac{\pi}2)$  \\
CC & $c(u)$ & $s(u)$ & $c(u+m\frac{\pi}2)$ & $s(u+n\frac{\pi}2)$  \\
SC & $1$ & $1$ & $1$ & $1$  \\
CS & $1$ & $1$ & $1$ & $1$ \\
&&& & \\[-0.3cm]
\end{tabular}
\captionof{table}{The weights for the four choices of boundary conditions.}\label{tab:lesPoids}
\end{center}
\end{figure}

The structure of the fusion hierarchy for the strip geometry is analogous to that of the periodic one (see the set of equations (5.11) in \cite{AMDPP19}): the shifts in the $f$ functions are identical and the recursion formula for the transfer tangle $\Db^{m,n}(u)$ involves the same tangles obtained at a previous step. However, a large number of boundary weights need to be included for the present case with a double-row transfer matrix with boundary operators. Here are the coefficients appearing in the recursive formulas:
\begin{subequations}
\begin{alignat}{2}
\Azero(u) &= w(u_{-3/2}) \wb(u_{1/2}) f(u_{-2}) \nonumber\\
\Aun(m,u) &= s(2u_{m-2}) s(2u_{2m-5})
w^{(m)}(u_{m-5/2}) \wb^{(-1)}(u_{2m-9/2}) f(u_{2m-5}) f(u_{2m-4}) \\
\Adeux(m,u) &= s(2u_{m-4}) s(2u_{2m-3})
w^{(m-1)}(u_{m-7/2}) \wb(u_{2m-7/2}) \nonumber\\
\Atrois(m,u) &= s(2u_{m-5}) s(2u_{2m-2})
\wb(u_{2m-9/2}) w(u_{2m-5/2}) f(u_{2m-2}) \nonumber\\[1ex]
\Bzero(m,u) &= f(u_{2m}) f(u_{2m+1}) \nonumber\\
\Bun(m,n,u) &= s(2u_{m+n-2}) s(2\slu_{2n-2}) f(u_{2m-2}) \nonumber\\
\Bdeux(m,n,u) &= s(2u_{m-2}) s(2\slu_{n-2})
w^{(m)}(u_{m-5/2}) w^{(n)}(\slu_{n-5/2}) \\
\Btrois(m,n,u) &= s(2u_{m-4}) s(2\slu_{n-4})
\wb^{(m)}(u_{m-7/2}) \wb^{(n)}(\slu_{n-7/2}) \nonumber\\ 
&\qquad\times 
\wb(u_{2m-3/2}) \wb(\slu_{2n-3/2}) w(u_{2m-1/2}) w(\slu_{2n-1/2}).\nonumber
\end{alignat}
\end{subequations}
Equation \eqref{eq:symCroiseD10} states the crossing symmetry of the fundamental tangle $\Db^{1,0}(u)$. As will be seen below, the fused tangles $\Db^{m,n}(u)$ satisfy a crossing symmetry extending \eqref{eq:symCroiseDmn}, with a shift that depends on the fusion indices. This symmetry will describe how $\Db^{m,n}(u)$ behaves under
\begin{equation} \label{def:slu} 
u \mapsto \slu = (4-2m-2n)\lambda-u\ . 
\ee
This new variable $\slu$ already appears in several of the functions $\beta^j$. The fusion hierarchy relations defining $\Db^{m,n}(u)$ are expressed in terms of the functions $\alpha^i$ and the $\beta^j$, and their form depends crucially on the position of the label $(m,n)$ in \cref{fig:maxP}, namely:\vspace{-0.3cm}
\begin{subequations} \label{eq:HF}
\begin{alignat}{3}
\intertext{(i) for the elements near the wedge of the region:}
\Aun(2,u) \Db^{2,0}_0 &= \Adeux(2,u) \Db^{1,0}_0 \Db^{1,0}_2
- \Atrois(2,u) \Azero(u_{-1}) \Db^{0,1}_0 \label{eq:HFa} \ ,\\
\Bun(1,1,u) \Db^{1,1}_0 &= \Bdeux(1,1,u) \Db^{1,0}_0 \Db^{0,1}_2
- \Btrois(1,1,u) \Bzero(1,u) \Bzero(1,\slu) \Ib \label{eq:HFb} \ , \\
\Aun(2,\slu) \Db^{0,2}_0 &= \Adeux(2,\slu) \Db^{0,1}_0 \Db^{0,1}_2
- \Atrois(2,\slu) \Azero(\slu_{-1}) \Db^{1,0}_2 \label{eq:HFc} \ ,
\intertext{(ii) for the elements at the boundary of the region:} 
\Aun(m,u) \Db^{m,0}_0 &= \Adeux(m,u) \Db^{m-1,0}_0 \Db^{1,0}_{2m-2}
- \Atrois(m,u) \Db^{m-2,1}_0 \label{eq:HFd} \ , \qquad &&m>2\ , \\ 
\Aun(n,\slu) \Db^{0,n}_0 &= \Adeux(n,\slu) \Db^{0,1}_{0} \Db^{0,n-1}_2 
- \Atrois(n,\slu) \Db^{1,n-2}_2 \label{eq:HFe} \ , \qquad &&n>2\ ,
\intertext{(iii) for the elements adjacent to a boundary elements:}
\Bun(m,1,u) \Db^{m,1}_0 &= \Bdeux(m,1,u) \Db^{m,0}_0 \Db^{0,1}_{2m}
- \Btrois(m,1,u) \Bzero(m,u) \Db^{m-1,0}_0 \label{eq:HFf} \ , \qquad &&m > 1\ , \\
\Bun(n,1,\slu) \Db^{1,n}_0 &= \Bdeux(n,1,\slu) \Db^{1,0}_0 \Db^{0,n}_{2}
 - \Btrois(n,1,\slu) \Bzero(n,\slu) \Db^{0,n-1}_4 \label{eq:HFg} \ , \qquad &&n > 1\ ,
\intertext{(iv) elsewhere:}
\Bun(m,n,u) \Db^{m,n}_0 &= \Bdeux(m,n,u) \Db^{m,0}_0 \Db^{0,n}_{2m}
- \Btrois(m,n,u) \Db^{m-1,0}_0 \Db^{0,n-1}_{2m+2} \ , \qquad &&m,n >1. \label{eq:HFh}
\end{alignat}
\end{subequations}
All matrices are evaluated at $u$ in \eqref{eq:HF}. The coefficients $\alpha^i$ apply to the boundary segments of the region shown in \cref{fig:maxP} and the coefficients $\beta^i$ are for every other fused tangles (anywhere inside the region). We note that the fusion hierarchy relations are divided in these many cases because of the different maximal degrees of the transfer tangles $\Db^{m,0}(u)$ and $\Db^{0,n}(u)$ compared to $\Db^{m,n}(u)$ with $m,n\ge 1$. The functions $\alpha^i$ and $\beta^j$ then compensate for these degrees accordingly: $\alpha^1$, $\alpha^2$, $\alpha^3$ and $\beta^0$ for the relations labeled by pairs $(m,n)$ near the boundary of the wedge in \cref{fig:maxP}, and $\Azero$ and $\Bzero$ near the corner.

In \eqref{eq:HF}, $\slu$ sometimes appears as the evaluation point for some functions $\alpha^i$. The integers $m$ and $n$ needed to compute $\slu$ are then the ones of the fused tangle created by the fusion hierarchy, that is, the one appearing on the left-hand side of each equation. The $m$ and $n$ in the definition of $\slu$ are thus the same throughout a given equation. The notation $\slu$ is handy, but somewhat dangerous as the operation $u\mapsto \slu$ does not commute with that of shifting $u\mapsto u_k=u+k\lambda$. Unless stated otherwise, the operation $u\mapsto \slu$ in these expressions is the last to be performed. For example, the function $\Azero(\slu_{-1})$ appears in \eqref{eq:HFc} where $m=0$ and $n=2$ and thus $\slu=-u$. This function for SS boundary conditions is
\begin{alignat}{2}
\Azero(\slu_{-1}) &= \Azero(u_{-1}) \big\rvert_{u\mapsto \slu=-u} 
\nonumber\\&
= \left(w(u_{-3/2}) \wb(u_{1/2}) f(u_{-2}) \right)_{-1} \big\rvert_{u\mapsto \slu=-u} 
\nonumber\\&
=s(u-\tfrac{5\lambda}{2}) c(u-\tfrac{\lambda}{2}) f(u-3\lambda) \big\rvert_{u\mapsto \slu=-u} \\&
=-s(u+\tfrac{5\lambda}{2}) c(u+\tfrac{\lambda}{2}) f(u+3\lambda) \ , \nonumber
\end{alignat}
where we used $f(-u) = f(u)$.

The weights $\beta^j$ appear in pairs invariant under $u\mapsto (4-2m+2n)\lambda-u$ and $m\leftrightarrow n$. For this reason, a relation between $\Db^{m,n}(u)$ and $\Db^{n,m}(\slu)$ becomes apparent, at least for the pair of equations \eqref{eq:HFf} and \eqref{eq:HFg}, and for the fused tangles defined with \eqref{eq:HFh}. This property, the generalization of the crossing symmetry \eqref{eq:symCroiseD10}, is proved in \cref{sub:proofCS}. Thus, any two tangles $\Db^{m,n}(u)$ and $\Db^{n,m}(u)$ that appear in \cref{fig:maxP} as mirror images through the horizontal line are related by crossing-symmetry.

\cref{fig:maxP} also gives the maximal power $\maxP$ of the fused tangles for identical boundary conditions (SS or CC) in the variable $z=\eE^{\iI u}$ introduced earlier. These maximal degrees should be contrasted with those of the periodic case obtained in \cite{AMDPP19}. In the latter case, fused matrices have different polynomial degrees: the ones with labels on a boundary of the region have $\maxP=2N$, and all the others have $\maxP=3N$. For the strip geometry, the cases of identical and mixed boundaries need to be distinguished. The mixed cases behave in a way similar to the periodic geometry: the maximal powers of the fused transfer matrices are respectively $4N$ and $6N$ on the boundary and inside the region. The identical cases do not behave as nicely, and the maximal power of $\Db^{m,n}(u)$ increases linearly with the fusion indices, as $4N+2m$ for the cases $(m,0)$ and $(0,m)$, and as $6N+2(m+n)+2$ for $(m,n)$ with $m,n \ge 1$.

The fused matrices satisfy the following properties for all fusion indices $m,n$:
\begin{subequations}
\begin{alignat}{3}
\text{crossing symmetry}&\qquad\qquad
&&\Db^{m,n}(u) = \Db^{n,m}((4-2m-2n)\lambda-u),\label{eq:symCroiseDmn}\\ 
\text{commutativity}&\qquad\qquad &&[\Db^{m,n}(u), \Db^{m',n'}(v)] = 0,\label{eq:commutDmn}\\[0.1cm]
\text{periodicity}&\qquad\qquad &&\Db^{m,n}(u) = \Db^{m,n}(u+\pi),\label{eq:piDmn}\\[0.1cm]
\text{conjugacy}&\qquad\qquad &&\Db^{0,n}(u) = \Db^{n,0}(u+\lambda),\label{eq:Dm0D0m}\\[0.1cm]
\text{polynomiality}& \qquad\qquad &&\text{The coefficients of $\Db^{m,n}(u)$ are Laurent polynomials in }z .
\end{alignat}
\end{subequations}
These equations generalize the properties \eqref{eq:symCroiseD10} to \eqref{eq:piD10} of the tangle $\Db^{1,0}(u)$. The first one is the crossing symmetry discussed above and its proof is given in \cref{sub:proofCS}. The second is the commutation that holds for any integers $m,n,m',n'$ and spectral parameters $u, v$. This property follows directly from \eqref{eq:commutD10}, since the fused transfer tangles are all expressible in terms of the fundamental transfer tangle evaluated at different values of $u$. The third one, the periodicity, follows similarly from \eqref{eq:piD10}. The proof of the conjugacy linking $\Db^{0,n}(u)$ and $\Db^{n,0}(u+\lambda)$ is non-trivial and relegated to \cref{sub:proofConjugacy}. Likewise, the polynomial properties $\Db^{m,n}(u)$ are not obvious and will be the topic of \cref{sec:polynomiality}.

%%%%%%%%%%%%%
\subsection{Diagrammatic definition of $\Db^{2,0}(u)$ and $\Db^{1,1}(u)$} \label{sub:fused.with.diagrams}
%%%%%%%%%%%%%

In this section, we focus on the first fused transfer matrices $\Db^{2,0}(u)$ and $\Db^{1,1}(u)$, for which the projectors $P^{2,0}$ and $P^{1,1}$ are known \cite{AMDPP19}. This allows us to give a diagrammatic definition for these two transfer matrices:
\begingroup
\allowdisplaybreaks
\begin{subequations}
\label{def:D20D11FullDiag}
\begin{alignat}{2} 
\label{def:D20FullDiag}
\Db^{2,0}(u) &= \frac1{Z^{2,0}(u)}\ \
\psset{unit=1.2cm}
\begin{pspicture}[shift=-1.8](-2.5,0)(6,4)
\rput(0,2){
\facegrid{(0,0)}{(4,2)}
\rput(2.5,1.5){$\ldots$}
\rput(2.5,0.5){$\ldots$}
\rput(0,1){\rput(0.5,0.5){$_{u_2+\xi_{(1)}}$}
\psarc[linewidth=0.025]{-}(1,0){0.16}{90}{180}}
\rput(1,1){\rput(0.5,0.5){$_{u_2+\xi_{(2)}}$}
\psarc[linewidth=0.025]{-}(1,0){0.16}{90}{180}}
\rput(3,1){\rput(0.5,0.5){$_{u_2+\xi_{(N)}}$}
\psarc[linewidth=0.025]{-}(1,0){0.16}{90}{180}}
\rput(0,0){
\rput(0.5,0.5){$_{u_0+\xi_{(1)}}$}
\psarc[linewidth=0.025]{-}(1,0){0.16}{90}{180}}
\rput(1,0){\rput(0.5,0.5){$_{u_0+\xi_{(2)}}$}
\psarc[linewidth=0.025]{-}(1,0){0.16}{90}{180}}
\rput(3,0){\rput(0.5,0.5){$_{u_0+\xi_{(N)}}$}
\psarc[linewidth=0.025]{-}(1,0){0.16}{90}{180}}
\psline[linecolor=blue,linewidth=1.5pt,linestyle=dashed,dash=2pt 2pt](4,1.5)(5.5,1.5)
\psline[linecolor=blue,linewidth=1.5pt,linestyle=dashed,dash=2pt 2pt](4,0.5)(5.5,0.5)
\rput(5,0){\triangleg \rput(0.65,1){${u_2}$}}
}
\rput(0,0){
\facegrid{(0,0)}{(4,2)}
\rput(2.5,1.5){$\ldots$}
\rput(2.5,0.5){$\ldots$}
\rput(0,1){\rput(0.5,0.5){$_{u_2-\xi_{(1)}}$}
\psarc[linewidth=0.025]{-}(0,0){0.16}{0}{90}}
\rput(1,1){\rput(0.5,0.5){$_{u_2-\xi_{(2)}}$}
\psarc[linewidth=0.025]{-}(0,0){0.16}{0}{90}}
\rput(3,1){\rput(0.5,0.5){$_{u_2-\xi_{(N)}}$}
\psarc[linewidth=0.025]{-}(0,0){0.16}{0}{90}}
\rput(0,0){\rput(0.5,0.5){$_{u_0-\xi_{(1)}}$}
\psarc[linewidth=0.025]{-}(0,0){0.16}{0}{90}}
\rput(1,0){\rput(0.5,0.5){$_{u_0-\xi_{(2)}}$}
\psarc[linewidth=0.025]{-}(0,0){0.16}{0}{90}}
\rput(3,0){\rput(0.5,0.5){$_{u_0-\xi_{(N)}}$}
\psarc[linewidth=0.025]{-}(0,0){0.16}{0}{90}}
\psline[linecolor=blue,linewidth=1.5pt,linestyle=dashed,dash=2pt 2pt](4,1.5)(5.5,1.5)
\psline[linecolor=blue,linewidth=1.5pt,linestyle=dashed,dash=2pt 2pt](4,0.5)(5.5,0.5)
\rput(5,0){\triangleg \rput(0.65,1){${u_0}$}}
}
\rput(-2.5,0){
\psline[linecolor=blue,linewidth=1.5pt,linestyle=dashed,dash=2pt 2pt](0,3.5)(2.5,3.5)
\psline[linecolor=blue,linewidth=1.5pt,linestyle=dashed,dash=2pt 2pt](0,2.5)(2.5,2.5)
\psline[linecolor=blue,linewidth=1.5pt,linestyle=dashed,dash=2pt 2pt](0,1.5)(2.5,1.5)
\psline[linecolor=blue,linewidth=1.5pt,linestyle=dashed,dash=2pt 2pt](0,0.5)(2.5,0.5)
\rput(0,2){\triangled \rput(0.45,1){\footnotesize${3\lambda\!-\!u_2}$}}
\rput(0,1){\losange\rput(1,1){$2u-\lambda$}}
\triangled \rput(0.45,1){\footnotesize$3\lambda\!-\!u_0$}
}
\rput(4,1){\losange\rput(1,1){$\lambda-2u$}}
\rput(-0.3,0){\projectorTwo\rput(0.18,1){\rput{90}(0,0){$_{2,0}$}}}
\end{pspicture}
\ \ ,
\\[0.2cm]
\label{def:D11FullDiag}
\Db^{1,1}(u) &= \frac1{Z^{1,1}(u) } \ \
\psset{unit=1.2cm}
\begin{pspicture}[shift=-1.8](-2.5,0)(6,4)
\rput(0,2){
\facegrid{(0,0)}{(4,2)}
\rput(2.5,1.5){$\ldots$}
\rput(2.5,0.5){$\ldots$}
\rput(0,1){\rput(0.5,0.5){$_{u_3+\xi_{(1)}}$}
\psarc[linewidth=0.025]{-}(1,0){0.16}{90}{180}}
\rput(1,1){\rput(0.5,0.5){$_{u_3+\xi_{(2)}}$}
\psarc[linewidth=0.025]{-}(1,0){0.16}{90}{180}}
\rput(3,1){\rput(0.5,0.5){$_{u_3+\xi_{(N)}}$}
\psarc[linewidth=0.025]{-}(1,0){0.16}{90}{180}}
\rput(0,0){\rput(0.5,0.5){$_{u_0+\xi_{(1)}}$}
\psarc[linewidth=0.025]{-}(1,0){0.16}{90}{180}}
\rput(1,0){\rput(0.5,0.5){$_{u_0+\xi_{(2)}}$}
\psarc[linewidth=0.025]{-}(1,0){0.16}{90}{180}}
\rput(3,0){\rput(0.5,0.5){$_{u_0+\xi_{(N)}}$}
\psarc[linewidth=0.025]{-}(1,0){0.16}{90}{180}}
\psline[linecolor=blue,linewidth=1.5pt,linestyle=dashed,dash=2pt 2pt](4,1.5)(5.5,1.5)
\psline[linecolor=blue,linewidth=1.5pt,linestyle=dashed,dash=2pt 2pt](4,0.5)(5.5,0.5)
\rput(5,0){\triangleg \rput(0.65,1){${u_3}$}}
}
\rput(0,0){
\facegrid{(0,0)}{(4,2)}
\rput(2.5,1.5){$\ldots$}
\rput(2.5,0.5){$\ldots$}
\rput(0,1){\rput(0.5,0.5){$_{u_3-\xi_{(1)}}$}
\psarc[linewidth=0.025]{-}(0,0){0.16}{0}{90}}
\rput(1,1){\rput(0.5,0.5){$_{u_3-\xi_{(2)}}$}
\psarc[linewidth=0.025]{-}(0,0){0.16}{0}{90}}
\rput(3,1){\rput(0.5,0.5){$_{u_3-\xi_{(N)}}$}
\psarc[linewidth=0.025]{-}(0,0){0.16}{0}{90}}
\rput(0,0){\rput(0.5,0.5){$_{u_0-\xi_{(1)}}$}
\psarc[linewidth=0.025]{-}(0,0){0.16}{0}{90}}
\rput(1,0){\rput(0.5,0.5){$_{u_0-\xi_{(2)}}$}
\psarc[linewidth=0.025]{-}(0,0){0.16}{0}{90}}
\rput(3,0){\rput(0.5,0.5){$_{u_0-\xi_{(N)}}$}
\psarc[linewidth=0.025]{-}(0,0){0.16}{0}{90}}
\psline[linecolor=blue,linewidth=1.5pt,linestyle=dashed,dash=2pt 2pt](4,1.5)(5.5,1.5)
\psline[linecolor=blue,linewidth=1.5pt,linestyle=dashed,dash=2pt 2pt](4,0.5)(5.5,0.5)
\rput(5,0){\triangleg \rput(0.65,1){$u_0$}}
}
\rput(-2.5,0){
\psline[linecolor=blue,linewidth=1.5pt,linestyle=dashed,dash=2pt 2pt](0,3.5)(2.5,3.5)
\psline[linecolor=blue,linewidth=1.5pt,linestyle=dashed,dash=2pt 2pt](0,2.5)(2.5,2.5)
\psline[linecolor=blue,linewidth=1.5pt,linestyle=dashed,dash=2pt 2pt](0,1.5)(2.5,1.5)
\psline[linecolor=blue,linewidth=1.5pt,linestyle=dashed,dash=2pt 2pt](0,0.5)(2.5,0.5)
\rput(0,2){\triangled \rput(0.45,1){\footnotesize$3\lambda\!-\!u_3$}}
\rput(0,1){\losange\rput(1,1){$2u$}}
\triangled \rput(0.45,1){\footnotesize$3\lambda\!-\!u_0$}
}
\rput(4,1){\losange\rput(1,1){$-2u$}}
\rput(-0.3,0){\projectorTwo\rput(0.18,1){\rput{90}(0,0){$_{1,1}$}}}
\end{pspicture}
\ \ ,
\end{alignat}
\end{subequations}
\endgroup
where 
\begin{subequations}
\label{eq:Z20Z11}
\begin{alignat}{2}
\label{def:Z20}
Z^{2,0}(u) &=
4 s(2u_{-1}) s(2u_{0}) \delta_L(u_{-3/2}) \delta_L(u_{-1/2}) \delta_R(u_{-1/2}) \delta_R(u_{1/2}) f(u_{-1}) f(u_0) \nonumber\\
&
\times\left\{
\begin{array}{cc}
-1 & \textrm{SS and CC,}\\[0.1cm]
\delta_L(u_{-5/2}) \delta_L(u_{1/2}) \delta_R(u_{-3/2}) \delta_R(u_{3/2})
& \textrm{SC and CS,}
\end{array}\right.
\\[0.15cm]
 \label{def:Z11}
Z^{1,1}(u) &=
4 s(2u_0)^2 \delta_L(\tfrac{3\lambda}{2}-u) \delta_R(\tfrac{3\lambda}{2}+u) f(u_0)\nonumber\\
&
\times\left\{
\begin{array}{cc}
-1 & \textrm{SS and CC,}\\[0.1cm]
\delta_L(\tfrac{\lambda}{2}+u) \delta_R(\tfrac{\lambda}{2}-u) \delta_L(\tfrac{3\lambda}{2}+u) \delta_R(\tfrac{3\lambda}{2}-u) \delta_L(\tfrac{5\lambda}{2}-u) \delta_R(\tfrac{5\lambda}{2}+u) 
& \textrm{SC and CS.}\\[0.1cm]
\end{array}\right.
\end{alignat}
\end{subequations}
Here we add subscripts $L$ and $R$ to the functions $\delta(u)$ to distinguish contributions coming from the left and right boundaries, thus allowing us to treat all four choices of boundary conditions simultaneously. 

We now show that these definitions are equivalent to those presented in \cref{sub:fused.with.relations}. In each case, expanding the projector as in \eqref{def:P20P11explicite} produces two terms:
\be
Z^{2,0}(u) \Db^{2,0}(u) = \Ab_{\mathrm{I}} - \Ab_{\mathrm{II}} \ ,
\qquad
Z^{1,1}(u) \Db^{1,1}(u) = \Bb_{\mathrm{I}} - \Bb_{\mathrm{II}} \ .
\ee
Using the notation
\begin{equation}
\psset{unit=1.05cm}
\begin{pspicture}[shift=-0.4](2,1)
\pspolygon[fillstyle=solid,fillcolor=lightlightblue](0,0)(2,0)(2,1)(0,1)\rput(1,0.5){$u-\xi$}
\psarc[linewidth=0.025]{-}(0,0){0.16}{0}{90}
\end{pspicture}
\ = \ \begin{pspicture}[shift=-0.4](4,1)
\facegrid{(0,0)}{(4,1)}
\rput(0,0){\rput(0.5,0.5){$_{u-\xi_{(1)}}$}
\psarc[linewidth=0.025]{-}(0,0){0.16}{0}{90}}
\rput(1,0){\rput(0.5,0.5){$_{u-\xi_{(2)}}$}
\psarc[linewidth=0.025]{-}(0,0){0.16}{0}{90}}
\rput(2,0){\rput(0.5,0.5){$\cdots$}}
\rput(3,0){\rput(0.5,0.5){$_{u-\xi_{(N)}}$}
\psarc[linewidth=0.025]{-}(0,0){0.16}{0}{90}}
\end{pspicture}
\ \ ,
\qquad 
\begin{pspicture}[shift=-0.4](2,1)
\pspolygon[fillstyle=solid,fillcolor=lightlightblue](0,0)(2,0)(2,1)(0,1)\rput(1,0.5){$u+\xi$}
\psarc[linewidth=0.025]{-}(2,0){0.16}{90}{180}
\end{pspicture}
\ = \ \begin{pspicture}[shift=-0.4](4,1)
\facegrid{(0,0)}{(4,1)}
\rput(0,0){\rput(0.5,0.5){$_{u+\xi_{(1)}}$}
\psarc[linewidth=0.025]{-}(1,0){0.16}{90}{180}}
\rput(1,0){\rput(0.5,0.5){$_{u+\xi_{(2)}}$}
\psarc[linewidth=0.025]{-}(1,0){0.16}{90}{180}}
\rput(2,0){\rput(0.5,0.5){$\cdots$}}
\rput(3,0){\rput(0.5,0.5){$_{u+\xi_{(N)}}$}
\psarc[linewidth=0.025]{-}(1,0){0.16}{90}{180}}
\end{pspicture}
\ \ ,
\end{equation}
we rewrite $\Ab_{\mathrm{I}}$ as
\begingroup
\allowdisplaybreaks
\begin{alignat}{2}
\Ab_{\mathrm{I}} &= \ 
\begin{pspicture}[shift=-1.8](-2,0)(4,4)
\rput(0,2){
\rput(0,0){\pspolygon[fillstyle=solid,fillcolor=lightlightblue](0,0)(2,0)(2,1)(0,1)\rput(1,0.5){$u_0+\xi$}
\psarc[linewidth=0.025]{-}(2,0){0.16}{90}{180}}
\rput(0,1){
\pspolygon[fillstyle=solid,fillcolor=lightlightblue](0,0)(2,0)(2,1)(0,1)\rput(1,0.5){$u_2+\xi$}\psarc[linewidth=0.025]{-}(2,0){0.16}{90}{180}}
\psline[linecolor=blue,linewidth=1.5pt,linestyle=dashed,dash=2pt 2pt](2,1.5)(3.5,1.5)
\psline[linecolor=blue,linewidth=1.5pt,linestyle=dashed,dash=2pt 2pt](2,0.5)(3.5,0.5)
\rput(3,0){\triangleg \rput(0.65,1){$u_2$}}
}
\rput(0,0){
\rput(0,0){
\pspolygon[fillstyle=solid,fillcolor=lightlightblue](0,0)(2,0)(2,1)(0,1)\rput(1,0.5){$u_0-\xi$}\psarc[linewidth=0.025]{-}(0,0){0.16}{0}{90}}
\rput(0,1){
\pspolygon[fillstyle=solid,fillcolor=lightlightblue](0,0)(2,0)(2,1)(0,1)\rput(1,0.5){$u_2-\xi$}\psarc[linewidth=0.025]{-}(0,0){0.16}{0}{90}}
\psline[linecolor=blue,linewidth=1.5pt,linestyle=dashed,dash=2pt 2pt](2,1.5)(3.5,1.5)
\psline[linecolor=blue,linewidth=1.5pt,linestyle=dashed,dash=2pt 2pt](2,0.5)(3.5,0.5)
\rput(3,0){\triangleg \rput(0.65,1){$u_0$}}
}
\rput(-2,0){
\psline[linecolor=blue,linewidth=1.5pt,linestyle=dashed,dash=2pt 2pt](0,3.5)(2,3.5)
\psline[linecolor=blue,linewidth=1.5pt,linestyle=dashed,dash=2pt 2pt](0,2.5)(2,2.5)
\psline[linecolor=blue,linewidth=1.5pt,linestyle=dashed,dash=2pt 2pt](0,1.5)(2,1.5)
\psline[linecolor=blue,linewidth=1.5pt,linestyle=dashed,dash=2pt 2pt](0,0.5)(2,0.5)
\rput(0,2){\triangled \rput(0.45,1){$_{3\lambda-u_2}$}}
\triangled \rput(0.45,1){$_{3\lambda-u_0}$}
}
\rput(2,1){\losange\rput(1,1){$\lambda-2u$}}
\rput(-2,1){\losange\rput(1,1){$2u-\lambda$}}
\end{pspicture}
\ = \
\begin{pspicture}[shift=-1.8](-4,0)(3,4)
\rput(0,2){
\rput(0,0){\pspolygon[fillstyle=solid,fillcolor=lightlightblue](0,0)(2,0)(2,1)(0,1)\rput(1,0.5){$u_2-\xi$}\psarc[linewidth=0.025]{-}(0,0){0.16}{0}{90}}
\rput(0,1){
\pspolygon[fillstyle=solid,fillcolor=lightlightblue](0,0)(2,0)(2,1)(0,1)\rput(1,0.5){$u_2+\xi$}\psarc[linewidth=0.025]{-}(2,0){0.16}{90}{180}}
\psline[linecolor=blue,linewidth=1.5pt,linestyle=dashed,dash=2pt 2pt](2,1.5)(2.5,1.5)
\psline[linecolor=blue,linewidth=1.5pt,linestyle=dashed,dash=2pt 2pt](2,0.5)(2.5,0.5)
\rput(2,0){\triangleg \rput(0.65,1){$u_2$}}
}
\rput(0,0){
\rput(0,0){
\pspolygon[fillstyle=solid,fillcolor=lightlightblue](0,0)(2,0)(2,1)(0,1)\rput(1,0.5){$u_0-\xi$}\psarc[linewidth=0.025]{-}(0,0){0.16}{0}{90}}
\rput(0,1){
\pspolygon[fillstyle=solid,fillcolor=lightlightblue](0,0)(2,0)(2,1)(0,1)\rput(1,0.5){$u_0+\xi$}
\psarc[linewidth=0.025]{-}(2,0){0.16}{90}{180}}
\psline[linecolor=blue,linewidth=1.5pt,linestyle=dashed,dash=2pt 2pt](2,1.5)(2.5,1.5)
\psline[linecolor=blue,linewidth=1.5pt,linestyle=dashed,dash=2pt 2pt](2,0.5)(2.5,0.5)
\rput(2,0){\triangleg \rput(0.65,1){$u_0$}}
}
\rput(-4,0){
\psline[linecolor=blue,linewidth=1.5pt,linestyle=dashed,dash=2pt 2pt](0,3.5)(4,3.5)
\psline[linecolor=blue,linewidth=1.5pt,linestyle=dashed,dash=2pt 2pt](0,2.5)(4,2.5)
\psline[linecolor=blue,linewidth=1.5pt,linestyle=dashed,dash=2pt 2pt](0,1.5)(4,1.5)
\psline[linecolor=blue,linewidth=1.5pt,linestyle=dashed,dash=2pt 2pt](0,0.5)(4,0.5)
\rput(0,2){\triangled \rput(0.45,1){$_{3\lambda-u_2}$}}
\triangled \rput(0.45,1){$_{3\lambda-u_0}$}
}
\rput(-2,1){\losange\rput(1,1){$\lambda-2u$}}
\rput(-4,1){\losange\rput(1,1){$2u-\lambda$}}
\end{pspicture}
\nonumber\\[0.2cm]
& \ = \rho_8(2u-\lambda) \rho_8(\lambda-2u) \
\begin{pspicture}[shift=-1.8](-1,0)(3,4)
\rput(0,2){
\rput(0,0){\pspolygon[fillstyle=solid,fillcolor=lightlightblue](0,0)(2,0)(2,1)(0,1)\rput(1,0.5){$u_2-\xi$}\psarc[linewidth=0.025]{-}(0,0){0.16}{0}{90}}
\rput(0,1){
\pspolygon[fillstyle=solid,fillcolor=lightlightblue](0,0)(2,0)(2,1)(0,1)\rput(1,0.5){$u_2+\xi$}\psarc[linewidth=0.025]{-}(2,0){0.16}{90}{180}}
\psline[linecolor=blue,linewidth=1.5pt,linestyle=dashed,dash=2pt 2pt](2,1.5)(2.5,1.5)
\psline[linecolor=blue,linewidth=1.5pt,linestyle=dashed,dash=2pt 2pt](2,0.5)(2.5,0.5)
\rput(2,0){\triangleg \rput(0.65,1){$u_2$}}
}
\rput(0,0){
\rput(0,0){
\pspolygon[fillstyle=solid,fillcolor=lightlightblue](0,0)(2,0)(2,1)(0,1)\rput(1,0.5){$u_0-\xi$}\psarc[linewidth=0.025]{-}(0,0){0.16}{0}{90}}
\rput(0,1){
\pspolygon[fillstyle=solid,fillcolor=lightlightblue](0,0)(2,0)(2,1)(0,1)\rput(1,0.5){$u_0+\xi$}
\psarc[linewidth=0.025]{-}(2,0){0.16}{90}{180}}
\psline[linecolor=blue,linewidth=1.5pt,linestyle=dashed,dash=2pt 2pt](2,1.5)(2.5,1.5)
\psline[linecolor=blue,linewidth=1.5pt,linestyle=dashed,dash=2pt 2pt](2,0.5)(2.5,0.5)
\rput(2,0){\triangleg \rput(0.65,1){$u_0$}}
}
\rput(-1,0){
\psline[linecolor=blue,linewidth=1.5pt,linestyle=dashed,dash=2pt 2pt](0,3.5)(1,3.5)
\psline[linecolor=blue,linewidth=1.5pt,linestyle=dashed,dash=2pt 2pt](0,2.5)(1,2.5)
\psline[linecolor=blue,linewidth=1.5pt,linestyle=dashed,dash=2pt 2pt](0,1.5)(1,1.5)
\psline[linecolor=blue,linewidth=1.5pt,linestyle=dashed,dash=2pt 2pt](0,0.5)(1,0.5)
\rput(0,2){\triangled \rput(0.45,1){$_{3\lambda-u_2}$}}
\triangled \rput(0.45,1){$_{3\lambda-u_0}$}
}
\end{pspicture}
\ = \rho_8(2u-\lambda) \rho_8(\lambda-2u) \Dbt^{1,0}_0 \Dbt^{1,0}_2 \ ,
\end{alignat}
\endgroup
where we first applied the Yang-Baxter equation and then the inversion identity. For $\Ab_{\mathrm{II}}$, we write
\begingroup
\allowdisplaybreaks
\begin{alignat}{2} 
\Ab_{\mathrm{II}} &= \ \begin{pspicture}[shift=-1.8](-3,0)(4,4)
\rput(0,2){
\rput(0,0){\pspolygon[fillstyle=solid,fillcolor=lightlightblue](0,0)(2,0)(2,1)(0,1)\rput(1,0.5){$u_0+\xi$}\psarc[linewidth=0.025]{-}(2,0){0.16}{90}{180}}
\rput(0,1){\pspolygon[fillstyle=solid,fillcolor=lightlightblue](0,0)(2,0)(2,1)(0,1)\rput(1,0.5){$u_2+\xi$}\psarc[linewidth=0.025]{-}(2,0){0.16}{90}{180}}
\psline[linecolor=blue,linewidth=1.5pt,linestyle=dashed,dash=2pt 2pt](2,1.5)(3.5,1.5)
\psline[linecolor=blue,linewidth=1.5pt,linestyle=dashed,dash=2pt 2pt](2,0.5)(3.5,0.5)
\rput(3,0){\triangleg \rput(0.65,1){$u_2$}}
}
\rput(0,0){
\rput(0,0){\pspolygon[fillstyle=solid,fillcolor=lightlightblue](0,0)(2,0)(2,1)(0,1)\rput(1,0.5){$u_0-\xi$}\psarc[linewidth=0.025]{-}(0,0){0.16}{0}{90}}
\rput(0,1){\pspolygon[fillstyle=solid,fillcolor=lightlightblue](0,0)(2,0)(2,1)(0,1)\rput(1,0.5){$u_2-\xi$}\psarc[linewidth=0.025]{-}(0,0){0.16}{0}{90}}
\psline[linecolor=blue,linewidth=1.5pt,linestyle=dashed,dash=2pt 2pt](2,1.5)(3.5,1.5)
\psline[linecolor=blue,linewidth=1.5pt,linestyle=dashed,dash=2pt 2pt](2,0.5)(3.5,0.5)
\rput(3,0){\triangleg \rput(0.65,1){$u_0$}}
}
\rput(-3,0){
\psline[linecolor=blue,linewidth=1.5pt,linestyle=dashed,dash=2pt 2pt](0,3.5)(3,3.5)
\psline[linecolor=blue,linewidth=1.5pt,linestyle=dashed,dash=2pt 2pt](0,2.5)(3,2.5)
\psline[linecolor=blue,linewidth=1.5pt,linestyle=dashed,dash=2pt 2pt](0,1.5)(3,1.5)
\psline[linecolor=blue,linewidth=1.5pt,linestyle=dashed,dash=2pt 2pt](0,0.5)(3,0.5)
\rput(0,2){\triangled \rput(0.45,1){$_{3\lambda-u_2}$}}
\rput(0,1){\losange\rput(1,1){$2u-\lambda$}}
\triangled \rput(0.45,1){$_{3\lambda-u_0}$}
}
\rput(2,1){\losange\rput(1,1){$\lambda-2u$}}
\rput(-2,0)\triangleCNg
\rput(-1,0){\triangled\rput(0.4,1){1}\psarc[linewidth=0.025]{-}(1,1){0.16}{135}{225}}
\end{pspicture}
\ = \ell(u) \
\begin{pspicture}[shift=-1.8](-2,0)(4,4)
\rput(0,2){
\rput(0,0){\pspolygon[fillstyle=solid,fillcolor=lightlightblue](0,0)(2,0)(2,1)(0,1)\rput(1,0.5){$u_0+\xi$}\psarc[linewidth=0.025]{-}(2,0){0.16}{90}{180}}
\rput(0,1){\pspolygon[fillstyle=solid,fillcolor=lightlightblue](0,0)(2,0)(2,1)(0,1)\rput(1,0.5){$u_2+\xi$}\psarc[linewidth=0.025]{-}(2,0){0.16}{90}{180}}
\psline[linecolor=blue,linewidth=1.5pt,linestyle=dashed,dash=2pt 2pt](2,1.5)(3.5,1.5)
\psline[linecolor=blue,linewidth=1.5pt,linestyle=dashed,dash=2pt 2pt](2,0.5)(3.5,0.5)
\rput(3,0){\triangleg \rput(0.65,1){$u_2$}}
}
\rput(0,0){
\rput(0,0){\pspolygon[fillstyle=solid,fillcolor=lightlightblue](0,0)(2,0)(2,1)(0,1)\rput(1,0.5){$u_0-\xi$}\psarc[linewidth=0.025]{-}(0,0){0.16}{0}{90}}
\rput(0,1){\pspolygon[fillstyle=solid,fillcolor=lightlightblue](0,0)(2,0)(2,1)(0,1)\rput(1,0.5){$u_2-\xi$}\psarc[linewidth=0.025]{-}(0,0){0.16}{0}{90}}
\psline[linecolor=blue,linewidth=1.5pt,linestyle=dashed,dash=2pt 2pt](2,1.5)(3.5,1.5)
\psline[linecolor=blue,linewidth=1.5pt,linestyle=dashed,dash=2pt 2pt](2,0.5)(3.5,0.5)
\rput(3,0){\triangleg \rput(0.65,1){$u_0$}}
}
\rput(-2,0){
\psline[linecolor=blue,linewidth=1.5pt,linestyle=dashed,dash=2pt 2pt](1.5,3.5)(2,3.5)
\psline[linecolor=blue,linewidth=1.5pt,linestyle=dashed,dash=2pt 2pt](1.5,2.5)(2,2.5)
\psline[linecolor=blue,linewidth=1.5pt,linestyle=dashed,dash=2pt 2pt](1.5,1.5)(2,1.5)
\psline[linecolor=blue,linewidth=1.5pt,linestyle=dashed,dash=2pt 2pt](1.5,0.5)(2,0.5)
\psline[linecolor=blue,linewidth=1.5pt,linestyle=dashed,dash=2pt 2pt](0.5,2.5)(1,3)
\psline[linewidth=1.5pt,linecolor=blue,linestyle=dashed,dash=2pt 2pt](0.5,1.5)(1,1)
\rput(1,2){\triangleCNd}
\rput(0,1){\triangled\rput(0.45,1){$_{3\lambda-u_1}$}}
\rput(1,0){\triangled\rput(0.4,1){1}\psarc[linewidth=0.025]{-}(1,1){0.16}{135}{225}}
}
\rput(2,1){\losange\rput(1,1){$\lambda-2u$}}
\end{pspicture}
\nonumber\\[0.2cm] & \hspace{-0.84cm}= \ell(u)t(u) \
\begin{pspicture}[shift=-1.8](-2,0)(4.5,4)
\rput(0,2){
\rput(0,0){\pspolygon[fillstyle=solid,fillcolor=lightlightblue](0,0)(2,0)(2,1)(0,1)\rput(1,0.5){$u_1+\xi$}\psarc[linewidth=0.025]{-}(2,0){0.16}{90}{180}}
\psline[linecolor=blue,linewidth=1.5pt,linestyle=dashed,dash=2pt 2pt](2,0.5)(2.5,1.0)
}
\rput(0,0){
\rput(0,0){\pspolygon[fillstyle=solid,fillcolor=lightlightblue](0,0)(2,0)(2,1)(0,1)\rput(1,0.5){$u_0-\xi$}\psarc[linewidth=0.025]{-}(0,0){0.16}{0}{90}}
\rput(0,1){\pspolygon[fillstyle=solid,fillcolor=lightlightblue](0,0)(2,0)(2,1)(0,1)\rput(1,0.5){$u_2-\xi$}\psarc[linewidth=0.025]{-}(0,0){0.16}{0}{90}}
\psline[linecolor=blue,linewidth=1.5pt,linestyle=dashed,dash=2pt 2pt](2,1.5)(4.5,1.5)
\psline[linecolor=blue,linewidth=1.5pt,linestyle=dashed,dash=2pt 2pt](2,0.5)(4.5,0.5)
}
\rput(-2,0){
\psline[linecolor=blue,linewidth=1.5pt,linestyle=dashed,dash=2pt 2pt](0.2,2.5)(2,2.5)
\psline[linecolor=blue,linewidth=1.5pt,linestyle=dashed,dash=2pt 2pt](1.5,1.5)(2,1.5)
\psline[linecolor=blue,linewidth=1.5pt,linestyle=dashed,dash=2pt 2pt](1.5,0.5)(2,0.5)
\psline[linewidth=1.5pt,linecolor=blue,linestyle=dashed,dash=2pt 2pt](0.5,1.5)(1,1)
\rput(0,1){\triangled\rput(0.45,1){$_{3\lambda-u_1}$}}
\rput(1,0){\triangled\rput(0.4,1){1}\psarc[linewidth=0.025]{-}(1,1){0.16}{135}{225}}
}
\rput(2.5,0){
\psline[linecolor=blue,linewidth=1.5pt,linestyle=dashed,dash=2pt 2pt](0.5,3.5)(1.5,3.5)
\rput(0,2){\triangleCNd}
\rput(1,2){\triangleg \rput(0.65,1){$u_2$}}
\rput(0,1){\losange\rput(1,1){$\lambda-2u$}}
\rput(1,0){\triangleg \rput(0.65,1){$u_0$}}}
\end{pspicture}\
= \ell(u)t(u)r(u) \ 
\begin{pspicture}[shift=-1.4](-2,0)(4,3)
\rput(0,2){
\rput(0,0){\pspolygon[fillstyle=solid,fillcolor=lightlightblue](0,0)(2,0)(2,1)(0,1)\rput(1,0.5){$u_1+\xi$}\psarc[linewidth=0.025]{-}(2,0){0.16}{90}{180}}
}
\rput(0,0){
\rput(0,0){\pspolygon[fillstyle=solid,fillcolor=lightlightblue](0,0)(2,0)(2,1)(0,1)\rput(1,0.5){$u_0-\xi$}\psarc[linewidth=0.025]{-}(0,0){0.16}{0}{90}}
\rput(0,1){\pspolygon[fillstyle=solid,fillcolor=lightlightblue](0,0)(2,0)(2,1)(0,1)\rput(1,0.5){$u_2-\xi$}\psarc[linewidth=0.025]{-}(0,0){0.16}{0}{90}}
}
\rput(-2,0){
\psline[linecolor=blue,linewidth=1.5pt,linestyle=dashed,dash=2pt 2pt](0.2,2.5)(2,2.5)
\psline[linewidth=1.5pt,linecolor=blue,linestyle=dashed,dash=2pt 2pt](0.5,1.5)(1,1)
\psline[linecolor=blue,linewidth=1.5pt,linestyle=dashed,dash=2pt 2pt](1,1.5)(2,1.5)
\psline[linecolor=blue,linewidth=1.5pt,linestyle=dashed,dash=2pt 2pt](1,0.5)(2,0.5)
\psline[linewidth=1.5pt,linecolor=blue,linestyle=dashed,dash=2pt 2pt](5.5,1.5)(5,1)
\rput(0,1){\triangled\rput(0.45,1){$_{3\lambda-u_1}$}}
\rput(1,0){\triangled\rput(0.4,1){1}\psarc[linewidth=0.025]{-}(1,1){0.16}{135}{225}}
}
\rput(2,0){
\psline[linecolor=blue,linewidth=1.5pt,linestyle=dashed,dash=2pt 2pt](0,2.5)(1.5,2.5)
\psline[linewidth=1.5pt,linecolor=blue,linestyle=dashed,dash=2pt 2pt](0.5,1.5)(1,1)
\psline[linecolor=blue,linewidth=1.5pt,linestyle=dashed,dash=2pt 2pt](0,1.5)(0.5,1.5)
\psline[linecolor=blue,linewidth=1.5pt,linestyle=dashed,dash=2pt 2pt](0,0.5)(0.5,0.5)
\rput(1,1){\triangleg\rput(0.65,1){$u_1$}}
\rput(0,0)\triangleCNg}
\end{pspicture}
\nonumber\\[0.2cm] & \hspace{-0.84cm}= \ell(u) t(u) r(u) b(u) \
\begin{pspicture}[shift=-1.4](-3.75,0)(3,3)
\rput(0,1){
\rput(0,0){\pspolygon[fillstyle=solid,fillcolor=lightlightblue](0,0)(2,0)(2,1)(0,1)\rput(1,0.5){$u_1+\xi$}\psarc[linewidth=0.025]{-}(2,0){0.16}{90}{180}}
}
\rput(0,0){
\rput(0,0){\pspolygon[fillstyle=solid,fillcolor=lightlightblue](0,0)(2,0)(2,1)(0,1)\rput(1,0.5){$u_1-\xi$}\psarc[linewidth=0.025]{-}(0,0){0.16}{0}{90}}
}
\psline[linecolor=blue,linewidth=1.5pt,linestyle=dashed,dash=2pt 2pt](-3.5,2.5)(-1,2.5)
\psarc[linewidth=1.5pt,linecolor=blue,linestyle=dashed,dash=2pt 2pt](-1,2){0.5}{5}{85}
\psarc[linewidth=1.5pt,linecolor=blue,linestyle=dashed,dash=2pt 2pt](0,2){0.5}{180}{270}
\psarc[linewidth=1.5pt,linecolor=blue,linestyle=dashed,dash=2pt 2pt](-0.5,0.75){0.25}{10}{80}
\psarc[linewidth=1.5pt,linecolor=blue,linestyle=dashed,dash=2pt 2pt](0,0.75){0.25}{180}{270}
\psline[linecolor=blue,linewidth=1.5pt,linestyle=dashed,dash=2pt 2pt](-1,1)(-0.5,1)
\rput(-3.75,0){
\psline[linewidth=1.5pt,linecolor=blue,linestyle=dashed,dash=2pt 2pt](0.5,1.5)(1,1)
\psline[linecolor=blue,linewidth=1.5pt,linestyle=dashed,dash=2pt 2pt](1,1.5)(3,1.5)
\psline[linecolor=blue,linewidth=1.5pt,linestyle=dashed,dash=2pt 2pt](1,0.5)(3,0.5)
\rput(0,1){\triangled\rput(0.45,1){$_{3\lambda-u_1}$}}
\rput(1,0){\triangled\rput(0.4,1){1}\psarc[linewidth=0.025]{-}(1,1){0.16}{135}{225}}
\rput(2,0)\triangleCNg}
\rput(2,0){
\psline[linecolor=blue,linewidth=1.5pt,linestyle=dashed,dash=2pt 2pt](0,1.5)(0.5,1.5)
\psline[linecolor=blue,linewidth=1.5pt,linestyle=dashed,dash=2pt 2pt](0,0.5)(0.5,0.5)
\rput(0,0){\triangleg\rput(0.65,1){$u_1$}}}
\end{pspicture}
\ = \ell(u) t(u) r(u) b(u) \Dbt^{0,1}_0.
\end{alignat}
\endgroup
Here we applied the push-through property $N$ times on the top and bottom of the diagram, at the third and fifth equality. The resulting functions $t(u)$ and $b(u)$ are  obtained from \eqref{rel:PTbulkTCN} and satisfy $t(u)b(u) = f(u_{-3})f(u_2)$. Similarly, the factors $\ell(u)$ and $r(u)$ arose at the second and fourth equality from the application of the boundary reflection property on the left and right boundary, respectively, and are read off from \eqref{rel:PTfrontiereTCN}: 
\be
\ell(u) = 2 s(2u_{-3})\delta_L(u_{-5/2})\bar\delta_L(\tfrac \lambda 2-u)\delta_L(u_{1/2}), \qquad
r(u) = \ell(\lambda-u)\big|_{\delta_L \to \delta_R}. 
\ee
Finally, we used \eqref{eq:spider.ids} at the last step and obtained $\Dbt^{1,0}_1 = \Dbt^{0,1}_0$. The final result is an identity relating $\Db^{2,0}_0$, $\Db^{1,0}_0\Db^{1,0}_2$ and $\Db^{0,1}_0$ which precisely reproduces \eqref{eq:HFa} after simplification of the trigonometric prefactors.

The same arguments can be applied to compute $\Bb_{\mathrm{I}}$ and $\Bb_{\mathrm{II}}$, leading to
\be
\Bb_{\mathrm{I}} = \rho_8(2u) \rho_8(-2u) \Dbt^{1,0}_0 \Dbt^{1,0}_3,
\qquad
\Bb_{\mathrm{II}} = \tilde\ell(u) \tilde t(u) \tilde r(u) \tilde b(u) \Ib,
\ee
with $\tilde t(u)\tilde b(u) = f(u_{-3})f(u_{-2})f(u_2)f(u_3)$ and the functions $\tilde\ell(u)$ and $\tilde r(u)=\tilde \ell(-u)|_{\delta_L \to \delta_R}$ read off from~\eqref{rel:PTfrontiereArc}. After simplification of the prefactors, the resulting relation is found to be exactly \eqref{eq:HFb}. This then confirms that the recursive and diagrammatic definitions are equivalent, for both $\Db^{2,0}(u)$ and $\Db^{1,1}(u)$.

We stress that we still have not proven that $\Db^{2,0}(u)$ and $\Db^{1,1}(u)$ are Laurent polynomials in $z = \eE^{\iI u}$. Indeed, both the recursive and diagrammatic definitions involve denominators written in terms of trigonometric functions of $u$ that could potentially lead to singularities. Showing the polynomiality of the fused transfer matrices is a non-trivial task which is addressed in \cref{sec:polynomiality}. Finally, we note that it is also possible to give diagrammatic definitions of $\Db^{m,0}(u)$ for $m>2$ using the projectors in the Appendix A of \cite{AMDPP19}, however these expressions are not needed in this paper.

%%%%%%%%%%%%%
\subsection{Determinantal form of the fused tangles} \label{sub:det}
%%%%%%%%%%%%%
The fusion relations define recursively the fused transfer tangles as polynomials in the elementary transfer tangles $\Db^{1,0}(u)$. In this section, we write $\Db^{m,n}(u)$ in terms of a formal determinant and use it to rewrite the fusion hierarchy \eqref{eq:HF}. The proof of these determinant formulas is sketched at the end the section. 

The determinants are expressed in terms of the following quantities:
\begin{subequations}
\begin{alignat}{2}
\Dbm(u) &= s(2u_{-4}) s(2u_{1})
\wb(u_{-3/2}) w^{(1)}(u_{-3/2}) \Db^{1,0}_0(u) = \det(1,0)(u) \ , \\
\fbm(u) &= s(2u_{-4}) s(2u_{3}) w(u_{-5/2}) w(u_{3/2})
\wb(u_{-3/2}) \wb(u_{1/2}) f(u_{-3}) f(u_{2}) \Ib \ .
\end{alignat}
We often use the shorthand notation $\Dbm_k=\Dbm(u_k)$ and $\fbm_k=\fbm(u_k)$. These new functions satisfy
\begin{alignat}{2} 
\label{eq:symDbm} \Dbm(u) &= \Dbm(3\lambda-u) \ , \qquad &\Dbm(u) &= \Dbm(u+\pi) \ , \\
\label{eq:symFbm} \fbm(u) &= \fbm(\lambda-u) \ , \qquad &\fbm(u) &= \fbm(u+\pi) \ .
\end{alignat}
\end{subequations}

The fused transfer matrices will be written in terms of two families of determinants. The first is for fused tangles with indices $(m,0)$ and $(0,n)$ corresponding to the boundary of the domain depicted in \cref{fig:maxP}: 
\begin{subequations}
\begin{alignat}{2} \label{def:detFormel(m,0)}
\det(m,0)(u) &= \begin{vmatrix}
\Dbm_{2m-2} & \Dbm_{2m-3} & \fbm_{2m-5} & 0 & 0 & 0 \\
\fbm_{2m-4} & \Dbm_{2m-4} & \Dbm_{2m-5} & \fbm_{2m-7} & 0 & 0\vphantom{\ddots} \\
0 & \fbm_{2m-6} & \Dbm_{2m-6} & \Dbm_{2m-7} & \ddots & 0 \\
0 & 0 & \ddots & \ddots & \ddots & \fbm_{1} \\
0 & 0 & 0 & \fbm_2 & \Dbm_2 & \Dbm_1\vphantom{\ddots} \\
0 & 0 & 0 & 0 & \fbm_0 & \Dbm_0\vphantom{\ddots}
\end{vmatrix} \ , \\
\det(0,n)(u) &= \det(n,0)(4\lambda-2n\lambda-u) \ . \label{def:detFormel(0,n)}
\end{alignat}
\end{subequations}
The matrix in the determinant $\det(m,0)_0$ has size $m\times m$. Like for the functions $\fbm(u)$ and $\Dbm(u)$, we will use the notation $\det(m,0)_k=\det(m,0)(u_k)$. 

The first determinant is tied to the fused matrix $\Db^{m,0}(u)$ by
\begin{alignat}{2} \label{def:detExplicite(m,0)}
\det(m,& 0)_0 = s(2u_{-4}) s(2u_{2m-1}) w^{(m)}(u_{m-5/2}) \Db^{m,0}(u)\\
& \times \Big( \prod^{m-2}_{j=0} s(2u_j) \Big)
\Big( \prod^{2m-5}_{j=m-3} s(2u_j) \Big)
\Big( \prod^{2m-4}_{l=-1} f(u_l) \Big) 
\Big( \prod^{m-1}_{k=0} \wb(u_{2k-3/2}) \Big) 
\Big( \prod^{m-2}_{k=0} w(u_{2k-1/2}) \Big) \ .\notag
\end{alignat}
We postpone the (sketch of the) proof of this statement until the end of the section. 

The definition $\det(0,n)(u)$ as $\det(n,0)(\slu)$, with $\slu=4\lambda-2n\lambda-u$, ensures that the crossing symmetry \eqref{eq:symCroiseDmn} is realized. Moreover, this first family of determinants has the property
\begin{equation} \label{def:detPratique(0,n)}
\det(0,n)_0 = \det(n,0)_1 \ .
\end{equation}
The proof is given in \cref{sub:proofConjugacy}. The property $\Db^{0,n}_0=\Db^{n,0}_1$ for $n>1$ is also proved in \cref{sub:proofConjugacy}.

The second family of determinants is
\begin{subequations}
\begin{alignat}{2} \label{def:detFormel(m,n)}
\det(m,n)(u) & = \left\rvert \begin{array}{ccc:cccc}
\ddots & \ddots & \fbm_{2m+2} & 0 & 0 & 0 & 0 \\
\ddots & \ddots & \Dbm_{2m+2} & 0 & 0 & 0 & 0 \\[1ex]
0 & \fbm_{2m+1} & \Dbm_{2m+1} & \fbm_{2m-1} & 0 & 0 & 0 \\[1ex]
\hdashline
&&&&&& \\[-2ex]
0 & 0 & \fbm_{2m-2} & \Dbm_{2m-2} & \Dbm_{2m-3} & \fbm_{2m-5} & 0 \\
0 & 0 & 0 & \fbm_{2m-4} & \ddots & \ddots & \ddots
\end{array} \right\rvert \\
&= \left|
\begin{matrix} \ \ \ 
\begin{pspicture}(0,0)(0,0)
\pspolygon(-0.3,0.2)(2.7,0.2)(2.7,-1)(-0.3,-1)
\rput(1.2,-0.4){$\det(n,0)_{2m+1}$}
\end{pspicture} &  &  & 0 & 0 & 0 \\ 
 &  &  & 0 & 0 & 0 \\
 &  &  & \qquad \fbm_{2m-1}\quad & 0 & 0 \\
0 & 0 & \quad \fbm_{2m-2}\qquad  & 
 &  &  \\
0 & 0 & 0 &  &  &  \\
0 & 0 & 0 &  &  & 
\begin{pspicture}(0,0)(0,0)
\rput(-2.4,1){\pspolygon(-0.3,0.2)(2.7,0.2)(2.7,-1)(-0.3,-1)\rput(1.2,-0.4){$\det(m,0)_0$}}
\end{pspicture} \ \ \ \\
\end{matrix}\ \right|\ .
\end{alignat}
\end{subequations}
It applies to all fusion pairs $(m,n)$ with $m,n >1$, corresponding to the 
positions in \cref{fig:maxP} that do not touch the edge of the wedge. The bottom-right and the top-left blocks are of size $m\times m$ and $n\times n$ respectively. The determinant \eqref{def:detFormel(m,n)} and the fused matrix $\Db^{m,n}(u)$ are related by
\begin{alignat}{2}\label{def:detExplicite(m,n)}
s(2u_{2m-4}&) s(2\slu_{2n-4})\det(m,n)(u) = \\
& s(2u_{-4}) s(2\slu_{-4}) s(2u_{m-3}) s(2\slu_{n-3})
s(2u_{m+n-2}) s(2\slu_{2n-2}) f(u_{2m-2}) \Db^{m,n}(u) \notag \\
&\times\Big( \prod^{2m-3}_{j=0} s(2u_j) \Big)
\Big( \prod^{m-1}_{k=0} \wb(u_{2k-3/2}) \Big)
\Big( \prod^{m-2}_{k=0} w(u_{2k-1/2}) \Big)
\Big( \prod^{2m-4}_{l=-1} f(u_l) \Big)\notag \\
&\times \Big( \prod^{2n-3}_{j=0} s(2\slu_j) \Big)
\Big( \prod^{n-1}_{k=0} \wb(\slu_{2k-3/2}) \Big)
\Big( \prod^{n-2}_{k=0} w(\slu_{2k-1/2}) \Big)
\Big( \prod^{2n-4}_{l=-1} f(\slu_l) \Big) \ , \notag
\end{alignat}
with $\slu=(4-2m-2n)\lambda-u$. We note that the right-hand side of \eqref{def:detExplicite(m,n)} always contains factors $s(2u_{2m-4}) s(2\slu_{2n-4})$, even for $m=1$ or $n=1$. It is however easier to leave them as prefactors on the left-hand side instead of identifying in which factor they appear in the right-hand side, as this depends on $m$ and $n$.

These determinants satisfy a number of recurrence relations. The following ones express the fusion hierarchies in a unified manner, namely for all four choices of boundary conditions and independently of the position of the label $(m,n)$ in \cref{fig:maxP}: 
\begin{subequations}
\label{eq:HFDet}
\begin{alignat}{2}
\label{eq:HFdDet} \det(m,0)_0 &= \Dbm_{2m-2} \det(m-1,0)_0
- \fbm_{2m-4} \det(m-2,1)_0  \ , \\% [0.8ex]
\label{eq:HFeDet} \det(0,n)_0 &= \Dbm_1 \det(0,n-1)_2 - \fbm_1 \det(1,n-2)_2 \ , \\%[0.8ex]
\label{eq:HFhDet} \det(m,n)_0 &= \det(m,0)_0 \det(0,n)_{2m} \\
& \qquad - \fbm_{2m-2} \fbm_{2m-1} \det(m-1,0)_0 \det(0,n-1)_{2m+2} \ . \notag
\end{alignat}
\end{subequations}
These relations are valid for $m,n\ge 1$, with the convention
\begin{equation}
\det(0,0)_k = 1 \ , \qquad \det(m,-1)_k = \det(-1,n)_k = 0 \ . 
\end{equation}
The relations \eqref{eq:HFdDet} and \eqref{eq:HFeDet} are respectively obtained by expanding $\det(m,0)_0$ and $\det(0,n)_0$ along the first row and the last column. Expanding with respect to the first column after the dashed line leads to \eqref{eq:HFhDet}. Two additional useful recurrence relations are
\begin{subequations} \label{eq:HFDet3}
\begin{alignat}{2}
\label{eq:HFdDet3}\det(m,0)_0 &= \Dbm_{2m-2} \det(m-1,0)_0 
\notag\\ & \qquad 
- \fbm_{2m-4} \big( \Dbm_{2m-3} \det(m-2,0)_0
- \fbm_{2m-6} \fbm_{2m-5} \det(m-3,0)_0 \big), \\[0.1cm]
\label{eq:HFeDet3}
 \det(m,0)_0 &= \Dbm_0 \det(m-1,0)_2
- \fbm_0 \big( \Dbm_1 \det(m-2,0)_4
- \fbm_1 \fbm_2 \det(m-3,0)_6 \big),
\end{alignat}
\end{subequations}
with $m \ge 2$. The first is a rewriting of \eqref{eq:HFdDet} and the second is obtained from the determinant \eqref{def:detFormel(m,0)} by an expansion along the last column.

The symmetry properties of the determinants are analogous to those satisfied by the fused transfer matrices. These are expressed in a form that is independent of the choice of boundary conditions:
\begin{subequations}
\begin{alignat}{3}
\label{eq:symCroiseDETmn}\text{crossing symmetry} &\qquad\qquad
&&\det(m,n)(u) = \det(n,m)((4-2m-2n)\lambda-u) \ ,\\ 
\label{eq:commutDETmn}\text{commutativity} &\qquad\qquad
&&[ \, \det(m,n)(u), \det(m',n')(v) \, ] = 0 \ ,\\
\label{eq:piDETmn}\text{periodicity} &\qquad\qquad
&&\det(m,n)(u) = \det(m,n)(u+\pi) \ ,\\
\label{eq:conjugDETmn}\text{conjugacy} &
\qquad\qquad &&\det(0,n)(u) = \det(n,0)(u+\lambda) \ .
\end{alignat}
\end{subequations}

We end this section by sketching the proof of the relations \eqref{def:detExplicite(m,0)} and \eqref{def:detExplicite(m,n)} that relate the formal determinants and the fused transfer transfer matrices. For \eqref{def:detExplicite(m,0)}, the proof is done by induction on~$m$ using \eqref{eq:HFdDet3}. The seed cases $m=1$ and $m=2$ are first checked separately. For $m=1$, \eqref{def:detExplicite(m,0)} follows directly from the definition of $\Dbm(u)$. For $m=2$, we check that \eqref{eq:HFdDet} reduces to \eqref{eq:HFa} using \eqref{def:detExplicite(m,0)}. For $m\ge3$, the proof proceeds using the induction hypothesis \eqref{def:detExplicite(m,0)} for $m'<m$. With \eqref{eq:HFd} and \eqref{eq:HFf}, one first derives a new relation that ties $\Db^{m,0}(u)$, $\Db^{m-1,0}(u)$, $\Db^{m-2,0}(u)$ and $\Db^{m-3,0}(u)$. Then we must show that \eqref{eq:HFdDet3} reduces to this relation upon applying \eqref{def:detExplicite(m,0)}. Each step is straigthforward and tedious, and amounts to writing the prefactors for each fused transfer matrix, cancelling the common terms, and then checking that the remaining ones precisely match those in the new relation. We omit the details of this calculation, which in the end proves \eqref{def:detExplicite(m,0)}. The similar result for $m,n\ge1$, namely the relation \eqref{def:detExplicite(m,n)}, is proved using the same idea. We use \eqref{eq:HFhDet} to expand $\det(m,n)_0$, and then \eqref{def:detExplicite(m,0)} to express all the determinants in terms of the fused transfer matrices. After the cancellation of many common terms, one obtains \eqref{eq:HFf}, \eqref{eq:HFg} or \eqref{eq:HFh} depending on the fusion indices $(m,n)$.

%%%%%%%%%%%%%
\subsection{Reduction relations} \label{sub:reductions}
%%%%%%%%%%%%%

In this subsection, we study properties of the fused transfer matrices $\Db^{m,0}(u)$ at specific values $u = \widehat u$ of the spectral parameter. We find that $\Db^{m,0}(\widehat u)$ is proportional to a fused transfer matrix with the index $m$ reduced by $3$. We refer to these identities as {\it reduction relations}. Applying these relations repeatedly, we find that $\Db^{m,0}(\widehat u)$ is proportional to the unit $\Ib$. To proceed, we first write down the functional relations
\begin{subequations}
\label{eq:Dm0munu}
\begin{alignat}{2}
\mu^1(m,u)\Db^{m,0}_0 &= \mu^2(m,u)\Db^{m-1,0}_0\Db^{1,0}_{2m-2} - \mu^3(m,u)\Db^{m-2,0}_0 \Db^{1,0}_{2m-3}+\mu^4(m,u) \Db^{m-3,0}_0,
\label{eq:Dm0mu}\\[0.1cm]
\nu^1(m,u)\Db^{m,0}_0 &= \nu^2(m,u)\Db^{1,0}_{0}\Db^{m-1,0}_2 - \nu^3(m,u)\Db^{1,0}_{1}\Db^{m-2,0}_4 +\nu^4(m,u) \Db^{m-3,0}_6,
\label{eq:Dm0nu}
\end{alignat}
\end{subequations}
where
\begin{subequations}
\begin{alignat}{2}
&\hspace{-0.2cm}\mu^1(m,u)=w(u_{2m-9/2})w^{(m)}(u_{m-5/2})s(2u_{m-3})s(2u_{m-2})s(2u_{2m-6})s(2u_{2m-5})f_{2m-6}f_{2m-5}f_{2m-4},
\\[0.1cm]
&\hspace{-0.2cm}\mu^2(m,u)=w^{(1)}(u_{2m-7/2})w^{(m-1)}(u_{m-7/2})s(2u_{m-4})s(2u_{m-3})s(2u_{2m-6})s(2u_{2m-3})f_{2m-6},
\\[0.1cm]
&\hspace{-0.2cm}\mu^3(m,u)=w(u_{2m-5/2})\bar w(u_{2m-9/2})w^{(1)}(u_{2m-9/2})w^{(m-2)}(u_{m-9/2})s(2u_{m-5})s(2u_{m-4})
\nonumber\\ &\hspace{1.7cm} s(2u_{2m-5})s(2u_{2m-2})f_{2m-2},
\\[0.1cm]
&\hspace{-0.2cm}\mu^4(m,u)=w(u_{2m-15/2})w(u_{2m-9/2})w(u_{2m-7/2})w(u_{2m-5/2})\bar w(u_{2m-13/2})\bar w(u_{2m-11/2})\bar w(u_{2m-9/2})\nonumber\\ 
& \hspace{1.7cm}w^{(m-3)}(u_{m-11/2})s(2u_{m-6})s(2u_{m-5})s(2u_{2m-3})s(2u_{2m-2})f_{2m-4}f_{2m-3}f_{2m-2},
\end{alignat}
\end{subequations}
and
\begin{subequations}
\begin{alignat}{2}
\nu^1(m,u)&=w(u_{-1/2})w^{(m)}(u_{m-5/2})s(2u_{0})s(2u_{1})s(2u_{m-3})s(2u_{m-2})f_{-1}f_{0}f_{1},
\\[0.1cm]
\nu^2(m,u)&=w^{(1)}(u_{-3/2})w^{(m-1)}(u_{m-3/2})s(2u_{-2})s(2u_{1})s(2u_{m-2})s(2u_{m-1})f_{1},
\\[0.1cm]
\nu^3(m,u)&=w(u_{-5/2})\bar w(u_{-1/2})w^{(1)}(u_{-1/2})w^{(m-2)}(u_{m-1/2})s(2u_{-3})s(2u_{0})
s(2u_{m-1})s(2u_{m})f_{-3}, 
\\[0.1cm]
\nu^4(m,u)&=w(u_{-5/2})w(u_{-3/2})w(u_{-1/2})w(u_{5/2})\bar w(u_{-1/2})\bar w(u_{1/2})\bar w(u_{3/2})w^{(m-3)}(u_{m+1/2})\nonumber\\ 
& \hspace{0.5cm}s(2u_{-3})s(2u_{-2})s(2u_{m})s(2u_{m+1})f_{-3}f_{-2}f_{-1}.
\end{alignat}
\end{subequations}
These are obtained by two possible methods: (i) by combining the fusion hierarchy relations for $\Db^{m,0}(u)$, $\Db^{m-2,1}(u)$ and $\Db^{1,m-2}(u)$, or (ii) by simplifying the common prefactors in \eqref{eq:HFDet3}. The above equations hold for $m \ge 4$, but also for $m = 2,3$ with the identifications $\Db^{0,0}_k \mapsto f_{k-2}f_{k-3}\Ib$ and $\Db^{-1,0}_k \mapsto 0$. Moreover, we remark that the relations \eqref{eq:Dm0mu} and \eqref{eq:Dm0nu} are related by a crossing symmetry linking the evaluations at $u$ and $(5-2m)\lambda-u$.

These relations allow us to derive {\it reduction relations} for the fused transfer tangles. For instance, we note that
\be
\begin{array}{cl}
\mu^2(m,\widehat u) = \mu^3(m,\widehat u)=0 \quad &\textrm{for}\quad \widehat u = (4-m)\lambda + \tfrac{r \pi}2,\\[0.15cm]
\nu^2(m,\widehat v) = \nu^3(m, \widehat v)=0 \quad &\textrm{for}\quad \widehat v = (1-m)\lambda + \tfrac{r \pi}2,
\end{array}
 \qquad r \in \mathbb Z.
\ee 
Moreover, for $m\geq 3$, $\mu^1(m,\widehat u)$ and $\mu^4(m,\widehat u)$ are non-zero for generic values of $\lambda$, and likewise for $\nu^1(m,\widehat v)$ and $\nu^4(m,\widehat v)$. We therefore have
\begin{subequations}
\label{eq:Dm0.reds}
\begin{alignat}{2}
\Db^{m,0}_0(\widehat u) &= \frac{\mu^4\big(m,\widehat u\big)}{\mu^1\big(m,\widehat u\big)} \Db^{m-3,0}_0(\widehat u) 
= \frac{\mu^4\big(m,\widehat u\big)}{\mu^1\big(m,\widehat u\big)} \Db^{m-3,0}_0\big((7-m)\lambda + \tfrac{r \pi}2\big)
= \frac{\mu^4\big(m,\widehat u\big)}{\mu^1\big(m,\widehat u\big)} \Big[\Db^{m,0}_0(\widehat u)\big|_{m \to m-3}\Big],\label{eq:Dm0.red.nu}
\\
\Db^{m,0}_0(\widehat v) &= \frac{\nu^4\big(m,\widehat v\big)}{\nu^1\big(m,\widehat v\big)} \Db^{m-3,0}_6(\widehat v) = \frac{\nu^4\big(m,\widehat v\big)}{\nu^1\big(m,\widehat v\big)} \Db^{m-3,0}_0((4-m)\lambda + \tfrac{r \pi}2) = \frac{\nu^4\big(m,\widehat v\big)}{\nu^1\big(m,\widehat v\big)}\Big[\Db^{m,0}_0(\widehat v)\big|_{m \to m-3}\Big],\label{eq:Dm0.red.mu}
\end{alignat}
\end{subequations}
where we used crossing symmetry at the second equalities. Writing $m = x+3y$ with $x \in \{0,1,2\}$ and $y \in \mathbb N$, we apply the first reduction relation repeatedly to write
\be
\Db^{m,0}(\widehat u) = \Db^{x,0}\big((4-x)\lambda+\tfrac {r \pi}2\big) \prod_{k=0}^{y-1} g^r(m-3k), 
\qquad 
g^r(m)=\frac{\mu^4\big(m,(4-m)\lambda+\tfrac{r\pi}2\big)}{\mu^1\big(m,(4-m)\lambda+\tfrac{r\pi}2\big)}.
\ee
This relation also holds for $y=0$ provided that the product is replaced by the factor $1$. Crucially, one can write each of the tangles for $m = 0,1,2$ in terms of one of the following objects:
\begin{subequations}
\label{eq:D012.ids}
\begin{alignat}{2}
\label{eq:D00.id}
&\Db^{0,0}(4\lambda+\tfrac{r\pi}2) \mapsto f(2\lambda+\tfrac{r\pi}2) f(\lambda+\tfrac {r\pi} 2) \Ib,\\[0.2cm]
\label{eq:D10.id}
&\Db^{1,0}(3\lambda+\tfrac{r\pi}2)
= \Db^{1,0}(\tfrac{r\pi}2) 
= f(2\lambda+\tfrac{r\pi}2) f(3\lambda+\tfrac{r\pi}2) \frac{s(6\lambda)}{s(2\lambda)}
\frac{w(\frac{5\lambda}2+\tfrac{r\pi}2) w(-\frac{3\lambda}2+\tfrac{r\pi}2) \wb(-\frac{\lambda}2+\tfrac{r\pi}2)}{w^{(1)}(\frac{3\lambda}2+\tfrac{r\pi}2)} \Ib,\\[0.2cm]
&\Db^{2,0}(2\lambda+\tfrac{r\pi}2) = -\frac{\alpha^3(2,2\lambda+\tfrac{r\pi}2)\alpha^0(\lambda+\tfrac{r\pi}2)}{\alpha^1(2,2\lambda+\tfrac{r\pi}2)}\Db^{1,0}(3\lambda+\tfrac{r\pi}2).
\label{eq:D20.id1}
\end{alignat}
\end{subequations}
The relation \eqref{eq:D00.id} follows directly from our convention for $\Db^{0,0}(u)$ used in \eqref{eq:Dm0munu}, and \eqref{eq:D10.id} was given in \eqref{eq:DI}. Finally for \eqref{eq:D20.id1}, we used the fact that $\Adeux(2,u)$ in \eqref{eq:HFa} has a factor $s(2u_{-2})$. 

This proves that $\Db^{m,0}(\widehat u)$ is proportional to the identity element $\Ib$, for all $m \ge 0$. The overall prefactor is a non-trivial product of trigonometric factors, which can be simplified to  
\begin{alignat}{2}
\Db^{m,0}(\widehat u) &= \Ib \times f(\widehat u_{2m-3}) f(\widehat u_{2m-2}) 
\frac{s(2\widehat u_{2m-3}) s(2\widehat u_{2m-2})}{s(2\widehat u_{m-3}) s(2\widehat u_{m-2})}
\frac{w(\widehat u_{2m-5/2})}{w^{(m)}(\widehat u_{m-5/2})}
\prod_{j=0}^{m-1} w(-\widehat u_{2j-3/2}) \wb(-\widehat u_{2j-5/2}).
\label{eq:simpler.D.reduction}
\end{alignat}
This formula is valid for all four boundary conditions. It can be checked first for $m = 0,1,2$ and then proved inductively on $m$ by steps of $3$ using \eqref{eq:Dm0.red.nu}.

%%%%%%%%%%%%%
\subsection{Braid limits} \label{sub:braid}
%%%%%%%%%%%%%

The \textit{bulk braid operators} are defined as the following limits of the face operator:
\begin{equation}
\begin{pspicture}[shift=-0.4](1,1)
\facegrid{(0,0)}{(1,1)}
\psarc[linewidth=0.025]{-}(0,0){0.16}{0}{90}
\rput(.5,.5){$\pm\infty$}
\end{pspicture}
\ =\lim_{u\rightarrow \pm \iI\infty} \frac{\eE^{\mp \iI(\pi-2\lambda)}}{\rho_8(u)} \
\begin{pspicture}[shift=-0.4](1,1)
\facegrid{(0,0)}{(1,1)}
\psarc[linewidth=0.025]{-}(0,0){0.16}{0}{90}
\rput(.5,.5){$u$}
\end{pspicture}
\ = \ \begin{pspicture}[shift=-0.4](1,1)
\loopa
\end{pspicture}
\ + \ \begin{pspicture}[shift=-0.4](1,1)
\loopf
\end{pspicture}
\ + \ \begin{pspicture}[shift=-0.4](1,1)
\loopg
\end{pspicture}
\ -\eE^{\pm 2\iI\lambda} \
\begin{pspicture}[shift=-0.4](1,1)
\looph
\end{pspicture}
\ -\eE^{\mp 2\iI\lambda} \
\begin{pspicture}[shift=-0.4](1,1)
\loopi
\end{pspicture} \ .
\end{equation}
Diagrammatic properties then follow from the ones of the face operator:
\begin{equation} \label{rel:relationsBulkTresse}
\begin{pspicture}[shift=-0.4](1,1)
\facegrid{(0,0)}{(1,1)}
\rput(0,0){\psarc[linewidth=0.025]{-}(0,0){0.16}{0}{90}\rput(.5,.5){$\pm\infty$}}
\end{pspicture}
= \begin{pspicture}[shift=-0.4](1,1)
\facegrid{(0,0)}{(1,1)}
\rput(0,0){\psarc[linewidth=0.025]{-}(1,0){0.16}{90}{180}\rput(.5,.5){$\mp\infty$}}
\end{pspicture} \ , \qquad
\begin{pspicture}[shift=-0.6](2.8,1.4)
\psline[linecolor=blue,linewidth=1.5pt,linestyle=dashed,dash=2pt 2pt](1,1.06)(2,1.06)
\psline[linecolor=blue,linewidth=1.5pt,linestyle=dashed,dash=2pt 2pt](1,0.35)(2,0.35)
\rput(0.71,0){\rput{45}{\loopa \psarc[linewidth=0.025]{-}(0,1){0.16}{-90}{0}}
\rput(0,0.71){$\pm\infty$}}
\rput(2.12,0){\rput{45}{\loopa \psarc[linewidth=0.025]{-}(0,1){0.16}{-90}{0}}
\rput(0,0.71){$\mp\infty$}}
\end{pspicture}
= \begin{pspicture}[shift=-0.6](1.4,1.4)
\rput{45}(0.71,0){\loopej}
\end{pspicture} \ , \qquad
\begin{pspicture}[shift=-0.7](2.4,1.6)
\psset{unit=0.8cm,linewidth=0.8\pslinewidth}
\psline[linecolor=blue,linewidth=1.5pt,linestyle=dashed,dash=2pt 2pt](1.5,0.5)(2,0.5)
\psline[linecolor=blue,linewidth=1.5pt,linestyle=dashed,dash=2pt 2pt](1.5,1.5)(2,1.5)
\losange \rput(1,1){$\pm\infty$}
\rput(2,0){
\facegrid{(0,0)}{(1,2)}
\psarc[linewidth=0.025]{-}(0,1){0.16}{0}{90}
\psarc[linewidth=0.025]{-}(0,0){0.16}{0}{90}
\rput(.5,1.5){$\mp\infty$}
\rput(.5,0.5){$\pm\infty$}}
\end{pspicture}
= \begin{pspicture}[shift=-0.7](2.4,1.6)
\psset{unit=0.8cm,linewidth=0.8\pslinewidth}
\psline[linecolor=blue,linewidth=1.5pt,linestyle=dashed,dash=2pt 2pt](1,0.5)(1.5,0.5)
\psline[linecolor=blue,linewidth=1.5pt,linestyle=dashed,dash=2pt 2pt](1,1.5)(1.5,1.5)
\rput(1,0){\losange \rput(1,1){$\mp\infty$}}
\facegrid{(0,0)}{(1,2)}
\psarc[linewidth=0.025]{-}(0,1){0.16}{0}{90}
\psarc[linewidth=0.025]{-}(0,0){0.16}{0}{90}
\rput(.5,1.5){$\pm\infty$}
\rput(.5,0.5){$\mp\infty$}
\end{pspicture} \ .
\end{equation}
The bulk braid operators also have push-trough properties for the arc and the vacancies:
\begin{equation} \label{rel:PTtresse}
\begin{pspicture}[shift=-0.4](2,1.5)
\rput(1,1){\psarc[linecolor=blue,linewidth=1.5pt]{-}(0,0){0.5}{0}{180}}
\facegrid{(0,0)}{(2,1)}
\rput(1,0){\psarc[linewidth=0.025]{-}(0,0){0.16}{0}{90}\rput(.5,.5){$\pm\infty$}}
\rput(0,0){\psarc[linewidth=0.025]{-}(0,0){0.16}{0}{90}\rput(.5,.5){$\pm\infty$}}
\end{pspicture}
\ = \ \begin{pspicture}[shift=-0.4](2,1.5)
\facegrid{(0,0)}{(2,1)}
\rput(1,1){\psarc[linecolor=blue,linewidth=1.5pt,linestyle=dashed,dash=2pt 2pt]
{-}(0,0){0.5}{0}{180}}
\rput(0,1){\psarc[linecolor=blue,linewidth=1.5pt,linestyle=dashed,dash=2pt 2pt]
{-}(0,0){0.5}{270}{360}}
\rput(2,1){\psarc[linecolor=blue,linewidth=1.5pt,linestyle=dashed,dash=2pt 2pt]
{-}(0,0){0.5}{180}{270}}
\rput(1,0){\psarc[linecolor=blue,linewidth=1.5pt]{-}(0,0){0.5}{0}{180}}
\end{pspicture} \ , \qquad
\begin{pspicture}[shift=-0.4](1,1)
\facegrid{(0,0)}{(1,1)}
\rput(0,0){\psarc[linewidth=0.025]{-}(0,0){0.16}{0}{90}\rput(.5,.5){$\pm\infty$}}
\rput(0.5,1){\pscircle[fillstyle=solid,fillcolor=black]{0.05}}
\end{pspicture}
\ = \ \begin{pspicture}[shift=-0.4](1,1)
\facegrid{(0,0)}{(1,1)}
\psline[linecolor=blue,linewidth=1.5pt,linestyle=dashed,dash=2pt 2pt](0,0.5)(1,0.5)
\rput(0.5,1){\pscircle[fillstyle=solid,fillcolor=black]{0.05}}
\rput(0.5,0){\pscircle[fillstyle=solid,fillcolor=black]{0.05}}
\end{pspicture} \ .
\end{equation}
The \textit{boundary braid operators} are defined as
\begin{equation} \label{def:tresseFrontiere}
\begin{pspicture}[shift=-0.9](1,2)
\psline[linecolor=blue,linewidth=1.5pt,linestyle=dashed,dash=2pt 2pt](0.5,1.5)(1,1.5)
\psline[linecolor=blue,linewidth=1.5pt,linestyle=dashed,dash=2pt 2pt](0.5,0.5)(1,0.5)
\triangled \rput(0.40,1){$\pm\infty$}
\end{pspicture}
= \lim_{u\rightarrow \pm \iI\infty} \frac{1}{\delta(u)}
\begin{pspicture}[shift=-0.9](-0.1,0)(1.1,2)
\psline[linecolor=blue,linewidth=1.5pt,linestyle=dashed,dash=2pt 2pt](0.5,1.5)(1,1.5)
\psline[linecolor=blue,linewidth=1.5pt,linestyle=dashed,dash=2pt 2pt](0.5,0.5)(1,0.5)
\triangled \rput(0.45,1){$u$}
\end{pspicture} 
\ =
\vartheta\, \eE^{\pm 3 \iI\lambda/2} \ \ 
\begin{pspicture}[shift=-0.65](0.9,1.5)
\psset{unit=0.75cm,linewidth=0.75\pslinewidth}
\triangled
\psline[linecolor=blue,linewidth=1.5pt](0.74,1.5)(1.25,1.5)
\psline[linecolor=blue,linewidth=1.5pt](0.74,0.5)(1.25,0.5)
\psarc[linecolor=blue,linewidth=1.5pt](0.75,1){0.50}{90}{270}
\end{pspicture}
+ \eE^{\mp 3 \iI \lambda/2} \ \
\begin{pspicture}[shift=-0.65](0.9,1.5)
\psset{unit=0.75cm,linewidth=0.75\pslinewidth}
\triangled
\pscircle[fillstyle=solid,fillcolor=black](0.5,0.5){0.05}
\pscircle[fillstyle=solid,fillcolor=black](0.5,1.5){0.05}
\end{pspicture}\ ,
\qquad
\vartheta = \left\{\begin{array}{cc}
-1 & \textrm{S} \\
1 & \textrm{C}
\end{array}\right..
\end{equation}
These satisfy the crossing symmetry relation
\begin{equation} \label{rel:symCroiseFrontiereTresse}
\begin{pspicture}[shift=-0.9](3,2)
\psline[linecolor=blue,linewidth=1.5pt,linestyle=dashed,dash=2pt 2pt](0.5,1.5)(1.5,1.5)
\psline[linecolor=blue,linewidth=1.5pt,linestyle=dashed,dash=2pt 2pt](0.5,0.5)(1.5,0.5)
\triangled \rput(0.45,1){$\pm\infty$}
\rput(1,0){\losange \rput(1,1){$\pm\infty$}}
\end{pspicture}
= \eE^{\mp 3\iI\lambda} \ 
\begin{pspicture}[shift=-0.9](1,2)
\triangled \rput(0.4,1){$\mp\infty$}
\end{pspicture} \ .
\end{equation}

The \textit{fundamental braid transfer tangle} is defined from the limits of the operators composing $\Db^{1,0}(u)$:
\be \label{def:D10tresse}
\Db_{\infty}= \Db^{1,0}_{\infty}
= \ \begin{pspicture}[shift=-0.9](-1,0)(6,2)
\facegrid{(0,0)}{(5,2)}
\rput(2.5,1.5){$\ldots$}\rput(3.5,1.5){$\ldots$}
\rput(2.5,0.5){$\ldots$}\rput(3.5,0.5){$\ldots$}
\rput(0,1){\rput(0.5,0.5){$\mp \infty$}\psarc[linewidth=0.025]{-}(0,0){0.16}{0}{90}}
\rput(1,1){\rput(0.5,0.5){$\mp \infty$}\psarc[linewidth=0.025]{-}(0,0){0.16}{0}{90}}
\rput(4,1){\rput(0.5,0.5){$\mp \infty$}\psarc[linewidth=0.025]{-}(0,0){0.16}{0}{90}}
\rput(0,0){\rput(0.5,0.5){$\pm \infty$}\psarc[linewidth=0.025]{-}(0,0){0.16}{0}{90}}
\rput(1,0){\rput(0.5,0.5){$\pm \infty$}\psarc[linewidth=0.025]{-}(0,0){0.16}{0}{90}}
\rput(4,0){\rput(0.5,0.5){$\pm \infty$}\psarc[linewidth=0.025]{-}(0,0){0.16}{0}{90}}
\psline[linecolor=blue,linewidth=1.5pt,linestyle=dashed,dash=2pt 2pt](5,1.5)(5.5,1.5)
\psline[linecolor=blue,linewidth=1.5pt,linestyle=dashed,dash=2pt 2pt](5,0.5)(5.5,0.5)
\rput(5,0){\triangleg \rput(0.62,1){$\pm \infty$}}
\psline[linecolor=blue,linewidth=1.5pt,linestyle=dashed,dash=2pt 2pt](0,1.5)(-1,1.5)
\psline[linecolor=blue,linewidth=1.5pt,linestyle=dashed,dash=2pt 2pt](0,0.5)(-1,0.5)
\rput(-1,0){\triangled \rput(0.40,1){$\mp \infty$}}
\end{pspicture}
= \lim_{u\rightarrow \pm \iI\infty}
\frac{\eE^{\pm 4\iI\lambda N}\Db^{1,0}(u)}{\gamma(u) f_{-3} f_{-2}} \ ,
\ee
where $\gamma(u)$ depends on the choice of boundary condition:
\be
\label{eq:gamma}
\text{identical: } \quad\gamma(u) = -w(u) w(3\lambda-u) \ ,
\qquad\qquad\text{mixed: }\quad \gamma(u) = 1 \ .
\ee
Using the braid inversion relation, the Yang-Baxter equation and the crossing symmetry at the boundary, one can show that
\be
\begin{pspicture}[shift=-0.9](-1,0)(5,2)
\facegrid{(0,0)}{(4,2)}
\rput(2.5,1.5){$\ldots$}\rput(3.5,1.5){$\ldots$}
\rput(2.5,0.5){$\ldots$}\rput(3.5,0.5){$\ldots$}
\rput(0,1){\rput(0.5,0.5){$-\infty$}\psarc[linewidth=0.025]{-}(0,0){0.16}{0}{90}}
\rput(1,1){\rput(0.5,0.5){$-\infty$}\psarc[linewidth=0.025]{-}(0,0){0.16}{0}{90}}
\rput(3,1){\rput(0.5,0.5){$-\infty$}\psarc[linewidth=0.025]{-}(0,0){0.16}{0}{90}}
\rput(0,0){\rput(0.5,0.5){$+\infty$}\psarc[linewidth=0.025]{-}(0,0){0.16}{0}{90}}
\rput(1,0){\rput(0.5,0.5){$+\infty$}\psarc[linewidth=0.025]{-}(0,0){0.16}{0}{90}}
\rput(3,0){\rput(0.5,0.5){$+\infty$}\psarc[linewidth=0.025]{-}(0,0){0.16}{0}{90}}
\psline[linecolor=blue,linewidth=1.5pt,linestyle=dashed,dash=2pt 2pt](4,1.5)(4.5,1.5)
\psline[linecolor=blue,linewidth=1.5pt,linestyle=dashed,dash=2pt 2pt](4,0.5)(4.5,0.5)
\rput(4,0){\triangleg \rput(0.62,1){$+\infty$}}
\psline[linecolor=blue,linewidth=1.5pt,linestyle=dashed,dash=2pt 2pt](0,1.5)(-1,1.5)
\psline[linecolor=blue,linewidth=1.5pt,linestyle=dashed,dash=2pt 2pt](0,0.5)(-1,0.5)
\rput(-1,0){\triangled \rput(0.40,1){$-\infty$}}
\end{pspicture}
\ = \
\begin{pspicture}[shift=-0.9](-1,0)(5,2)
\facegrid{(0,0)}{(4,2)}
\rput(2.5,1.5){$\ldots$}\rput(3.5,1.5){$\ldots$}
\rput(2.5,0.5){$\ldots$}\rput(3.5,0.5){$\ldots$}
\rput(0,1){\rput(0.5,0.5){$+\infty$}\psarc[linewidth=0.025]{-}(0,0){0.16}{0}{90}}
\rput(1,1){\rput(0.5,0.5){$+\infty$}\psarc[linewidth=0.025]{-}(0,0){0.16}{0}{90}}
\rput(3,1){\rput(0.5,0.5){$+\infty$}\psarc[linewidth=0.025]{-}(0,0){0.16}{0}{90}}
\rput(0,0){\rput(0.5,0.5){$-\infty$}\psarc[linewidth=0.025]{-}(0,0){0.16}{0}{90}}
\rput(1,0){\rput(0.5,0.5){$-\infty$}\psarc[linewidth=0.025]{-}(0,0){0.16}{0}{90}}
\rput(3,0){\rput(0.5,0.5){$-\infty$}\psarc[linewidth=0.025]{-}(0,0){0.16}{0}{90}}
\psline[linecolor=blue,linewidth=1.5pt,linestyle=dashed,dash=2pt 2pt](4,1.5)(4.5,1.5)
\psline[linecolor=blue,linewidth=1.5pt,linestyle=dashed,dash=2pt 2pt](4,0.5)(4.5,0.5)
\rput(4,0){\triangleg \rput(0.62,1){$-\infty$}}
\psline[linecolor=blue,linewidth=1.5pt,linestyle=dashed,dash=2pt 2pt](0,1.5)(-1,1.5)
\psline[linecolor=blue,linewidth=1.5pt,linestyle=dashed,dash=2pt 2pt](0,0.5)(-1,0.5)
\rput(-1,0){\triangled \rput(0.40,1){$+\infty$}}
\end{pspicture} \ .
\ee
It thus follows that $\Db^{1,0}_{\infty}$ does not depend on the choice of limit $u\rightarrow\pm \iI\infty$ in its definition. 

The \textit{fused braid transfer tangles} are defined as
\begin{subequations} \label{def:tresseFused} 
\begin{alignat}{2}
\Db^{m,0}_{\infty} \label{def:tresse(m,0)}
&= \lim_{u\rightarrow \pm \iI\infty}
\frac{\eE^{\pm 4\iI\lambda mN} \Db^{m,0}(u)}
{f(u_{-3}) f(u_{-2}) \prod_{j=0}^{m-1} \gamma(u_{2j})} \ , \\
\Db^{0,n}_{\infty} &= \Db^{n,0}_{\infty} \ , \\
\Db^{m,n}_{\infty} \label{def:tresse(m,n)}
&= \lim_{u\rightarrow \pm \iI\infty}
\frac{\eE^{\pm 4\iI\lambda (m+n)N} \Db^{m,n}(u)}
{w^{(m)}(u_{m-5/2}) w^{(n)}(\slu_{n-5/2}) f(u_{-3}) f(u_{-2}) f(u_{2m-1})} 
\frac{1}{\prod_{j=0}^{m-1} \gamma(u_{2j}) \prod_{j=0}^{n-1} \gamma(\slu_{2j})} \ ,
\end{alignat} \end{subequations}
for $m,n>1$. The definition \eqref{def:tresseFused} is completed with $\Db^{0,0}_\infty=\Ib$. For the mixed boundary cases, these definitions are similar to those used for the periodic geometry \cite{AMDPP19}. For all choices of boundary condition, the fusion hierarchy relations in the limit $u\rightarrow\pm \iI\infty$ become:
\begin{subequations} \label{eq:HFtresse} 
\begin{alignat}{2}
\Db^{m,0}_{\infty} \label{eq:HFtresse(m,0)}
&= \Db^{m-1,0}_{\infty} \Db^{1,0}_{\infty} - \Db^{m-2,1}_{\infty} \ , \\
\Db^{m,n}_{\infty} \label{eq:HFtresse(m,n)}
&= \Db^{m,0}_{\infty} \Db^{0,n}_{\infty}
- \Db^{m-1,0}_{\infty} \Db^{0,n-1}_{\infty} \ .
\end{alignat} 
\end{subequations}
It immediately follows that each fused braid tangle is a polynomial in $\Db_{\infty}$ that can be expressed as a determinant:
\be
\label{tresseDet(m,0)} \Db^{m,0}_{\infty} = \begin{vmatrix}
\Db_{\infty} & \Db_{\infty} & \Ib & 0 & 0 & 0 \\[0.5ex]
\Ib & \Db_{\infty} & \Db_{\infty} & \Ib & 0 & 0 \\[0.5ex]
0 & \ddots & \ddots & \ddots & \ddots & 0 \\
0 & 0 & \Ib & \Db_{\infty} & \Db_{\infty} & \Ib \\[0.5ex]
0 & 0 & 0 & \Ib & \Db_{\infty} & \Db_{\infty} \\[0.5ex]
0 & 0 & 0 & 0 & \Ib & \Db_{\infty}
\end{vmatrix}, 
\qquad
\begin{pspicture}[shift=-0.1](0.15,1.5) 
$ \Db^{m,n}_{\infty} = \begin{vmatrix}
\pnode[0,1.5ex]{n1} & & & 0 & 0 & 0\hphantom{x} \\
& n \times n & & 0 & 0 & 0\hphantom{x} \\
& & \pnode[0, 0]{n2} & \Ib & 0 & 0\hphantom{x} \\
\hphantom{x}0 & 0 & \Ib & \pnode[0, 1.5ex]{m1} & \\
\hphantom{x}0 & 0 & 0 & & m \times m & \\
\hphantom{x}0 & 0 & 0 & & & \pnode[0, 0]{m2}
\end{vmatrix}$
\psframe(n1)(n2)
\psframe(m1)(m2)
\end{pspicture}.
\ee
As previously in \eqref{def:detFormel(m,n)}, the upper block is associated to $\Db^{0,n}_\infty$ and the lower block to $\Db^{m,0}_{\infty}$.

%%%%%%%%%%%%%
\subsection{$T$-system and $Y$-system}\label{sub:TY.systems}
%%%%%%%%%%%%%

The $T$-system is a set of quadratic equations satisfied by the fused transfer matrices, that follows from the fusion hierarchy relations. For the $\Atwotwo$ model on the strip, it can be written in a uniform way using the determinantal forms given in \cref{sub:det}, namely
\be 
\det(m,0)_0 \det(m-k,0)_{2k+2}
= \Big( \prod^{m-1}_{j=k} \fbm_{2j} \Big) \det(k,m-k)_0 
+ \det(m+1,0)_0 \det(m-k-1,0)_{2k+2} \label{Tsystem}
\ee
where $0 \leq k < m$ and $\det(0,0)=\Ib$. The proof, given in \cref{sub:proof.T-system}, is completely analogous to the one given for the periodic case in \cite{AMDPP19}. Of course, the $T$-system can be written directly in terms of the fused transfer matrices $\Db^{m,n}(u)$. For all boundary conditions, the $T$-system equation reads
\begin{alignat}{2}
& s(2u_{m-3}) s(2u_{m+k}) w^{(m)}(u_{m-5/2}) w^{(m-k)}(u_{m+k-1/2})
\Db^{m,0}_0 \Db^{m-k,0}_{2k+2} \nonumber\\
&= \Big( \prod^{m-1}_{j=k} \wb(-u_{2j-1/2}) \Big) \Big( \prod^{m-2}_{j=k} w(-u_{2j+1/2}) \Big)
s(2u_{k-3}) s(2u_{2m}) w(u_{2m-1/2})
f_{2m} \Db^{k,m-k}_0 \nonumber\\
&+ s(2u_{m-1}) s(2u_{m+k-2}) w^{(m+1)}(u_{m-3/2}) w^{(m-k-1)}(u_{m+k-3/2})
\Db^{m+1,0}_0 \Db^{m-k-1,0}_{2k+2} \  \label{TsystemInt}
\end{alignat}
for $0<k<m$, with $\Db^{0,0}_\ell = f_{\ell-3} f_{\ell-2} \Ib$ used for the case $k=m-1$. A similar result holds for $k=0$ with the first term of the right-hand side replaced by
\begin{alignat}{2}
\Big( \prod^{m-1}_{j=0} &\wb(-u_{2j-1/2}) \Big)\Big( \prod^{m-2}_{j=0} w(-u_{2j+1/2}) \Big)
\nonumber\\& \times 
s(2u_{-3}) s(2u_{2m}) w(u_{-5/2}) w(u_{2m-1/2}) w^{(m)}(-u_{m-3/2}) f_{-3} f_{2m}\Db^{0,m}_0 \ .
\end{alignat}

It is clearly more convenient to work with the determinantal expressions. To express this in terms of a $Y$-system, we first define the functions $\db^m(u)$ as
\be
\db^m(u) = \frac{\det(m+1,0)_0\det(m-1,0)_2}{
\Big(\prod_{j=0}^{m-1}\fbm_{2j}\Big) \det(m,0)_1}, \qquad m \ge 0,
\ee
with $\db^0(u)=0$. It then directly follows from \eqref{Tsystem} that
\be
\Ib+\db^m_0 = \frac{\det(m,0)_0\det(m,0)_2}{
\Big(\prod_{j=0}^{m-1}\fbm_{2j}\Big) \det(m,0)_1}.
\ee
The starting point to derive the $Y$-system equation is to compute the product $\db^m_0\db^m_2$:
\begin{alignat}{2}
\db^m_0\db^m_2 &= \frac{\det(m-1,0)_2 \det(m-1,0)_4}{\prod_{j=0}^{m-1} \fbm_{2j}\prod_{j=1}^{m} \fbm_{2j}} \frac{\det(m+1,0)_0\det(m+1,0)_2}{\det(m,0)_1\det(m,0)_3}
\nonumber\\[0.1cm] & = \frac{\prod_{j=0}^{m}\fbm_{2j}\prod_{j=1}^{m-1}\fbm_{2j}}{\prod_{j=0}^{m-1}\fbm_{2j}\prod_{j=1}^{m}\fbm_{2j}} \frac{(\Ib+\db^{m-1}_2)(\Ib+\db^{m+1}_0)}{(\Ib+\db^m_1)} \frac{\det(m+1,0)_1\det(m-1,0)_3}{\det(m,0)_2 \prod_{j=0}^{m-1}\fbm_{2j+1}}
\nonumber\\[0.1cm] & = \frac{(\Ib+\db^{m-1}_2)(\Ib+\db^{m+1}_0)}{(\Ib+\db^m_1)}\, \db^m_1.
\end{alignat}
The $Y$-system is therefore identical to the one obtained for periodic boundary conditions:
\be 
\label{eq:ddd.Y.system}
\frac{\db^m_0 \db^m_2}{\db^m_1} =
\frac{ \big( \Ib + \db^{m-1}_2 \big) \big( \Ib + \db^{m+1}_0 \big)}
{\Ib+\db^m_1} \ , \qquad m \ge 1 \ .
\ee

%%%%%%%%%%%%%%%%%%%%%%%%%%%%%%%%%%%%%%%%%%%%%%%
%
\section{Polynomiality of the fused transfer matrices}\label{sec:polynomiality}
%
%%%%%%%%%%%%%%%%%%%%%%%%%%%%%%%%%%%%%%%%%%%%%%%

In this section, we prove an important property of the fused transfer matrices $\Db^{m,n}(u)$, namely that they are Laurent polynomials in $z = \eE^{\iI u}$. The proof is divided in three cases. In \cref{sub:poly.D20.D11}, the polynomiality of $\Db^{2,0}(u)$ and $\Db^{1,1}(u)$ is established using the diagrammatic definitions of these transfer matrices given in \cref{sub:fused.with.diagrams}. Then in \cref{sub:poly.Dm0,sub:poly.Dmn}, we respectively prove the polynomiality of $\Db^{m,0}(u)$ and $\Db^{m,n}(u)$ for $m,n >1$, by cleverly using the fusion hierarchy relations. Throughout the section, the arguments are done for generic values of $\lambda$. By continuity, they then hold at all values of $\lambda$, including those for which $q$ is a root of unity. A reader uninterested in these technical proofs may wish to accept this result and skip forward to \cref{sec:closure}.

%%%%%%%%%%%%%
\subsection{Polynomiality of $\Db^{2,0}(u)$ and $\Db^{1,1}(u)$} \label{sub:poly.D20.D11}
%%%%%%%%%%%%%

In this section, we prove that $\Db^{2,0}(u)$ and $\Db^{1,1}(u)$ are Laurent polynomials in $z = \eE^{\iI u}$. Let us first recall that we have two equivalent definitions for these objects. The first definitions in \eqref{eq:HF} express $\Db^{2,0}(u)$ and $\Db^{1,1}(u)$ as functions of $\Db^{1,0}(u)$ and its shifts, whereas the second definitions in \eqref{def:D20D11FullDiag} are diagrammatic and written in terms of projectors. For $\Db^{2,0}(u)$, the potential poles are the zeros of the functions $\alpha^1(2,u)$ and $Z^{2,0}(u)$, or more precisely the common zeros of these two functions:
\begin{alignat}{3}
\textrm{(i)}\ &u =\tfrac {r \pi}2,\qquad
&\textrm{(ii)}\ &u = \lambda+\tfrac {r \pi}2, \qquad
&& \textrm{(iii)}\ u = \tfrac \lambda 2 + 
\left\{\begin{array}{cc}
r\pi & \textrm{SS,}\\[0.1cm]
(r+\tfrac12)\pi & \textrm{CC,}
\end{array}\right. \nonumber\\
\textrm{(iv)}\ &u = \lambda\pm \xi_{(j)}+ r \pi, \qquad
&\textrm{(v)}\ &u = \pm \xi_{(j)} + r \pi, \qquad 
&&\textrm{for}\quad j = 1, \dots, N, \quad r \in \mathbb Z.
\label{eq:polesofD20}
\end{alignat} 
For the mixed cases, there are no potential poles of types (iii). We also note that the functions $\alpha^1(2,u)$ and $Z^{2,0}(u)$ have simple zeros for (i), (ii), (iv) and (v), and double zeros for (iii) for the boundary conditions SS and CC. Likewise the potential poles of $\Db^{1,1}(u)$ are the common zeros of $\beta^1(1,1,u)$ and $Z^{1,1}(u)$, which are the values of $u$ in (i) and (v). In this case, the zeros are double for (i) and simple for (v), and this holds for all boundary conditions.

To prove that $\Db^{2,0}(u)$ and $\Db^{1,1}(u)$ are regular at the points, we note that their diagrammatic definitions can be rewritten as
\begingroup
\allowdisplaybreaks
\begin{subequations}
\label{def:D20D11FullDiagProjs}
\begin{alignat}{2} 
\label{def:D20FullDiagProjs}
\Db^{2,0}(u) &= \frac1{Z^{2,0}(u)} \ 
\psset{unit=1.2cm}
\begin{pspicture}[shift=-1.8](-2.5,0)(7.4,4)
\psline[linecolor=blue,linewidth=1.5pt,linestyle=dashed,dash=2pt 2pt](4.9,3.5)(7,3.5)
\psline[linecolor=blue,linewidth=1.5pt,linestyle=dashed,dash=2pt 2pt](4.9,2.5)(7,2.5)
\psline[linecolor=blue,linewidth=1.5pt,linestyle=dashed,dash=2pt 2pt](4.9,1.5)(7,1.5)
\psline[linecolor=blue,linewidth=1.5pt,linestyle=dashed,dash=2pt 2pt](4.9,0.5)(7,0.5)
\rput(-2.5,0){
\psline[linecolor=blue,linewidth=1.5pt,linestyle=dashed,dash=2pt 2pt](0,3.5)(2.5,3.5)
\psline[linecolor=blue,linewidth=1.5pt,linestyle=dashed,dash=2pt 2pt](0,2.5)(2.5,2.5)
\psline[linecolor=blue,linewidth=1.5pt,linestyle=dashed,dash=2pt 2pt](0,1.5)(2.5,1.5)
\psline[linecolor=blue,linewidth=1.5pt,linestyle=dashed,dash=2pt 2pt](0,0.5)(2.5,0.5)
\rput(0,2){\triangled \rput(0.45,1){\footnotesize$3\lambda\!-\!u_2$}}
\rput(0,1){\losange\rput(1,1){$2u-\lambda$}}
\triangled \rput(0.45,1){\footnotesize$3\lambda\!-\!u_0$}
\rput{-135}(0.28,4.15){\projectorBdryTwo\rput(0.18,1.32){\rput{90}(0,0){$_{2,0}$}}}
}
\rput(0,2){
\facegrid{(0,0)}{(1,2)}
\rput(0,1){\rput(0.5,0.5){$_{u_2+\xi_{(1)}}$}
\psarc[linewidth=0.025]{-}(1,0){0.16}{90}{180}}
\rput(0,0){\rput(0.5,0.5){$_{u_0+\xi_{(1)}}$}
\psarc[linewidth=0.025]{-}(1,0){0.16}{90}{180}}
\rput{-180}(1.3,2){\projectorTwo\rput(0.18,1){\rput{90}(0,0){$_{2,0}$}}}
}
\rput(1.3,2){
\facegrid{(0,0)}{(1,2)}
\rput(0,1){\rput(0.5,0.5){$_{u_2+\xi_{(2)}}$}
\psarc[linewidth=0.025]{-}(1,0){0.16}{90}{180}}
\rput(0,0){\rput(0.5,0.5){$_{u_0+\xi_{(2)}}$}
\psarc[linewidth=0.025]{-}(1,0){0.16}{90}{180}}
\rput{180}(1.3,2){\projectorTwo\rput(0.18,1){\rput{90}(0,0){$_{2,0}$}}}
}
\rput(2.6,2){
\facegrid{(0,0)}{(1,2)}
\rput(0.5,1.5){$\ldots$}
\rput(0.5,0.5){$\ldots$}
\rput{180}(1.3,2){\projectorTwo\rput(0.18,1){\rput{90}(0,0){$_{2,0}$}}}
}
\rput(3.9,2){
\facegrid{(0,0)}{(1,2)}
\rput(0,1){\rput(0.5,0.5){$_{u_2+\xi_{(N)}}$}
\psarc[linewidth=0.025]{-}(1,0){0.16}{90}{180}}
\rput(0,0){\rput(0.5,0.5){$_{u_0+\xi_{(N)}}$}
\psarc[linewidth=0.025]{-}(1,0){0.16}{90}{180}}
\rput{180}(1.3,2){\projectorTwo\rput(0.18,1){\rput{90}(0,0){$_{2,0}$}}}
}
\rput(0,0){
\facegrid{(0,0)}{(1,2)}
\rput(0,1){\rput(0.5,0.5){$_{u_2-\xi_{(1)}}$}
\psarc[linewidth=0.025]{-}(0,0){0.16}{0}{90}}
\rput(0,0){\rput(0.5,0.5){$_{u_0-\xi_{(1)}}$}
\psarc[linewidth=0.025]{-}(0,0){0.16}{0}{90}}
\rput(-0.3,0){\projectorTwo\rput(0.18,1){\rput{90}(0,0){$_{2,0}$}}}
}
\rput(1.3,0){
\facegrid{(0,0)}{(1,2)}
\rput(0,1){\rput(0.5,0.5){$_{u_2-\xi_{(2)}}$}
\psarc[linewidth=0.025]{-}(0,0){0.16}{0}{90}}
\rput(0,0){\rput(0.5,0.5){$_{u_0-\xi_{(2)}}$}
\psarc[linewidth=0.025]{-}(0,0){0.16}{0}{90}}
\rput(-0.3,0){\projectorTwo\rput(0.18,1){\rput{90}(0,0){$_{2,0}$}}}
}
\rput(2.6,0){
\facegrid{(0,0)}{(1,2)}
\rput(0.5,1.5){$\ldots$}
\rput(0.5,0.5){$\ldots$}
\rput(-0.3,0){\projectorTwo\rput(0.18,1){\rput{90}(0,0){$_{2,0}$}}}
}
\rput(3.9,0){
\facegrid{(0,0)}{(1,2)}
\rput(0,1){\rput(0.5,0.5){$_{u_2-\xi_{(N)}}$}
\psarc[linewidth=0.025]{-}(0,0){0.16}{0}{90}}
\rput(0,0){\rput(0.5,0.5){$_{u_0-\xi_{(N)}}$}
\psarc[linewidth=0.025]{-}(0,0){0.16}{0}{90}}
\rput(-0.3,0){\projectorTwo\rput(0.18,1){\rput{90}(0,0){$_{2,0}$}}}
}
\rput(5.4,0){
\rput(1,2){\triangleg \rput(0.65,1){$u_2$}}
\rput(0,1){\losange\rput(1,1){$\lambda-2u$}}
\rput(1,0){\triangleg \rput(0.65,1){$u_0$}}
\rput{45}(1.73,-0.16){\projectorBdryTwo\rput(0.18,1.32){\rput{90}(0,0){$_{2,0}$}}}
}
\end{pspicture}
\ \ ,
\\[0.3cm]
\label{def:D11FullDiagProjs}
\Db^{1,1}(u) &=\frac1{Z^{1,1}(u)} \ 
\psset{unit=1.2cm}
\begin{pspicture}[shift=-1.8](-2.5,0)(7.4,4)
\psline[linecolor=blue,linewidth=1.5pt,linestyle=dashed,dash=2pt 2pt](4.9,3.5)(7,3.5)
\psline[linecolor=blue,linewidth=1.5pt,linestyle=dashed,dash=2pt 2pt](4.9,2.5)(7,2.5)
\psline[linecolor=blue,linewidth=1.5pt,linestyle=dashed,dash=2pt 2pt](4.9,1.5)(7,1.5)
\psline[linecolor=blue,linewidth=1.5pt,linestyle=dashed,dash=2pt 2pt](4.9,0.5)(7,0.5)
\rput(-2.5,0){
\psline[linecolor=blue,linewidth=1.5pt,linestyle=dashed,dash=2pt 2pt](0,3.5)(2.5,3.5)
\psline[linecolor=blue,linewidth=1.5pt,linestyle=dashed,dash=2pt 2pt](0,2.5)(2.5,2.5)
\psline[linecolor=blue,linewidth=1.5pt,linestyle=dashed,dash=2pt 2pt](0,1.5)(2.5,1.5)
\psline[linecolor=blue,linewidth=1.5pt,linestyle=dashed,dash=2pt 2pt](0,0.5)(2.5,0.5)
\rput(0,2){\triangled \rput(0.45,1){\footnotesize$3\lambda\!-\!u_3$}}
\rput(0,1){\losange\rput(1,1){$2u$}}
\triangled \rput(0.45,1){\footnotesize$3\lambda\!-\!u_0$}
\rput{-135}(0.28,4.15){\projectorBdryTwo\rput(0.18,1.32){\rput{90}(0,0){$_{1,1}$}}}
}
\rput(0,2){
\facegrid{(0,0)}{(1,2)}
\rput(0,1){\rput(0.5,0.5){$_{u_3+\xi_{(1)}}$}
\psarc[linewidth=0.025]{-}(1,0){0.16}{90}{180}}
\rput(0,0){\rput(0.5,0.5){$_{u_0+\xi_{(1)}}$}
\psarc[linewidth=0.025]{-}(1,0){0.16}{90}{180}}
\rput{180}(1.3,2){\projectorTwo\rput(0.18,1){\rput{90}(0,0){$_{1,1}$}}}
}
\rput(1.3,2){
\facegrid{(0,0)}{(1,2)}
\rput(0,1){\rput(0.5,0.5){$_{u_3+\xi_{(2)}}$}
\psarc[linewidth=0.025]{-}(1,0){0.16}{90}{180}}
\rput(0,0){\rput(0.5,0.5){$_{u_0+\xi_{(2)}}$}
\psarc[linewidth=0.025]{-}(1,0){0.16}{90}{180}}
\rput{180}(1.3,2){\projectorTwo\rput(0.18,1){\rput{90}(0,0){$_{1,1}$}}}
}
\rput(2.6,2){
\facegrid{(0,0)}{(1,2)}
\rput(0.5,1.5){$\ldots$}
\rput(0.5,0.5){$\ldots$}
\rput{180}(1.3,2){\projectorTwo\rput(0.18,1){\rput{90}(0,0){$_{1,1}$}}}
}
\rput(3.9,2){
\facegrid{(0,0)}{(1,2)}
\rput(0,1){\rput(0.5,0.5){$_{u_3+\xi_{(N)}}$}
\psarc[linewidth=0.025]{-}(1,0){0.16}{90}{180}}
\rput(0,0){\rput(0.5,0.5){$_{u_0+\xi_{(N)}}$}
\psarc[linewidth=0.025]{-}(1,0){0.16}{90}{180}}
\rput{180}(1.3,2){\projectorTwo\rput(0.18,1){\rput{90}(0,0){$_{1,1}$}}}
}
\rput(0,0){
\facegrid{(0,0)}{(1,2)}
\rput(0,1){\rput(0.5,0.5){$_{u_3-\xi_{(1)}}$}
\psarc[linewidth=0.025]{-}(0,0){0.16}{0}{90}}
\rput(0,0){\rput(0.5,0.5){$_{u_0-\xi_{(1)}}$}
\psarc[linewidth=0.025]{-}(0,0){0.16}{0}{90}}
\rput(-0.3,0){\projectorTwo\rput(0.18,1){\rput{90}(0,0){$_{1,1}$}}}
}
\rput(1.3,0){
\facegrid{(0,0)}{(1,2)}
\rput(0,1){\rput(0.5,0.5){$_{u_3-\xi_{(2)}}$}
\psarc[linewidth=0.025]{-}(0,0){0.16}{0}{90}}
\rput(0,0){\rput(0.5,0.5){$_{u_0-\xi_{(2)}}$}
\psarc[linewidth=0.025]{-}(0,0){0.16}{0}{90}}
\rput(-0.3,0){\projectorTwo\rput(0.18,1){\rput{90}(0,0){$_{1,1}$}}}
}
\rput(2.6,0){
\facegrid{(0,0)}{(1,2)}
\rput(0.5,1.5){$\ldots$}
\rput(0.5,0.5){$\ldots$}
\rput(-0.3,0){\projectorTwo\rput(0.18,1){\rput{90}(0,0){$_{1,1}$}}}
}
\rput(3.9,0){
\facegrid{(0,0)}{(1,2)}
\rput(0,1){\rput(0.5,0.5){$_{u_3-\xi_{(N)}}$}
\psarc[linewidth=0.025]{-}(0,0){0.16}{0}{90}}
\rput(0,0){\rput(0.5,0.5){$_{u_0-\xi_{(N)}}$}
\psarc[linewidth=0.025]{-}(0,0){0.16}{0}{90}}
\rput(-0.3,0){\projectorTwo\rput(0.18,1){\rput{90}(0,0){$_{1,1}$}}}
}
\rput(5.4,0){
\rput(1,2){\triangleg \rput(0.65,1){$u_3$}}
\rput(0,1){\losange\rput(1,1){$-2u$}}
\rput(1,0){\triangleg \rput(0.65,1){$u_0$}}
\rput{45}(1.73,-0.16){\projectorBdryTwo\rput(0.18,1.32){\rput{90}(0,0){$_{1,1}$}}}
}
\end{pspicture}
\ \ .
\end{alignat}
\end{subequations}
\endgroup
We justify this claim as follows. For $\Db^{2,0}(u)$, let us choose any of the projectors $P^{2,0}$ in \eqref{def:D20FullDiagProjs} and expand it using \eqref{def:P20P11explicite}. The black triangle of the second term in this decomposition points leftwards if we selected a projector in the bottom part of the diagram, and to the right if it belongs to the top part. This triangle is then either pushed through a pair of face operators using \eqref{rel:PTbulkTCN}, or it reflects on the boundary using \eqref{rel:PTfrontiereTCN}. In all cases, this black triangle then connects with another $P^{2,0}$ projector, yielding a zero result due to the third identity in \eqref{eq:spider.ids}. Only the first term of the decomposition \eqref{def:P20P11explicite} survives, and the final result is a diagram identical to the original one, but with this projector now absent. This argument is repeated until all projectors are removed except for one. The same argument applies for $\Db^{1,1}(u)$ and its $P^{1,1}$ projectors.

These expressions with many projectors allow us to investigate the polynomiality of $\Db^{2,0}(u)$ and $\Db^{1,1}(u)$ in terms of some of their constituting parts. First, let us consider the combinations
\begin{equation} \label{eq:zerosBulk(2,0)}
\frac1{s(u_{-1}-\xi_{(j)}) s(u_0-\xi_{(j)})} \ 
\psset{unit=1.2}
\begin{pspicture}[shift=-0.9](-0.3,0)(1,2.0)
\facegrid{(0,0)}{(1,2)}
\rput(0,1){\rput(0.5,0.5){$_{u_2-\xi_{(j)}}$}
\psarc[linewidth=0.025]{-}(0,0){0.16}{0}{90}}
\rput(0,0){\rput(0.5,0.5){$_{u_0-\xi_{(j)}}$}
\psarc[linewidth=0.025]{-}(0,0){0.16}{0}{90}}
\rput(-0.3,0){\projectorTwo\rput(0.18,1){\rput{90}(0,0){$_{2,0}$}}}
\end{pspicture} \ \ , 
\qquad\quad
\frac1{s(u_0-\xi_{(j)})} \ 
\begin{pspicture}[shift=-0.9](-0.3,0)(1,2)
\facegrid{(0,0)}{(1,2)}
\rput(0,1){\rput(0.5,0.5){$_{u_3-\xi_{(j)}}$}
\psarc[linewidth=0.025]{-}(0,0){0.16}{0}{90}}
\rput(0,0){\rput(0.5,0.5){$_{u_0-\xi_{(j)}}$}
\psarc[linewidth=0.025]{-}(0,0){0.16}{0}{90}}
\rput(-0.3,0){\projectorTwo\rput(0.18,1){\rput{90}(0,0){$_{1,1}$}}}
\end{pspicture} \ \ .
\end{equation}
Each of these combinations is a Laurent polynomial in $z=\eE^{\iI u}$. To show this, we note that 
\begin{subequations}
\begin{alignat}{2}
\label{eq:vanishing.with.P20}
\psset{unit=1.2}
\begin{pspicture}[shift=-0.9](-0.3,0)(1.2,2.4)
\facegrid{(0,0)}{(1,2)}
\rput(0,1){\rput(0.5,0.5){$_{u_2-\xi_{(j)}}$}
\psarc[linewidth=0.025]{-}(0,0){0.16}{0}{90}}
\rput(0,0){\rput(0.5,0.5){$_{u_0-\xi_{(j)}}$}
\psarc[linewidth=0.025]{-}(0,0){0.16}{0}{90}}
\rput(-0.3,0){\projectorTwo\rput(0.18,1){\rput{90}(0,0){$_{2,0}$}}}
\end{pspicture}
\Big\rvert_{u=\xi_{(j)}}
&= s(\lambda) s^2(2\lambda) s(3\lambda) \
\begin{pspicture}[shift=-0.9](-0.3,0)(1.2,2.4)
\rput(0,1){ \loopa \psline[linewidth=1pt](0,1)(1,0)
\pspolygon[fillstyle=solid,fillcolor=black](0.65,0.65)(0.75,0.65)(0.75,0.75)(0.65,0.75)
\pspolygon[fillstyle=solid,fillcolor=black](0.35,0.35)(0.25,0.35)(0.25,0.25)(0.35,0.25)}
\rput(0,0)\loopid
\rput(-0.3,0){\projectorTwo\rput(0.18,1){\rput{90}(0,0){$_{2,0}$}}}
\end{pspicture}
= 0 \ ,
\\
\psset{unit=1.2}
\begin{pspicture}[shift=-0.9](-0.3,0)(1.2,2.4)
\facegrid{(0,0)}{(1,2)}
\rput(0,1){\rput(0.5,0.5){$_{u_3-\xi_{(j)}}$}
\psarc[linewidth=0.025]{-}(0,0){0.16}{0}{90}}
\rput(0,0){\rput(0.5,0.5){$_{u_0-\xi_{(j)}}$}
\psarc[linewidth=0.025]{-}(0,0){0.16}{0}{90}}
\rput(-0.3,0){\projectorTwo\rput(0.18,1){\rput{90}(0,0){$_{1,1}$}}}
\end{pspicture}
\Big\rvert_{u=\xi_{(j)}}
&= s^2(2\lambda) s^2(3\lambda) \
\begin{pspicture}[shift=-0.9](-0.3,0)(1.2,2.4)
\rput(0,1){\loopej}
\rput(0,0){\loopid}
\rput(-0.3,0){\projectorTwo\rput(0.18,1){\rput{90}(0,0){$_{1,1}$}}}
\end{pspicture}
= 0 \ ,
\end{alignat}
\end{subequations}
where we used \eqref{rel:evalRemarquable0et3}, \eqref{rel:evalRemarquable1et2} and \eqref{eq:spider.ids}. Similar identities holds with $u$ shifted by $\pi$, and likewise for \eqref{eq:vanishing.with.P20} with $u = \lambda + \xi_{(j)} + r \pi$. The combinations \eqref{eq:zerosBulk(2,0)} appear in the bottom bulk sections of the diagrams in \eqref{def:D20D11FullDiagProjs}, which are now understood to be Laurent polynomials. The same holds for the top bulk parts of these diagrams, as this follows from repeating the  argument with $\xi_{(j)} \mapsto - \xi_{(j)}$, using combinations similar to those in \eqref{eq:zerosBulk(2,0)} but with projectors attached to the right. This ends the proof that $\Db^{2,0}(u)$ is regular at the points (iv) and (v), and likewise that $\Db^{1,1}(u)$ is regular at the points (v).

The proofs of the regularity at the point (i), (ii) and (iii) are instead related to the operators on the boundaries. In this case, we consider the combinations
\begin{equation} \label{eq:second.combinations}
\frac1{s(2u_{-1}) \delta_L(u_{-3/2}) \delta_L(u_{-1/2})} \  
\psset{unit=1.2}
\begin{pspicture}[shift=-1.9](0,0)(2,4)
\rput(0,2){\triangled \rput(0.45,1){$_{3\lambda-u_2}$}}
\rput(0,1){\losange\rput(1,1){$2u-\lambda$}}
\triangled \rput(0.45,1){$_{3\lambda-u_0}$}
\rput{-135}(0.28,4.15){\projectorBdryTwo\rput(0.18,1.32){\rput{90}(0,0){$_{2,0}$}}}
\end{pspicture}
\ \ ,
\qquad \quad
\frac1{s(2u_0) \delta_L(u_{-3/2})} \  
\begin{pspicture}[shift=-1.9](0,0)(2,4)
\rput(0,2){\triangled \rput(0.45,1){$_{3\lambda-u_3}$}}
\rput(0,1){\losange\rput(1,1){$2u$}}
\triangled \rput(0.45,1){$_{3\lambda-u_0}$}
\rput{-135}(0.28,4.15){\projectorBdryTwo\rput(0.18,1.32){\rput{90}(0,0){$_{1,1}$}}}
\end{pspicture}\ \  .
\end{equation}
These are also Laurent polynomials. The proof uses the same arguments, namely one shows using \eqref{eq:bdy.identities.type1}, \eqref{eq:bdy.identities.type2a} and \eqref{eq:bdy.identities.type2b} that the diagrams evaluate to zero if $u$ is specified to a value where the denominator vanishes. Similar identities hold on the right boundary. The relevant combinations are obtained from \eqref{eq:second.combinations} by rotating the diagram, replacing $\delta_L$ by $\delta_R$, and changing $u \mapsto \lambda - u$ and $u \mapsto -u$ for the first and second combinations, respectively. This allows us to conclude that the expressions \eqref{def:D20D11FullDiagProjs} for $\Db^{2,0}(u)$ and $\Db^{1,1}(u)$ are regular at a number of values of $u$ that includes all the points of type (i), (ii) and (iii). This is also true in the cases of double zeros, as in these cases, one zero is accounted for in the combination corresponding to the left boundary and the second for the one in the right boundary. This ends the proof of the polynomiality of $\Db^{2,0}(u)$ and $\Db^{1,1}(u)$.

%%%%%%%%%%%%%
\subsection{Polynomiality of $\Db^{m,0}(u)$} \label{sub:poly.Dm0}
%%%%%%%%%%%%%

\paragraph{Idea of the inductive proof.}
The polynomiality of $\Db^{m,0}(u)$ is already established for $m=1$ and $m=2$. This section proves that this also holds for $m\ge 3$, using induction on $m$. Dividing \eqref{eq:Dm0mu} and \eqref{eq:Dm0nu} by $\mu^1(m,u)$ and $\nu^1(m,u)$ respectively gives two different expressions for $\Db^{m,0}(u)$ in terms of fused transfer matrices $\Db^{m',0}(u)$ with $m'<m$. By the inductive hypothesis, these transfer matrices with lower fusion indices are assumed to be Laurent
polynomials in $z = \eE^{\iI u}$. Therefore the only potential poles for $\Db^{m,0}(u)$ are the zeros of $\mu^1(m,u)$ and $\nu^1(m,u)$, or more precisely the common zeros of these two functions. Upon inspection, we see that they share the factor $s(2u_{m-3})s(2u_{m-2})w^{(m)}(u_{m-5/2})$. Therefore $\Db^{m,0}(u)$ potentially has poles at the following values of $u$:
\be
\textrm{(i)} \  u = (2-m)\lambda + \tfrac {r \pi}2, 
\qquad
\textrm{(ii)} \  u = (3-m)\lambda + \tfrac {r \pi}2, 
\qquad
\textrm{(iii)} \  u = (\tfrac52-m)\lambda + r \pi + 
\left\{\begin{array}{cc}
m \tfrac \pi 2 & \textrm{SS,}\\
(m+1) \tfrac \pi 2 & \textrm{CC,}\\
\end{array}\right. 
\ee
with $r \in \mathbb Z$.
Of course, there is no potential pole of type (iii) for the mixed cases SC and CS. For $m \ge 4$, all the other zeros of $\mu^1(m,u)$ are not zeros of $\nu^1(m,u)$, so the above list exhausts all the possible problematic values of $u$. Let us thus first consider $m \ge 4$. (The case $m=3$ is discussed at the end of the section). Below we prove that $\Db^{m,0}(u)$ is regular at each of these points.

\paragraph{Proof for (i) and (ii).} From the periodicity and crossing symmetry, we have
\be
\Db^{m,0}\big((2-m)\lambda + \tfrac{r \pi}2\big) = \Db^{m,0}\big((3-m)\lambda + \tfrac{r \pi}2\big).
\ee
As a result, the proof of regularity for (i) will imply the same result for (ii). We thus focus on (i) and define $\widehat u = (2-m)\lambda + \frac{r \pi}2$. The function $\nu^1(m,u)$ has a zero of degree one at $u = \widehat u$. To show that $\Db^{m,0}(\widehat u)$ is non-singular, we must show that
\be
\Kb(m)= \lim_{u \to \widehat u} \left(\nu^2(m,u) \Db^{1,0}_0 \Db^{m-1,0}_2 - \nu^3(m,u) \Db^{1,0}_1 \Db^{m-2,0}_4 + \nu^4(m,u) \Db^{m-3,0}_6\right) = 0.
\ee
We readily observe that $\nu^2(m,u)=0$ at $u = \widehat u$ because it has a factor of $s(2u_{m-2})$. In contrast, $\nu^3(m,\widehat u)$ and $\nu^4(m,\widehat u)$ are non-zero. We thus have
\be
\Kb(m) = - \nu^3(m,\widehat u) \Db^{1,0}_1(\widehat u) \Db^{m-2,0}_4(\widehat u) + \nu^4(m,\widehat u) \Db^{m-3,0}_6(\widehat u).
\ee
As a first step, we note that 
\be \label{eq:casm=4}
\Db^{m-3,0}_6(\widehat u) = \frac{\mu^2(m-3,\widehat u_6)}{\mu^1(m-3,\widehat u_6)} \Db^{m-4,0}_6(\widehat u)\Db^{1,0}_{2m-2}(\widehat u),
\ee
with the other two terms absent due to zeros of $\mu^3(m-3,\widehat u_6)$ and $\mu^4(m-3,\widehat u_6)$. 
This holds for $m \ge 4$.\footnote{For $m=4$, $\Db^{m-3,0}_{6}(\widehat u)=\Db^{1,0}_{6}(\widehat u)$. The relation \eqref{eq:Dm0mu} holds trivially as
$\mu^1(1,u)\Db^{1,0}(u) = \mu^2(1,u)\Db^{0,0}(u)\Db^{1,0}(u)$ and all other terms vanish due to negative fusion indices. In \eqref{eq:casm=4}, each $\mu^i(m-3,\widehat u_6)$ vanishes, but one can check that the equality still holds by using a limit and understanding $\frac{\mu^2(m-3,\widehat u_6)}{\mu^1(m-3,\widehat u_6)}$ as $\lim_{u \to \widehat u}\frac{\mu^2(m-3, u_6)}{\mu^1(m-3, u_6)}$.}
From crossing symmetry and periodicity, we have $\Db^{1,0}_{2m-2}(\widehat u)=\Db^{1,0}_{1}(\widehat u)$ and
\be
\label{eq:KmJm}
\Kb(m) = - \nu^3(m,\widehat u) \Db^{1,0}_1(\widehat u) \Big( \underbrace{\Db^{m-2,0}_4(\widehat u) -\frac{\nu^4(m,\widehat u)}{\nu^3(m,\widehat u)} \frac{\mu^2(m-3,\widehat u_6)}{\mu^1(m-3,\widehat u_6)} \Db^{m-4,0}_6(\widehat u)}_{=\boldsymbol{J}(m)} \Big).
\ee
The factor in front of the parenthesis is non-zero leaving us to show that the content of the parenthesis, $\Jb(m)$, vanishes. We check the cases $m=4,5,6$ explicitly:
\begin{subequations}
\label{eq:J4J5J6}
\begin{alignat}{3}
\Jb(4) &= \Db^{2,0}_4(\widehat u) -\frac{\nu^4(4,\widehat u)}{\nu^3(4,\widehat u)} \frac{\mu^2(1,\widehat u_6)}{\mu^1(1,\widehat u_6)} \Db^{0,0}_6(\widehat u), \qquad && \widehat u = -2\lambda + \tfrac{r \pi}2, 
\\
\Jb(5) &= \Db^{3,0}_4(\widehat u) -\frac{\nu^4(5,\widehat u)}{\nu^3(5,\widehat u)} \frac{\mu^2(2,\widehat u_6)}{\mu^1(2,\widehat u_6)} \Db^{1,0}_6(\widehat u),\qquad && \widehat u = -3\lambda + \tfrac{r \pi}2,
\\
\Jb(6) &= \Db^{4,0}_4(\widehat u) -\frac{\nu^4(6,\widehat u)}{\nu^3(6,\widehat u)} \frac{\mu^2(3,\widehat u_6)}{\mu^1(3,\widehat u_6)} \Db^{2,0}_6(\widehat u),\qquad && \widehat u = -4\lambda + \tfrac{r \pi}2.
\end{alignat}
\end{subequations}
All six transfer tangles in \eqref{eq:J4J5J6} turn out to be proportional to the identity $\Ib$. Indeed, using the crossing symmetry, periodicity and conjugacy of the transfer matrices as well as the reduction relations \eqref{eq:Dm0.reds}, one can write each of these tangles in terms of the objects in \eqref{eq:D012.ids}. Verifying that $\Jb(4)$, $\Jb(5)$ and $\Jb(6)$ vanish is then straightforward, as it amounts to checking three equalities satisfied by finite products of trigonometric factors.

Returning to the cases $m\ge7$ with $\widehat u = (2-m)\lambda + \frac{r \pi}2$, we apply \eqref{eq:Dm0.reds} to both transfer matrices in $\Jb(m)$ and find
\begin{alignat}{2}
\Jb(m) &= \frac{\mu_4(m-2,\widehat u_4)}{\mu_1(m-2,\widehat u_4)}\Db^{m-5,0}_4(\widehat u) - \frac{\nu^4(m,\widehat u)}{\nu^3(m,\widehat u)}\frac{\mu^2(m-3,\widehat u_6)}{\mu^1(m-3,\widehat u_6)}\frac{\mu^4(m-4,\widehat u_6)}{\mu^1(m-4,\widehat u_6)}\Db^{m-7,0}_6(\widehat u)\nonumber\\
&=\frac{\mu_4(m-2,\widehat u_4)}{\mu_1(m-2,\widehat u_4)}\Jb(m-3).
\end{alignat}
The last equality stems from the definition \eqref{eq:KmJm} of $\Jb(m-3)$ and the relations
\begin{subequations}
\begin{alignat}{2}
&\frac{\nu^4(m,\widehat u)}{\nu^3(m,\widehat u)}\frac{\mu^1(m-2,\widehat u_4)}{\mu^4(m-2,\widehat u_4)}\frac{\mu^2(m-3,\widehat u_6)}{\mu^1(m-3,\widehat u_6)}\frac{\mu^4(m-4,\widehat u_6)}{\mu^1(m-4,\widehat u_6)} = \frac{\nu^4(m-3,\widehat v)}{\nu^3(m-3,\widehat v)}\frac{\mu^2(m-6,\widehat v_6)}{\mu^1(m-6,\widehat v_6)},
\label{eq:mu.nu.ratio}\\[0.15cm]
\Db^{m-5,0}_4&(\widehat u) = \Db^{m-5,0}_4(\widehat v), \qquad 
\Db^{m-7,0}_6(\widehat u) = \Db^{m-7,0}_6(\widehat v), \qquad \widehat v = \widehat u \big|_{m\to m-3} = (5-m)\lambda + \tfrac{r \pi}2,
\end{alignat}
valid for $m\ge7$.\footnote{This also works in the case $m=7$, in which case the functions $\mu^1(m-6,\widehat v_6)$ and $\mu^2(m-6,\widehat v_6)$ both vanish and their ratio in \eqref{eq:mu.nu.ratio} should be interpreted as $\lim_{v \to \widehat v} \frac{\mu^2(m-6,v_6)}{\mu^1(m-6, v_6)}$.} By the induction hypothesis, $\Db^{m-3,0}\big((5-m)\lambda+\frac{r\pi}2\big)$ is finite and therefore $\Jb(m-3)=0$. Because $\mu_1(m-2,\widehat u_4) \neq 0$ we deduce that $\Jb(m)=0$. This ends the proof that $\Db^{m,0}(u)$ is regular at $u=(2-m)\lambda + \frac{r \pi}2$ and $(3-m)\lambda + \frac{r \pi}2$.
\end{subequations}

\paragraph{Proof for (iii).} This case applies only to SS and CC boundary conditions. We start by deriving a new functional equation. Recall that a section of the matrix used to defined $\det(m,0)(u)$ takes the form
\be
\det(m,0)(u) = \begin{vmatrix}[rccc:cccc]
\ddots & \ddots & \ddots & 0 &  \\
\ddots & \Dbm_{2k+4} & \Dbm_{2k+3} & \fbm_{2k+1} & 0 &  &  &  \\
0 & \fbm_{2k+2} & \Dbm_{2k+2} & \Dbm_{2k+1} & \fbm_{2k-1} & 0 &  \\
& 0 & \fbm_{2k} & \Dbm_{2k} & \Dbm_{2k-1} & \fbm_{2k-3} & 0 &  \\\cdashline{2-7}
& & 0 & \fbm_{2k-2} & \Dbm_{2k-2} & \Dbm_{2k-3} & \fbm_{2k-5} & 0 \\
& & & 0 & \fbm_{2k-4} & \Dbm_{2k-4} & \Dbm_{2k-5} & \ddots \\
& & & & 0 & \fbm_{2k-6} & \Dbm_{2k-6} & \ddots \\
& & & & & 0 & \ddots & \ddots \\
\end{vmatrix}.
\ee
Using the determinant minor expansion, we obtain the five-term relation
\begin{alignat}{2}
\det(m,0)_0 &= \det(m-k,0)_{2k}\det(k,0)_0 + \fbm_{2k-2}\fbm_{2k-1}\fbm_{2k} \det(m-k-2,0)_{2k+4}\det(k-1,0)_0 \nonumber\\&- \fbm_{2k-2} \Dbm_{2k-1} \det(m-k-1,0)_{2k+2}\det(k-1,0)_0 \nonumber\\&+ \fbm_{2k-4}\fbm_{2k-3}\fbm_{2k-2} \det(m-k-1,0)_{2k+2}\det(k-2,0)_0,
\qquad k = 1, \dots, m-1.
\label{eq:5term.rec}
\end{alignat}
Rewriting this equation in terms of the fused transfer tangles, we find
\begin{alignat}{2}
\label{eq:Dm.with.taus}
\tau^1(m,k,u) \Db^{m,0}_0 &= 
\tau^2(m,k,u) \Db^{m-k,0}_{2k}\Db^{k,0}_0
+\tau^3(m,k,u) \Db^{m-k-2,0}_{2k+4}\Db^{k-1,0}_0
\nonumber\\&+\tau^4(m,k,u) \Db^{m-k-1,0}_{2k+2}\Db^{1,0}_{2k-1}\Db^{k-1,0}_0
+\tau^5(m,k,u) \Db^{m-k-1,0}_{2k+2} \Db^{k-2,0}_0, 
\end{alignat}
where $k = 2, \dots, m-2$ and
\begingroup \allowdisplaybreaks
\begin{subequations}
\begin{alignat}{2}
\tau^1(m,k,u) &= s(2u_{m-2})s(2u_{m-3})s(2u_{2k-4})s(2u_{2k-3})s(2u_{2k-2})s(2u_{2k-1})
 w(u_{2k-5/2})w^{(m)}(u_{m-5/2}) \nonumber\\&\times f(u_{2k-4})f(u_{2k-3})f(u_{2k-2})f(u_{2k-1}),
\\[0.1cm]
\tau^2(m,k,u) &= s(2u_{k-3})s(2u_{k-2})s(2u_{2k-4})s(2u_{2k-1})s(2u_{m+k-3})s(2u_{m+k-2})
w^{(m-k)}(u_{m+k-5/2})\nonumber\\&\times w^{(k)}(u_{k-5/2}) f(u_{2k-4})f(u_{2k-1}),
\\[0.1cm]
\tau^3(m,k,u) &=s(2u_{k-4})s(2u_{k-3})s(2u_{2k-4})s(2u_{2k-3})s(2u_{m+k-1})s(2u_{m+k})
w(u_{2k-7/2}) w(u_{2k-5/2}) \nonumber\\&\times w(u_{2k+1/2})\bar w(u_{2k-5/2})\bar w(u_{2k-3/2})\bar w(u_{2k-1/2}) w^{(m-k-2)}(u_{m+k-1/2})w^{(k-1)}(u_{k-7/2}) \nonumber\\&\times f(u_{2k-4})f(u_{2k-3}),
\\[0.1cm]
\tau^4(m,k,u) &= -s(2u_{k-4})s(2u_{k-3})s(2u_{2k-3})s(2u_{2k-2})s(2u_{m+k-2})s(2u_{m+k-1})
\bar w(u_{2k-5/2})\nonumber\\&\times w^{(m-k-1)}(u_{m+k-3/2})w^{(1)}(u_{2k-5/2})w^{(k-1)}(u_{k-7/2}),
\\[0.1cm]
\tau^5(m,k,u) &= s(2u_{k-5})s(2u_{k-4})s(2u_{2k-2})s(2u_{2k-1})s(2u_{m+k-2})s(2u_{m+k-1})
w(u_{2k-11/2})w(u_{2k-5/2}) \nonumber\\&\times w(u_{2k-3/2}) \bar w(u_{2k-9/2})\bar w(u_{2k-7/2})\bar w(u_{2k-5/2})w^{(m-k-1)}(u_{m+k-3/2})w^{(k-2)}(u_{k-9/2}) \nonumber\\&\times f(u_{2k-2})f(u_{2k-1}).
\end{alignat}
\end{subequations}
\endgroup
The determinant relations \eqref{eq:5term.rec} for $k=1$ and $k=m-1$ correspond to \eqref{eq:Dm0mu} and \eqref{eq:Dm0nu}. For all values of $k$, the function $\tau^1(m,k,u)$ contains the problematic factor $w^{(m)}(u_{m-5/2})$, which may produce potential poles for $\Db^{m,0}(u)$ for SS and CC boundary conditions. To show that this is not the case, we treat separately the cases of $m$ odd and even. 

Starting with $m$ odd, we set $\widehat u = (\frac52 -m ) \lambda + (r+\frac12) \pi$ for SS boundary conditions and $\widehat u = (\frac52 -m ) \lambda + r \pi$ for CC boundary conditions. We use the relation \eqref{eq:Dm.with.taus} with $k = \frac{m+1}2$, noting that $\tau^5(m,\frac{m+1}2,\widehat u) = 0$ and that the degree of the zero of $\tau^1(m,\frac{m+1}2,u)$ at $u = \widehat u$ is one. Thus to prove that $\Db^{m,0}(\widehat u)$ is non-singular, we must show that  
\begin{alignat}{2}
&\tau^2(m,\tfrac{m+1}2,\widehat u) \Db^{(m-1)/2,0}_{m+1}(\widehat u)\Db^{(m+1)/2,0}_0(\widehat u)
+\tau^3(m,\tfrac{m+1}2,\widehat u) \Db^{(m-5)/2,0}_{m+5}(\widehat u) \Db^{(m-1)/2,0}_0(\widehat u)
\nonumber\\
&+\tau^4(m,\tfrac{m+1}2,\widehat u) \Db^{(m-3)/2,0}_{m+3}(\widehat u)\Db^{1,0}_{m}(\widehat u)\Db^{(m-1)/2,0}_0(\widehat u)
 = 0.
 \label{eq:must.be.zero}
\end{alignat}
With the function $\mu^1(\frac{m+1}2,\widehat u)$ non-zero, we rewrite $\Db^{(m+1)/2,0}(\widehat u)$ as
\be
\Db^{(m+1)/2,0}_0(\widehat u) = - \frac{\mu^3(\frac{m+1}2,\widehat u)}{\mu^1(\frac{m+1}2,\widehat u)}\Db^{(m-3)/2,0}_0(\widehat u) \Db^{1,0}_{m-2}(\widehat u)+\frac{\mu^4(\frac{m+1}2,\widehat u)}{\mu^1(\frac{m+1}2,\widehat u)} \Db^{(m-5)/2,0}_0(\widehat u).
\ee
This expresses the left side of \eqref{eq:must.be.zero} as a sum of four terms. With crossing symmetry, we find that two of them are proportional to $\Db^{(m-5)/2,0}_{m+5}(\widehat u) \Db^{(m-1)/2,0}_0(\widehat u)$, whereas the other two are proportional to $\Db^{(m-3)/2,0}_{m+3}(\widehat u)\Db^{1,0}_{m}(\widehat u)\Db^{(m-1)/2,0}_0(\widehat u)$. It is then straightforward to check that the sum of their coefficients vanishes. This also holds for the special case $m=5$ where $\Db^{(m-5)/2,0}_0(\widehat u)\mapsto f_{-3}f_{-2}(\widehat u)$, and using $f_{-3}(5\lambda-u)f_{-2}(5\lambda-u)=f_{-3}(u)f_{-2}(u)$. This ends the proof for $m$ odd.

For $m$ even, we define $\widehat u = (\tfrac52-m)\lambda + r \pi$ for SS boundary conditions and $\widehat u = (\tfrac52-m)\lambda + (r+\frac12) \pi$ for CC boundary conditions. In this case, we use the relation \eqref{eq:Dm.with.taus} with $k = \frac{m}2$. We note that $\tau^j(m,\frac m2,u)$ vanishes at $u = \widehat u$ for $j=1,3,5$, and that $\tau^1(m,\frac m2,u)$ has a double zero whereas $\tau^3(m,\frac m2,u)$ and $\tau^5(m,\frac m2,u)$ have simple zeros. To show that $\Db^{m,0}(u)$ is regular at $u = \widehat u$, we show that 
\be
\Lb = \lim_{u \to \widehat u}\frac{\tau^1(m,\frac m2,u)\Db^{m,0}(u)}{w^{(m)}(u_{m-5/2})} = 0,
\ee
or equivalently
\begin{alignat}{2}
\Lb = \lim_{u \to \widehat u}&\frac1{w^{(m)}(u_{m-5/2})}\bigg[
\tau^2(m,\tfrac m2,u) \Db^{m/2,0}_{m}\Db^{m/2,0}_0
+\tau^3(m,\tfrac m2,u) \Db^{m/2-2,0}_{m+4}\Db^{m/2-1,0}_0
\nonumber\\
&\hspace{1cm}+\tau^4(m,\tfrac m2,u) \Db^{m/2-1,0}_{m+2}\Db^{1,0}_{m-1}\Db^{m/2-1,0}_0
+\tau^5(m,\tfrac m2,u) \Db^{m/2-1,0}_{m+2} \Db^{m/2-2,0}_0\bigg] = 0.
\end{alignat}
Because $\tau^3(m,\frac m2,\widehat u)=\tau^5(m,\frac m2,\widehat u)=0$, the limits of the second and fourth terms are evaluated directly and their sum is found to vanish due to the relations
\be
\label{eq:some.relations}
\Db^{m/2-j,0}_{m+2j}(\widehat u) = \Db^{m/2-j,0}_0(\widehat u),
\qquad
\lim_{u \to \widehat u}\frac{\tau^3(m,\frac m2,u)}{w^{(m)}(u_{m-5/2})}
=-\lim_{u \to \widehat u}\frac{\tau^5(m,\frac m2,u)}{w^{(m)}(u_{m-5/2})},
\ee
the first of which is obtained by crossing symmetry. 
For the remaining terms, we expand $\Db^{m/2,0}_{m}$ and $\Db^{m/2,0}_0$ as
\begin{subequations}
\label{eq:two.Dmu.relations}
\begin{alignat}{2}
\Db^{m/2,0}_0 &= \frac{\mu^2(\frac m2,u)}{\mu^1(\frac m2,u)}\Db^{m/2-1,0}_0\Db^{1,0}_{m-2} - \frac{\mu^3(\frac m2,u)}{\mu^1(\frac m2,u)}\Db^{m/2-2,0}_0 \Db^{1,0}_{m-3}+\frac{\mu^4(\frac m2,u)}{\mu^1(\frac m2,u)} \Db^{m/2-3,0}_0,
\\
\Db^{m/2,0}_m &= \frac{\nu^2(\frac m2,u_m)}{\nu^1(\frac m2,u_m)}\Db^{1,0}_{m}\Db^{m/2-1,0}_{m+2} - \frac{\nu^3(\frac m2,u_m)}{\nu^1(\frac m2,u_m)}\Db^{1,0}_{m+1}\Db^{m/2-2,0}_{m+4} +\frac{\nu^4(\frac m2,u_m)}{\nu^1(\frac m2,u_m)} \Db^{m/2-3,0}_{m+6}.
\end{alignat}
\end{subequations}
The functions $\mu^3(\frac m2,u)$, $\mu^4(\frac m2,u)$, $\nu^3(\frac m2,u_m)$ and $\nu^4(\frac m2,u_m)$ vanish at $u = \widehat u$ with simple zeros. Expanding $\Db^{m/2,0}_{m}\Db^{m/2,0}_0$ with \eqref{eq:two.Dmu.relations} yields nine terms, four of which contain two of these functions. Dividing  by $w^{(m)}(u_{m-5/2})$ and taking the limit $u \to \widehat u$, these four terms vanish. Four more terms contain exactly one of the functions $\mu^3(\frac m2,u)$, $\mu^4(\frac m2,u)$, $\nu^3(\frac m2,u_m)$ and $\nu^4(\frac m2,u_m)$. After the division by $w^{(m)}(u_{m-5/2})$, the limit of each term is well-defined and nonzero, and involves fused transfer matrices evaluated at $u=\widehat u$ and shifts thereof. These four terms then cancel pairwise due to the crossing-symmetry relations in \eqref{eq:some.relations} and 
\be
\mu^2(\tfrac m2,\widehat u) = (-1)^{m/2} \nu^2(\tfrac m2,\widehat u_m), \qquad 
\lim_{u \to \widehat u}\frac{\mu^j(\frac m2,u)}{w^{(m)}(u_{m-5/2})}
=(-1)^{m/2-1}\lim_{u \to \widehat u}\frac{\nu^j(\frac m2,u_m)}{w^{(m)}(u_{m-5/2})}, \quad j = 3,4.
\ee
The remaining terms combine to 
\begin{alignat}{2}
\label{eq:L.final.limit}
\Lb = \lim_{u \to \widehat u} \frac1{w^{(m)}(u_{m-5/2})} \bigg[
\tau^2(m,\tfrac m2,u)&\frac{\mu^2(\frac m2,u)}{\mu^1(\frac m2,u)}\frac{\nu^2(\frac m2,u_m)}{\nu^1(\frac m2,u_m)}\Db^{1,0}_{m}\Db^{1,0}_{m-2}\\
&+\tau^4(m,\tfrac m2,u) \Db^{1,0}_{m-1}\bigg] \Db^{m/2-1,0}_0\Db^{m/2-1,0}_{m+2}.\nonumber
\end{alignat}
With 
\begin{equation}
\gamma = \frac{\tau^2(m,\frac m2,u) \mu^2(\frac m2,u) \nu^2(\frac m2,u_m)}{\mu^1(\frac m2,u) \nu^1(\frac m2,u_m) \Adeux(2,u_{m-2})}
= -\frac{\tau^4(m,\frac m2,u)}{\Atrois(2,u_{m-2}) \Azero(u_{m-3})},
\end{equation}
it follows that
\begin{equation}
\Lb = \lim_{u \to \widehat u} \frac{\gamma}{w^{(m)}(u_{m-5/2})}
\Big[ \Aun(2,u_{m-2})\Db^{2,0}_{m-2}\Big] \Db^{m/2-1,0}_0\Db^{m/2-1,0}_{m+2}.
\end{equation}
The function $\Aun(2,u_{m-2})$ has a double zero at $u = \widehat u$, so after dividing by $w^{(m)}(u_{m-5/2})$, we find that~$\Lb$ vanishes, ending the proof that $\Db^{m,0}(u)$ is regular at $u = \widehat u$ for $m$ even.

\paragraph{The special case $m=3$.} For $m=3$, \eqref{eq:HFd} and \eqref{eq:HFe} provide two equations for $\Db^{3,0}(u)$. The  results of \cref{sub:poly.D20.D11} confirm that the right-hand sides of these equations are Laurent polynomials in $z = \eE^{\iI u}$. The  coefficients of $\Db^{3,0}(u)$ in these equations are respectively
\begin{subequations}
\begin{alignat}{2}
\Aun(3,u)&=s(2u_1)^2 w^{(3)}(u_{1/2})\bar w^{(-1)}(u_{3/2})f_1f_2,\\
\Aun(3,\slu-\lambda)&=s(2u_0)^2 w^{(3)}(-u_{1/2}) \bar w^{(-1)}(-u_{-1/2})f_0f_{-1}.
\end{alignat}
\end{subequations}
For the mixed boundary conditions, these two have no common zeroes for generic inhomogeneities $\xi_{(j)}$ and $\Db^{3,0}(u)$ is thus a polynomial. For the boundary conditions SS and CC, the only common zeroes are those arising from $w^{(3)}(u_{1/2})$ which is equal up to a possible sign to $w^{(3)}(-u_{1/2})$. Thus the only zeroes to be studied are 
\begin{equation}
\widehat u=-\frac{\lambda}2+
\left\{\begin{array}{cc} 
(r+\frac12)\pi& \text{SS},\\[0.1cm]
r\pi & \text{CC}.\end{array}\right.
\end{equation}
To show that $\Db^{3,0}(u)$ is regular at this value, we use \eqref{eq:Dm0mu} and note that $\mu^1(3,u)$ and $\mu^4(3,u)$ have simple zeros at $u = \widehat u$. We must then show that 
\be
\mu^2(3,\widehat u)\Db^{2,0}_0(\widehat u)\Db^{1,0}_{4}(\widehat u) - \mu^3(3,\widehat u)\Db^{1,0}_0(\widehat u) \Db^{1,0}_{3}(\widehat u)=0.
\ee
This follows directly from the crossing symmetry of $\Db^{1,0}(u)$ and the relation
\be
\Db^{2,0}(\widehat u) = -\frac{\alpha^3(2,\widehat u)\alpha^0(\widehat u_{-1})}{\alpha^1(2,\widehat u)} \Db^{1,0}_1(\widehat u).
\ee
This ends the proof of the polynomiality of $\Db^{3,0}(u)$, and of $\Db^{m,0}(u)$ by the inductive argument.

%%%%%%%%%%%%%
\subsection{Polynomiality of $\Db^{m,n}(u)$} \label{sub:poly.Dmn}
%%%%%%%%%%%%%

In this section, we show the polynomiality of $\Db^{m,n}(u)$ for $m,n \ge 1$, assuming that it holds for $\Db^{m,0}(u)$. Recall also that this property was already established for $\Db^{1,1}(u)$ in \cref{sub:poly.D20.D11}. From \eqref{eq:HFh}, the potential poles of $\Db^{m,n}(u)$ are at the values of $u$ where
\begin{equation}
\Bun(m,n,u) = -s(2u_{m+n-2}) s(2u_{2m-2}) f(u_{2m-2}) = 0\ .
\end{equation}

We write down a second formula for $\Db^{m,n}(u)$ starting from the $T$-system relation \eqref{Tsystem} with $(m,k) \mapsto (m+n,m)$:
\begin{equation} \label{eq:TsystemPourPoly}
\Big( \prod^{m+n-1}_{j=m} \fbm_{2j} \Big) \det(m,n)_0
=\det(m+n,0)_0 \det(n,0)_{2m+2}
- \det(m+n+1,0)_0 \det(n-1,0)_{2m+2} \ .
\end{equation}
Reformulated in terms of fused transfer matrices, it reads
\begin{equation} \label{eq:TsystemExpli}
\eta^1(m,n,u) \Db^{m,n}_0
= \eta^2(m,n,u) \Db^{m+n,0}_0 \Db^{n,0}_{2m+2}
- \eta^3(m,n,u) \Db^{m+n+1,0}_0 \Db^{n-1,0}_{2m+2} \ ,
\end{equation}
with
\begin{subequations}
\begin{alignat}{2}
\eta^1(m,n,u) &=
\Big( \prod^{n-1}_{j=0} \wb(\slu_{2j-3/2}) \Big) \Big( \prod^{n-2}_{j=0} w(\slu_{2j-1/2}) \Big)
s(2u_{m-3}) s(2u_{2m+2n}) w(u_{2m+2n-1/2}) f_{2m+2n}  \ , \\[-0.5ex]
\eta^2(m,n,u) &= s(2u_{m+n-3}) s(2u_{2m+n})
w^{(m+n)}(u_{m+n-5/2}) w^{(n)}(u_{2m+n-1/2}) \ , \\[1ex]
\eta^3(m,n,u) &= s(2u_{m+n-1}) s(2u_{2m+n-2})
w^{(m+n+1)}(u_{m+n-3/2}) w^{(n-1)}(u_{2m+n-3/2}) \ .
\end{alignat}
\end{subequations}
This relation holds for $m,n\geq 1$ with the convention $\Db^{0,0}_k \mapsto f_{k-3} f_{k-2} \Ib$ used for the special case $n=1$. Having established in \cref{sub:poly.Dm0} the polynomiality of the transfer matrices $\Db^{m,0}(u)$ with $m\geq 1$, we conclude that the right side of \eqref{eq:TsystemExpli} is also a Laurent polynomial. It can then be checked that the intersection of the zeros of $\Bun(m,n,u)$ and $\eta^1(m,n,u)$ is empty for generic values of $\lambda$, for all $m,n \ge 1$. This ends the proof that $\Db^{m,n}(u)$ is a Laurent polynomial in $z = \eE^{\iI u}$.

%%%%%%%%%%%%%%%%%%%%%%%%%%%%%%%%%%%%%%%%%%%%%%%
%
\section{Closure at roots of unity}\label{sec:closure}
%
%%%%%%%%%%%%%%%%%%%%%%%%%%%%%%%%%%%%%%%%%%%%%%%

For generic values of $\lambda$, the fusion hierarchy, $T$-system and $Y$-system are infinite systems of equations. In this section, we fix the crossing parameter to rational multiples of $\pi$
\begin{equation} \label{def:lambdaRacineUnite}
\lambda = \lambda_{a,b} = \frac{\pi(b-a)}{2b} \ ,
\qquad \gcd(a,b)=1 \ ,
\end{equation}
for which these systems close finitely. The main result of this section, stated in \cref{sub:results.and.prelims}, is the closure relation for the fusion hierarchy. There, we also give the strategy of the proof, which is then the topic of \cref{sec:closure.symmetries,sec:finite.evaluations,sec:infinite.evaluation}. Finally, \cref{sec:closure.Y.system} shows that the $Y$-system also closes finitely at roots of unity.

%%%%%%%%%%%%%
\subsection{Results and skeleton of the proof}\label{sub:results.and.prelims}
%%%%%%%%%%%%%

A closure relation for the fusion hierarchy is a linear relation between fused transfer matrices. The following theorem gives these relations for the $\Atwotwo$ loop model on the strip. 
\begin{Theorem} 
For $\lambda = \lambda_{a,b}$ with $b\geq2$, the fusion hierarchy is finite, namely $\Db^{b,0}(u)$ can be expressed as a linear combination of $\Db^{m,n}(u)$ with $m+n<b$. For $b>2$, the closure relation reads
\begin{subequations}
\label{eq:closure}
\begin{alignat}{2}
&w^{(a)}(u_{-5/2}) f_{-1} \Db^{b,0}_0 - w^{(1)}(u_{-5/2}) \Db^{b-2,1}_2 
\nonumber\\[0.1cm]
&+ w(u_{-7/2}) w(u_{-5/2}) \wb(u_{-5/2}) \wb(u_{-3/2}) \wb(u_{-1/2})
\wb^{(a)}(u_{-3/2}) \wb^{(-1)}(u_{1/2}) f_{-3} \Db^{b-3,0}_4
\label{eq:general.closure}\\[0.1cm]
&= w(u_{-5/2})(\leU -2\leV) f_{-3} f_{-2} f_{-1} \,\kappa\, \Ib,
\nonumber
\end{alignat}
and for $b=2$
\begin{alignat}{2}
\label{eq:closure.b=2}
w^{(a)}(u_{-5/2}) \Db^{2,0}_0 - w^{(1)}(u_{-5/2}) w^{(1)}(-u_{3/2}) w(u_{-1/2}) \Db^{0,1}_2
= w(u_{-5/2})(\leU-2\leV) f_{-3} f_{-2} \,\kappa\, \Ib,
\end{alignat}
where 
\be  
\label{eq:definitionUV}
\leU(u) = \prod_{j=0}^{b-1} \wb(u_{2j+1/2}) w(u_{2j+3/2}),\qquad
\leV(u) = \prod_{j=0}^{b-1} \wb(u_{2j+3/2}) w(u_{2j+1/2}), 
\ee
and
\be
\kappa = \left\{\begin{array}{cc}
(-1)^{a-1}\big(2+(-1)^b\big) & \mathrm{identical}, \\[0.15cm]
-3 & \mathrm{mixed}.
\end{array}\right.
\ee
\end{subequations}
The relation \eqref{eq:closure} holds for $b=3$ with $\Db^{0,0}_{4} \mapsto f_{1} f_{2} \Ib $.
\end{Theorem}
\noindent This result thus holds for all four choices of boundary conditions. 
A proof of the theorem follows in \cref{sec:closure.symmetries,sec:finite.evaluations,sec:infinite.evaluation}. The rest of this section presents the skeleton of this proof.  

First we note that \eqref{eq:general.closure} is an equality between centered Laurent polynomials in $z = \eE^{\iI u}$ of maximal degree $6N+2b+1$ and $6N$, for identical and mixed boundary conditions respectively. Using the strategy outlined at the end of \cref{sub:propertiesBoundary}, it thus suffices to check that the identity holds for $12N+4b+3$ and $12N+1$ values of $z$, respectively. 

An important simplification arises from the fact that the closure relation can be written in terms of the determinants as 
\be
\det(b,0)_0 -\fbm_{-2} \det(b-2,1)_2
+ \fbm_{-2} \fbm_{-1} \fbm_{0} \det(b-3,0)_4 
=  \Lambda (\leU - 2\leV) \,\kappa\, \Ib
\ee
with
\be\label{eq:grandLambda}
\Lambda(u) = \prod_{j=0}^{2b-1} s(2u_j) f_{j}
\prod_{\ell=0}^{b-1} \wb(u_{2\ell+1/2}) w(u_{2\ell+3/2}).
\ee
We then define
\be
\label{eq:PJ}
\Pbm_0 = \det(b,0)_0 -\fbm_{-2} \det(b-2,1)_2
+ \fbm_{-2} \fbm_{-1} \fbm_{0} \det(b-3,0)_4,
\qquad
\Jbm_0  = \Lambda (\leU - 2\leV) \,\kappa\, \Ib,
\ee
so that the closure equation reads $\Pbm_0 = \Jbm_0$. This equation is also an equality between Laurent polynomials, whose degrees are however larger than the ones given above for \eqref{eq:general.closure}. We will show in \cref{sec:closure.symmetries} that for $\lambda = \lambda_{a,b}$, $\Pbm(u)$ and $\Jbm(u)$ satisfy the periodicity and crossing symmetries
\be
\label{eq:PJ.symmetries}
\Pbm(u) = \Pbm(u+2 \lambda) = \Pbm(5 \lambda-2b \lambda - u), \qquad
\Jbm(u) = \Jbm(u+2 \lambda) = \Jbm(5 \lambda-2b \lambda - u).
\ee
These symmetries turn out to be useful to reduce the number of points where \eqref{eq:general.closure} needs to be evaluated. Indeed, dividing both sides of \eqref{eq:general.closure} by $w(u_{-5/2}) f_{-3} f_{-2} f_{-1}$, we see that the equation we want to prove reduces to
\begin{alignat}{2}
&\frac1{w(u_{-5/2})f_{-3} f_{-2} f_{-1}}\Big[w^{(a)}(u_{-5/2})f_{-1}\Db^{b,0}_0 - w^{(1)}(u_{-5/2}) \Db^{b-2,1}_2 
\nonumber\\[0.1cm]&
+ w(u_{-7/2})w(u_{-5/2})\bar w(u_{-5/2})\bar w(u_{-3/2})\bar w(u_{-1/2}) \bar w^{(a)}(u_{-3/2}) \bar w^{(-1)}(u_{1/2})f_{-3} \Db^{b-3,0}_4\Big]
\label{eq:renormalized.closure}\\[0.1cm]&
=(\leU-2\leV)  \,\kappa\, \Ib.
\nonumber
\end{alignat}
We readily observe that the right-hand side is invariant under shifts of $2\lambda$. The same applies to the left-hand side, as it can be expressed as $\Pbm(u)/\Lambda(u)$, with both factors individually invariant under shifts of $2\lambda$.

The left-hand side of \eqref{eq:renormalized.closure} has a non-trivial denominator, letting us believe that it may have poles at values of $u$ where $w(u_{-5/2})f_{-3} f_{-2} f_{-1}=0$. But it is clear that it has no poles at values of $u$ where $w(u_{-1/2}) f_{0} f_{1}=0$. Once \eqref{eq:PJ.symmetries} is established, we know that the left-hand side is periodic in $u$ with period $2\lambda$, and thus deduce that it has no poles at all, including at the zeros of the denominator $w(u_{-5/2})f_{-3} f_{-2} f_{-1}$. This equivalently implies that \eqref{eq:general.closure} holds at all values where $w(u_{-5/2})f_{-3} f_{-2} f_{-1}=0$. We conclude that \eqref{eq:renormalized.closure} is an equality of Laurent polynomials of degree width $4b$ for SS and CC boundary conditions, and of degree width $0$ for SC and CS boundary conditions. To prove the equation, we must show that it holds for $4b+1$ and $1$ values of $z$, for identical and mixed boundary conditions, respectively. In \cref{sec:finite.evaluations}, we will prove, for the identical cases, that \eqref{eq:renormalized.closure} holds at $4b$ finite values of $u$ corresponding to $4b$ distinct values of $z = \eE^{\iI u}$. The last evaluation point, needed for both the identical and mixed cases, is the braid limit $u \to \iI \infty$, corresponding to $z = 0$. This will be the topic of \cref{sec:infinite.evaluation}, and will then end the proof of the theorem, for all four boundary conditions. 

%%%%%%%%%%%%%
\subsection{Symmetries at roots of unity}\label{sec:closure.symmetries}
%%%%%%%%%%%%%
 
In this section, we establish the periodicity and crossing properties \eqref{eq:PJ.symmetries} of the functions $\Pbm(u)$ and $\Jbm(u)$. The function $\Jbm(u)$ is equal to $\Ib$ times a function involving only simple trigonometric functions, and verifying that it satisfies the two symmetry properties is straightforward. We therefore focus on the function $\Pbm(u)$. We will use repeatedly the symmetries $\Dbm_{2b+k}= \Dbm_{k}$ and $\fbm_{2b+k}= \fbm_{k}$. Using \eqref{eq:HFhDet} with $(m,n) = (b-2,1)$, we express $\Pbm(u)$ as
\begin{alignat}{2}
\label{eq:P0.refined}
\Pbm(u) &= \det(b,0)_0 - \fbm_{-2} \Dbm_{-1} \det(b-2,0)_2 
+ \fbm_{-4} \fbm_{-3} \fbm_{-2} \det(b-3,0)_2 \nonumber\\&
+ \fbm_{-2} \fbm_{-1} \fbm_{0} \det(b-3,0)_4 \ .
\end{alignat}

We now show the crossing symmetry of $\Pbm(u)$. It directly follows from \eqref{eq:symCroiseDETmn} and \eqref{eq:conjugDETmn} that $\det(b,0)(u)=\det(b,0)(5\lambda-2b\lambda-u)$, so the first term in \eqref{eq:P0.refined} is invariant under $u \mapsto 5\lambda-2b\lambda-u$. For the second term, we have
\begin{alignat}{2} 
\fbm(5\lambda-2b\lambda-u-2\lambda) &= \fbm(u-2\lambda+2b\lambda) = \fbm_{-2} \ , \nonumber\\
\Dbm(5\lambda-2b\lambda-u-\lambda) &= \Dbm(u-\lambda+2b\lambda)=  \Dbm_{-1} \ ,
\\\nonumber
\det(b-2,0)(5\lambda-2b\lambda-u+2\lambda)
&=\det(b-2,0)(5\lambda-2(b-2)\lambda-(u+2\lambda)) 
=\det(b-2,0)_2 \ ,
\end{alignat}
where we used \eqref{eq:symDbm} and \eqref{eq:symFbm}. We thus see that this second term is also invariant. This is in contrast with the last two terms of \eqref{eq:P0.refined}, which are mapped to one another:
\be
\det(b-3,0)(5\lambda-2b\lambda-u+2\lambda)
=\det(b-3,0)(5\lambda-2(b-3)\lambda-(u+4\lambda)) =\det(b-3,0)_4(u) \ .
\ee
The same is seen to apply to the prefactors, namely we have
\be
\Big(\fbm_{-4} \fbm_{-3} \fbm_{-2} \det(b-3,0)_2 \Big)(5\lambda-2b\lambda-u)
=\Big( \fbm_{-2} \fbm_{-1} \fbm_{0} \det(b-3,0)_4 \Big)(u) \ ,
\ee
ending the proof of the crossing symmetry.

To prove the periodicity, we compute $\Pbm_0-\Pbm_2$. The terms proportional to $\det(b-3,0)_4$ immediately cancel, and we are left with
\begin{alignat}{2}
\Pbm_0-\Pbm_2 &= \det(b,0)_0 - \fbm_{-2} \Dbm_{-1} \det(b-2,0)_2
+ \fbm_{-4} \fbm_{-3} \fbm_{-2} \det(b-3,0)_2 \nonumber\\
&- \det(b,0)_2 + \fbm_{0} \Dbm_{1} \det(b-2,0)_4
- \fbm_{0} \fbm_{1} \fbm_{2} \det(b-3,0)_6 \ .
\end{alignat}
We now expand $\det(b,0)_0$ using \eqref{eq:HFeDet3}. The terms proportional to $\det(b-2,0)_4$ and $\det(b-3,0)_6$ cancel, resulting in
\begin{alignat}{2}
\Pbm_0 - \Pbm_2
&= \Dbm_0 \det(b-1,0)_2 -  \fbm_{-2} \Dbm_{-1} \det(b-2,0)_2 
\nonumber\\&
+  \fbm_{-4} \fbm_{-3} \fbm_{-2} \det(b-3,0)_2 - \det(b,0)_2 = 0\ ,
\end{alignat}
where at the last step we rewrote $\det(b,0)_2$ using \eqref{eq:HFdDet3}. This ends the proof of the periodicity of $\Pbm(u)$.

%%%%%%%%%%%%%
\subsection{Evaluations at finite points}\label{sec:finite.evaluations}
%%%%%%%%%%%%%

In this section, we focus on the SS and CC boundary conditions. Our goal is to prove that the closure relation holds at $4b$ distinct values of $u$ that are real and in the interval $[0,2 \pi)$.

Let us suppose that we are able to prove it for some finite value $u=\widehat u$. By the periodicity and crossing properties \eqref{eq:PJ.symmetries} of $\Pbm(u)$ and $\Jbm(u)$, we will also have proved the relation for $u=\widehat u+2k \lambda$, $\widehat u+2k \lambda + \pi$, $-\widehat u+(2k+1) \lambda$ and $-\widehat u+(2k+1) \lambda + \pi$, with $k = 0, \dots, b-1$. Depending on $\widehat u$ and $\lambda$, these may or may not lead to a full set of $4b$ distinct values of $z = \eE^{\iI u}$. This turns out to depend on the parity of $b-a$. For $b-a$ odd, choosing $\widehat u = \ell \lambda$ for a given $\ell \in \mathbb Z$ does in fact lead to $4b$ distinct values of $z$. In contrast, for $b-a$ even, choosing $\widehat u = \ell \lambda$ for any $\ell \in \mathbb Z$ only gives $2b$ distinct values. These can then be combined to $2b$ more distinct values obtained from $\widehat u = \ell' \lambda+\frac \pi 2$ for some $\ell' \in \mathbb Z$, leading to a full set of $4b$ values.

Thus for the proof, we will show that the closure relation holds for $\widehat u=(4-b)\lambda+\frac{r\pi}2$ with $r \in \{0,1\}$. The cases $b=2$ and $b=3$ are treated separately in \cref{sub:proof.closure.b23}. For $b \ge 4,$ we keep $\lambda$ generic for now and note that\footnote{
We note that \eqref{eq:intermediate.reductions.2} also works for $b=4$, with $\Db^{0,0}_4 \mapsto f_1 f_{2} \Ib$ and $\frac{\mu^2(b-3,\widehat u_4)}{\mu^1(b-3,\widehat u_4)}$ understood as $\lim_{u \to \widehat u}\frac{\mu^2(b-3, u_4)}{\mu^1(b-3,u_4)}$.
} 
\begin{subequations}
\label{eq:intermediate.reductions}
\begin{alignat}{2}
\label{eq:intermediate.reductions.1}
\Db^{b-2,1}_2(\widehat u) &= \frac{\beta^2(b-2,1,\widehat u_2)}{\beta^1(b-2,1,\widehat u_2)} \Db^{1,0}_{2b-1}(\widehat u)\Db^{b-2,0}_2(\widehat u), 
\\[0.1cm]
\label{eq:intermediate.reductions.2}
\Db^{b-3,0}_4(\widehat u) &= \frac{\mu^2(b-3,\widehat u_4)}{\mu^1(b-3,\widehat u_4)}\Db^{1,0}_{2b-4}(\widehat u)\Db^{b-4,0}_4(\widehat u).
\end{alignat}
\end{subequations}
The three remaining fused transfer matrices, $\Db^{b,0}_0(\widehat u)$, $\Db^{b-2,0}_2(\widehat u)$ and $\Db^{b-4,0}_4(\widehat u)$, are precisely of the form $\Db^{m,0}_0(\widehat v)$ with $\widehat v = (4-m)\lambda + \tfrac{r \pi}2$, for $m = b$, $b-2$, and $b-4$ respectively. By \eqref{eq:simpler.D.reduction}, these tangles are all multiples of $\Ib$. Let us define 
\begin{subequations}\label{eq:lesgrandsR}
\begin{alignat}{2}
\Rb^1(u)&=w^{(a)}(u_{-5/2})f_{-1}\Db^{b,0}_0,
\\[0.1cm]
\Rb^2(u)&= - w^{(1)}(u_{-5/2}) \Db^{b-2,1}_2,
\\[0.1cm]
\Rb^3(u)&= w(u_{-7/2})w(u_{-5/2})\bar w(u_{-5/2})\bar w(u_{-3/2})\bar w(u_{-1/2}) \bar w^{(a)}(u_{-3/2}) \bar w^{(-1)}(u_{1/2})f_{-3} \Db^{b-3,0}_4,
\\[0.1cm]
\Rb^4(u)&=w(u_{-5/2})\leU f_{-3} f_{-2} f_{-1}\, \Ib,
\\[0.1cm]
\Rb^5(u)&=w(u_{-5/2})\leV f_{-3} f_{-2} f_{-1}\, \Ib.
\end{alignat}
\end{subequations}
Using \eqref{eq:intermediate.reductions} and \eqref{eq:simpler.D.reduction}, we obtain simple expressions for $\Rb^1(\widehat u)$, $\Rb^2(\widehat u)$ and $\Rb^3(\widehat u)$: $\Rb^1(\widehat u)$ is proportional to $\Ib$, whereas $\Rb^2(\widehat u)$ and $\Rb^3(\widehat u)$ are respectively proportional to $\Db^{1,0}_{2b-1}(\widehat u)$ and $\Db^{1,0}_{2b-4}(\widehat u)$. This is true for all $\lambda$. Setting $\lambda = \lambda_{a,b}$, we remark that
\be
\Db^{1,0}_{2b-1}(\widehat u) = \Db^{1,0}_{2b-4}(\widehat u) = \Db^{1,0}_0\big((b-a+r)\tfrac \pi 2\big),
\ee
which follows from periodicity and crossing symmetry. From \eqref{eq:D10.id}, we see that $\Db^{1,0}_{2b-1}(\widehat u)$ and $\Db^{1,0}_{2b-4}(\widehat u)$ are scalar multiples of $\Ib$ at $\lambda = \lambda_{a,b}$. 

The evaluation of the closure relation at $\widehat u$ then reads 
\be
\label{eq:final.closure}
\Rb^1(\widehat u)+ \Rb^2(\widehat u) + \Rb^3(\widehat u) = \kappa \big(\Rb^4(\widehat u)-2\Rb^5(\widehat u)\big).
\ee
and checking it amounts to comparing the prefactors of $\Ib$ in $\Rb^1(\widehat u)$, $\Rb^2(\widehat u)$ and $\Rb^3(\widehat u)$ with those in $\Rb^4(\widehat u)$ and $\Rb^5(\widehat u)$. Evaluating the prefactors, one finds
\be
\label{eq:R.equalities}
\Rb^1(\widehat u) = \Rb^2(\widehat u)= \Rb^3(\widehat u) = -\Rb^4(\widehat u ) = (-1)^{b+1}\Rb^5(\widehat u),\qquad \text{for } b\ge 4.
\ee
For $b-a$ even, each $\Rb^i(\widehat u)$ in fact vanishes, and thus the above equalities hold trivially in this case. In contrast, for $b-a$ odd, none of the $\Rb^i(\widehat u)$ vanish. For either parity of $b-a$, the proof of \eqref{eq:final.closure} is trivial using \eqref{eq:R.equalities}.

The proof of \eqref{eq:R.equalities} is lengthy but straightforward. To show how it works, let us prove that $\Rb^4(\widehat u) = (-1)^b\Rb^5(\widehat u)$. From the definition \eqref{eq:lesgrandsR}, we have
\be
\label{eq:R4R5}
U(\widehat u) = \prod_{j=0}^{b-1} \bar w_{-b+2j+9/2}\, w_{-b+2j+11/2}\ ,
\qquad
V(\widehat u) = \prod_{j=0}^{b-1} \bar w_{-b+2j+11/2}\, w_{-b+2j+9/2}\ ,
\ee
where we use the short-hand notation
\be
\label{eq:short.hand.wk}
w_k = w(k \lambda+\tfrac{r \pi}2), \qquad \bar w_k = \bar w(k \lambda+\tfrac{r \pi}2).
\ee
The proof rests on the identities
\be
\label{eq:easyIdentities}
w_{k+2b} = (-1)^{b-a} w_{k}, 
\qquad
\bar w_{k+2b} = (-1)^{b-a} \bar w_{k},
\qquad
w_{-k} \bar w_{-\ell} = -w_{k} \bar w_{\ell}, 
\ee
valid for both SS and CC and all $k,\ell$. The first two of these equations hold for $\lambda=\lambda_{a,b}$, whereas the last one holds for all~$\lambda$. The first two identities allow us to write
\be
\prod_{j=0}^{b-1} \bar w_{2j+x}\, w_{2j+y} = \prod_{j'=1}^{b} \bar w_{2(j'-1)+x}\, w_{2(j'-1)+y} = \prod_{j'=0}^{b-1} \bar w_{2j'+x-2}\, w_{2j'+y-2}\ ,
\ee
for all $x,y$. Thus we may freely shift the indices of all functions in the products by two. For the products in \eqref{eq:R4R5}, this yields
\begin{subequations}
\begin{alignat}{2}
U(\widehat u) &= \prod_{j=0}^{b-1} \bar w_{-b+2j+1/2}\, w_{-b+2j+3/2}\ ,
\\
V(\widehat u) &= (-1)^b \prod_{j=0}^{b-1} \bar w_{b-2j-11/2}\, w_{b-2j-9/2}
= (-1)^b \prod_{j'=0}^{b-1} \bar w_{b-2(b-1-j')-11/2}\, w_{b-2(b-1-j')-9/2}
\nonumber\\&
= (-1)^b \prod_{j'=0}^{b-1} \bar w_{-b+2j'-7/2}\, w_{-b+2j'-5/2}
= (-1)^b \prod_{j'=0}^{b-1} \bar w_{-b+2j'+1/2}\, w_{-b+2j'+3/2}\ .
\end{alignat}
\end{subequations}
The two products are thus equal up to the sign $(-1)^b$, confirming that $\Rb^4(\widehat u) = (-1)^b\Rb^5(\widehat u)$. Showing that $\Rb^1(\widehat u), \Rb^2(\widehat u)$ and $\Rb^3(\widehat u)$ are equal to $-\Rb^4(\widehat u)$ then follows three similar paths. For each, one can show, using the identities $f(-u) = f(u) = f(u+\pi)$ and $s(-u) = -s(u) = s(u+\pi)$, that all functions $f(u_k)$ and $s(2u_k)$ appearing in the functions $\Rb^1(\widehat u)$, $\Rb^2(\widehat u)$ and $\Rb^3(\widehat u)$ simplify up to a potential sign. A finer analysis is needed for the factors $w$ and $\bar w$, but one can check that they always coincide with those of $\Rb^4(\widehat u)$, again up to potential signs. These lengthy checks complete the proof of \eqref{eq:R.equalities}. Finally, the fact that each $\Rb^i(\widehat u)$ vanishes for $b-a$ even follows from an analysis of the products of $w,\bar w$ functions: either the argument of a sine function is an integer multiple of $\pi$ or the argument of a cosine function is an odd integer multiple of~$\tfrac{\pi}2$. This ends the proof of the closure relation at $4b$ separate values, for both parities of $b-a$.

%%%%%%%%%%%%%
\subsection{Evaluation at infinity}\label{sec:infinite.evaluation}
%%%%%%%%%%%%%

This section finishes the proof of the closure relations by taking the braid limit of the four terms in \eqref{eq:general.closure}. These are all polynomials in $z=\eE^{\iI u}$ of the same degree. Multiplying them with a properly chosen rational function of $\eE^{\iI u}$ gives a finite result in the limit $u \to \iI \infty$. This rational function will be chosen as
\begin{equation}
r(u)=\frac{\eE^{4\iI\lambda bN}}{w^{(a)}(u_{-5/2})f_{-3}f_{-2}f_{-1}\prod_{j=0}^{b-1}\gamma(u_{2j})},
\end{equation}
where $\gamma(u)$ defined in \eqref{eq:gamma}. This choice ensures that the first term is 
\be
\lim_{u\to \iI \infty}r(u)w^{(a)}(u_{-5/2})\Db^{b,0}_0(u)=\Db^{b,0}_\infty,
\ee
as introduced in \eqref{def:tresse(m,0)}, and thus that the limit of the closure relation is non-trivial. With this rational function $r(u)$, the limit of the other terms are
\begin{alignat}{2}
\lim_{u\to i\infty}& - r(u) w^{(1)}(u_{-5/2})\Db^{b-2,1}_2(u)=-\Db^{b-2,1}_\infty,\nonumber\\
\lim_{u\to i\infty}& r(u) w(u_{-7/2})w(u_{-5/2})\bar w(u_{-5/2})\bar w(u_{-3/2})\bar w(u_{-1/2})\bar w^{(a)}(u_{-3/2})\bar w^{(-1)}(u_{1/2})f_3\Db^{b-3,0}_4(u)=\Db^{b-3,0}_\infty,\nonumber\\
\lim_{u\to i\infty}& r(u)w(u_{-5/2}) (\leU-2\leV) f_{-3}f_{-2}f_{-1}\kappa\, \Ib=\Delta^b=\Ib \times
   \left\{\begin{array}{cc}
   1+2(-1)^b& \text{identical,}\\[0.1cm]
   3&           \text{mixed.}
   \end{array}\right.
\end{alignat}
The computation of these limits proceeds as in the proof leading to the identities \eqref{eq:HFtresse} for the fusion hierarchy of braid transfer tangles. The problem is thus reduced to proving that the braid transfer tangles satisfy
\be\label{eq:closureAtInfinity}
\Db^{b,0}_\infty-\Db^{b-2,1}_\infty+\Db^{b-3,0}_\infty-\Delta^b=0, \qquad b \ge 2.
\ee
(For the special cases $b=2$ and $b=3$, this is also true with the conventions $\Db^{0,0}_\infty = \Ib$ and $\Db^{-1,0}_\infty = 0$.) The proof of this identity proceeds in several steps: (i)~we introduce families of polynomials $p_m(x)$ and $q_m(x)$ related to Chebyshev polynomials and establish a link with the fused transfer matrices; (ii)~we show that one of these polynomials is equal, up to an additive constant, to the left-hand side of \eqref{eq:closureAtInfinity} with the variable $x$ set to $\Db_\infty-\Ib$; (iii)~we recall that the tangle $\Db_\infty$ is a central element of $\dtl_N(\beta)$ and discuss its minimal polynomial in the regular representation of this algebra; (iv)~we show that the determinant obtained in (ii) has this minimal polynomial as a factor, thus confirming that the left-hand side of \eqref{eq:closureAtInfinity} is equal to zero.

Let $p_m(x)$ with $m\ge 1$ and $q_m(x)$ with $m \ge 3$ be the polynomials in $x$ defined as the following $m\times m$ determinants: 
\be
p_m(x)=\left|\begin{matrix}x & 1 & 0 & \dots & 0 & 0\\
                           1 & x & 1 & \dots & 0 & 0\\
                           0 & 1 & x & \dots & 0 & 0\\
                           \vdots &\vdots & \vdots & \ddots & \vdots & \vdots\\
                           0 & 0 & 0 & \dots & x & 1\\
                           0 & 0 & 0 & \dots & 1 & x\end{matrix}\right|\,,
                           \qquad
q_m(x)=\left|\begin{matrix}x & 1 & 0 & \dots & 0 & 1\\
                           1 & x & 1 & \dots & 0 & 0\\
                           0 & 1 & x & \dots & 0 & 0\\
                           \vdots &\vdots & \vdots & \ddots & \vdots & \vdots\\
                           0 & 0 & 0 & \dots & x & 1\\
                           1 & 0 & 0 & \dots & 1 & x\end{matrix}\right|. 
\ee
By expanding the determinants along the first line, they are seen to satisfy the recurrence relations 
\be
\label{eq:rec.UT}
p_m(x)=xp_{m-1}(x)-p_{m-2}(x), \qquad 
q_m(x)=p_m(x)-p_{m-2}(x)-2(-1)^m, \qquad
m \ge 3.
\ee  
The solutions are written in terms of Chebyshev polynomials as 
\begin{equation}
p_m(x)= U_m(\tfrac x2), \qquad
q_m(x)=2\,T_m(\textstyle{\frac x2})-2(-1)^m.
\end{equation}
This then allows us to extend the definition of $q_m(x)$ to $m=1$ and $m=2$.

The determinant form of $q_m(x)$ was crucial in the study of the dense loop models (see Section 7 of \cite{AMDPPJR14}). We now relate the fused braid transfer matrices of the dilute $\Atwotwo$ model to the polynomials $p_m(x)$. First, we note the matrix identity
\begin{alignat}{2}
&\begin{pmatrix}
1 & 1 & 0 & \dots & 0 & 0\\
0 & 1 & 1 & \dots & 0 & 0\\
0 & 0 & 1 & \dots & 0 & 0\\
\vdots &\vdots & \vdots & \ddots & \vdots & \vdots\\
0 & 0 & 0 & \dots & 1 & 1\\
0 & 0 & 0 & \dots & 0 & 1
\end{pmatrix}
\cdot
\begin{pmatrix}
x & 1 & 0 & \dots & 0 & 0\\
1 & x & 1 & \dots & 0 & 0\\
0 & 1 & x & \dots & 0 & 0\\
\vdots &\vdots & \vdots & \ddots & \vdots & \vdots\\
0 & 0 & 0 & \dots & x & 1\\
0 & 0 & 0 & \dots & 1 & x
\end{pmatrix}
\nonumber\\&\label{eq:matrix.identity}
= 
\begin{pmatrix}
x+1 & x+1 & 1 & 0 & \dots & 0 & 0 & 0\\
1 & x+1 & x+1 & 1 &\dots & 0 & 0 & 0\\
0 & 1 & x+1 & x+1 &\dots & 0 & 0 & 0\\
0 & 0 & 1 & x+1 &\dots & 0 & 0 & 0\\
\vdots &\vdots & \vdots & \ddots & \vdots & \vdots & \vdots\\
0 & 0 & 0 & 0 & \dots & x+1 & x+1 & 1\\
0 & 0 & 0 & 0 &\dots & 1 & x+1 & x+1\\
0 & 0 & 0 & 0 & \dots & 0 & 1 & x
\end{pmatrix}.
\end{alignat}
Except for the bottom right entry, this matrix with $x+1 \to \Db_\infty$ is identical to matrix in the expression \eqref{tresseDet(m,0)} for $\Db^{m,0}_\infty$. Moreover, the first matrix in \eqref{eq:matrix.identity} clearly has a unit determinant. By expanding the determinants of the other two matrices along the last row, we find the simple identity
\be
\Db^{m,0}_\infty = \Db^{m-1,0}_\infty + p_m(\Db_\infty-\Ib).
\ee  
This completes step (i).

Step (ii) aims at tying the polynomials $q_m(x)$ to the left-hand side of \eqref{eq:closureAtInfinity}. The relation is remarkably simple:
\begin{equation}\label{eq:closureQm}
\Db^{m,0}_\infty-\Db^{m-2,1}_\infty+\Db^{m-3,0}_\infty-\Delta^m=q_m(\Db^{1,0}_\infty-\Ib)
+
\left\{\begin{array}{cc}
0&	\text{identical,}\\[0.1cm]
              2\big((-1)^m-1\big)\Ib&\text{mixed,}
   \end{array}\right.
   \quad m \ge 2,
\end{equation}
where we set $\Db^{-1,0}_\infty \mapsto 0$ for $m=2$. This relation is easily checked for $m=2$ and $m=3$. The proof for $m\ge 4$ is by induction:
\begin{alignat}{2}
\Db^{m,0}_\infty-\Db^{m-2,1}_\infty+\Db^{m-3,0}_\infty &= \Db^{m,0}_\infty-\Db_\infty\Db^{m-2,0}_\infty+2\Db^{m-3,0}_\infty 
\nonumber\\&= \Db^{m-1,0}_\infty-\Db_\infty\Db^{m-3,0}_\infty+2\Db^{m-4,0}_\infty + p_m(x) -(1+x) p_{m-2}(x) + 2 p_{m-3}(x)
\nonumber\\&
= q_{m-1}(x) + (1-2(-1)^m)\Ib + p_m(x) - (1+x) p_{m-2}(x) + 2 p_{m-3}(x) 
\nonumber\\&
= q_{m}(x) + (1+2(-1)^m)\Ib,
\end{alignat}
where $x = \Db_\infty-\Ib$. The inductive hypothesis was used at the third equality, and the relations \eqref{eq:rec.UT} at the fourth. This ends the proof of \eqref{eq:closureQm}, and therefore of step (ii).

The first two steps translated the statement \eqref{eq:closureAtInfinity} into one involving only the fundamental braid transfert tangle $\Db_\infty$ defined in \eqref{def:D10tresse}. Defining the polynomial 
\begin{equation}\label{eq:polyQ}
\mathbb Q_m(x)=
q_m(x-1)
+
\left\{\begin{array}{cc}
	0&\text{identical,}\\[0.1cm]
         2\big((-1)^m-1\big)&\text{mixed,}
   \end{array}\right.
\end{equation}
we must then show that $\mathbb Q_b(x)$ vanishes for $x=\Db_\infty$ with $\lambda = \lambda_{a,b}$. It will be so if this polynomial contains, as a factor, the minimal polynomial of $\Db_\infty$ or, more precisely, the minimal polynomial $\mathbb P_{a,b}(x)$ of its action on the algebra $\dtl_N(\beta)$ with $\beta = -2 \cos 4\lambda_{a,b}$. In other words, we need to prove that $\mathbb P_{a,b}\ |\ \mathbb Q_b$. Step (iii) consists in finding all factors that can possibly appear in this minimal polynomial $\mathbb P_{a,b}(x)$. A key observation is that $\Db_\infty$ is central. A proof of this for a slightly different element of $\dtl_N(\beta)$ is given in Appendix B of \cite{JBYSA13}. It rests upon relations similar to \eqref{rel:PTtresse} and goes through without change for $\Db_\infty$. Since it is central, it defines a homomorphism $d:\mathsf M\to\mathsf M$ on any module $\mathsf M$ by multiplication $\mu\mapsto \Db_\infty \mu$, for $\mu\in\mathsf M$. Such a homomorphism can have only one eigenvalue on an indecomposable module. Computing this eigenvalue $\Upsilon_N^d$ on the standard module $\wW_N^d$ is then a good idea. Indeed, these standard modules are indecomposable and their quotients by their radicals give a complete family of isomorphic classes of simple modules. In other words, the eigenvalues of the action of $\Db_\infty$ on any module $\mathsf M$ will be among those on the $\wW_N^d$ with $0\leq d\leq N$. Similar computations were done before. For example, Section~3.1 of \cite{AMDYSA11} does such a computation for the action of the fundamental braid transfer matrix for the ordinary Temperley-Lieb algebra $\tl_N(\beta)$ on its standard modules. The computation for $\Db_\infty$ is similar and gives
\begin{equation}\label{eq:D10onStandard}
\Upsilon_N^d=\left.\Db_\infty\right|_{\wW_N^d}=1+2\epsilon \cos\big(4\lambda(1+d)\big), \qquad
\epsilon=
\left\{\begin{array}{cc}
	-1 &	\text{identical,}\\[0.1cm]
         1 &\text{mixed.}
\end{array}\right.
\end{equation}
Can any of these eigenvalues have multiplicities in the minimal polynomial of $\Db_\infty$ in the regular representation? For roots of unity, the answer is yes \cite{JBYSA13}. For any $d$ that is not critical, there is a non-zero nilpotent homomorphism from the block of $\dtl_N(\beta)$, where $\Db_\infty$ has the eigenvalue $\Upsilon_N^d$, to itself. The square of this homomorphism is zero and no other such homomorphisms $\rho$ can be nilpotent of a higher order, that is, there exists no homomorphism $\rho$ such that $\rho^i=0$ but $\rho^{i-1}\neq0$ with $i>2$. For $d$ critical, any indecomposable module on which $\Db_\infty$ has the eigenvalue $\Upsilon_N^d$ is in fact irreducible and the minimal polynomial of $\Db_\infty$ on these is simply $(x-\Upsilon_N^d)$. 
Summing up, the minimal polynomial of $\Db_\infty$ may only contain factors $(x-\Upsilon_N^d)^{1}$ for $d$ critical and $(x-\Upsilon_N^d)^{2}$ for $d$ non-critical, but none with higher powers. In conclusion, the minimal polynomial $\mathbb P_{a,b}$ of $\Db_\infty$ divides the polynomial 
\begin{equation}\label{eq:minimalD10}
\mathbb R_{a,b}(x)=\prod_{d\in D}(x-\Upsilon_N^d)^{n_d}, \qquad 
n_d = \left\{\begin{array}{cl}
1 & d \textrm{ critical},
\\[0.1cm]
2 & d \textrm{ non-critical},
\end{array}\right.
\end{equation}
where $\lambda = \lambda_{a,b}$ and $D$ is a subset of $\{0, 1,\dots, N\}$ such that $\{\Upsilon_N^d\ |\ d\in D\}$ contains all possible eigenvalues in $\{\Upsilon_N^d\ |\ 0\leq d\leq N\}$ once and only once. (For $N<2b-1$, the orbit of some non-critical $d$ will only contain $d$. In this case, $n_d=1$ and $\mathbb P_{a,b}$ divides $\mathbb R_{a,b}$ without being equal to it. If $N\geq 2b-1$, then $\mathbb P_{a,b}=\mathbb R_{a,b}$.)

The goal of step (iv) is to establish whether the polynomial $\mathbb Q_b(x)$ specified to $\lambda = \lambda_{a,b}$ contains the polynomial $\mathbb R_{a,b}(x)$ as a factor. If it does, then $\mathbb P_{a,b}\ |\ \mathbb R_{a,b} \ |\ \mathbb Q_b$ 
and thus $\mathbb P_{a,b}\ |\ \mathbb Q_b$ as desired. We rewrite the polynomial $\mathbb Q_b$ using \eqref{eq:closureQm} as
\begin{equation}\label{eq:dernierPoly}
\mathbb Q_b(x) =
2\big(T_b\big(\tfrac12(x-1)\big) - \epsilon^b \big).
\end{equation}
We must thus check that this polynomial vanishes for $x$ replaced by the eigenvalue $\Upsilon_N^d$, and show that the multiplicity of such a zero is greater than or equal to its multiplicity $n_d$ in $\mathbb R_{a,b}(x)$. Because of the expression \eqref{eq:D10onStandard} for $\Upsilon_N^d$, it is natural to introduce the change of variables $x=1+2\epsilon \cos\theta$, thus rewriting
\begin{equation}
\mathbb Q_b(x)=2(T_b(\epsilon \cos \theta)-\epsilon^b)
=
2\epsilon^b(T_b(\cos\theta)-1)=-4 \epsilon^b \sin^2(\textstyle{\frac12}b\theta).
\end{equation}
The property $T_b(- x) = (- 1)^b T_b(x)$ was used at the second equality and the trigonometric formula $T_b(\cos \theta)=\cos(b\theta)$ at the third one. The evaluation at $x = \Upsilon_N^d$, namely at $\theta=4\lambda_{a,b}(1+d)$, gives
\be
\mathbb Q_b(\Upsilon_N^d)=-4\,\epsilon^b\sin^2((b-a)(1+d)\pi),
\ee 
which is clearly zero for any integer $d$. 

The multiplicity of these zeros must be computed for $\mathbb Q_b(x)$ seen as a polynomial in $x$. Care must be taken to ensure that the change of variables $x\mapsto \theta$ has not introduced spurious multiplicities. To get the correct multiplicities, we compute 
\be
\label{eq:dQ.dx}
\frac{\dd}{\dd x}\mathbb Q_b(x) = \frac{\dd \theta}{\dd x}\frac{\dd}{\dd \theta}\mathbb Q_b(1+2 \epsilon \cos \theta) = b\, \epsilon^{b-1} \frac{\sin b \theta}{\sin \theta}.
\ee
The multiplicity of a zero $x_0$ is larger than one if and only if $\mathbb Q_b'(x_0)=0$. We see from \eqref{eq:dQ.dx} that this occurs for $\theta_0 = 4\lambda(1+d)$ if $\theta_0\neq n \pi$ for $n \in \mathbb Z$. In these cases, taking the second derivative of $\mathbb Q_b(x)$ reveals that the multiplicity is precisely $2$. In contrast, for $\theta_0 = n \pi$, $\mathbb Q_b(x_0)'\neq 0$ and the multiplicity of the zero is $1$. Recalling that $q=\eE^{\iI(4\lambda-\pi)}$ and $\theta_0 = 2\pi (\frac{b-a}b)(1+d)$, we see that the condition $\theta_0=n\pi$ can be written as $q^{2(d+1)}=1$. This is precisely the condition for criticality, for which the multiplicities $n_d$ in \eqref{eq:minimalD10} are also equal to one. All zeros of $\mathbb R_{a,b}$ are thus among those of $\mathbb Q_b$ (also counting the multiplicities), and thus $\mathbb R_{a,b}$ divides $\mathbb Q_b$. This completes the proof of the closure relation for all four boundary conditions.

%%%%%%%%%%%%%
\subsection{Closure of the $Y$-system}\label{sec:closure.Y.system}
%%%%%%%%%%%%%

In this section, we use the closure of the fusion hierarchy to derive the closure relations of the $Y$-system. The first step towards this goal is to derive extra closure relations in the fusion hierarchy, satisfied by the transfer matrices $\Db^{b,k}(u)$ and $\Db^{k,b}(u)$, for $k = 1, \dots, b-1$. For this, it turns out to be simpler to work with the determinants. The result reads:\begin{subequations}
\begin{alignat}{2}
\det(b,k)_0 &= \fbm_{-2}\det(b-2,k+1)_2 - \bigg(\prod_{j=-2}^{2k} \fbm_j \bigg) \det(b-3-k,0)_{2k+4} + \Jbm_0
\det(0,k)_0, \label{eq:closurebk} \\
\det(k,b)_0 &= \fbm_{2k-1}\det(k+1,b-2)_0 - \bigg(\prod_{j=-3}^{2k-1} \fbm_j \bigg) \det(0,b-3-k)_{2k+2} + \Jbm_1
\det(k,0)_0,\label{eq:closurekb}
\end{alignat}
\end{subequations}
where we recall the definition \eqref{eq:PJ} of $\Jbm(u)$. The closure relations for $\det(b,k)_0$ are proven inductively from the same relation for $\det(b,0)_0$ following the arguments of Proposition B.3 of \cite{AMDPP19}. The closure relations \eqref{eq:closurekb} are then directly obtained by applying the crossing symmetry to \eqref{eq:closurebk}.

The second step is to derive the following quartic relation:
\begin{alignat}{2}
&\big(\det(b-1,0)_0\det(0,b-1)_0 - \fbm_{-2}\fbm_{-3}\det(b-2,0)_2\det(0,b-2)_0\big) \nonumber\\
&\times \big(\det(b-1,0)_2\det(0,b-1)_0 - \fbm_{-2}\fbm_{-1}\det(b-2,0)_2\det(0,b-2)_2\big)\nonumber\\ 
&= \Big( \prod_{j=0}^{b-2} \fbm_{2j} \Big) \det\!{}^3(0,b-1)_0
+ \Jbm_0
\det\!{}^2(0,b-1)_0 \det(b-2,0)_2
\\&+ \Jbm_1
\fbm_{-2} \det(0,b-1)_0 \det\!{}^2(b-2,0)_2
+ (\fbm_{-2})^2 \Big(\prod_{j=0}^{b-1} \fbm_{2j+1}\Big) \det\!{}^3(b-2,0)_2 \ . \nonumber
\end{alignat}
\noindent Its proof follows the same arguments as those in Proposition B.4 of \cite{AMDPP19}.
This relation can be rewritten as
\be
(\Ib+\db^{b-1}_0)(\Ib+\yb_0)(\Ib+\yb_1) = \Ib + \frac{\Jbm_0 \xb_0}{\prod_{j=0}^{b-1} \fbm_{2j}}  + \frac{\Jbm_1 (\xb_0)^2}{\prod_{j=0}^{b-1} \fbm_{2j}} + \frac{{\prod_{j=0}^{b-1} \fbm_{2j+1}}}{\prod_{j=0}^{b-1} \fbm_{2j}} (\xb_0)^3
\ee
where
\be
\xb(u) = \frac{\fbm_{-2} \det(b-2,0)_2}{\det(b-1,0)_1}, \qquad \yb(u) = - \xb_{-1}\xb_0.
\ee
To factorize the right-hand side, we recall that $\Jbm_0=\kappa\, \Lambda_0(\leU-2\leV)$, with $\leU(u)$, $\leV(u)$ and $\kappa$ introduced in \eqref{eq:definitionUV} and $\Lambda(u)$ in \eqref{eq:grandLambda}, and use the equalities
\be
\prod_{j=0}^{b-1} \fbm_{2j} = \sigma\, \Lambda_0 \leU, 
\qquad 
\prod_{j=0}^{b-1} \fbm_{2j+1} = \Lambda_1 \leV,
\qquad
\Lambda_1 \, \leU =
\sigma \Lambda_0 \, \leV ,
\qquad 
\sigma = \left\{\begin{array}{cc}
(-1)^{b-a} & \textrm{identical,}\\[0.1cm]
1 & \textrm{mixed,}
\end{array}\right.
\ee
and
\be
\frac{\leV}{\leU} = \left\{\begin{array}{cl}
1 & \textrm{SS and CC with $a$ odd,}  \\[0.1cm]
-\tan^2(b u + (b-a)\tfrac \pi 4) & \textrm{SS with $a$ even,}\\[0.1cm]
-\cot^2(b u + (b-a)\tfrac \pi 4) & \textrm{CC with $a$ even,}\\[0.1cm]
1 & \textrm{SC and CS.}  \\[0.1cm]
\end{array}
\right.
\ee
With this information, we treat separately the identical and mixed boundary conditions, as well as the different parities of $a$ and $b$, and find that the above equation can be nicely rewritten as
\be
\label{eq:1+d.closure}
\Ib+\db^{b-1}_0=\frac{(\Ib+\xb_0)(\Ib+\iota\frac{\leV}{\leU}\xb_0)^2}{(\Ib+\yb_0)(\Ib+\yb_1)},
\qquad 
\iota = \left\{\begin{array}{cc}
(-1)^{b} & \textrm{identical,}\\[0.1cm]
1 & \textrm{mixed,}
\end{array}\right.
\ee
for all parities of $a$ and $b$. We also have the relation
\be
\label{eq:x.closure}
\frac{\xb_0\xb_2}{\xb_1} = \frac {\leU^2}{\leV^2} \frac{\Ib+\db^{b-2}_2}{\Ib+\db^{b-1}_1} = \frac {\leU^2}{\leV^2} \frac{(\Ib+\db^{b-2}_2)(\Ib+\yb_1)(\Ib+\yb_2)}{(\Ib+\xb_1)(\Ib+\iota \frac{\leU}{\leV} \xb_1)^2}.
\ee
Thus the closed $Y$-system for $\lambda = \lambda_{a,b}$ consists of the relations \eqref{eq:ddd.Y.system} for $m =1, \dots, b-2$, along with the relations \eqref{eq:1+d.closure} and \eqref{eq:x.closure}.

%%%%%%%%%%%%%%%%%%%%%%%%%%%%%%%%%%%%%%%%%%%%%%%
%
\section{Free energies}\label{sec:free.energies}
%
%%%%%%%%%%%%%%%%%%%%%%%%%%%%%%%%%%%%%%%%%%%%%%%

In this section, we show how the fusion hierarchy relation for $\Db^{1,1}(u)$ can be used to compute the bulk and boundary free energies of the homogeneous transfer matrix (whereby $\xi_{(j)}=0$), following the method of Baxter \cite{B82}. This relation reads
\begin{equation}
\beta^2(1,1,u)\Db^{1,0}_0\Db^{1,0}_3 = \beta^1(1,1,u)\Db^{1,1}_0 + \beta^0(1,u)\beta^0(1,-u) \beta^3(1,1,u)\Ib. 
\end{equation}
The idea consists in noticing that the function $\beta^1(1,1,u)$ has a zero of degree $2N+2$ at $u=0$, thus rendering the first term on the right side exponentially negligible compared to the second in the neighbourhood of $u=0$. The bulk and boundary free energies for the leading eigenvalues $D(u)$ of $\Db^{1,0}(u)$ can then be computed from the relation
\begin{alignat}{2}
D(u)D(u+3\lambda)& \simeq \beta^0(1,u)\beta^0(1,-u)\frac{\beta^3(1,1,u)}{\beta^2(1,1,u)} \nonumber\\ &= \Big(s(u+2\lambda)s(u-2\lambda)s(u+3\lambda)s(u-3\lambda)\Big)^{2N} \\&\times
\frac{s(6\lambda-2u)s(6\lambda+2u)}{s(2\lambda-2u)s(2\lambda+2u)}\frac{\bar w(\frac\lambda2+u)\bar w(\frac\lambda2-u)}{\bar w(\frac{3\lambda}2+u)\bar w(\frac{3\lambda}2-u)}w(\tfrac{5 \lambda}2+u)w(\tfrac{5 \lambda}2-u)w(\tfrac{3 \lambda}2+u)w(\tfrac{3 \lambda}2-u).
\nonumber
\end{alignat}
Writing $\log D(u) \simeq -2N f_{\textrm{bulk}}(u)-f_{\textrm{bdy}}(u)$, we identify together the different orders in $N$  and write
\begin{subequations}
\begin{alignat}{2}
f_{\textrm{bulk}}(u) + f_{\textrm{bulk}}(u+3\lambda) = -\log \big(&s(u+2\lambda)s(u-2\lambda)s(u+3\lambda)s(u-3\lambda)\big),
\\
f_{\textrm{bdy}}(u) + f_{\textrm{bdy}}(u+3\lambda) = -\log\bigg[&\frac{s(6\lambda-2u)s(6\lambda+2u)}{s(2\lambda-2u)s(2\lambda+2u)}\frac{\bar w(\frac\lambda2+u)\bar w(\frac\lambda2-u)}{\bar w(\frac{3\lambda}2+u)\bar w(\frac{3\lambda}2-u)}\label{eq:bdy.functional.relation}\\
&\times w(\tfrac{5 \lambda}2+u)w(\tfrac{5 \lambda}2-u)w(\tfrac{3 \lambda}2+u)w(\tfrac{3 \lambda}2-u)\bigg].\nonumber
\end{alignat}
\end{subequations}

We may now solve these relations using Fourier transforms and assuming that the function $D(u)$ is analytic and non-zero in the strip $0\le \textrm{Re}(u)\le 3\lambda$. For the bulk free energy, this calculation was done previously by Warnaar, Batchelor and Nienhuis \cite{WBN92}. For $0<\lambda<\frac \pi 3$, the final expression is
\be
\label{eq:fbulk}
f_{\textrm{bulk}}(u) = -\log \sin(2\lambda) \sin(3\lambda) - 2 \int_{-\infty}^\infty \dd k\,  \frac{\sinh[uk]\sinh[(3\lambda-u)k]\cosh[(\pi-5 \lambda)k]\cosh[\lambda k]}{k \sinh(\pi k) \cosh(3 \lambda k)}.
\ee

Let us now solve \eqref{eq:bdy.functional.relation} for $f_{\textrm{bdy}}(u)$, starting with the mixed boundary conditions for which $w(u)=\bar w(u)=1$. We first study the asymptotic behavior of $\log D(u)$ as $u \to \iI \infty$. From the form of $D(u)$ as a centered Laurent polynomial in $\eE^{\iI u}$, we deduce that its behavior for $u \to \iI \infty$ is given by 
\begin{alignat}{2}
\log D(u) &=  \log \big(\alpha_{-4N} \eE^{-4N \iI u}+\alpha_{-(4N-2)} \eE^{-(4N-2) \iI u}+ \dots \big)
\simeq  -4 \iI u N + \log \alpha_{-4N} + \dots 
\end{alignat}
with the next terms going to zero in the limit. We thus find that\footnote{For SS and CC boundary conditions, the lowest power of $D(u)$ as a Laurent polynomial is $\eE^{-(4N+2)\iI u}$, and in this case only the second derivative of $f_{\textrm{bdy}}(u)$ vanishes, with the first derivative instead given by $\lim_{u\to \iI \infty}f'_{\textrm{bdy}}(u) = 2\iI$.}
\be
\label{eq:zero.derivatives}
\lim_{u\to\iI \infty} f''_{\textrm{bulk}}(u) = \lim_{u\to \iI\infty} f'_{\textrm{bdy}}(u) = 0.
\ee

As a next step, we define the function
\be
g(y) = f_{\textrm{bdy}}(\iI y + \tfrac{3\lambda}2),
\ee
for which
\be
g''(y+\tfrac{3 \iI \lambda}2) + g''(y-\tfrac{3 \iI \lambda}2)
= \frac4{\sinh^2(2y+6 \iI \lambda)} + \frac4{\sinh^2(2y-6 \iI \lambda)}
- \frac4{\sinh^2(2y+2 \iI \lambda)} - \frac4{\sinh^2(2y-2 \iI \lambda)}.
\ee
We define the Fourier transform of $g''(y)$:
\be
\label{eq:Gk.shifts}
G(k) = \frac1{2\pi} \int_{-\infty}^\infty \dd y\, \eE^{-\iI k y} g''(y), \qquad
g''(y) = \int_{-\infty}^\infty \dd k\, \eE^{\iI k y} G(k).
\ee
With the assumption that $f_{\textrm{bdy}}(u)$ has no zeros in the analyticity strip $0\le \textrm{Re}(u)\le 3 \lambda$, we find
\be
\frac1{2\pi} \int_{-\infty}^\infty \dd y\, \eE^{-\iI k y} g''(y \pm \tfrac{3 \iI \lambda}2) = \eE^{\mp 3 k \lambda/2} G(k). 
\ee
As a result, we have
\begin{alignat}{2}\label{eq:G.intermediate}
2 \cosh(\tfrac{3\lambda k}2) G(k) = \frac1{2\pi} \int_{-\infty}^\infty \dd y \, \eE^{-\iI k y} \bigg(&\frac4{\sinh^2(2y+6 \iI \lambda)} + \frac4{\sinh^2(2y-6 \iI \lambda)}\\
& - \frac4{\sinh^2(2y+2 \iI \lambda)} - \frac4{\sinh^2(2y-2 \iI \lambda)} \bigg).\nonumber
\end{alignat}
We can compute these integrals using the identity
\be\label{eq:int.identity}
\frac 1{2\pi} \int_{-\infty}^\infty \dd y\, \eE^{-\iI k y}\frac 1{\sinh^2(y-\iI \gamma)} =- \frac{k\, \eE^{k\gamma}}{\eE^{\pi k}-1}, \qquad \gamma \in (0,\pi),
\ee
which is easily obtained from the residue theorem. With $\gamma \mapsto \pi - \gamma$, we then also have 
\be
\frac 1{2\pi} \int_{-\infty}^\infty \dd y\, \eE^{-\iI k y}\frac 1{\sinh^2(y+\iI \gamma)} =- \frac{k\, \eE^{k(\pi-\gamma)}}{\eE^{\pi k}-1}, \qquad \gamma \in (0,\pi).
\ee
We also give the identity
\be\label{eq:int.identity.cosh}
\frac 1{2\pi} \int_{-\infty}^\infty \dd y\, \eE^{-\iI k y}\frac 1{\cosh^2(y\pm\iI \gamma)} = \frac{k\, \eE^{k(\frac{\pi}2\mp \gamma)}}{\eE^{\pi k}-1}, \qquad \gamma \in (-\tfrac{\pi}2,\tfrac{\pi}2),
\ee
as it is needed for the boundary conditions SS and CC. Applying this to the four terms in \eqref{eq:G.intermediate}, we find after some simplifications
\be
G(k) = \frac k{2\cosh(\frac{3 \lambda k}2)}\frac{\cosh[(\frac \pi 4-\lambda)k]-\cosh[(\frac \pi 4-3\lambda)k]}{\sinh(\frac{\pi k}4)}.
\ee
The inverse transform is then given by 
\begin{alignat}{2}
g''(y) &= \int_{-\infty}^\infty \dd k\, \frac {k\,\eE^{\iI k y}}{2\cosh(\frac{3 \lambda k}2)}\frac{\cosh[(\frac \pi 4-\lambda)k]-\cosh[(\frac \pi 4-3\lambda)k]}{\sinh(\frac{\pi k}4)} 
\nonumber\\
&= \int_{-\infty}^\infty \dd k\, \frac {k\cos(ky)}{2\cosh(\frac{3 \lambda k}2)}\frac{\cosh[(\frac \pi 4-\lambda)k]-\cosh[(\frac \pi 4-3\lambda)k]}{\sinh(\frac{\pi k}4)},
\end{alignat}
where we used the symmetry of the integrand under $k \to -k$.
Integrating once with respect to $y$ yields
\be
g'(y) = \int_{-\infty}^\infty \dd k\, \frac {\sin(ky)}{2\cosh(\frac{3 \lambda k}2)}\frac{\cosh[(\frac \pi 4-\lambda)k]-\cosh[(\frac \pi 4-3\lambda)k]}{\sinh(\frac{\pi k}4)} + C.
\ee
We fix the constant $C$ by imposing that $\lim_{y\to \infty} g'(y) = 0$, see \eqref{eq:zero.derivatives}. 
Indeed, we have
\be
\lim_{y\to \infty} g'(y) - C = 
\lim_{y\to \infty} \int_{-\infty}^\infty \dd k'\, \frac {\sin k'}{2y\cosh(\frac{3 \lambda k'}{2y})}\frac{\cosh[(\frac \pi 4-\lambda)\frac{k'}y]-\cosh[(\frac \pi 4-3\lambda)\frac{k'}y]}{\sinh(\frac{\pi k'}{4y})} = 0, \ee
as the integrand behaves as $y^{-2}$. We thus conclude that $C = 0$. Integrating once more, we find 
\be
g(y) = -\int_{-\infty}^\infty \dd k\,  \frac {\cos(ky)}{2k\cosh(\frac{3 \lambda k}2)}\frac{\cosh[(\frac \pi 4-\lambda)k]-\cosh[(\frac \pi 4-3\lambda)k]}{\sinh(\frac{\pi k}4)} + K.
\ee
The value of $g(\frac{3i\lambda}2)$ can be read off directly from \eqref{eq:DI}: $g(\frac{3 \iI \lambda}2) = -\log\big[\frac{\sin(6 \lambda)}{\sin(2\lambda)}\big]$. This can also be obtained from \eqref{eq:bdy.functional.relation} by noting that $g(\frac{3 \iI \lambda}2)$ and $g(-\frac{3 \iI \lambda}2)$ are equal due to crossing symmetry. This fixes the value of $K$. After simplification, this yields
\be
g(y) = -\log\bigg[\frac{\sin(6 \lambda)}{\sin(2\lambda)}\bigg] + \int_{-\infty}^\infty \dd k\, \frac {\sinh\big[(\frac{3\lambda}2+\iI y)\frac{k}2\big]\sinh\big[(\frac{3\lambda}2-\iI y)\frac{k}2\big]}{k\cosh(\frac{3 \lambda k}2)}\frac{\cosh[(\frac \pi 4-\lambda)k]-\cosh[(\frac \pi 4-3\lambda)k]}{\sinh(\frac{\pi k}4)}
\ee
and finally,
\be
\label{eq:fbdySC}
f_{\textrm{bdy}}(u) =-\log\bigg[\frac{\sin(6 \lambda)}{\sin(2\lambda)}\bigg] + \int_{-\infty}^\infty \dd k\, \frac {\sinh\big[\frac{uk}2\big]\sinh\big[(3 \lambda -u)\frac{k}2\big]}{k\cosh(\frac{3 \lambda k}2)}\frac{\cosh[(\frac \pi 4-\lambda)k]-\cosh[(\frac \pi 4-3\lambda)k]}{\sinh(\frac{\pi k}4)}.
\ee
In deriving this result, we applied \eqref{eq:int.identity} with $\gamma = 6 \lambda$, so this result holds for $0 < \lambda < \frac \pi 6$. In \cref{fig:SC.energies}, we compare these results with the groundstate eigenvalue $D(u)$ (namely the largest eigenvalue at $u=\frac{3 \lambda}2$), for SC boundary conditions and with $\lambda = \frac{\pi}{10}$. We see from the pattern of zeros of $D(u)$ that it has no zeros in the strip $0\le \textrm{Re}(u) \le 3\lambda$. We also see that the integral formulas for $f_{\textrm{bulk}}(u)$ and $f_{\textrm{bdy}}(u)$ nicely match the numerical data. We find a similar agreement numerically for the CS boundary condition.
\begin{figure}
\centering
\begin{pspicture}(-3,-2.2)(3,1.8)
\rput(0,0){\includegraphics[width=.38\textwidth]{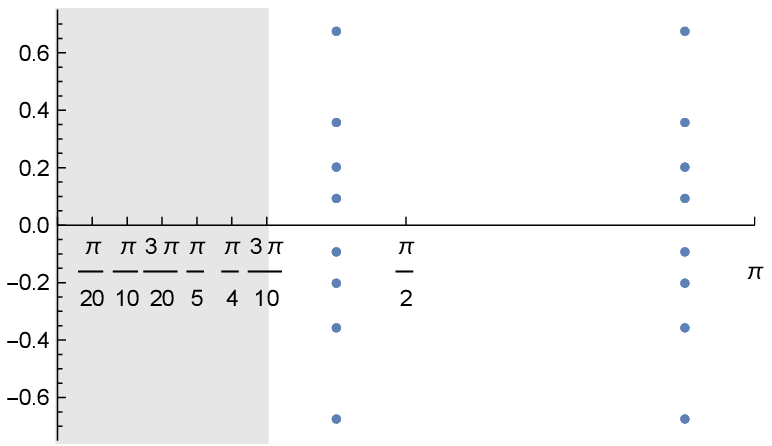}}
\end{pspicture}
\\[0.5cm]
\begin{pspicture}(-3,-2.2)(3,1.8)
\rput(0,0){\includegraphics[width=.38\textwidth]{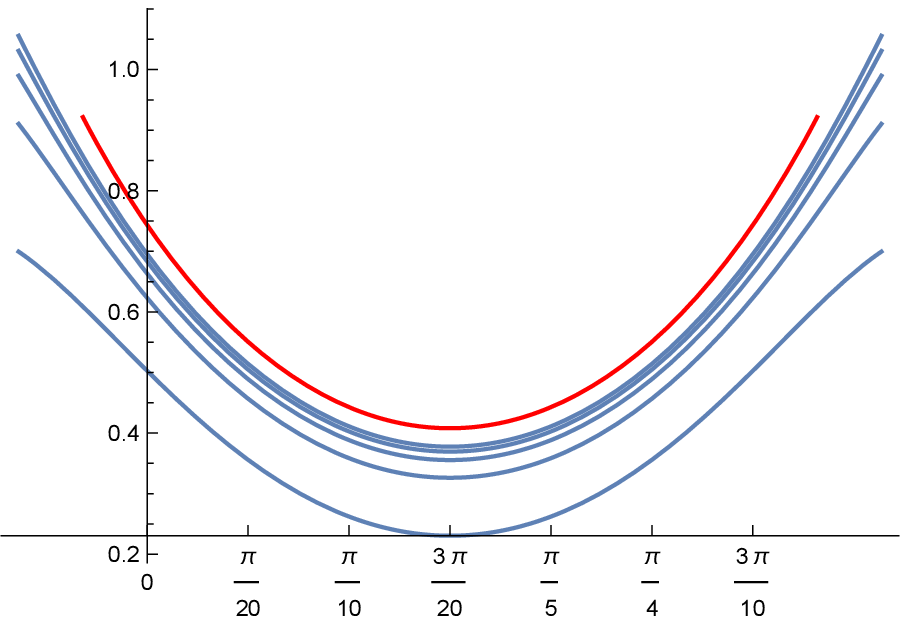}}
\rput(3,-1.5){$u$}
\rput(-1.5,1.75){$f_{\textrm{bulk}}(u)$}
\end{pspicture}
\hspace{3cm}
\begin{pspicture}(-3,-2.2)(3,1.8)
\rput(0,0){\includegraphics[width=.38\textwidth]{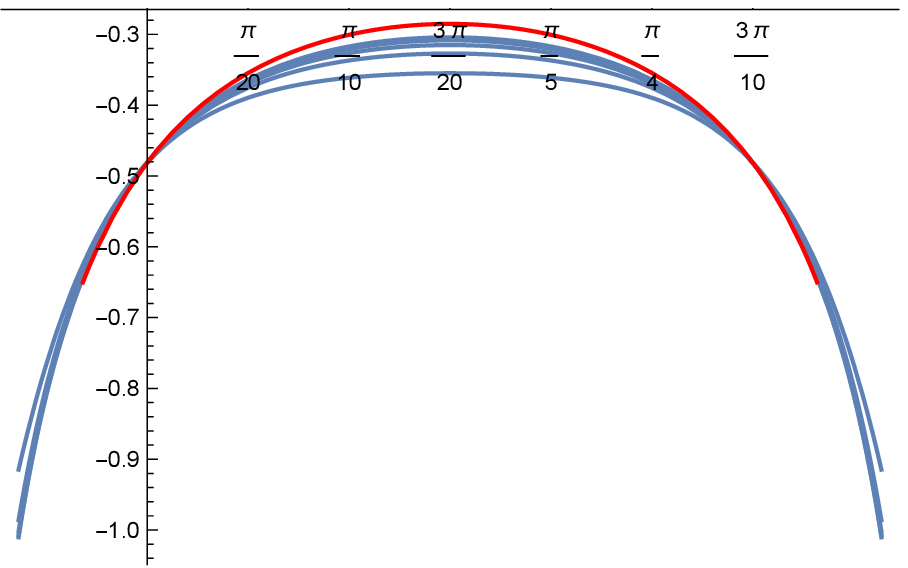}}
\rput(3.1,1.9){$u$}
\rput(-1.45,-1.65){$f_{\textrm{bdy}}(u)$}
\end{pspicture}
\caption{Zeros and free energies for SC boundary conditions with $\lambda = \frac \pi {10}$. \textit{Top:} the position of the zeros of $D(u)$ in the complex $u$-plane for $N=4$, with the analyticity strip colored in gray. \textit{Left:} the integral formula \eqref{eq:fbulk} for $f_{\textrm{bulk}}(u)$ in red and its finite-size approximations $-\frac1{2N}\log D(u)$ for $N=1, \dots, 5$ in blue. \textit{Right:} the integral formula \eqref{eq:fbdySC} for $f_{\textrm{bdy}}(u)$ in red and its finite-size approximations $-\log D(u)-2N f_{\textrm{bulk}}(u)$ for $N=1, \dots, 5$ in blue.}
\label{fig:SC.energies}
\end{figure}

For SS boundary conditions, the linear functional relation for $f_{\textrm{bdy}}(u)$ involves more terms on the right-hand side:
\begin{alignat}{2}
f_{\textrm{bdy}}(u) + f_{\textrm{bdy}}(u+3\lambda) = -\log\bigg[&\frac{s(6\lambda-2u)s(6\lambda+2u)}{s(2\lambda-2u)s(2\lambda+2u)}\frac{ c(\frac\lambda2+u)c(\frac\lambda2-u)}{c(\frac{3\lambda}2+u)c(\frac{3\lambda}2-u)}\\
&\times s(\tfrac{5 \lambda}2+u)s(\tfrac{5 \lambda}2-u)s(\tfrac{3 \lambda}2+u)s(\tfrac{3 \lambda}2-u)\bigg].
\nonumber
\end{alignat}
From this expression, the steps are similar to those for the mixed boundary conditions. We provide only the key steps. The second derivative of $g(y)$ is given by
\begin{alignat}{2}
g''(y)=\int_{-\infty}^\infty \dd k& \ \frac{k \cos(ky)}{2\cosh{(\tfrac{3\lambda k}2})}\left[\frac{\cosh[(\tfrac{\pi}4-\lambda)k]-\cosh[(\tfrac{\pi}4-3\lambda)k]}{\sinh(\tfrac{\pi k}4)}\right. \\
&+\left. \frac{\cosh(\tfrac{3\lambda k}2)-\cosh(\tfrac{\lambda k}2)-\cosh[(\tfrac{\pi-5\lambda}2)k]-\cosh[(\tfrac{\pi-3\lambda}2)k]}{\sinh(\tfrac{\pi k}2)}\right].\notag
\end{alignat}
As before, the first integration on $y$ replaces the first numerator $k\cos(ky)$ by $\sin(ky)$ and introduces a new constant of integration $C$ which is obtained by computing
\begin{alignat}{2}
\lim_{y\to \infty} g'(y) - C = 
\lim_{y\to \infty} &\int_{-\infty}^\infty \dd k'\, \frac {\sin k'}{2y\cosh(\frac{3 \lambda k'}{2y})}
\Bigg[
\frac{\cosh[(\frac \pi 4-\lambda)\frac{k'}y]-\cosh[(\frac \pi 4-3\lambda)\frac{k'}y]}{\sinh(\frac{\pi k'}{4y})} \\ 
&+
\frac{\cosh(\frac{3\lambda} 2 \frac{k'}y)-\cosh(\frac{\lambda} 2 \frac{k'}y)-
\cosh[(\frac{\pi-5\lambda}2)\frac{k'}y]-
\cosh[(\frac{\pi-3\lambda}2)\frac{k'}y]
}{\sinh(\frac{\pi k'}{2y})}
\Bigg]= -2. \notag
\end{alignat} 
The value of $g'(y) = \iI f'_{\textrm{bdy}}(u)$ in the limit $u \to \iI \infty$ is $-2$
by the above footnote, and $C$ must be set to~$0$. The final result for the these boundary conditions is thus
\begin{alignat}{2}
\label{eq:fbdySS}
f_{\textrm{bdy}}(u) &=-\log\bigg[-\frac{\sin(6 \lambda)}{\sin(2\lambda)}\frac{\cos(\frac\lambda2)\sin(\frac{5\lambda}2)\sin(\frac{3\lambda}2)}{\cos(\frac{3\lambda}2)}\bigg] \\
&+ \int_{-\infty}^\infty \dd k\, \frac {\sinh(\frac{uk}2)\sinh\big[(3 \lambda -u)\frac{k}2\big]}{k\cosh(\frac{3 \lambda k}2)}\bigg[\frac{\cosh[(\frac \pi 4-\lambda)k]-\cosh[(\frac \pi 4-3\lambda)k]}{\sinh(\frac{\pi k}4)}\nonumber\\
&\hspace{6cm}+\frac{\cosh(\frac{3\lambda k}2)-\cosh(\frac{\lambda k}2)-\cosh[\frac{(\pi-5\lambda) k}2]-\cosh[\frac{(\pi-3\lambda) k}2]}{\sinh(\frac{\pi k}2)}\bigg].\nonumber
\end{alignat}
It again holds for $0<\lambda< \frac \pi 6$. As illustrated in \cref{fig:SS.energies}, our numerical investigations for small~$N$ confirm the analyticity assumptions for the groundstate eigenvalue $D(u)$ and show a good match between the integral formulas for the free energies and the numerical data. We also note that, for $0<\lambda< \frac \pi 6$, the logarithm on the first line of \eqref{eq:fbdySS} has a negative argument, which implies that $f_{\textrm{bdy}}(u)$ has a non-zero imaginary part equal to an odd multiple of $\pi$. This is a simple artifact of our choice of normalisation for $\Db(u)$, causing it to be negative for $0<u<3\lambda$ for SS boundary conditions, see for instance \eqref{eq:DI}. As a result, we only plot the real part $f_{\textrm{bdy}}(u)$ in the right panel of \cref{fig:SS.energies}.
\begin{figure}
\centering
\begin{pspicture}(-3,-2.0)(3,1.8)
\rput(0,0){\includegraphics[width=.38\textwidth]{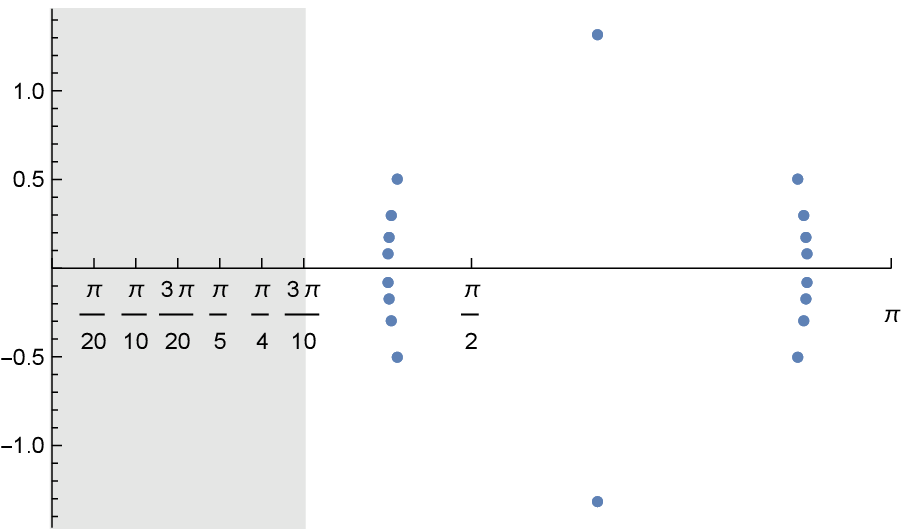}}
\end{pspicture}
\\[0.5cm]
\begin{pspicture}(-3,-2.2)(3,1.8)
\rput(0,0){\includegraphics[width=.38\textwidth]{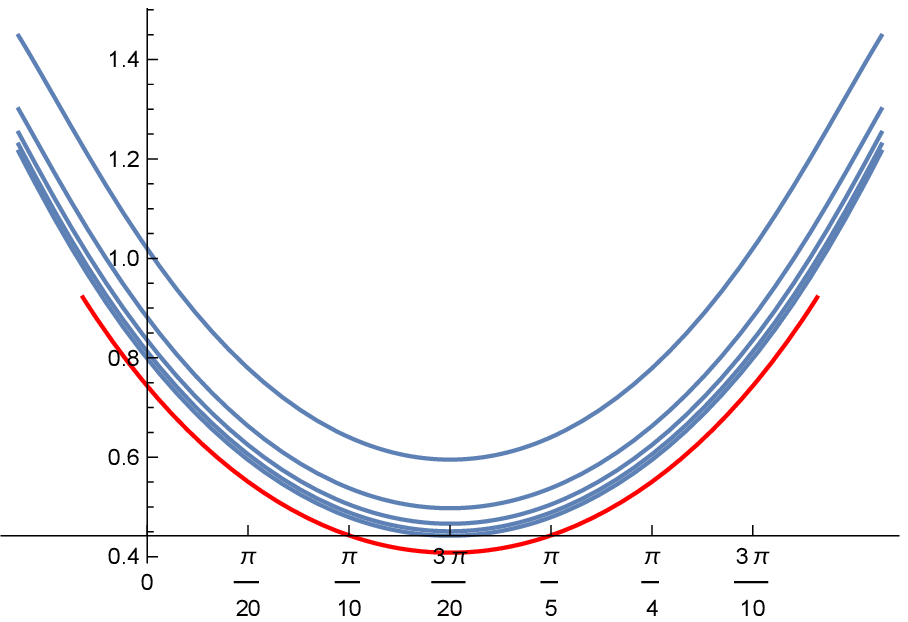}}
\rput(3,-1.5){$u$}
\rput(-1.5,1.75){$f_{\textrm{bulk}}(u)$}
\end{pspicture}
\hspace{3cm}
\begin{pspicture}(-3,-2.2)(3,1.8)
\rput(3,-1.5){$u$}
\rput(-1.1,1.75){Re$\big(f_{\textrm{bdy}}(u)\big)$}
\rput(0,0){\includegraphics[width=.38\textwidth]{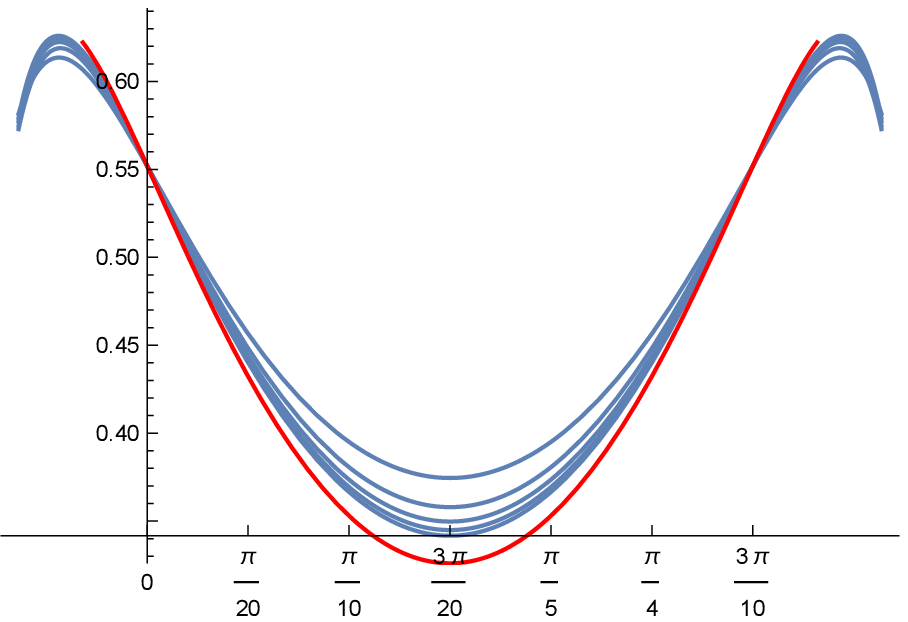}}
\end{pspicture}
\caption{Zeros and free energies for SS boundary conditions, with $\lambda = \frac \pi {10}$. \textit{Top:} the position of the zeros of $D(u)$ in the complex $u$-plane for $N=4$, with the analyticity strip colored in gray. \textit{Left:} the integral formula \eqref{eq:fbulk} for $f_{\textrm{bulk}}(u)$ in red and its finite-size approximations $-\frac1{2N}\log D(u)$ for $N=1, \dots, 5$ in blue. \textit{Right:} the integral formula \eqref{eq:fbdySS} for the real part of $f_{\textrm{bdy}}(u)$ in red and its finite-size approximations $-\log D(u)-2N f_{\textrm{bulk}}(u)$ for $N=1, \dots, 5$ in blue.}
\label{fig:SS.energies}
\end{figure}

\begin{figure}
\centering
\begin{pspicture}(-3,-2.0)(3,1.8)
\rput(0,0){\includegraphics[width=.38\textwidth]{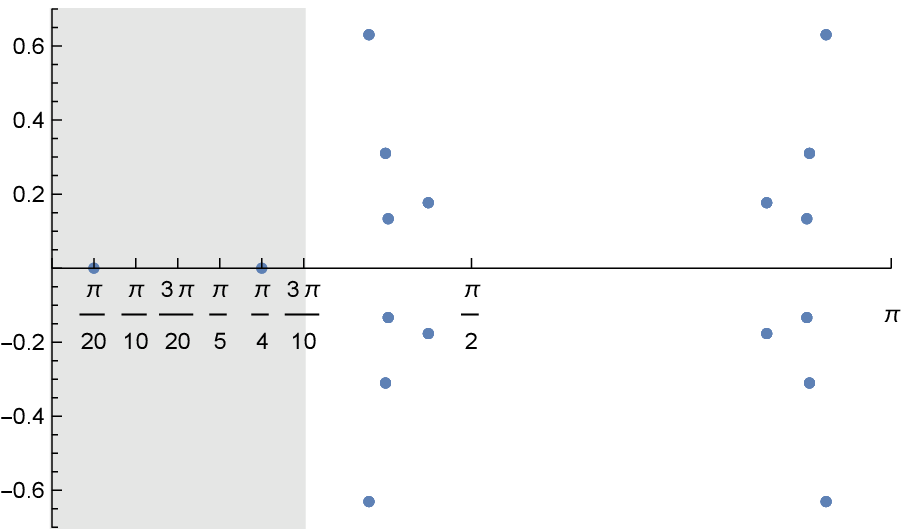}}
\end{pspicture}
\\[0.5cm]
\begin{pspicture}(-3,-2.2)(3,1.8)
\rput(3,-1.60){$u$}
\rput(0,1.95){$f_{\textrm{bulk}}(u)$}
\rput(0,0){\includegraphics[width=.38\textwidth]{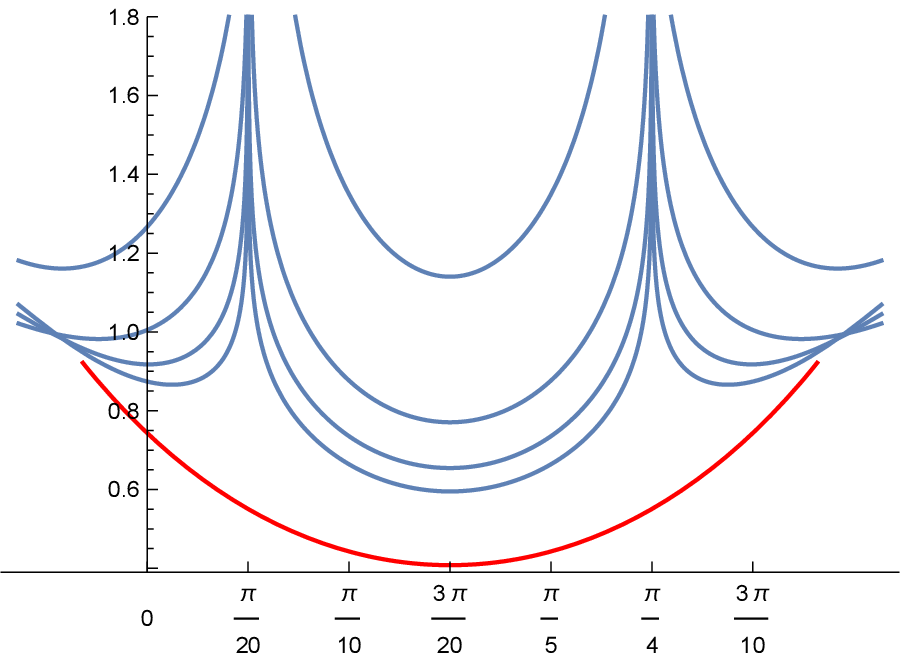}}
\end{pspicture}
\hspace{3cm}
\begin{pspicture}(-3,-2.2)(3,1.8)
\rput(3.2,-1.20){$u$}
\rput(0,1.75){Re$\big(f_{\textrm{bdy}}(u)\big)$}
\rput(0,0){\includegraphics[width=.38\textwidth]{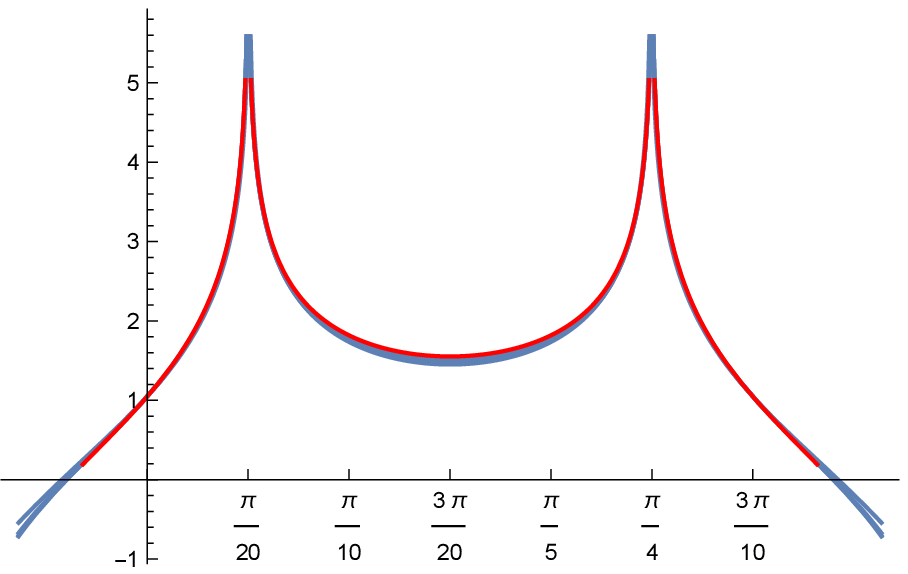}}
\end{pspicture}
\caption{Zeros and free energies for CC boundary conditions, with $\lambda = \frac \pi {10}$. \textit{Top:} the position of the zeros of $D(u)$ in the complex $u$-plane for $N=4$, with the analyticity strip colored in gray. \textit{Left:} the integral formula \eqref{eq:fbulk} for $f_{\textrm{bulk}}(u)$ in red and its finite-size approximations $-\frac1{2N}\log D(u)$ for $N=2, \dots, 5$ in blue. \textit{Right:} the integral formula \eqref{eq:fbdyCC} for the real part of $f_{\textrm{bdy}}(u)$ in red and its finite-size approximations $-\log D(u)-2N f_{\textrm{bulk}}(u)$ for $N=2, \dots, 5$ in blue.}
\label{fig:CC.energies}
\end{figure}

For CC boundary conditions, the functional relation \eqref{eq:bdy.functional.relation} is
\begin{alignat}{2}
f_{\textrm{bdy}}(u) + f_{\textrm{bdy}}(u+3\lambda) = -\log\bigg[&\frac{s(6\lambda-2u)s(6\lambda+2u)}{s(2\lambda-2u)s(2\lambda+2u)}\frac{s(\frac\lambda2+u)s(\frac\lambda2-u)}{s(\frac{3\lambda}2+u)s(\frac{3\lambda}2-u)}\\
&\times c(\tfrac{5 \lambda}2+u)c(\tfrac{5 \lambda}2-u)c(\tfrac{3 \lambda}2+u)c(\tfrac{3 \lambda}2-u)\bigg].\nonumber
\end{alignat}
While it is possible to solve this equation with the same arguments as before, the solution turns out not to match the numerical data. Indeed, for CC boundary conditions, our numerical explorations reveal that the groundstate eigenvalue $D(u)$ has a pair of real zeros very close to $u = \frac \lambda2$ and $u = \frac {5\lambda}2$. These thus lie inside the strip $0<\textrm{Re}(u)<3\lambda$, and the condition that $D(u)$ is free of zeros in this strip is violated. As a result, \eqref{eq:Gk.shifts} no longer holds, as the argument leading to this identity requires that we deform the integration path across an area that is free of poles, a condition that does not hold in the present case. The position of these zeros is not exactly at $u = \frac \lambda2, \frac {5\lambda}2$, however they appear to converge to these positions rapidly as $N$ is increased. To bypass this problem and compute $f_{\textrm{bdy}}(u)$ for CC boundary conditions, it is useful to consider the eigenvalue 
\be\label{eq:pourEnleverLesZeros}
\widehat D(u) = \frac{D(u)}{\sin(\frac{\lambda}2-u)\sin(\frac{5\lambda}2-u)},\ee
a function that has no zero in the analyticity strip, in the limit $N \to \infty$. Its corresponding boundary free energy satisfies the functional relation 
\be
\widehat f_{\textrm{bdy}}(u) + \widehat f_{\textrm{bdy}}(u+3\lambda) = -\log\bigg[\frac{s(6\lambda-2u)s(6\lambda+2u)}{s(2\lambda-2u)s(2\lambda+2u)}
\frac{c(\tfrac{5 \lambda}2+u)c(\tfrac{5 \lambda}2-u)c(\tfrac{3 \lambda}2+u)c(\tfrac{3 \lambda}2+u)}{s(\frac{3\lambda}2+u)s(\frac{3\lambda}2-u)s(\frac{5\lambda}2+u)s(\frac{5\lambda}2-u)}\bigg].
\ee
Defining $\widehat g(y) = \widehat f_{\textrm{bdy}}(\iI y + \frac{3\lambda}2)$, we find that its second derivative is
\begin{alignat}{2}
\widehat g''(y)=\int_{-\infty}^\infty \dd k& \ \frac{k \cos(ky)}{2\cosh{(\tfrac{3\lambda k}2})}\left[\frac{\cosh[(\tfrac{\pi}4-\lambda)k]-\cosh[(\tfrac{\pi}4-3\lambda)k]}{\sinh(\tfrac{\pi k}4)}\right.\\
&+\left. \frac{\cosh[(\tfrac{\pi-5\lambda}2)k]+\cosh[(\tfrac{\pi-3\lambda}2)k]-\cosh(\tfrac{5\lambda k}2)-\cosh(\tfrac{3\lambda k}2)}{\sinh(\tfrac{\pi k}2)}\right]\notag
\end{alignat}
and the integral formula for $f_{\textrm{bdy}}(u)$ for CC boundary conditions reads
\begin{alignat}{2}
\label{eq:fbdyCC}
f_{\textrm{bdy}}(u)
&=-\log \bigg[\frac{\sin(6 \lambda)}{\sin(2\lambda)}\frac{\cos(\frac{5\lambda}2)\cos(\frac{3\lambda}2)}{\sin(\frac{5\lambda}2)\sin(\frac{3\lambda}2)}\bigg]
- \log \big(\sin(\tfrac\lambda2-u)\sin(\tfrac{5\lambda}2-u)\big) \\
&+ \int_{-\infty}^\infty \dd k \, \frac {\sinh\big[\frac{uk}2\big]\sinh\big[(3 \lambda -u)\frac{k}2\big]}{k\cosh(\frac{3 \lambda k}2)}\bigg[\frac{\cosh[(\frac \pi 4-\lambda)k]-\cosh[(\frac \pi 4-3\lambda)k]}{\sinh(\frac{\pi k}4)}\nonumber\\
&\qquad\qquad +\frac{\cosh\big(\frac{(\pi-3\lambda) k}2\big)+\cosh\big(\frac{(\pi-5\lambda) k}2\big)-\cosh\big(\frac{5\lambda k}2\big)-\cosh\big(\frac{3\lambda k}2\big)}{\sinh(\frac{\pi k}2)}\bigg]\nonumber
\end{alignat}
and is valid for $0<\lambda<\frac \pi 6$. One can then check that $\lim_{u\to \iI \infty} f_{\textrm{bdy}}'(u) = 2\iI$ as needed. The second logarithm in \eqref{eq:fbdyCC} gives rise to an imaginary part for $u \in (\frac{\lambda}2,\frac{5\lambda}2)$. This is due to the presence of the poles in $\widehat D(u)$, see \eqref{eq:pourEnleverLesZeros}. Only the real part has a physical interpretation. The corresponding numerical data is given in \cref{fig:CC.energies}. In addition to the zeros of $D(u)$ lying inside the analyticity strip, we also see that the convergence of the numerical data for $f_{\textrm{bulk}}(u)$ to its integral formula is much slower in this case due to the two singularities of $f_{\textrm{bdy}}(u)$. The agreement for Re$\big(f_{\textrm{bdy}}(u)\big)$ is in contrast very convincing.

%%%%%%%%%%%%%%%%%%%%%%%%%%%%%%%%%%%%%%%%%%%%%%%
%
\section{Conclusion}\label{sec:conclusion}
%
%%%%%%%%%%%%%%%%%%%%%%%%%%%%%%%%%%%%%%%%%%%%%%%

In this paper, we studied the commuting transfer matrices of the $\Atwotwo$ model on the strip. The main results of the article are the fusion hierarchy relations \eqref{eq:HF} defining the fused transfer matrices, their $T$-system \eqref{Tsystem}, and for $\lambda/\pi \in \mathbb Q$ the closure relation \eqref{eq:closure} for the hierarchy and the corresponding the finite $Y$-systems, consisting of the relations \eqref{eq:ddd.Y.system}, \eqref{eq:1+d.closure} and \eqref{eq:x.closure}.
These formulas and their proofs turn out to have a much greater complexity than the similar results for the dense loop model or for the $\Atwotwo$ model with periodic boundary conditions. This can be traced back to the fact that the solutions of the boundary Yang-Baxter equation have a non-trivial dependence on the spectral parameter $u$, which then causes the fused transfer matrices to have polynomial degrees in $\eE^{\iI u}$ that increase with the fusion indices. A similar feature in fact arises for the dense loop model with loop segments that can be attached to the boundaries of the strip, in the context of the two-boundary Temperley-Lieb algebra. The closure relation in this case is conjectured in Section 5.2 of \cite{FrahmMDP2019}, but remains without proof. We expect that the methods developed here will apply to this case and lead to a proof of this result.

In contrast, here we provided complete proofs for all the results. These proofs, particularly the difficult ones for the polynomiality of the fused tangles and their closure relation, indicate how tightly constrained the definition of the fusion hierarchy is. Despite the substantial length of this text, it is worth mentioning that our first versions of many proofs were even longer. For example, an alternative proof of the polynomiality of $\Db^{m,0}(u)$ and $\Db^{m,1}(u)$ relies on their equivalent diagrammatic definitions with the projectors $P^{m,0}$ and $P^{m,1}$ defined in \cite{AMDPP19}. Moreover, we note that, like for the periodic $\Atwotwo$ dilute loop models, a diagrammatic definition of $\Db^{m,n}(u)$ for $m,n>1$ and of the corresponding projectors $P^{m,n}$ is still missing. Finally, we note that unifying the presentation of the hierarchies for the four boundary conditions (SS, CC, SC and CS) through the notation of \cref{tab:lesPoids} was not an easy task. It reduced the length of the proofs significantly.

We also computed the bulk and boundary free energies using the first fusion hierarchy relations and the analyticity properties of the groundstate eigenvalues. Their agreement with numerical data, obtained for small strip lengths $N$, is striking. In particular, the case of CC boundary conditions featured zeros inside the analyticity strip and was resolved convincingly. This, of course, raises the question of which conformal weights arise in the continuum limit of the loop models. The method of non-linear integral equations, originally developed by Kl\"umper and Pearce in the nineties \cite{PK91,KP91,KP92}, allows for the computation of the next leading term in the large-$N$ expansion of the eigenvalues. This term is proportional to $1/N$ and its coefficient involves the combination $c-24 \Delta$, where $c$ is the central charge and $\Delta$ is the conformal weight of the field that the corresponding eigenstate scales to in the limit. These techniques were recently applied to the groundstate of the loop models in the standard module with no defects, for the $\Aoneone$, $\Atwoone$ and $\Atwotwo$ models with periodic boundary conditions \cite{AMDAKPP21}. For $\Atwotwo$, these results were found to be consistent with the known values of the central charge  in Regimes I and II. The general formula is \cite{WBN92}
\be
c = \left\{\begin{array}{clc}
\displaystyle 1-\frac{3(\pi-4\lambda)^2}{\pi(\pi-2\lambda)},
&\quad\textrm{Regime I:}&0<\lambda<\frac{\pi}{2},
\\[0.3cm]
\displaystyle -1+\frac{6(\pi-2\lambda)^2}{\pi(\pi-\lambda)},
&\quad\textrm{Regime III:}&\frac{\pi}{2}<\lambda<\frac{2\pi}{3},
\\[0.3cm]
\displaystyle \frac32-\frac{3(3\pi-4\lambda)^2}{2\pi(\pi-\lambda)},
&\quad\textrm{Regime II:}&\frac{2\pi}{3}<\lambda<\pi.
\end{array}\right.
\ee
This work should be extended to the same models on the strip, for all four choices of boundary conditions. In the representation $\wW_{N}^{d=0}$, one can expect for the mixed boundary conditions a groundstate with a non-zero conformal weight $\Delta$, corresponding to the insertion at infinity of a field that changes the boundary condition from $S$ to $C$. For identical boundary conditions, we instead expect that the groundstate has the weight $\Delta=0$, since no boundary condition changing field must be inserted in this case. The technique of non-linear equations and Rogers dilogarithms should be applied to confirm this, and subsequently to compute the conformal weights of the groundstates in all the standard representations $\wW_{N}^d$. A more ambitious project would then be to apply this to the excited states of the $\Atwotwo$ loop model as well, and thus to compute the properly scaled transfer matrix traces and express them in terms of characters of representations of the Virasoro algebra. This program is currently being implemented for the special case of $\lambda = \frac \pi 3$ corresponding to the model of critical site percolation on the triangular lattice, with the results to appear soon \cite{MDKP22}.

%%%%%%%%%%%
%
% many thanks
%
%%%%%%%%%%%

\section*{Acknowledgements}

FB thanks the D\'epartement de physique of the Universit\'e de Montr\'eal for its financial support during her graduate studies. AMD is supported by the EOS Research Project, project number 30889451. He also acknowledges support from the Max-Planck-Institute f\"ur Mathematik in Bonn and the Leibniz Universit\"at Hannover, where parts of this work were done. YSA holds a grant from the Natural Sciences and Engineering Research Council of Canada. This support is gratefully acknowledged. AMD and YSA thank J.~Rasmussen for collaborative investigations of the dilute $\Atwotwo$ loop model that predated this project.

\appendix

%%%%%%%%%%%%%%%%%%%%%%%%%%%%%%%%%%%%%%%%%%%%%%%
%
\section{Proofs of properties of the fused transfer matrices}\label{app:proofs}
%
%%%%%%%%%%%%%%%%%%%%%%%%%%%%%%%%%%%%%%%%%%%%%%%

In this appendix, we collect the technical proofs of some of the properties of the fused transfer matrices.

%%%%%%%%%%%%%
\subsection{Proof of the crossing symmetry of $\Db^{m,n}(u)$ and $\det(m,n)(u)$} \label{sub:proofCS}
%%%%%%%%%%%%%

This section is devoted to proving the crossing symmetries \eqref{eq:symCroiseDmn} and \eqref{eq:symCroiseDETmn}. We start with the proof for $\Db^{m,0}(u)$. Together with certain properties of the functions $\beta^i(m,n,u)$, it will imply the crossing symmetry of both $\Db^{m,n}(u)$ and $\det(m,n)(u)$. 

The proof of the equality $\Db^{m,0}((4-2m)\lambda-u)=\Db^{0,m}(u)$ is by induction on $m$. The inductive argument requires that the cases $m-1$, $m-2$ and $m-3$ be already shown. We thus check the crossing symmetry for transfer tangles $\Db^{1,0}(u)$, $\Db^{2,0}(u)$ and $\Db^{3,0}(u)$ first. The case $m=1$ follows readily from \eqref{eq:symCroiseD10} and \eqref{def:D01}. For $m=2$, we use the definition \eqref{eq:HFa} for $\Db^{2,0}(u)$ and the evaluation point $\slu = (4-4)\lambda-u=-u$. We show that $\Db^{2,0}(\slu)=\Db^{2,0}(u)$ using the definitions of the weights in \eqref{eq:HFa} and \eqref{eq:HFc} and the crossing symmetry of $\Db^{1,0}(u)$. For $m=3$, proving the crossing symmetry for $\Db^{3,0}(u)$ requires that we understand the same property for $\Db^{1,1}(u)$. Its crossing symmetry follows from \eqref{eq:HFb}, the crossing symmetry of $\Db^{1,0}(u)$, and the fact that the three functions $\beta^i(1,1,u)$, $i=1, 2, 3$, are invariant under $u\mapsto \slu=-u$. Then, the symmetry of $\Db^{3,0}(u)$ is checked using the same arguments as for $\Db^{2,0}(u)$, with the property just proved for $\Db^{1,1}(u)$.

We now consider $m\geq 4$, for which $\slu=(4-2m)\lambda-u$. The fusion hierarchies \eqref{eq:HFd} and \eqref{eq:HFe} respectively define $\Db^{m,0}(u)$ and $\Db^{0,m}(u)$ and involve the fused transfer matrices $\Db^{m-2,1}(u)$ and $\Db^{1,m-2}(u_2)$, to which the inductive assumption does not apply. These are replaced by their definitions \eqref{eq:HFf} and \eqref{eq:HFg}:
\begin{subequations}
\begin{alignat}{2}
\Db^{m-2,1}(u) &= \frac{\Bdeux(m-2,1,u) \Db^{m-2,0}_0 \Db^{0,1}_{2(m-2)} \label{eq:m21}
- \Btrois(m-2,1,u) \Bzero(m-2,u) \Db^{m-3,0}_0}{\Bun(m-2,1,u)} \ , \\
\Db^{1,m-2}(u_2) &= \frac{\Bdeux(m-2,1,v) \Db^{1,0}_{2} \Db^{0,m-2}_{4} \label{eq:1m2}
- \Btrois(m-2,1,v) \Bzero(m-2,v) \Db^{0,m-3}_{6}}{\Bun(m-2,1,v)} \ ,
\end{alignat}
\end{subequations}
where, in the second equation, $v$ must be computed following the prescription  \eqref{eq:HFg}:
\begin{equation}
v = (4-2-2(m-2))\lambda-u_2 = (4-2m)\lambda-u = \slu \ .
\end{equation}
Equation \eqref{eq:1m2} provides an example where the operation $u\mapsto\slu$ is performed before the shift $u\mapsto u_2$. Thus, $v$ is precisely the combination appearing in the crossing symmetry for $\Db^{m,0}(u)$. The expressions for $\Db^{m,0}(u)$ and $\Db^{0,m}(u)$ become
\begin{subequations}
\begin{alignat}{2}
\Aun&(m,u) \Db^{m,0}(u) = \Big[\Adeux(m,u) \Db^{m-1,0}_0 \Db^{1,0}_{2m-2}\notag \\
&- \frac{\Atrois(m,u)}{\Bun(m-2,1,u)}\Big(\Bdeux(m-2,1,u) \Db^{m-2,0}_0 \Db^{0,1}_{2m-4} - \Btrois(m-2,1,u) \Bzero(m-2,u) \Db^{m-3,0}_0 \Big)\Big] \ ,\label{eq:proofCSm0} \\
\Aun&(m,\slu) \Db^{0,m}(u) =  \Big[\Adeux(m,\slu) \Db^{0,m-1}_2 \Db^{0,1}_{0} \notag \\
&- \frac{\Atrois(m,\slu)}{\Bun(m-2,1,\slu)} \Big(\Bdeux(m-2,1,\slu) \Db^{1,0}_{2}\Db^{0,m-2}_{4} - \Btrois(m-2,1,\slu) \Bzero(m-2,\slu) \Db^{0,m-3}_{6} \Big)\Big] \label{eq:proofCS0m} \ .
\end{alignat}
\end{subequations}
In both of these equations, the fused matrices on the right-hand side are evaluated at $u$. We now evaluate all terms in \eqref{eq:proofCSm0} at $\slu$. All weight functions become exactly equal to those of \eqref{eq:proofCS0m}. It remains to check that the transfer tangles transform correctly. Here are two examples:
\begin{alignat}{2}
&\Db^{m-1,0}_0(\slu)
= \Db^{m-1,0}_0((4-2m)\lambda-u)
= \Db^{m-1,0}_0((4-2(m-1))\lambda-(u+2\lambda))
= \Db^{0,m-1}_2(u) \ , \nonumber\\
&\Db^{1,0}_{2m-2}(\slu)
= \Db^{1,0}_0((4-2m)\lambda-u+(2m-2)\lambda)
= \Db^{1,0}_0((4-2)\lambda-u)
= \Db^{0,1}_0(u) \ .
\end{alignat}
The last equality in each line uses the induction hypothesis with $m'=m-1$ and $m'=1$ respectively. The proofs for the other matrices is similar. This completes the proof of the crossing symmetry for $\Db^{m,0}(u)$.

The proof of $\Db^{m,n}(u)=\Db^{n,m}((4-2m-2n)\lambda-u)$ for $m, n>0$ rests on the following property of the functions $\beta^i$: 
\begin{equation} \label{eq:symCroiseBeta}
\beta^i(m,n,\slu) = \beta^i(n,m,u), \qquad i=1, 2, 3\ .
\end{equation}
(We already used this fact to prove the crossing symmetry of $\Db^{1,1}(u)$.) This equality requires the identity $f(u)=f(-u)$ and is clear for all the other weights, as they appear in pairs.

We give the details of the proof for \eqref{eq:HFh} wherein both $m, n\geq 2$. The manipulations are similar for \eqref{eq:HFf} or \eqref{eq:HFg}. With \eqref{eq:symCroiseBeta}, \eqref{eq:HFh} evaluated at $\slu$ reads
\begin{alignat}{2}
\Db^{m,n}_0(\slu)
&= \frac{\Bdeux(m,n,\slu) \Db^{m,0}_0(\slu) \Db^{0,n}_{2m}(\slu)
- \Btrois(m,n,\slu) \Db^{m-1,0}_0(\slu) \Db^{0,n-1}_{2m+2}(\slu)}{\Bun(m,n,\slu)} \nonumber\\
&= \frac{\Bdeux(n,m,u) \Db^{m,0}_0(\slu) \Db^{0,n}_{2m}(\slu)
- \Btrois(n,m,u) \Db^{m-1,0}_0(\slu) \Db^{0,n-1}_{2m+2}(\slu)}{\Bun(n,m,u)} \ .
\end{alignat}
The remaining step is to check that the fused tangles in $\Db^{m,n}_0(\slu)$ are identical to those appearing in the definition of $\Db^{n,m}(u)$. Here are two examples, the two others being obtained with similar arguments:
\begin{alignat}{2}
&\Db^{m,0}_0(\slu)
= \Db^{m,0}_0((4-2m-2n)\lambda-u)
= \Db^{m,0}_0((4-2m)\lambda-(u+2n\lambda))
= \Db^{0,m}_{2n}(u)\ , \nonumber\\
&\Db^{0,n}_{2m}(\slu)
= \Db^{0,n}_0((4-2m-2n)\lambda-u+2m\lambda)
= \Db^{0,n}_0((4-2n)\lambda-u)
= \Db^{n,0}_{0}(u)\ .
\end{alignat}
The last equality in each line uses $\Db^{m',0}((4-2m')\lambda-u)=\Db^{0,m'}(u)$. This ends the proof of the crossing symmetry \eqref{eq:symCroiseDmn}.

The proof of the crossing symmetry of the determinantal form \eqref{def:detFormel(m,n)} follows easily. For $m,n\geq 1$, it consists in exchanging $m$ and $n$ in \eqref{def:detExplicite(m,n)} and evaluating the result at $\slu=(4-2m-2n)\lambda-u$. Clearly the two bottom lines of the right-hand side of \eqref{def:detExplicite(m,n)} simply get exchanged. So are all the other factors, except maybe the three factors $s(2u_{m+n-2}), s(2\slu_{2n-2})$ and $f(u_{2m-2})$. A direct computation shows that the latter remains invariant under the changes while the two former become $-s(2u_{m+n-2})$ and $-s(2\slu_{2m-2})$ respectively. The last step is to use the crossing symmetry of $\Db^{m,n}(u)$. This confirms the crossing symmetry \eqref{eq:symCroiseDETmn} of the determinantal form, for $m,n>1$. Note that, by definition of $\det(0,n)$ in \eqref{def:detFormel(0,n)}, the crossing symmetry when one of $m$ and $n$ is zero is actually built in, so nothing has to be proved. Still the relation \eqref{def:detExplicite(m,0)} between $\det(m,0)(u)$ and $\Db^{m,0}(u)$ enjoys an invariance similar to that of \eqref{def:detExplicite(m,n)} and it will play a role in the next section.

%%%%%%%%%%%%%
\subsection{Proof of the conjugacy of $\Db^{m,n}(u)$ and $\det(m,n)(u)$} \label{sub:proofConjugacy}
%%%%%%%%%%%%%

This section proves the conjugacy properties \eqref{eq:Dm0D0m} and \eqref{eq:conjugDETmn}. We start with the latter: $\det(0,m)(u) = \det(m,0)(u+\lambda)$.

By definition, we have $\det(0,m)(u)=\det(m,0)((4-2m)\lambda-u)$. It is thus natural to evaluate the entries in the determinant $\det(m,0)$ at $\slu=(4-2m)\lambda-u$. Let $k$ be an arbitrary shift of the spectral parameter and $v=u+\lambda$. The functions $\Dbm_k$ and $\fbm_k$ in this determinant become
\begin{alignat}{2}
\Dbm_k(u) \big\rvert_{u\mapsto(4-2m)\lambda-u}
&= \Dbm((4-2m)\lambda-u+k\lambda) 
= \Dbm(3\lambda - (u+(2m-k-1)\lambda)) \nonumber \\
& \stackrel{\eqref{eq:symDbm}}{=} \Dbm(u+(2m-k-1)\lambda) 
= \Dbm_{2m-2-k}(v) \ ,
\intertext{and}
\fbm_k(u) \big\rvert_{u\mapsto(4-2m)\lambda-u}
&= \fbm((4-2m)\lambda-u+k\lambda) 
= \fbm(\lambda - (u+(2m-k-3)\lambda)) 
\nonumber \\& 
\stackrel{\eqref{eq:symFbm}}{=} \fbm(u+(2m-k-3)\lambda) 
= \fbm_{2m-4-k}(v) \ .
\end{alignat}
It follows that
\begin{equation}
\det(m,0)(4\lambda-2m\lambda-u) = \begin{vmatrix}
\Dbm_{0}(v) & \Dbm_{1}(v) & \fbm_{1}(v) & 0 & 0 \\
\fbm_{0}(v) & \Dbm_{2}(v) & \Dbm_{3}(v) & \ddots & 0 \\
0 & \fbm_{2}(v) & \Dbm_{4}(v) & \ddots & \fbm_{2m-5}(v) \\
0 & 0 & \ddots & \ddots & \Dbm_{2m-3}(v) \\
0 & 0 & 0 & \fbm_{2m-4}(v) & \Dbm_{2m-2}(v)
\end{vmatrix} \ .
\end{equation}
This matrix is the transpose with respect to the anti-diagonal of the one appearing in $\det(m,0)(v)$. Their determinants are equal, and therefore
\begin{equation} \label{eq:proofConjugEquationDets}
\det(0,m)(u)
= \det(m,0)((4-2m)\lambda-u)
= \det(m,0)(v)
= \det(m,0)_1(u) \ .
\end{equation}

We now argue that the previous result implies the conjugacy property for the fused matrix, namely $\Db^{m,0}(u+\lambda)=\Db^{0,m}(u)$. From \eqref{eq:proofConjugEquationDets}, we have
\begin{equation} \label{eq:proofConjugDetSym}
\det(m,0)((5-2m)\lambda-u) = \det(m,0)(u) \ .
\end{equation}
The product of weights in the expression \eqref{def:detExplicite(m,0)} for $\det{(m,0)}_0(u)$ is invariant under $u\mapsto (5-2m)\lambda-u$. (This fact can be verified by direct computation. Some signs appear during this computation, but they all cancel at the end. This invariance is the property alluded to at the end of the previous section.) From \eqref{eq:proofConjugDetSym}, the identity $\Db^{m,0}((5-2m)\lambda-u) = \Db^{m,0}(u)$ follows and, by the crossing symmetry \eqref{eq:symCroiseDmn}, the desired conjugacy property: $\Db^{m,0}_1(u)=\Db^{0,m}_0(u)$.

%%%%%%%%%%%%%
\subsection{Proof of the $T$-system relations} \label{sub:proof.T-system}
%%%%%%%%%%%%%

The proof of the $T$-system relations \eqref{Tsystem} presented in this section is completely analogous to the same proof given for the periodic case in \cite{AMDPP19}. The proof is inductive on $m$ and requires that we first check the seed cases $m=1,2,3$ separately, for all the allowed values of $k$, namely for $k = 0, \dots, m-1$. For the proof, we will repeatedly use \eqref{eq:HFDet} as well as the relations 
\begin{subequations}
\begin{alignat}{2}
\det(m,n)_0 &= \Dbm_0 \det(m-1,n)_2 - \fbm_0 \Dbm_1 \det(m-2,n)_4 + \fbm_0 \fbm_1 \fbm_2 \det(m-3,n)_6\ ,\label{eq:dDevBas}\\[0.1cm]
\det(m,n)_0&= \Dbm_{2m+2n-1} \det(m,n-1)_0 - \fbm_{2m+2n-3} \Dbm_{2m+2n-2} \det(m,n-2)_0 \nonumber\\[0.1cm]&
+\fbm_{2m+2n-5} \fbm_{2m+2n-4} \fbm_{2m+2n-3} \det(m,n-3)_0\ .\label{eq:dDevHaut}
\end{alignat}
\end{subequations}
These are obtained by expanding the determinant \eqref{def:detFormel(m,n)} along the first row and the last column, respectively. We also recall that $\det(1,0)_k = \Dbm_k$ and $\det(0,0)_k=\Ib$. 

It is easy to see that the $T$-system relation for $(m,k) = (m,m-1)$ is directly equivalent to the fusion hierarchy relation \eqref{eq:HFdDet} with $m \mapsto m+1$. As a result, for the seed cases, we need only check $(m,k)=(2,0)$, $(3,0)$ and $(3,1)$. For $(m,k)=(2,0)$, we have
\begin{alignat}{2}
\det(2,0)_0 \det(2,0)_{2}
&= \big( \Dbm_0 \Dbm_2 - \Dbm_1 \fbm_0 \big) \det(2,0)_{2}
= \Dbm_0 \Dbm_2 \det(2,0)_{2} - \Dbm_1 \fbm_0 \big( \Dbm_2 \Dbm_4 - \Dbm_3 \fbm_2 \big) \nonumber\\
& = \Dbm_2 \big( \Dbm_0 \det(2,0)_{2} - \fbm_0 \Dbm_1 \Dbm_4 \big)
+ \fbm_0 \fbm_2 (\Dbm_1 \Dbm_3) \nonumber\\
& = \Dbm_2 \big( \Dbm_0 \det(2,0)_{2} - \fbm_0 \Dbm_1 \Dbm_4 + \fbm_0 \fbm_1 \fbm_2 \big)
+ \fbm_0 \fbm_2 \det(2,0)_1 \nonumber\\
& = \Dbm_2 \det(3,0)_0 + \fbm_0 \fbm_2 \det(0,2)_0 \ ,
\end{alignat}
which is the desired result. For $(m,k)=(3,0)$, we use \eqref{eq:dDevBas} and find
\begin{alignat}{2}
\det(3,0)_0 \det(3,0)_2
&= \big( \Dbm_0 \det(2,0)_2 - \fbm_0 \Dbm_1 \Dbm_4 + \fbm_0 \fbm_1 \fbm_2 \big) \det(3,0)_{2} \nonumber\\
&= \det(2,0)_2 \Dbm_0 \det(3,0)_{2} - \fbm_0 \Dbm_1 \Dbm_4 \det(3,0)_{2}
+ \fbm_0 \fbm_1 \fbm_2 \det(3,0)_{2}\ .
\end{alignat}
Here we rewrite $\Dbm_4 \det(3,0)_{2}$ using the $T$-system relation for $(m,k)=(2,0)$ already proven above, and use \eqref{eq:HFdDet} for $\det(3,0)_{2}$ in the last term. This yields
\begin{alignat}{2}
\det(3,0)_0 \det(3,0)_2&= \det(2,0)_2 \big( \Dbm_0 \det(3,0)_{2} - \fbm_0 \Dbm_1 \det(2,0)_4
+ \fbm_0 \fbm_1 \fbm_2 \Dbm_6 \big) \nonumber\\
&+ \fbm_0 \fbm_2 \fbm_4 \big( \Dbm_1 \det(0,2)_2  - \fbm_1 \det(1,1)_2 \big)\nonumber\\
& = \det(2,0)_2 \det(4,0)_0 + \fbm_0 \fbm_2 \fbm_4 \det(0,3)_0
\ ,
\end{alignat}
which ends the proof in this case. For $(m,k)=(3,1)$, we have
\be
\det(3,0)_0 \det(2,0)_4
= \big( \Dbm_0 \det(2,0)_2 - \fbm_0 \Dbm_1 \Dbm_4 + \fbm_0 \fbm_1 \fbm_2 \big) \det(2,0)_4 \ .
\ee
We rewrite the first term using the $T$-system for $(m,k)=(2,0)$ proven above, and the second term using the fusion hierarchy relation \eqref{eq:HFdDet} for $\det(2,0)_4$:
\begin{alignat}{2}
\det(3,0)_0 \det(2,0)_4 &= \Dbm_0 \big( \Dbm_4 \det(3,0)_2 + \fbm_2 \fbm_4 \det(0,2)_2 \big)- \fbm_0 \Dbm_1 \Dbm_4 \det(2,0)_4 
\nonumber\\&
+ \fbm_0 \fbm_1 \fbm_2 \big( \Dbm_4 \Dbm_6 - \Dbm_5 \fbm_4 \big) 
\nonumber\\&= \Dbm_4 \big( \Dbm_0 \det(3,0)_2 - \fbm_0 \Dbm_1 \det(2,0)_4 + \fbm_0 \fbm_1 \fbm_2 \Dbm_6 \big)\ 
\nonumber\\&+ \fbm_2 \fbm_4 \big( \Dbm_0 \det(0,2)_2 - \fbm_0 \fbm_1 \Dbm_5 \big)
\nonumber\\&= \Dbm_4 \det(4,0)_0 +\fbm_2 \fbm_4 \det(1,2)_0\ ,
\end{alignat}
ending the proof.  

Having proven all the seed cases, we now move on to the general case $(m,k)$. We thus assume that \eqref{Tsystem} holds for $m-1, m-2, m-3$ and all corresponding values of $k$. We expand $\det(m,0)$ using \eqref{eq:dDevBas} and find
\begin{alignat}{2}
 \det(m,0) &\det(m-k,0)_{2k+2} = \Dbm_0 \det(m-1,0)_2 \det(m-k,0)_{2k+2} \nonumber\\[0.1cm]
&- \Dbm_1 \fbm_0 \det(m-2,0)_4 \det(m-k,0)_{2k+2}
+ \fbm_0 \fbm_1 \fbm_2 \det(m-3,0)_6 \det(m-k,0)_{2k+2}\ \nonumber\\[0.1cm]
&= \Dbm_0 \big( \det(m-1,0)_0 \det((m-1)-(k-1),0)_{2(k-1)+2} \big)_2 \nonumber\\[0.1cm]
&- \Dbm_1 \fbm_0 \big( \det(m-2,0)_0 \det((m-2)-(k-2),0)_{2(k-2)+2} \big)_4 \nonumber\\[0.1cm]
&+ \fbm_0 \fbm_1 \fbm_2 \big( \det(m-3,0)_0 \det((m-3)-(k-3),0)_{2(k-3)+2} \big)_6\ .
\end{alignat}
For each of the three terms, we use the induction hypothesis and apply the $T$-system relation for $(m,k)\mapsto (m-1,k-1)$, $(m-2,k-2)$ and $(m-3,k-3)$, respectively.
To avoid negative indices, we first consider the case $k\geq3$. In this case, we find
\begin{alignat}{2}
&\det(m,0) \det(m-k,0)_{2k+2}= \Dbm_0 \Big( \det(k-1,m-k)_0 \prod^{m-2}_{j=k-1} \fbm_{2j}
+ \det(m,0)_0 \det(m-k-1,0)_{2k} \Big)_2 \nonumber\\
&\hspace{0.5cm}- \Dbm_1 \fbm_0 \Big( \det(k-2,m-k)_0 \prod^{m-3}_{j=k-2} \fbm_{2j}
+ \det(m-1,0)_0 \det(m-k-1,0)_{2k-2} \Big)_4 \nonumber\\
&\hspace{0.5cm}+ \fbm_0 \fbm_1 \fbm_2 \Big( \det(k-3,m-k)_0 \prod^{m-4}_{j=k-3} \fbm_{2j}
+ \det(m-2,0)_0 \det(m-k-1,0)_{2k-4} \Big)_6
\nonumber\\
&= \det(k,m-k)_0 \prod^{m-1}_{j=k} \fbm_{2j} + \det(m+1,0)_0 \det(m-k-1,0)_{2k+2}\ .
\end{alignat}
At the last step, we applied \eqref{eq:dDevBas} with $(m,n) \mapsto (k,m-k)$ to combine the first, third and fifth term, and the same equation with $(m,n) \mapsto (m+1,0)$ for the other three.

It thus remains to check the cases $k=0$, $k=1$ and $k=2$. For $k=2$, we find
\begin{alignat}{2}
 \det(m,0)_0 \det(m-2,0)_{6}
&= \Dbm_0 \det(m-1,0)_2 \det(m-2,0)_{6} \nonumber\\[0.1cm]
&- \Dbm_1 \fbm_0 \det(m-2,0)_4 \det(m-2,0)_{6}
+ \fbm_0 \fbm_1 \fbm_2 \det(m-3,0)_6 \det(m-2,0)_{6}\nonumber\\&= \Dbm_0 \Big( \det(1,m-2)_0  \prod^{m-2}_{j=1} \fbm_{2j}
+ \det(m,0)_0 \det(m-3,0)_{4} \Big)_2 \nonumber\\
&- \Dbm_1 \fbm_0 \Big( \det(0,m-2)_0 \prod^{m-3}_{j=0} \fbm_{2j}
+ \det(m-1,0)_0 \det(m-3,0)_{2} \Big)_4 \nonumber\\
&+ \fbm_0 \fbm_1 \fbm_2 \det(m-3,0)_6 \det(m-2,0)_{6} \vphantom{\prod^{m}}
\nonumber\\
&\hspace{-3.0cm}= \prod^{m-1}_{j=2} \fbm_{2j}
\Big( \Dbm_0 \det(1,m-2)_2 - \Dbm_1 \fbm_0 \det(0,m-2)_4 \Big)
+ \det(m+1,0)_0 \det(m-3,0)_{6}\ ,
\end{alignat}
which is the announced result. The only difference with the general case is that we only used two $T$-system relations inductively at the second equality. For $k=1$, we have
\begin{alignat}{2}
& \det(m,0)_0 \det(m-1,0)_4
= \Dbm_0 \det(m-1,0)_2 \det(m-1,0)_{4} \nonumber\\[0.1cm]
&- \Dbm_1 \fbm_0 \det(m-2,0)_4 \det(m-1,0)_{4}
+ \fbm_0 \fbm_1 \fbm_2 \det(m-3,0)_6 \det(m-1,0)_{4} \nonumber\\
&= \Dbm_0 \Big( \det(0,m-1)_0 \prod^{m-2}_{j=0} \fbm_{2j}
+ \det(m,0)_0 \det(m-2,0)_{2} \Big)_2
- \Dbm_1 \fbm_0 \det(m-2,0)_4 \det(m-1,0)_{4} \nonumber\\
&+ \fbm_0 \fbm_1 \fbm_2 \Big( \det(m-2,0)_4 \det(m-2,0)_6
- \det(0,m-2)_4 \prod^{m-1}_{j=2} \fbm_{2j} \Big) \ ,
\end{alignat}
where we applied the $T$-system relation \eqref{Tsystem} inductively for the first and last terms, with $(m,n) \mapsto (m-1,0)$ and $(m-2,0)$ respectively.
Collecting the second, third and fourth terms, we apply \eqref{eq:dDevBas} and find
\begin{alignat}{2}
 \det(m,0)_0 \det(m-1,0)_4 &= \prod^{m-1}_{j=1} \fbm_{2j} \Big( \Dbm_0 \det(0,m-1)_2
- \fbm_0 \fbm_1 \det(0,m-2)_4 \Big) \nonumber\\&+ \det(m+1,0)_0 \det(m-2,0)_4 
\nonumber\\&= \det(1,m-1)_0 \prod^{m-1}_{j=1} \fbm_{2j}  + \det(m+1,0)_0 \det(m-2,0)_4\ ,
\end{alignat}
ending the proof for this case.

Finally, for $k=0$, we first use \eqref{eq:dDevBas} and find
\begin{alignat}{2}
 \det(m,0) \det(m,0)_{2}
&= \Dbm_0 \det(m-1,0)_2 \det(m,0)_{2} - \Dbm_1 \fbm_0 \det(m-2,0)_4 \det(m,0)_{2} 
\nonumber\\[0.1cm]&+ \fbm_0 \fbm_1 \fbm_2 \det(m-3,0)_6 \det(m,0)_{2} \ 
\nonumber\\[0.1cm]&= \Dbm_0 \det(m-1,0)_2 \det(m,0)_{2} \nonumber\\
&- \Dbm_1 \fbm_0 \Big( \det(m-1,0)_2 \det(m-1,0)_4
- \det(0,m-1)_2 \prod^{m-1}_{j=1} \fbm_{2j} \Big) \nonumber\\
&+ \fbm_0 \fbm_1 \fbm_2 \Big( \det(m-1,0)_2 \det(m-2,0)_6
- \det(1,m-2)_2 \prod^{m-1}_{j=2} \fbm_{2j} \Big) \ .
\end{alignat}
At the last equality, we used inductively the $T$-system relation for $(m,n) \mapsto (m-1,0)$ and $(m-1,1)$ to reexpress the second and third terms, respectively. Collecting the first, second and fourth terms, which all have a factor of $\det(m-1,0)_2$, we use \eqref{eq:dDevBas} and find
\begin{alignat}{2}
 \det(m,0) \det(m,0)_{2} &= \det(m+1,0)_0 \det(m-1,0)_2
+ \prod^{m-1}_{j=0} \fbm_{2j} \big( \Dbm_1 \det(0,m-1)_2 - \fbm_1 \det(1,m-2)_2 \big)
\nonumber\\ &
= \det(m+1,0)_0 \det(m-1,0)_2
+ \det(0,m)_0 \prod^{m-1}_{j=0} \fbm_{2j},
\end{alignat}
ending the proof.

%%%%%%%%%%%%%
\subsection{Closure relations for $b=2$ and $b=3$} \label{sub:proof.closure.b23}
%%%%%%%%%%%%%

In this section, we give more details on the closure relations and their proofs, for $b=2$ and $b=3$ with identical boundary conditions. We show that these relations hold at $u = \widehat u = (4-b)\lambda + \frac{r \pi}2$, thus completing the work of \cref{sec:finite.evaluations} where the proof is presented for $b \ge 4$. The arguments are given separately for (i) $b = 2$, (ii) $b=3$ with $a$ odd, and (iii) $b=3$ with $a$ even.

\paragraph{The case $b=2$.} In this case, $a$ is an odd integer and the loop weight is $\beta = 2$. Following the same idea as in \eqref{eq:lesgrandsR}, we denote the four terms in the closure relation \eqref{eq:closure.b=2} as $\Sb^1(u), \dots, \Sb^4(u)$ in such a way that it reads
\be
\label{eq:R.closure.b=2}
\Sb^1(u)+\Sb^2(u) = \kappa( \Sb^3(u) - 2 \Sb^4(u)), \qquad \kappa = 3.
\ee
At $\lambda = \lambda_{a,2}$, the various functions have the symmetries
\begin{subequations}
\begin{alignat}{3}
\Db^{m,n}(u) &= \Db^{m,n}(u_4) = \Db^{m,n}(u_2 + \tfrac \pi 2)\qquad &f(u) &= f(-u) = f(u_4) = f(u_2+\tfrac \pi 2),
\\[0.1cm]
w(u) &= - w(u_{4}),\qquad
&\bar w(u) &= \zeta \varrho\, w(u_{2}),
\\
w^{(1)}(u) &= \zeta\, w(u_{2}), 
&\bar w^{(1)}(u) &= -\varrho\, w(u),
\\[0.1cm]
w(-u) &= -\varrho\, w(u), \qquad
&\bar w(-u) &= \varrho\, \bar w(u), \qquad
\end{alignat}
\end{subequations}
where 
\be
\label{eq:zeta.varrho}
\zeta = (-1)^{(a-1)/2}, \qquad
\varrho = \left\{\begin{array}{cc}
1 &\textrm{SS},
\\[0.1cm]
-1 & \textrm{CC}.
\qquad
\end{array}\right.
\ee
We use this information to evaluate each term of the closure relation at $\widehat u = 2 \lambda + \frac{r\pi}2$. Using the notation \eqref{eq:short.hand.wk}, we find
\be
\Sb^1(\widehat u) = 2\, \Sb^2(\widehat u) = -2\,\Sb^3(\widehat u) = -2\,\Sb^4(\widehat u) = (-1)^{r+1}2\varrho\, w_{1/2}^3 w_{3/2}^2 f(\tfrac{r \pi}2) f(\lambda+\tfrac{r \pi}2) \Ib,
\ee
which we use to confirm that \eqref{eq:R.closure.b=2} holds at $u = \widehat u$.
In particular, to compute $\Sb^1(\widehat u)$, one must carefully use the result \eqref{eq:simpler.D.reduction} for generic $\lambda$ and then take the limit $\lambda \to \lambda_{a,2}$. Finally, we note that the closure relation for $b=2$ can be rewritten as the quadratic functional relation in $\Db(u)$
\begin{alignat}{2}
0&=w(u_{1/2})w(u_{5/2}) \Db_0 \Db_2 + \varrho\, w(u_{3/2})^4 f_1 f_2 \Db_1 - \varrho\, w(u_{7/2})^4 f_0 f_3 \Db_3 
\nonumber\\[0.15cm]&
+ 3\, w(u_{1/2})w(u_{3/2})^2w(u_{5/2})w(u_{7/2})^2 f_0 f_1 f_2 f_3 \Ib.
\end{alignat}

\paragraph{The case $b=3$ with $a$ odd.} 
In this case, the loop weight is $\beta = 1$ and the functions entering the closure relation~\eqref{eq:closure} have the symmetries
\begin{subequations}
\label{eq:b3.aodd.symmetries}
\begin{alignat}{3}
\Db^{m,n}(u) &= \Db^{m,n}(u_3), \qquad &f(u) &= f(-u) = f(u_3), \quad
\\[0.1cm]
w(u) &= -\zeta \, w (u_{3}),\qquad
&\bar w(u) &= -\zeta\, \bar w(u_{3}),&
\\
w^{(1)}(u) &= \varrho\, \bar w(u), 
&\bar w^{(1)}(u) &= -\varrho\, w(u),
\\[0.1cm]
w(-u) &= -\varrho\, w(u), \qquad
&\bar w(-u) &= \varrho\, \bar w(u), \qquad
\end{alignat}
\end{subequations}
where $\zeta$ and $\varrho$ are as in \eqref{eq:zeta.varrho}. Using this information, we find that all five terms in the closure relation vanish at $\widehat u=\lambda+\frac{r\pi}2$ with $r\in\{0,1\}$:
\be
\Rb^i(\widehat u) = 0, \qquad i = 1, \dots, 5.
\ee
To show this, we first note that exactly one of $w(\frac{3\lambda}2)$ or $\bar w(\frac{3\lambda}2)$ vanishes at $\lambda = \lambda_{a,3}$ with $a$ odd. Then, depending of the values of $\lambda$ and $r$ and the choice of boundary conditions, the fact that $\Rb^i(\widehat u) = 0$ may be easy to check due to the vanishing of one of the factors arising explicitly in the definition of this function. If this factor is $w^{(a)}(u_{-5/2})$ or $w^{(1)}(u_{-5/2})$, then one must use the symmetries \eqref{eq:b3.aodd.symmetries} to show that it indeed vanishes. In fact, $\Rb^3(\widehat u)$, $\Rb^4(\widehat u)$ and $\Rb^5(\widehat u)$ are always easy to compute in this sense. For $\Rb^1(\widehat u)$, when it is not easy to compute, one must instead rewrite $\Db^{3,0}(\widehat u)$ with \eqref{eq:simpler.D.reduction} to show that it vanishes. Similarly, in the cases where $\Rb^2(\widehat u)$ is not straightforward to compute, showing that it vanishes requires that we carefully apply the limit $\lambda \to \lambda_{a,3}$ to the fusion hierarchy relation \eqref{eq:HFb} for $\Db^{1,1}(\widehat u_2)$.

Note that, for $b=3$ and $a$ odd, the face operator \eqref{def:tuileDeBase} vanishes at $u = 0,\pi$. Moreover, the weights $\delta(u+\frac{3\lambda}2)$ and $\delta(u-\frac{3\lambda}2)$ that define the boundary operator in \eqref{def:tuileFrontiere} are equal up to a sign. This implies that the renormalised transfer matrix
\be
\Dbh(u) = \frac{\Db(u)}{f(u) w(u_{3/2})^2}
\ee
is a Laurent polynomial in $\eE^{\iI u}$. Its maximal power is $2N$.
In terms of this matrix, the closure relation is equivalent to the simple cubic functional relation for $\Dbh(u)$
\be
0=\varrho\, \Dbh_0\Dbh_1\Dbh_2 - f_0 \Dbh_0^2  - f_1 \Dbh_1^2 - f_2 \Dbh_2^2 + 4 f_0 f_1 f_2 \Ib.
\ee
This is precisely the functional relation studied in \cite{MDKP22} for $\lambda = \frac \pi 3$ corresponding to critical site percolation on the triangular lattice.

\paragraph{The case $b=3$ with $a$ even.} In this case, the loop weight is also
$\beta = 1$. We see from \eqref{eq:DI} that $\Db(u)$ vanishes at $u = 0,\frac \pi 2$. We therefore define the renormalised transfer matrix
\be
\label{eq:Dhat.a.even}
\Dbh(u) = \frac{\Db(u)}{\sin 2u}.
\ee
It is a Laurent polynomial in $\eE^{\iI u}$ of maximal power $4N$. Of course, these zeroes are particular to $b=3$ with $a$ even and, in the computations below, the evaluation on $\lambda$ must thus precede that on $u$. (This observation turns out to be crucial for the computation of $\Rb^2(\widehat u)$.) The various functions in \eqref{eq:closure} have the symmetries 
\begin{subequations}
\label{eq:b3.aeven.symmetries}
\begin{alignat}{3}
\Db^{m,n}(u) &= \Db^{m,n}(u_6), \qquad &f(u) &= f(-u) = f(u_6),
\\[0.1cm]
w(u) &= - w (u_{6}),\qquad
&\bar w(u) &= -\varrho\zeta\, w(u_{3}),&
\\
w^{(1)}(u) &= -\zeta\, w(u_{3}), 
&\bar w^{(1)}(u) &= -\varrho\, w(u),
\\[0.1cm]
w(-u) &= -\varrho\, w(u), \qquad
&\bar w(-u) &= \varrho\, \bar w(u), \qquad
\end{alignat}
\end{subequations}
where
\be
\label{eq:zeta.varrho.aEven}
\zeta = (-1)^{a/2}, \qquad
\varrho = \left\{\begin{array}{cc}
1 &\textrm{SS},
\\[0.1cm]
-1 & \textrm{CC}.
\qquad
\end{array}\right.
\ee
Using the short-hand notation \eqref{eq:short.hand.wk}, we find
\begin{subequations}
\begin{alignat}{2}
\Rb^1(\widehat u) &= -\Rb^4(\widehat u) = \Rb^5(\widehat u) = (-1)^r \zeta\, w_{1/2}^2 w_{3/2}^3 w_{5/2}^2 f(\tfrac{r \pi}2) f(\lambda+\tfrac{r \pi}2) f(2\lambda+\tfrac{r \pi}2)\Ib,
\\
\Rb^2(\widehat u) &= \zeta\, w_{3/2} \Db^{1,1}(3\lambda + \tfrac{r \pi} 2) = \frac{(-1)^{r+1} \zeta \, w_{3/2}^3}{f(3\lambda+\tfrac{r \pi}2)} \Big(s(2\lambda)^2 \Dbh(0)\Dbh(\tfrac \pi 2) - w_{1/2}^4 f(\tfrac{r \pi}2)^2 f(\lambda+\tfrac{r \pi}2)^2 \Ib\Big),
\\
\Rb^3(\widehat u) &= (-1)^r\zeta\, w_{3/2}^3 w_{5/2}^4 f(2\lambda+\tfrac{r\pi}2)^2 f(3\lambda+\tfrac{r\pi}2)\Ib.
\end{alignat}
\end{subequations}
Here we used \eqref{eq:simpler.D.reduction} for $\Rb^1(\widehat u)$ and simplified the terms. Likewise the product of factors for $\Rb^3(\widehat u)$, $\Rb^4(\widehat u)$ and $\Rb^5(\widehat u)$ were simplified using \eqref{eq:b3.aeven.symmetries}. Only $\Rb^2(\widehat u)$ is not yet in its final form. In \cref{lem:Dh(0)} below, we show that $\Dbh(0)$ and $\Dbh(\frac \pi 2)$ are equal and proportional to the unit $\Ib$, and we compute the corresponding prefactor. Inserting these results in \eqref{eq:final.closure}, we find that the identity indeed holds, ending the proof of the closure relation at $u = \widehat u$. Finally, we note that the closure relation in this case is equivalent to the cubic functional relation
\begin{alignat}{2}
0&=\varrho\, s(2u_0) s(2u_1) s(2u_2) \Dbh_0\Dbh_2\Dbh_4\\
&+ s(2u_0)s(2u_2)w(u_{5/2})^2 f_2 f_3\Dbh_1\Dbh_4
- s(2u_1)s(2u_2) w(u_{9/2})^2 f_4f_5\Dbh_0\Dbh_3
\nonumber\\[0.2cm]
&- s(2u_0)s(2u_1) w(u_{1/2})^2 f_0 f_1\Dbh_2\Dbh_5
+ w(u_{1/2})^2 w(u_{3/2})^2 w(u_{5/2})^2 f_0 f_1^2 f_2^2 f_3 \Ib
\nonumber\\[0.2cm] 
&+ w(u_{5/2})^2 w(u_{7/2})^2 w(u_{9/2})^2 f_2 f_3^2 f_4^2 f_5 \Ib
+ w(u_{1/2})^2 w(u_{9/2})^2 w(u_{11/2})^2 f_0^2 f_1 f_4 f_5^2 \Ib\\ 
&- \bigg[\prod_{j=0}^5 f_j \bigg] \Big( w(u_{3/2})^2 w(u_{7/2})^2 w(u_{11/2})^2 + 2\, w(u_{1/2})^2 w(u_{5/2})^2 w(u_{9/2})^2 \Big)\Ib.
\nonumber
\end{alignat}

\begin{Lemme}
\label{lem:Dh(0)}
The renormalized transfer matrix satisfies
\be
\label{eq:Dbh0}
\Dbh(0) = \Dbh(\tfrac{\pi}2) = \frac {\varrho}{s(2\lambda)} \big(w(\tfrac \lambda2)^2 f(0)f(\lambda) - w(\tfrac {5\lambda}2)^2f(2\lambda)f(3\lambda)\big) \Ib.
\ee
\end{Lemme}
\noindent {\scshape Proof.\ }
We give the proof for $\Dbh(0)$, noting that the result will also hold for $\Dbh(\tfrac{\pi}2)$ because of crossing symmetry. To compute $\Dbh(0)$, we use L'H\^opital's rule to write
\be\label{eq:leSigneDuD}
\Dbh(0) = \frac12 \frac{\dd}{\dd u}\Db(u) \Big|_{u=0}= -\frac12 \frac{\dd}{\dd u}\Dbt(u) \Big|_{u=0}
\ee
where $\Dbt(u)$ is defined in \eqref{def:D10tilde}. Let us then define
\be
\psset{unit=0.75cm}
\begin{pspicture}[shift=-0.4](1,1)
\facegrid{(0,0)}{(1,1)}
\psarc[linewidth=0.025]{-}(0,0){0.16}{0}{90}
\rput(.5,.5){$v'$}
\end{pspicture} 
\ = \frac{\dd}{\dd u}\ 
\begin{pspicture}[shift=-0.4](1,1)
\facegrid{(0,0)}{(1,1)}
\psarc[linewidth=0.025]{-}(0,0){0.16}{0}{90}
\rput(.5,.5){$u$}
\end{pspicture} \ \bigg|_{u = v}\ ,
\qquad \quad
\begin{pspicture}[shift=-0.9](0,0)(1.25,2)
\psline[linecolor=blue,linewidth=1.5pt,linestyle=dashed,dash=2pt 2pt](0,1.5)(1.25,1.5)
\psline[linecolor=blue,linewidth=1.5pt,linestyle=dashed,dash=2pt 2pt](0,0.5)(1.25,0.5)
\triangled \rput(0.35,1){$v'$}
\end{pspicture}
\ = \frac{\dd}{\dd u}\
\begin{pspicture}[shift=-0.9](0,0)(1.25,2)
\psline[linecolor=blue,linewidth=1.5pt,linestyle=dashed,dash=2pt 2pt](0,1.5)(1.25,1.5)
\psline[linecolor=blue,linewidth=1.5pt,linestyle=dashed,dash=2pt 2pt](0,0.5)(1.25,0.5)
\triangled \rput(0.35,1){$u$}
\end{pspicture}
\ \bigg|_{u = v}\ .
\ee
For $\lambda = \lambda_{a,3}$ with $a$ even, we have
\begin{subequations}
\begin{alignat}{2}
&
\psset{unit=0.75cm}
\begin{pspicture}[shift=-0.9](1.25,2)
\psline[linecolor=blue,linewidth=1.5pt,linestyle=dashed,dash=2pt 2pt](0,1.5)(1.25,1.5)
\psline[linecolor=blue,linewidth=1.5pt,linestyle=dashed,dash=2pt 2pt](0,0.5)(1.25,0.5)
\triangled \rput(0.35,1){\small$0$}
\end{pspicture}
\ = \delta(\tfrac{3\lambda}2)\
\begin{pspicture}[shift=-0.9](-0.1,0)(1.25,2)
\psline[linecolor=blue,linewidth=1.5pt,linestyle=dashed,dash=2pt 2pt](0.77,1.5)(1.25,1.5)
\psline[linecolor=blue,linewidth=1.5pt,linestyle=dashed,dash=2pt 2pt](0.77,0.5)(1.25,0.5)
\rput(-0.1,0){\triangled}
\psarc[linecolor=blue,linewidth=1.5pt,linestyle=dashed,dash=2pt 2pt](0.7,1){0.5}{90}{270}
\end{pspicture}\ ,
\qquad \quad
&&
\psset{unit=0.75cm}
\begin{pspicture}[shift=-0.9](-0.2,0)(1.25,2)
\psline[linecolor=blue,linewidth=1.5pt,linestyle=dashed,dash=2pt 2pt](0,1.5)(1.25,1.5)
\psline[linecolor=blue,linewidth=1.5pt,linestyle=dashed,dash=2pt 2pt](0,0.5)(1.25,0.5)
\triangled \rput(0.35,1){\small$0'$}
\end{pspicture} 
\ = -\varrho\zeta \, \delta(\tfrac{3\lambda}2)\
\begin{pspicture}[shift=-0.9](-0.1,0)(1.25,2)
\psline[linecolor=blue,linewidth=1.5pt,linestyle=dashed,dash=2pt 2pt](0.77,1.5)(1.25,1.5)
\psline[linecolor=blue,linewidth=1.5pt,linestyle=dashed,dash=2pt 2pt](0.77,0.5)(1.25,0.5)
\rput(-0.1,0){\triangled}
\psarc[linecolor=blue,linewidth=1.5pt,linestyle=dashed,dash=2pt 2pt](0.7,1){0.5}{90}{270}
\rput(0,1){\pspolygon[fillstyle=solid,fillcolor=white](0,-0.125)(0.45,-0.125)(0.45,0.125)(0,0.125)}
\end{pspicture}
\ ,
\\[0.2cm]
&
\psset{unit=0.75cm}
\begin{pspicture}[shift=-0.9](1.25,2)
\psline[linecolor=blue,linewidth=1.5pt,linestyle=dashed,dash=2pt 2pt](0,1.5)(1.25,1.5)
\psline[linecolor=blue,linewidth=1.5pt,linestyle=dashed,dash=2pt 2pt](0,0.5)(1.25,0.5)
\triangled \rput(0.35,1){\small$3\lambda$}
\end{pspicture} 
\ = \varrho\,\delta(\tfrac{3\lambda}2)\
\begin{pspicture}[shift=-0.9](-0.1,0)(1.25,2)
\psline[linecolor=blue,linewidth=1.5pt,linestyle=dashed,dash=2pt 2pt](0.77,1.5)(1.25,1.5)
\psline[linecolor=blue,linewidth=1.5pt,linestyle=dashed,dash=2pt 2pt](0.77,0.5)(1.25,0.5)
\rput(-0.1,0){\triangled}
\psarc[linecolor=blue,linewidth=1.5pt,linestyle=dashed,dash=2pt 2pt](0.7,1){0.5}{90}{270}
\rput(0,1){\pspolygon[fillstyle=solid,fillcolor=white](0,-0.125)(0.45,-0.125)(0.45,0.125)(0,0.125)}
\end{pspicture}\ ,
\qquad \quad
&&
\psset{unit=0.75cm}
\begin{pspicture}[shift=-0.9](-0.2,0)(1.25,2)
\psline[linecolor=blue,linewidth=1.5pt,linestyle=dashed,dash=2pt 2pt](0,1.5)(1.25,1.5)
\psline[linecolor=blue,linewidth=1.5pt,linestyle=dashed,dash=2pt 2pt](0,0.5)(1.25,0.5)
\triangled \rput(0.38,1){\small$3\lambda'$}
\end{pspicture} 
\ = \zeta\, \delta(\tfrac{3\lambda}2)\
\begin{pspicture}[shift=-0.9](-0.1,0)(1.25,2)
\psline[linecolor=blue,linewidth=1.5pt,linestyle=dashed,dash=2pt 2pt](0.77,1.5)(1.25,1.5)
\psline[linecolor=blue,linewidth=1.5pt,linestyle=dashed,dash=2pt 2pt](0.77,0.5)(1.25,0.5)
\rput(-0.1,0){\triangled}
\psarc[linecolor=blue,linewidth=1.5pt,linestyle=dashed,dash=2pt 2pt](0.7,1){0.5}{90}{270}
\end{pspicture}\ .
\end{alignat}
\end{subequations}
There are two push-through properties. The first, given in \eqref{rel:PTbulkArc}, has dashed arcs that propagate to the left. The second, specific to $\lambda = \lambda_{a,3}$, involves a dashed arc with a gauge operator which is moved to the right:
\be
\begin{pspicture}[shift=-0.9](-0.5,0)(1,2)
\psarc[linecolor=blue,linewidth=1.5pt,linestyle=dashed,dash=2pt 2pt](0,1){0.5}{90}{270}
\facegrid{(0,0)}{(1,2)}
\rput(0,1){\psarc[linewidth=0.025]{-}(0,0){0.16}{0}{90}\rput(.5,0.5){$u_3$}}
\rput(0,0){\psarc[linewidth=0.025]{-}(0,0){0.16}{0}{90}\rput(.5,0.5){$u_0$}}
\rput(-0.5,1){\pspolygon[fillstyle=solid,fillcolor=white](-0.3,-0.125)(0.3,-0.125)(0.3,0.125)(-0.3,0.125)}
\end{pspicture}
\ = s(u_{-1}) s(u_{1}) s(u_0)^2 \
\begin{pspicture}[shift=-0.9](-0.5,0)(1.0,2)
\facegrid{(0,0)}{(1,2)}
\psarc[linecolor=blue,linewidth=1.5pt,linestyle=dashed,dash=2pt 2pt](0,1){0.5}{90}{270}
\rput(0,1){
\psarc[linewidth=1.5pt,linecolor=blue,linestyle=dashed,dash=2pt 2pt](0,1){.5}{-90}{0}
\psarc[linewidth=1.5pt,linecolor=blue,linestyle=dashed,dash=2pt 2pt](1,0){0.5}{90}{180}}
\psarc[linewidth=1.5pt,linecolor=blue,linestyle=dashed,dash=2pt 2pt](0,0){0.5}{0}{90}
\psarc[linewidth=1.5pt,linecolor=blue,linestyle=dashed,dash=2pt 2pt](1,1){0.5}{180}{270}
\rput(0.5,1){\pspolygon[fillstyle=solid,fillcolor=white](-0.3,-0.125)(0.3,-0.125)(0.3,0.125)(-0.3,0.125)}
\end{pspicture}\ .
\ee
This relation is derived from \eqref{rel:PTbulkArc} using \eqref{rel:jaugePi} and the fact that $6\lambda = \pi \textrm{ mod }2\pi$. We apply the derivative to the right side of \eqref{def:D10tilde} and use the product rule. The derivative must be applied to each of the face and boundary operators, so this produces a total of $2N+2$ terms: two boundary terms and $2N$ bulk terms. For each, we use the two push-through properties repeatedly.
This yields
\begin{alignat}{2}
\Dbh(0) &=\frac12\left(f(2 \lambda)f(3 \lambda) \
\psset{unit=0.75}
\begin{pspicture}[shift=-0.9](2,2)
\psline[linecolor=blue,linewidth=1.5pt,linestyle=dashed,dash=2pt 2pt](0,1.5)(2,1.5)
\psline[linecolor=blue,linewidth=1.5pt,linestyle=dashed,dash=2pt 2pt](0,0.5)(2,0.5)
\triangled \rput(0.35,1){\small$3\lambda'$}
\rput(1,0){\triangleg \rput(0.65,1){\small$0$}}
\end{pspicture} 
\ - f(0)f(\lambda) \
\begin{pspicture}[shift=-0.9](2,2)
\psline[linecolor=blue,linewidth=1.5pt,linestyle=dashed,dash=2pt 2pt](0,1.5)(2,1.5)
\psline[linecolor=blue,linewidth=1.5pt,linestyle=dashed,dash=2pt 2pt](0,0.5)(2,0.5)
\triangled \rput(0.35,1){\small$3\lambda$}
\rput(1,0){\triangleg \rput(0.65,1){\small$0'$}}
\end{pspicture} \  
\right) \Ib_N
\nonumber\\[0.2cm]
& - \frac12 \sum_{j=1}^N \Bigg[\prod_{k=1}^{j-1} s(\xi_{(k)}+\lambda)s(\xi_{(k)}-\lambda)s(\xi_{(k)})^2\Bigg]
\Bigg[\prod_{\ell=j+1}^N s(\xi_{(\ell)}+2\lambda)s(\xi_{(\ell)}-2\lambda)s(\xi_{(\ell)}+3\lambda)s(\xi_{(\ell)}-3\lambda)\Bigg]
\nonumber\\[0.2cm]
&\hspace{1.5cm}\times \Ib_{j-1} \otimes \left(\ 
\psset{unit=0.75cm}
\begin{pspicture}[shift=-0.9](0,0)(3,2)
\psline[linecolor=blue,linewidth=1.5pt,linestyle=dashed,dash=2pt 2pt](0,1.5)(3,1.5)
\psline[linecolor=blue,linewidth=1.5pt,linestyle=dashed,dash=2pt 2pt](0,0.5)(3,0.5)
\triangled \rput(0.35,1){\small$3\lambda$}
\facegrid{(1,0)}{(2,2)}
\rput(1.5,0.5){\footnotesize$-\xi_{(j)}$}\rput(1.5,1.5){\footnotesize$\xi'_{(j)}$}
\rput(2,0){\triangleg \rput(0.65,1){\small$0$}}
\rput(1,0){\psarc[linewidth=0.025]{-}(0,0){0.16}{0}{90}}
\rput(2,1){\psarc[linewidth=0.025]{-}(0,0){0.16}{90}{180}}
\end{pspicture} 
\ + \ 
\begin{pspicture}[shift=-0.9](0,0)(3,2)
\psline[linecolor=blue,linewidth=1.5pt,linestyle=dashed,dash=2pt 2pt](0,1.5)(3,1.5)
\psline[linecolor=blue,linewidth=1.5pt,linestyle=dashed,dash=2pt 2pt](0,0.5)(3,0.5)
\triangled \rput(0.35,1){\small$3\lambda$}
\facegrid{(1,0)}{(2,2)}
\rput(1.5,0.5){\footnotesize$-\xi'_{(j)}$}\rput(1.5,1.5){\footnotesize$\xi_{(j)}$}
\rput(2,0){\triangleg \rput(0.65,1){\small$0$}}
\rput(1,0){\psarc[linewidth=0.025]{-}(0,0){0.16}{0}{90}}
\rput(2,1){\psarc[linewidth=0.025]{-}(0,0){0.16}{90}{180}}
\end{pspicture}
\ \right) \otimes \Ib_{N-j},
\label{eq:Dhat0sum}
\end{alignat}
where, to get the correct signs, one must remember the sign in \eqref{eq:leSigneDuD} and another coming from the derivative of the left boundary triangle. Moverover, we
use the notation $\Ib_N$ to denote the identity in $\dtl_N(\beta)$, and $c_1 \otimes c_2$ to denote the horizontal juxtaposition of two elements of $\dtl_{N_1}(\beta)$ and $\dtl_{N_2}(\beta)$ that forms an element of  $\dtl_{N_1+N_2}(\beta)$. The remaining diagrams are evaluated as
\begin{subequations}
\begin{alignat}{2}
&\psset{unit=0.75cm}
\begin{pspicture}[shift=-0.9](2,2)
\psline[linecolor=blue,linewidth=1.5pt,linestyle=dashed,dash=2pt 2pt](0,1.5)(2,1.5)
\psline[linecolor=blue,linewidth=1.5pt,linestyle=dashed,dash=2pt 2pt](0,0.5)(2,0.5)
\triangled \rput(0.35,1){\small$3\lambda'$}
\rput(1,0){\triangleg \rput(0.65,1){\small$0$}}
\end{pspicture}
\ = - \ 
\begin{pspicture}[shift=-0.9](2,2)
\psline[linecolor=blue,linewidth=1.5pt,linestyle=dashed,dash=2pt 2pt](0,1.5)(2,1.5)
\psline[linecolor=blue,linewidth=1.5pt,linestyle=dashed,dash=2pt 2pt](0,0.5)(2,0.5)
\triangled \rput(0.35,1){\small$3\lambda$}
\rput(1,0){\triangleg \rput(0.65,1){\small$0'$}}
\end{pspicture}
\ = \zeta\, \delta(\tfrac{3\lambda}2)^2(1+\beta) = \zeta,
\\[0.2cm]
&\psset{unit=0.75cm}
\begin{pspicture}[shift=-0.9](0,0)(3,2)
\psline[linecolor=blue,linewidth=1.5pt,linestyle=dashed,dash=2pt 2pt](0,1.5)(3,1.5)
\psline[linecolor=blue,linewidth=1.5pt,linestyle=dashed,dash=2pt 2pt](0,0.5)(3,0.5)
\triangled \rput(0.35,1){\small$3\lambda$}
\facegrid{(1,0)}{(2,2)}
\rput(1.5,0.5){$-\xi$}\rput(1.5,1.5){$\xi'$}
\rput(2,0){\triangleg \rput(0.65,1){\small$0$}}
\rput(1,0){\psarc[linewidth=0.025]{-}(0,0){0.16}{0}{90}}
\rput(2,1){\psarc[linewidth=0.025]{-}(0,0){0.16}{90}{180}}
\end{pspicture} 
\ + \ 
\begin{pspicture}[shift=-0.9](0,0)(3,2)
\psline[linecolor=blue,linewidth=1.5pt,linestyle=dashed,dash=2pt 2pt](0,1.5)(3,1.5)
\psline[linecolor=blue,linewidth=1.5pt,linestyle=dashed,dash=2pt 2pt](0,0.5)(3,0.5)
\triangled \rput(0.35,1){\small$3\lambda$}
\facegrid{(1,0)}{(2,2)}
\rput(1.5,0.5){$-\xi'$}\rput(1.5,1.5){$\xi$}
\rput(2,0){\triangleg \rput(0.65,1){\small$0$}}
\rput(1,0){\psarc[linewidth=0.025]{-}(0,0){0.16}{0}{90}}
\rput(2,1){\psarc[linewidth=0.025]{-}(0,0){0.16}{90}{180}}
\end{pspicture}
\ = \ \frac{\varrho}{s(2\lambda)} \big( s(\xi_{-3})s(\xi_{-2}) s(\xi_{2}) s(\xi_{3}) - s(\xi_{-1})s(\xi_0)^2 s(\xi_1) \big)\Ib_1.
\label{eq:sum.of.two.terms}
\end{alignat}
\end{subequations}
These two relations lead to a straightforward check of \eqref{eq:Dbh0} for $N=1$.
To show this, we use
\be
\tfrac{\varrho \zeta}2 s(2\lambda)+\tfrac12=w(\tfrac{\lambda}2)^2 ,\qquad \tfrac{\varrho \zeta}2 s(2\lambda)-\tfrac12=-w(\tfrac{5\lambda}2)^2.
\ee
For $N>1$, 
inserting \eqref{eq:sum.of.two.terms} into \eqref{eq:Dhat0sum}, we find that all terms in the sum cancel pairwise, except for one contribution from $j=1$ and another from $j=N$. This yields
\be
\Dbh(0) = \frac{\zeta}2\big(f(2\lambda)f(3\lambda)+f(0)f(\lambda)\big)\Ib -\frac{\varrho}{2s(2\lambda)}\big(f(2\lambda)f(3\lambda)-f(0)f(\lambda)\big)\Ib,
\ee
which is found to be equal to \eqref{eq:Dbh0} after simplification.
{\hfill \rule{0.5em}{0.5em}\medskip}

%%%%%%%%%%%%%%%%%%%%%%%%%%%%%%%%%%%%%%%%%%%%%%%%%%

\end{document}